\documentclass[twocolumn,showpacs,groupedaddress]{revtex4-1} 
\usepackage[utf8]{inputenc}
\usepackage{latexsym, amssymb, amsmath, graphicx, amsthm}
\usepackage{hyperref}   
\usepackage{color}
\usepackage{textcomp}

\begin{document}

\preprint{Review of Modern Physics - accepted.}

\title{White organic light-emitting diodes: Status and perspective}

\author{Sebastian Reineke}
\affiliation{Department of Electrical Engineering and Computer Science, Massachusetts Institute of Technology, 77 Massachusetts Avenue, Cambridge, MA 02139, USA.}
\email{reineke@mit.edu}

\author{Michael Thomschke}
\affiliation{Institut f\"ur Angewandte Photophysik, TU Dresden, George-B\"ahr-Str. 1, 01062 Dresden, Germany.}
\affiliation{Current address: Fraunhofer Research Institution for Organics, Materials and Electronic Devices COMEDD, Maria-Reiche-Str. 2,
01109 Dresden, Germany.}

\author{Bj\"orn L\"ussem}
\affiliation{Institut f\"ur Angewandte Photophysik, TU Dresden, George-B\"ahr-Str. 1, 01062 Dresden, Germany.}

\author{Karl Leo}
\affiliation{Institut f\"ur Angewandte Photophysik, TU Dresden, George-B\"ahr-Str. 1, 01062 Dresden, Germany.}

\date{\today}

\begin{abstract}
White organic light-emitting diodes (OLEDs) are ultra-thin, large-area light sources made from organic semiconductor materials. Over the last decades, much research has been spent on finding the suitable materials to realize highly efficient monochrome and white OLEDs. With their high efficiency, color-tunability, and color-quality, white OLEDs are emerging to become one of the next generation light sources. In this review, we discuss the physics of a variety of device concepts that are introduced to realize white OLEDs based on both polymer and small molecule organic materials. Owing to the fact that about 80\,\% of the internally generated photons are trapped within the thin-film layer structure, we put a second focus on reviewing promising concepts for improved light outcoupling.

\end{abstract}


\maketitle

\tableofcontents

\section{\label{Intro}Introduction}
In 1880, Thomas A. Edison introduced the incandescent bulb -- an epochal technical breakthrough that brought new light and comfort into peoples everyday life. Electricity is converted into a photon flux, so that artificial lighting became as versatile as never been before [\cite{Edison1880}]. Ever since, the energy demand of mankind has steadily increased to levels which clearly question our current dealing with natural energy resources. Besides the fact that fossil resources are limited, our current energy consumption will most likely harm the global ecosystem, calling for another revolution in the way energy is used. Today, it is one of the most important challenges to find efficient solutions for any energy-consuming process or application.

Edison's light bulb -- a Planckian radiator -- is still unmatched with respect to its color quality, because it is inherently best at resembling the natural sun light, which is most comfortable for the human perception of both light and, equally important, of the content that is illuminated. However, because incandescent lamps only convert roughly 5\,\% of the consumed electric power into light, new lighting solutions with higher luminous efficacies (LEs), given in lumen per watt (lm\,W$^{-1}$)  [\cite{Ohta2005}], need to be developed. At the same time, alternative technologies should match the light quality of incandescent bulbs. With luminous efficacies  up to 90 lm\,W$^{-1}$ [\cite{Steele2007}], fluorescent tubes  are widely used as energy efficient light sources, however, they lack a good color quality and contain toxic mercury. In the past decades, a new class of light sources has emerged, referred to as solid state lighting (SSL). In contrast to the gas discharge lamps, charge carriers (electrons and holes) are injected into semiconductor materials in their condensed phase, where they recombine under emission of photons.\footnote{Note that luminescence can also be generated without injecting charges, driven by an alternating electric field, in inorganic thin films [\cite{Rack1998}], quantum dots [\cite{Wood2011}], and organic systems [\cite{Perumal2012}]. However, to date, these technologies are not considered mainstream for SSL.} The first light-emitting diodes (LEDs) have been realized using inorganic semiconductors, where electroluminescence (EL) has initially been observed in silicon carbide by Oleg Vladimirovich Losev in 1928 [\cite{Zheludev2007}]. Similar to fluorescent tubes and compact fluorescent lamps (CFLs), inorganic white LEDs make use of phosphorous down-conversion layers excited by a blue LED to achieve white light [\cite{Ohno2004}]. White LEDs show remarkable efficiencies -- latest reported device efficiencies reach up to 169 lm\,W$^{-1}$ -- which is almost double the values of typical fluorescent tubes [\cite{Narukawa2006, Narukawa2008}]. Due to the crystallinity of the inorganic semiconductors used, these LEDs are point light sources with forward directed emission characteristics. 

The reports of \cite{Tang1987}  and later of \cite{Burroughes1990} on electroluminescence from thin organic films made of small molecular weight molecules (devices referred to as \emph{OLEDs}) and conducting polymers (\emph{PLEDs}), respectively, opened a new field of research. Early efforts following these pioneering works focused on the improvement of these devices with respect to their efficiency, stability and color tunability -- however, solely monochrome devices\footnote{Here, monochrome is used to describe EL devices, where emission stems from one type of emitter molecule only. This is, even though any organic semiconductor has a certain spectral distribution of its emitted spectrum typically with full width at half maximum in the range of 50 to 100\,nm [\cite{Pope1999}].} have been investigated. Roughly a decade later, first multi-color OLEDs [\cite{Kido1994}] and PLEDs [\cite{Wang1999}] have been reported, demonstrating that LEDs based on organic materials may become an alternative for general lighting applications.

It is our objective to review the topic of white organic light-emitting diodes\footnote{Starting here, the abbreviation OLED is jointly used for small molecule and polymer organic LEDs, simply because both material classes belong to the organic chemistry. The context will clarify, whether OLEDs or PLEDs are discussed.} made of small molecules and/or polymers. We discuss different concepts that enable white light emission. We will first focus on the \textsl{\textbf{internal efficiency of converting charges into photons}} -- the most important prerequisite for highly efficient white OLEDs. The discussion of various concepts will show that OLEDs have the potential to reach very high internal efficiencies approaching unity, however a large fraction of the light generated within the device cannot escape the thin film layer structure. This results in low light outcoupling efficiencies -- typically in the order of 20\% [\cite{Greenham1994}] -- bearing great potential for improvements. Therefore, we will put a second focus on \textsl{\textbf{key concepts for outcoupling enhancement}}  known to date. Despite the fact that material and device stability still is one of the most important challenges for OLEDs to become a mature technology, we will not include the discussion of stability related issues herein because this topic itself deserves a comprehensive review [e.g. \cite{So2010}].

This review is organized as follows. Within this Section \ref{Intro}, fundamentals of organic light-emitting diodes (Sec. \ref{OLEDs}) and their characterization (Sec. \ref{Quantification}) will be discussed. This is followed by sections discussing concepts for white OLEDs based on polymers (Sec. \ref{whitePOLED}) and small molecules (Sec. \ref{whiteSMOLED}).  Concepts for enhanced light outcoupling will be jointly discussed for polymer and small molecule based devices in Section \ref{Outcoupling}. The review is closed with outlining the perspective of white OLEDs with regard to their device efficiency in Section \ref{Perspective}. 
	
	\subsection{\label{OLEDs}Working principle of OLEDs}

		\subsubsection{\label{Layouts}Device configurations and white light generation}
Organic light-emitting diodes are ultra-thin, large-area light sources made of thin film organic semiconductors sandwiched between two electrodes. State-of-the-art small molecule based OLEDs consist of various different layers -- each layer having a distinct functionality. These films are prepared by thermal evaporation in high vacuum or organic vapor phase deposition [\cite{Baldo1997, Zhou2005, Forrest1997, Forrest2004}]. In contrast, polymer OLEDs are typically processed by spin-on or spray-coating techniques [\cite{Forrest2004, Friend1999}], where the solvent is removed by annealing steps. Polymer OLEDs are limited in their complexity owing to the fact that the solvents used often harm the underlying layers. In order to improve the general complexity of wet-processed devices, tremendous efforts are spent on improving polymer processing. These efforts include the use of cross-linking polymers to enable deposition of sequential layers from solution [\cite{Rehmann2008, Zuniga2011}], cross-linking in connection with direct photolithography to achieve patterned polymer layers [\cite{Gather2007a}], and the laser-induced forward transfer of individual device pixels [\cite{Stewart2012}]. Besides these uniform coating techniques, inkjet-printing can be used to process polymer-based devices.

\begin{figure}[h]
\includegraphics[width=8.5cm]{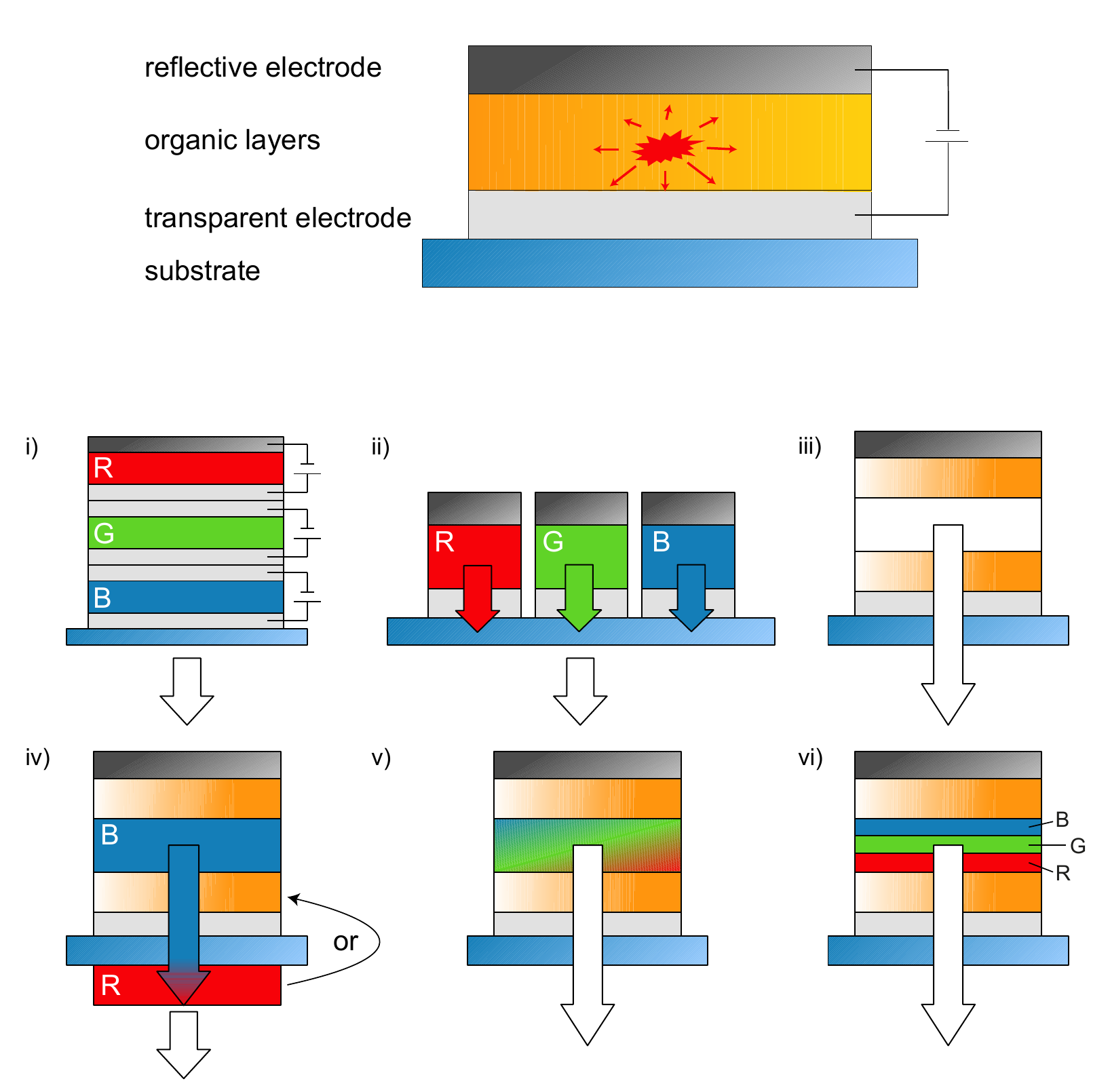}
\caption{\label{_OLED_layout}(color online) Top: Schematic cross-section of a bottom-emitting OLED. Bottom: Various device layouts to realize white light emission. i) Vertically stacked OLEDs, ii) pixelated monochrome OLEDs, iii) single emitter based white OLEDs, iv) blue OLEDs with down-conversion layers, v) multiple doped emission layers (EMLs), and vi) single OLEDs with a sub-layer EML design. For iii) to vi): orange layers represent optional functional layers, e.g. transport layers [not shown for i) and ii) for better visibility]. R, G, and B stand for red, green, and blue, respectively.} 

\end{figure}

The general architecture of an OLED is shown in a cross-section in Figure \ref{_OLED_layout} (Top). The conventional bottom-emitting device comprises a transparent electrode on top of a glass substrate, followed by one or more layers of organic material and capped with a highly reflective metal electrode. By altering the optical properties of the electrodes, top-emitting [\cite{Kanno2005, Huang2006c, Riel2003}] and transparent [\cite{Bulovic1996}] OLEDs can be fabricated. As organic materials have emission bands with  50-100\,nm full width at half maximum (FWHM) [\cite{Shimizu2010, Thompson2007, Pope1999}], typically more than one emitting material is necessary to realize white light. Figure \ref{_OLED_layout} (Bottom) schematically shows the common concepts to realize white light emission in OLEDs. Stacked OLEDs [cf. Fig. \ref{_OLED_layout} i)], where each unit can host different emitters, can be realized with additional [\cite{Burrows1996}] and without [\cite{Kanno2006, Kanno2006a}] electrodes. Here, optical optimization is challenging because the emitters placed far apart within the optical cavity must all be located at their respective field antinode for efficient outcoupling [\cite{Mladenovski2009a, Lin2006}]. Alternatively, the individual device units emitting red, green and blue can be independently designed in a pixelated approach as shown in Figure \ref{_OLED_layout} ii), however this approach has major drawbacks because it involves comparably complicated structuring processes and higher current density for each color, which accelerates degradation. Apart from these concepts where high technological efforts are necessary, all other approaches for white OLEDs are based on a single device unit [cf. Fig. \ref{_OLED_layout} iii-vi)]. These are: single emitter based devices [cf. Fig. \ref{_OLED_layout} iii), \cite{Kalinowski2007, Williams2007, DAndrade2002, Adamovich2002, Cocchi2007, Tsai2003}], blue OLEDs with external [\cite{Krummacher2006}, \cite{Gohri2011}] or internal [\cite{Schwab2011}] down-conversion layers as depicted in Figure \ref{_OLED_layout} iv) , OLEDs with single emission layers (EMLs) comprising all emitter molecules [cf. Fig. \ref{_OLED_layout} v), \cite{DAndrade2004, Kido1995a, Liu2005}], and single white OLEDs comprising an EML containing different sub-layers for red, green and blue [cf. Fig. \ref{_OLED_layout} vi), \cite{Reineke2009a, Schwartz2006, Sun2007, Kido1995, Chao1998}]. All these concepts will be discussed in detail in Sections \ref{whitePOLED} and \ref{whiteSMOLED}.

		\subsubsection{\label{Layers}Functional layers}
\begin{figure*}[t]
\includegraphics[width=13cm]{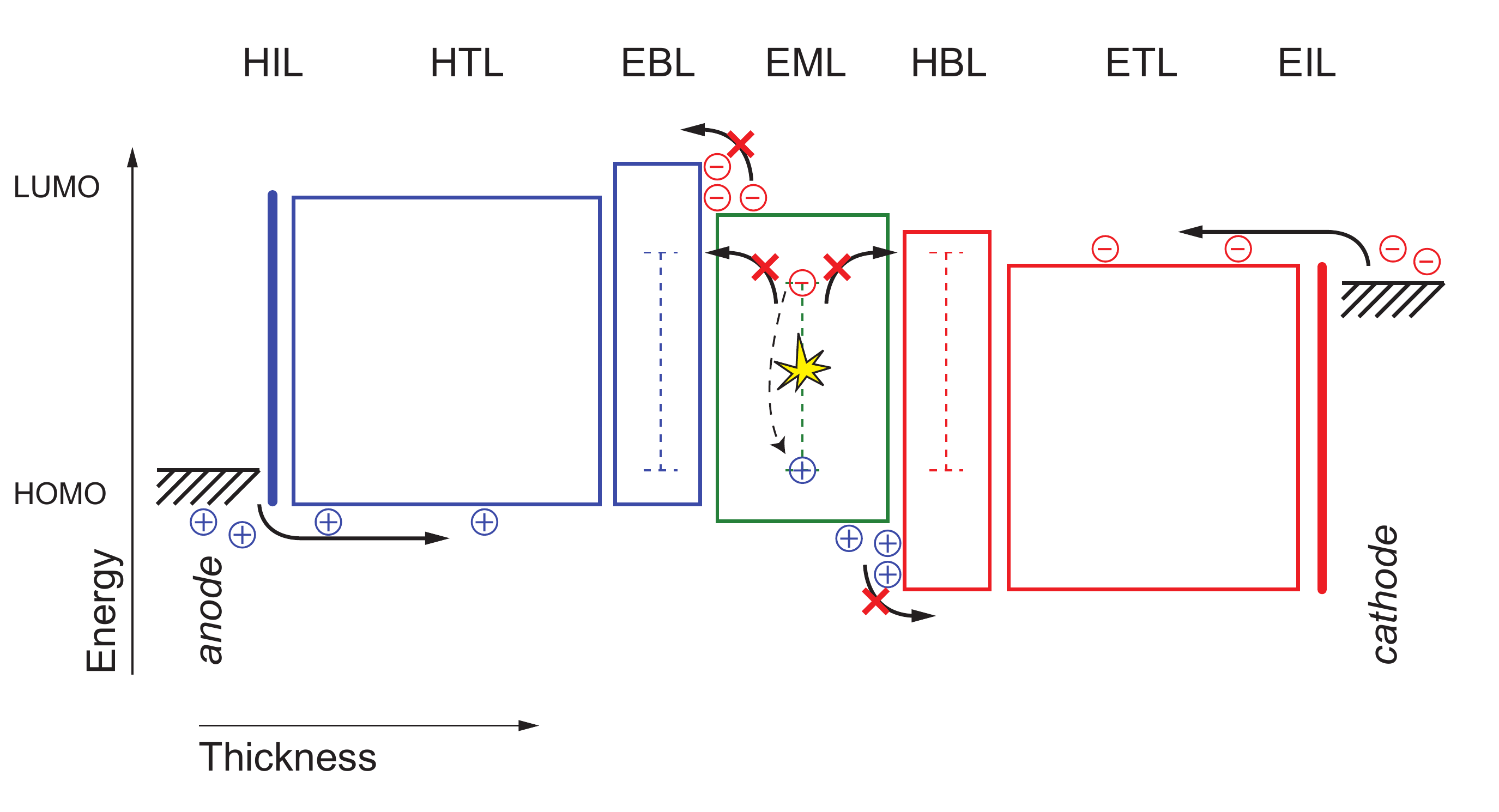}
\caption{\label{_Functional_layers}(color online) Energy diagram of a typical multilayer OLED. Note that in many devices some of the layers depicted are redundant, because different functions may be combined in one layer. From anode to cathode there are: hole injection layer (HIL), hole transport layer (HTL), electron blocking layer (EBL), emission layer (EML), hole blocking layer (HBL), electron transport layer (ETL), and electron injection layer (EIL). Boxes indicate HOMO and LUMO levels of the materials. The dashed lines in the EBL, EML, and HBL are the desired triplet energies of the materials in case of phosphorescent OLEDs.}
\end{figure*}

Electroluminescence occurs as a consequence of charges -- i.e. electrons and holes -- being injected into a semiconductor material where they meet and recombine under the emission of photons. Originally observed in anthracene crystals [\cite{Helfrich1965}], efficient EL has been reported by  \cite{Tang1987} [later by \cite{Burroughes1990} for polymers]. Since then, highly efficient OLEDs have become complex multilayer systems, where various functions like charge transport, recombination, etc. are separated to reach maximum device efficiency. Figure \ref{_Functional_layers} illustrates a multilayer OLED sequence with its functional layers.

The emission layer is located in the center of the device, where charges meet to form excited molecular states -- the so-called excitons. In order to reach the EML, holes are injected from a high work function metal [in bottom-emission layout mostly the transparent conductor indium tin oxide (ITO)] into the highest occupied molecular orbital (HOMO) of an organic semiconductor [the hole transport layer (HTL)], itself having a comparably high hole mobility. Injection can be improved by an ultra-thin hole injection layer (HIL), which works for both small molecule [\cite{Koch2005}] and polymer [\cite{Guo2006}] devices. The interested reader shall be referred to reviews of  \cite{book_Kahn2001} or \cite{Koch2012}, discussing the details of interface physics between both metal/organic as well as organic/organic interfaces. A similar effect can be achieved using metal oxides (like tungsten-, vanadium, or molybdenum-oxide), enabling ohmic injection into a wide variety of organic semiconductors [\cite{Greiner2012}]. These oxides provide very good injection properties with a high degree of flexibility, especially because the actual oxide can be exchanged easily to meet specific needs. Electrical doping of the HTL can be applied to simultaneously improve injection and transport [small molecules: \cite{Walzer2007, Blochwitz1998}, polymers: \cite{Yamamori1998}]. Before the holes reach the EML, they have to pass another (optional) layer -- the electron blocking layer (EBL). This blocking layer is often important to reach high efficiencies and has three key functions (cf. Fig. \ref{_Functional_layers}): (i) prevent leakage of the opposite charge carrier type (here the electrons) from the EML into the HTL making use of a large step in the lowest unoccupied molecular orbital (LUMO) levels forming an energy barrier, (ii) spatially separate the excitons from the HTL in case of doping (because the dopants are effective luminescence quenchers) [\cite{Zhou2001}], and (iii) realize exciton confinement. The latter function is especially important in case of phosphorescent emitters, because their long excited state lifetimes enhance their migration within the film. Here, the requirement is to use blocker materials having a higher triplet level than the emitter molecule to suppress energy transfer [\cite{DAndrade2003, Goushi2004, Chin2007}]. 
The injection and transport of electrons to the EML follows the same principles (cf. Fig. \ref{_Functional_layers}), with the only difference that they migrate on the LUMO levels of the respective materials. As cathode materials, low work function materials like aluminum, silver, and magnesium are typically used [\cite{Tang1987, Reineke2009a, Adachi2001, He2004a}]. 

In polymer OLEDs, the device architecture is less complex with respect to the number of layers used, which is a consequence of solvents involved in the preparation process [\cite{Forrest2004}]. Often, polymer OLEDs consist only of a single active layer sandwiched between the electrodes, where various materials are blended having different functionality (e.g. host and transport materials, chromophores, and even small molecule emitter materials) [\cite{Anthopoulos2003, Chuang2007, Gong2004, Huang2006b, Huang2006a, Kim2004, Kim2006, Lee2005, Liu2007a, Liu2007, Liu2006, Luo2007, Tasch1997, Tu2006, Tu2004, Wu2008, Wu2006, Xu2004}]. It is worth noting that devices are still seen as single layer diodes despite having electrodes (both anode and cathode [\cite{Huang2006b}]) that comprise a multilayer design mainly to improve charge injection. Well-known is the combination of poly(3,4- ethylene dioxythiophene):poly(styrene sulfonic acid) (PEDOT:PSS) on top of the transparent ITO anode [\cite{Gong2005, Huang2009, Kawamura2002, Niu2006, Niu2007, Xu2005}], which is widely used as heterolayer anode in polymer OLEDs. State-of-the-art wet-processing techniques allow a higher degree of complexity, for instance by carefully choosing orthogonal solvents for subsequent layers [\cite{Chen2002, Gong2005, Granstrom1996, Hu2002, Huang2009, Niu2006, Niu2004, Niu2007, Noh2003, Pschenitzka1999, Xu2005}]. Cross-linking polymer networks introduce an additional route to design multilayer polymer systems [\cite{Kohnen2010, Rehmann2008}]. A compromise is often found in the combination of solution processes and thermal evaporation to achieve multilayer OLEDs that partially consist of layers that comprise solely small molecules [\cite{Granstrom1996, Kawamura2002, Kim2006, Niu2006, Niu2007, Niu2004, Noh2003, Wang1999}]. Finally, similar to small molecule OLEDs, the preparation of the highly reflective cathode requires thermal evaporation in high vacuum.

		\subsubsection{\label{FluoPhos}Fluorescent and phosphorescent electroluminescence}
\begin{figure*}[t]
\includegraphics[width=12cm]{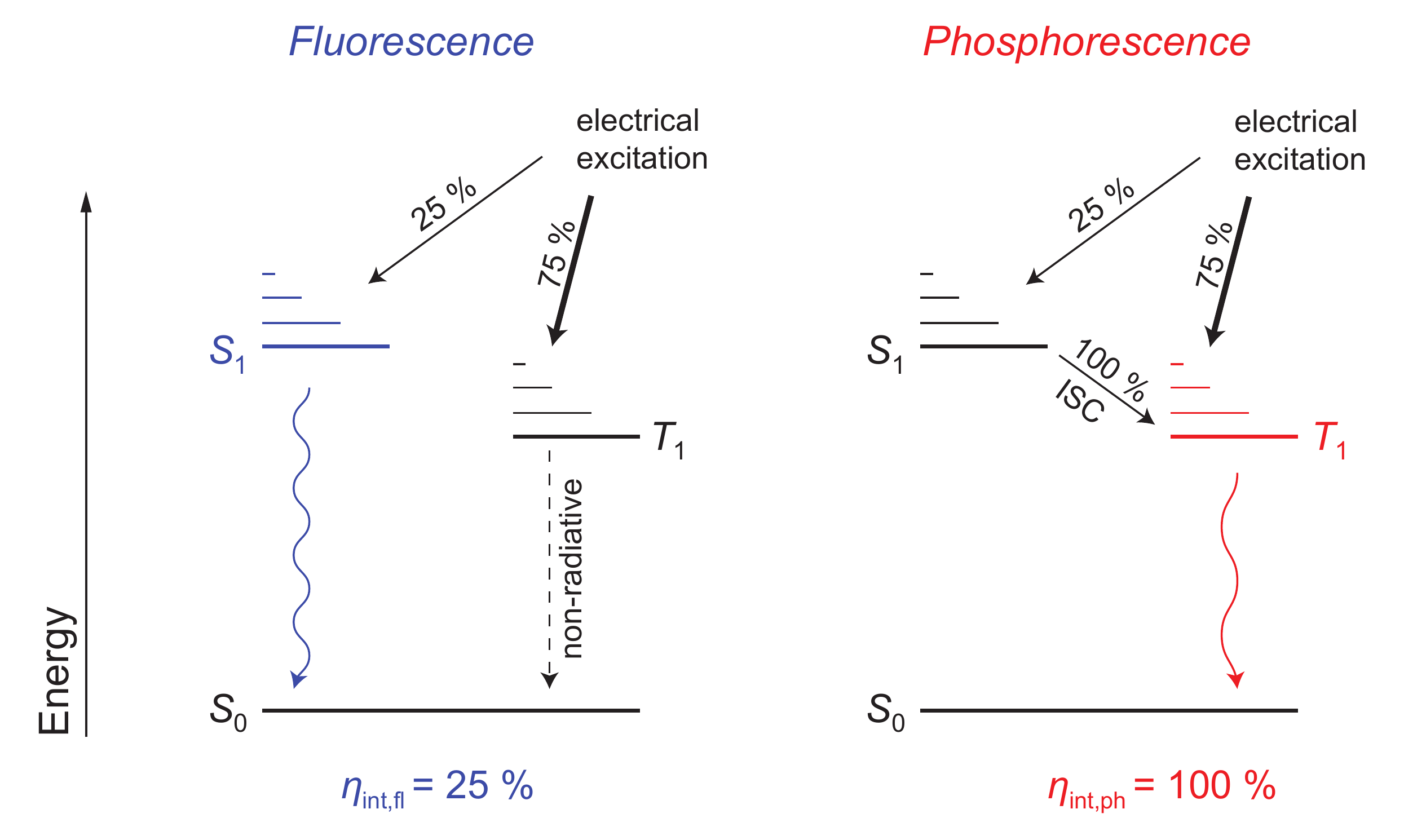}
\caption{\label{_Fluorescence_Phosphorescence}(color online) Population scheme of singlet and triplet level of the organic molecules under electrical excitation. For phosphorescent emitter materials, the singlet excitons created are efficiently transferred to the triplet state via intersystem crossing (ISC). Additionally given are the theoretical limits for the internal quantum efficiency $\eta_\text{int}$.}
\end{figure*}

The majority of organic semiconductors form -- in contrast to their inorganic counterparts -- amorphous, disordered films [\cite{Pope1999}], where Van-der-Waals forces determine the structure on the nanoscale. As a consequence, charges are injected statistically with respect to their electron spin, finally determining the formation of singlet and triplet excited states. Because the triplet state has a multiplicity of three [\cite{Pope1999}], on average 75\,\% of the excitons formed are triplet states, with the remaining 25\,\% being singlets. The work of \cite{Segal2003} showed slightly smaller values for the singlet fraction in both small molecule and polymeric systems [$\chi_\text S = (20\pm1) \,\%$ and $(20\pm4)\,\%$, respectively], which is in rather good agreement  with this simple statistical picture [for further details also see \cite{Baldo1999b}]. The low singlet fraction causes OLEDs based on fluorescent emitter molecules to be rather inefficient with an upper limit of the internal quantum efficiency of $\eta_\text{int,fl}=25\,\%$, because emission solely occurs in its singlet manifold as shown in Figure \ref{_Fluorescence_Phosphorescence}.

The efficiency of OLEDs was drastically improved with the introduction of phosphorescent emitter molecules [\cite{Baldo1998, Ma1998, Reineke2012}]. These materials are organometallic complexes comprising a heavy metal atom like iridium, platinum, palladium, etc. in the molecular core. Making use of this heavy metal effect, the spin-orbit coupling is strongly enhanced, ultimately weakening the selection rules for previously forbidden, radiative transitions in the triplet manifold of the molecule [\cite{Yersin2004, Yersin2002, Thompson2007}]. Simultaneous to realizing a highly efficient emissive triplet state in a molecule, the heavy metal effect strongly enhances the intersystem crossing (ISC) rates between singlet and triplet manifold [\cite{Yersin2002, Yersin2004}]. Thus, the fraction of singlet excitons that are created under electrical excitation are efficiently converted into triplet states before they can recombine radiatively. ISC is close to unity in various phosphorescent systems [\cite{Kawamura2005, Kawamura2006}]. Therefore, phosphorescent materials in OLEDs can lead to internal EL efficiencies of $\eta_\text{int,ph}=100\,\%$ (cf. Fig. \ref{_Fluorescence_Phosphorescence}). Furthermore, state-of-the-art emitters are especially optimized for having short excited state lifetimes -- typical values are in the order of microseconds -- in order to reduce bimolecular quenching processes limiting the photoluminescence quantum yield at high excitation levels [\cite{Baldo2000a, Reineke2007}]. Furthermore, the emitter lifetimes need to be compatible with the $RC$-time of the OLED to avoid emitter saturation effects [\cite{Thompson2007}]. In contrast to fluorescence, where emission originates from the lowest excited singlet state, phosphorescent EL induces thermalization losses in the order of the singlet-triplet splitting [\cite{Schwartz2009}] for every exciton that is captured by it via energy transfer from host materials or ISC [\cite{Sun2006}]. The loss might be circumvented if the excitons are resonantly generated on the emitter dopant.

	\subsubsection{\label{Exotic}Exotic types of electroluminescence}
Much research effort is spent on finding alternative concepts to phosphorescence that surpass the limit of $\eta_\text{int,fl}=25\,\%$ in case of fluorescence, because phosphorescence is accompanied with a serious efficiency decrease at high excitation levels [\cite{Reineke2009b, Staroske2007, Reineke2007, Baldo2000a}] (for more details see Sec. \ref{Rolloff}).

As mentioned in the previous section, the vast majority of excitons is created as triplets where, in case of fluorescence, the excited triplet state is long lived [\cite{Pope1999}]. Thus, the triplet exciton density in fluorescent OLEDs will be comparably high. A concept to improve the internal quantum efficiency of fluorescent EL makes use of this high density via \emph{delayed fluorescence} [\cite{Pope1999}]. Here, the interaction of two triplet states -- called \emph{triplet-triplet annihilation} (TTA) -- will create delayed singlet excitons: $T_1+T_1 \rightarrow S_0+S_\text n$ [\cite{Kepler1963}]. Based on this non-linear process, an internal electron-photon conversion efficiency of unity cannot be reached. The device data of \cite{Okumoto2006} showing a twofold improvement to the $\eta_\text{int,fl}=25\,\%$ limit [nearly 10\,\% external quantum efficiency (EQE)] suggests this process to take place. \cite{Kondakov2007} gives experimental evidence that delayed fluorescence substantially contributes to the internal efficiency of fluorescent OLEDs, however the author suggests that this process cannot be the only reason for the very high efficiency of \cite{Okumoto2006}.

\cite{Endo2009} suggest an alternative concept -- \emph{thermally activated delayed fluorescence} (TADF) -- to feed the singlet state of a molecule with its triplet excitons. By tailoring molecules with a small singlet-triplet splitting, reverse intersystem crossing (RISC) will occur with an increased probability, because this process is thermally activated: $T_1 + E[\sim k_\text B T] \rightarrow S_1$ [see also \cite{Endo2011}]. \cite{Deaton2010} have reported on very high efficiency devices based on TADF, reaching 16\,\% EQE. Based on this general idea, \cite{Goushi2012} reported on efficient triplet-to-singlet back-conversion in OLEDs where the emissive state is an interfacial exciplex formed between two organic layers. Devices based on this concept showed a very high RISC efficiency of 86.5\,\% and a external quantum efficiency beyond the fluorescence limit of 5\,\%. Very recently, \cite{Uoyama2012} reported very promising OLED performance data based on this TADF concept. With a specially designed novel class of organic materials, the exchange splitting could be reduced to approx. 80\,meV, giving rise to very effective reverse intersystem crossing. These materials possess a very high rate of delayed fluorescence in the order of 10$^{-6}$ seconds - which is comparable to the radiative rates of phosphorescent emitters [\cite{Thompson2007}]. In their report, OLEDs very discussed reaching 19\,\% EQE, which is on par with the currently used phosphorescent emitter technology. It is interesting to see, how this concept develops in future. For the first time, a promising, very general concept has matured to a serious alternative to the heavy metal complex based phosphorescence.

Finally, \emph{extrafluorescence} has been introduced by \cite{Segal2007}. This concept makes use of an anomaly in the energetic order of singlet and triplet charge-transfer (CT) states [the precursor states in the exciton formation process \cite{Pope1999, Segal2003}] of a molecule. By having a higher lying triplet CT-state, the rates of singlet exciton formation can be significantly increased, leading to a singlet fraction of as high as $\chi_\text S = (0.84\pm0.03)$ [\cite{Segal2007}].

It is worth mentioning that neither of these concepts find application in white OLED concepts to date, mainly owing to the fact that the underlying working principles are not yet fully understood. 

Even though LEDs based on colloidal quantum-dots (QDs) have inorganic lumophores, these QD-LEDs share to a large extend the general device layout with OLEDs.  \cite{Coe2002} have shown that a single monolayer of QDs can be incorporated into an OLED architecture solely acting as luminescent center of the device. It also has been shown that by incorporating differently emitting QDs into the monolayer, multicolor and white QD-LEDs can be fabricated [\cite{Anikeeva2007}]. One important difference to OLEDs is the comparably small FWHM ($<40$\,nm [\cite{Anikeeva2007}]) of the QD's luminescence, directly affecting the color rendition properties of white QD-LEDs. Another important distinction from organic lumophors is the fact that QDs are not affected by the spin statistics as observed in OLEDs. QDs are quantum systems, where spin-singlet and spin-triplet character states are mixed very effectively [\cite{Coe2002}]. However, QDs also have 'bright' and 'dark' excitonic band edge states that are spin allowed and forbidden, respectively [\cite{Shirasaki2013}]. In very efficient QDs, their energetic splitting can be as small as 25 meV, thus, the dark states are effectively thermally activated to the bright states (state-of-the-art CdSe QDs can harness virtually 100\,\% of the excitation in the 'bright' state [\cite{Shirasaki2013}]. It is worth to note that much effort is spent to replace the organic functional layers in a QD-LED by inorganic ones to benefit from the robustness of the latter [\cite{Caruge2008}].


\subsubsection{Intermolecular energy transfer}\label{Energytransfer}
White OLEDs are highly complex, multi-component luminescent systems which greatly rely on various energy transfer mechanisms. These in turn lead to the distribution of the excitation to the desired emitter molecules. Thus, it is necessary to briefly review the possible energy transfers that can happen between different molecular species, which are referred to as \emph{donor} $D$ or \emph{acceptor} $A$ whenever they yield or accept energy, respectively. Furthermore, their multiplicities will be denoted with preceding superscripts, i.e. 1 or 3 for singlets and triplets, respectively. Furthermore, asterisks will mark excited states and double asterisks levels higher than the lowest possible electronic excitation, correspondingly. 

Trivially, energy can be transferred as a two-step process (radiative energy transfer) that involves the emission and absorption of a photon $h\nu$ having a frequence $\nu$:
\begin{eqnarray}
D^*\rightarrow D\quad+ \quad& h\nu &  \\
& h\nu &\quad + \quad A\rightarrow A^*. 
\end{eqnarray}
This energy exchange is often referred to as reabsorption. Because the two steps are completely decoupled and solely depend on the specific properties of $D$ and $A$, it is not necessary to distinguish between singlets and triplets. Reabsorption in OLEDs plays only a minor role, because most organic materials possess a significant Stokes-shift between absorption and emission bands so that devices become transparent for the emitted wavelength - which is a great advantage of OLED compared to inorganic LED. However, reabsorbing photons is of importance in white OLED concepts that make use of an external down-conversion layer [cf. Fig. \ref{_OLED_layout} iv)]. In contrast to the following energy transfer types, reabsorption can overcome macroscopic distances between $D$ and $A$ states.

Non-radiative energy transfers are of central importance in OLEDs. Such transfers conserve the initial donor energy and are proportional to the number of transitions in the emission band of the donor $I_D(\nu)$ and in the absorption band of the acceptor $\varepsilon_A(\nu)$ that are equal in energy [\cite{b_Klessinger1989}]. This is quantified in the spectral overlap integral $J$ which reads:
\begin{equation}\label{eq:4_spectral_overlap}
J=\int \limits_0^\infty \bar I_D(\nu)\bar\varepsilon_A( \nu)\text d \nu,
\end{equation}
where $\bar I_D$ and $\bar \varepsilon_A$ represent normalized intensities with respect to the integrated band. Without going further into detail, there are two distinctive energy transfer mechanisms,  introduced by and named after \cite{Foerster1948} and \cite{Dexter1953}, which can be ascribed to Coulomb and exchange interactions, respectively [for more details on the quantum mechanical description, the reader is referred to the book of \cite{b_Klessinger1989} or equivalents]. 

In the F\"orster framework [\cite{Foerster1948}], the rate constant for the dominating dipole-dipole interaction can be written as [\cite{Braslavsky2008}]:
\begin{equation}\label{eq:4_Foerster_rate}
k_\text{F}=k_{\not{A}}\frac{9 (\ln 10) \kappa^2 \mathit{\Phi_D}}{128 \pi^5 N_\text{A} n^4}\cdot J\cdot \frac{1}{R_{DA}^6}=k_{\not{A}}\left\lbrack{\frac{R_0}{R_{DA}}}\right\rbrack^6,
\end{equation}
for distances exceeding orbital overlap interactions. Here, $k_{\not{A}}$ is the rate constant of the excited donor in absence of an acceptor, $\kappa$ an orientation factor, $n$ the refractive index of the medium in the range of spectral overlap, $N_\text A$ the Avogadro constant, $\mathit{\Phi_D}$ the luminescence quantum yield  of the donor emission, and $R_{DA}$ the intermolecular distance between donor and acceptor. Here, the various parameters merge to become the F\"orster radius $R_0$. This transfer can only occur if both $D$ and $A$ transitions are allowed [\cite{b_Klessinger1989}], which leads to the following allowed energy transfer reactions:
\begin{eqnarray}
{}^1D^* + {}^1A & \longrightarrow& {}^1D + {}^1A^* \label{Eq:singlet_foerster} \\
{}^1D^* + {}^3A & \longrightarrow& {}^1D + {}^3A^* \label{Eq:triplet_transfer}.
\end{eqnarray}
Note that there are examples of molecules having a triplet ground state configuration [\cite{Tanaka2006, b_Reinhold2004}], giving rise to ${}^3A$ on the left side of Equation \eqref{Eq:triplet_transfer}. A transfer of a triplet to a singlet state, i.e. ${}^3D^* + {}^1A \rightarrow {}^1D + {}^3A^*$, is strictly forbidden in the F\"orster theory as it would require two simultaneous intersystem crossing steps. This picture changes if a phosphorescent donor is incorporated. Here, the recombination in the triplet manifold is enhanced due to spin-orbit coupling. The following additional transfers are thus possible [\cite{b_Klessinger1989, b_Reinhold2004}]:
\begin{eqnarray}
    {}^{3}D^{*}+{}^{1}A & \longrightarrow & {}^{1}D + {}^{1}A^{*}
    \label{eq:4_phos_donor1} \\
    {}^{3}D^{*}+{}^{3}A & \longrightarrow & {}^{1}D + {}^{3}A^{*}.
\end{eqnarray}
Even though the transition from ${}^{3}D^{*}$ to ${}^{1}D$ requires intersystem crossing and thus has a lower rate than starting from ${}^{1}D^*$, they may have a similar probability, because the lifetime of the triplets is correspondingly longer. Note, for the process in Equation \eqref{eq:4_phos_donor1}, two different types of molecules have to be involved in order to excite the energetically higher singlet state of $A$ with the triplet $D$ state energy. The F\"orster energy transfer can efficiently overcome distances of up to $10\,\text{nm}$, which is much larger than typical molecular dimensions [\cite{Pope1999}].

Dexter energy transfer in contrast, is mediated by exchange interactions, which requires orbital overlap of $D$ and $A$, resulting in a decrease of this interaction with increasing intermolecular distance [\cite{Dexter1953}]. Dexter-type energy exchange obeys the \emph{Wigner-Witmer} spin conservation rules, requiring the total spin of the configuration to be conserved throughout the reaction [\cite{Widmer1928}]. The resulting energy transfer reactions read:
\begin{eqnarray}
{}^1D^* + {}^1A & \longrightarrow& {}^1D + {}^1A^* \label{Eq:singlet_dexter} \\
{}^3D^* + {}^1A & \longrightarrow& {}^1D + {}^3A^*, 
\end{eqnarray}
and
\begin{equation}
    {}^{3}D^{*}+{}^{3}A^{*}  \longrightarrow  {}^{1}D + {}^{1,3,5}A^{*},
\end{equation}
for triplet-triplet annihilation [\cite{Kepler1963, Suna1970}]. In the latter equation, $A$ can be in its singlet, triplet, or quintet configuration [\cite{b_Klessinger1989}]. Since the singlet-singlet interaction is a very efficient F\"orster-type transfer [cf. Eqn. \eqref{Eq:singlet_foerster}], it is rarely observed based on exchange interactions. In contrast, the triplet-triplet energy transfer is of great importance as it provides the basis for efficient triplet excited state migration in organic materials. The corresponding rate constant reads [\cite{Dexter1953}]:
\begin{equation}
k_\text{D}=\frac{2\pi}{\hbar}K^2\cdot J\cdot e^{-2R_{DA}/L},
\end{equation}
where $K$ is a constant in units of energy [\cite{Murphy2004}]. The exponential dependence on the intermolecular distance $R_{DA}$ accounts for the necessity of molecular orbital overlap. Accordingly, Dexter transfers are short distance interactions, typically reaching up to $1\,\text{nm}$ (cf. up to 10\,nm for F\"orster-type energy exchange).

Both F\"orster- and Dexter-type energy transfers enable excitons to migrate throughout organic solids. Here, the net charge carried is zero. The driving force of this exciton motion is a gradient in the exciton concentration $\nabla n(\vec r,t)$ leading to a series of uncorrelated hopping steps from molecule to molecule (\emph{random walk}). Particle diffusion is described by Fick's $2^{\text{nd}}$ law [\cite{Fick1995}]. Neglecting higher order processes and applying it to excitons, it reads:
\begin{equation}
\frac{\partial n(\vec r,t)}{\partial t}=G(\vec r,t)-\frac{n(\vec r,t)}{\tau}+D\nabla^2 n(\vec r,t).
\end{equation}
$G(\vec r,t)$ is the exciton generation, $D$ the diffusion constant, and $\tau$ the excited state lifetime. 

Exciton diffusion plays a key role in the working principle of OLEDs, especially in white OLEDs that need to distribute excitons to different emitters. Under electrical excitation, excitons are often formed close to an interface between different materials, usually with a generation width small compared to the total layer thickness [\cite{Reineke2007, Wuensche2010, Rosenow2010, Sun2006}]. Thus, it is often adequate to model the exciton generation to be a delta-function in space, i.e. $G(x,t)=G\cdot\delta(x=x_0,t)$. This solves to the steady-state ($\partial n/\partial t = 0$) solution of Fick's $2{^\text{nd}}$ law [\cite{Zhou2007, Giebink2006, Baldo2000b, Wuensche2010}]:
\begin{equation}
n(x)=n_0\cdot \text{e}^{-x/L_x} \quad \text{with}\quad L_x=\sqrt{D\tau},
\end{equation}
where $L_x$ is the diffusion length and $n_0$ is the exciton density at the interface.\footnote{Because exciton motion is typically isotropic and the systems are planar structures, diffusion can be reduced to one dimension in space, e.g. $L_x$.}

		\subsubsection{\label{Modes}Where the light goes}
After discussing the fundamentals of light generation and exciton transfer in the previous sections, we will now briefly discuss, where the photons -- created in the emission layer -- will propagate with respect to the important question: What fraction of photons is able to escape to air (here: the far-field, defined as the photons that leave the device to the forward hemisphere). 

\begin{figure*}[t]
\includegraphics[width=17.0cm]{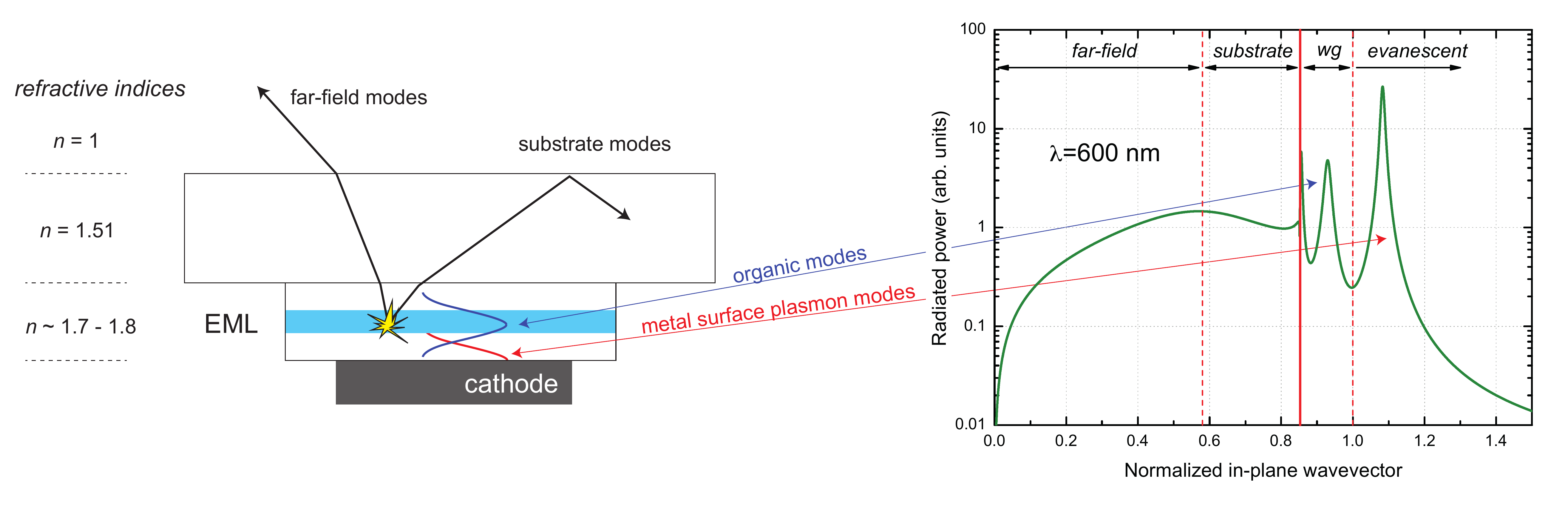}
\caption{\label{_Power_Spectrum}(color online) Left: cross-section of an OLED with indication of different light modes. Right: typical power spectrum of the internally generated light shown as a function of the in-plane wavevector. Vertical lines separate the various possible light modes. Both, waveguide (wg) and evanescent modes (thick line) cannot be accessed with external light outcoupling techniques, thus they dissipate within the layer structure. Model calculation for a bottom-emitting device as discussed in \cite{Meerheim2008}.}
\end{figure*}

Figure \ref{_Power_Spectrum} shows a cross-section of a conventional bottom-emission OLED, additionally indicating various light propagation modes. They are mainly determined by the thin-film structure of the device and the respective optical properties (i.e. refractive indices and absorption coefficients) [\cite{Greiner2007}]. Organic materials and ITO (the latter depending on its composition) typically have refractive indices in the range of $n\sim1.7-1.9$ and conventional glass substrates of $n=1.51$. Thus, in first approximation\footnote{Because (i) the organic materials all show different, distinct wavelength dependencies and (ii) slight changes in $n$ are observable for every two organic materials compared.}, two optical interfaces, i.e. the organic/substrate and the substrate/air interfaces, are formed, where total internal reflection may occur [\cite{Mladenovski2009a, Reineke2009a, Krummacher2009}]. The refractive index difference at the organic/substrate interface causes a large fraction of light to be trapped inside the organic layer stack, forming the so-called organic or waveguide (wg) modes (cf. Fig. \ref{_Power_Spectrum}). Based on similar considerations, only a fraction of the light that is originally coupled to the substrate can escape the device (the so-called far-field, air, or outcoupled modes). Additionally, the emitting molecules can directly couple to surface plasmons of the highly reflective electrode (here: cathode) -- a process that is very efficient for short distances between EML and cathode and strongly decreases with increasing spacing [\cite{Lin2006, Reineke2009a, Mladenovski2009a, Krummacher2009}].

Figure \ref{_Power_Spectrum} additionally shows a power spectrum, obtained from model calculations, of a conventional monochrome bottom-emitting OLED [\cite{Meerheim2008, Meerheim2010}], plotted as a function of the in-plane wavevector [\cite{Meerheim2010, Furno2010}]. In such a power spectrum, the modes discussed above can easily be attributed to different ranges of the in-plane wavevector, indicated by the vertical lines in Figure \ref{_Power_Spectrum}. Here, the fraction of photons that directly leaves the device (far-field) typically is in the range of only 20\,\% [\cite{Meerheim2010, Furno2010, Mladenovski2009a, Gaertner2008, Adachi2001b, Krummacher2009}]. More light can be extracted to the far-field by applying modifications of the substrate/air interface (e.g. periodic, shaped substrates) by converting substrate into air modes [\cite{Reineke2009a, Moller2002, Greiner2007, Sun2008}; for details see Sec. \ref{Outcoupling}]. On the contrary, as indicated by the thick solid line in Figure \ref{_Power_Spectrum}, modes with larger in-plane wavevector, i.e. waveguide and evanescent surface plasmon modes, cannot be outcoupled by external techniques. 

Concepts for improved light outcoupling, including approaches to reduce waveguide and surface plasmon modes, will be discussed in Section \ref{BottomOut}. Similarly, top-emitting OLEDs, which have significantly different optical properties compared to bottom-emitting devices that largely  influence the outcoupling efficiency, will be introduced and analyzed in Section \ref{TopOut}.

		\subsubsection{\label{Rolloff}Efficiency roll-off}	
Even though state-of-the-art phosphors have excited state lifetimes down to $1\,\text{\textmu s}$, the lifetime is still about two orders of magnitude longer than their fluorescent counterparts, which is the main reason for different electroluminescent properties of fluorescent and phosphorescent emitters [\cite{Pope1999}]. The following calculation illustrates the difference in the respective excited state properties. Representative fluorescent and phosphorescent excited state lifetimes are set to 10\,ns and $1\,\text{\textmu s}$, respectively [\cite{Sokolik1996, Reineke2007}]. The brightness $L$ of an OLED is proportional to the excited state density $n$ and inversely proportional to the excited state lifetime $\tau$ [\cite{Reineke2007}]: $L \sim n/\tau$. Considering the spin statistics discussed in Section \ref{FluoPhos}, one derives for a fluorescent (fl) and a phosphorescent (ph) system, respectively:

\begin{equation}\label{eq:3_04}
       		L \sim \frac{n_\text{fl}}{4 \cdot \tau_\text{fl}} \quad \text{and} \quad L \sim \frac{n_\text{ph}}{\tau_\text{ph}},
\end{equation}
with corresponding subscripts. This leads to an expression for the ratio between the excited state densities $n_\text{ph}/n_\text{fl}$:

\begin{equation}\label{eq:3_05}
       		\frac{n_\text{ph}}{n_\text{fl}} \sim \frac{\tau_\text{ph}}{4 \cdot \tau_\text{fl}} = \left[\frac{1\,\text{\textmu s}}{4\cdot 10\,\text{ns}} = \frac{\bf 25}{\bf 1}\right].
\end{equation}

\begin{figure}[h]
\includegraphics[width=8.5cm]{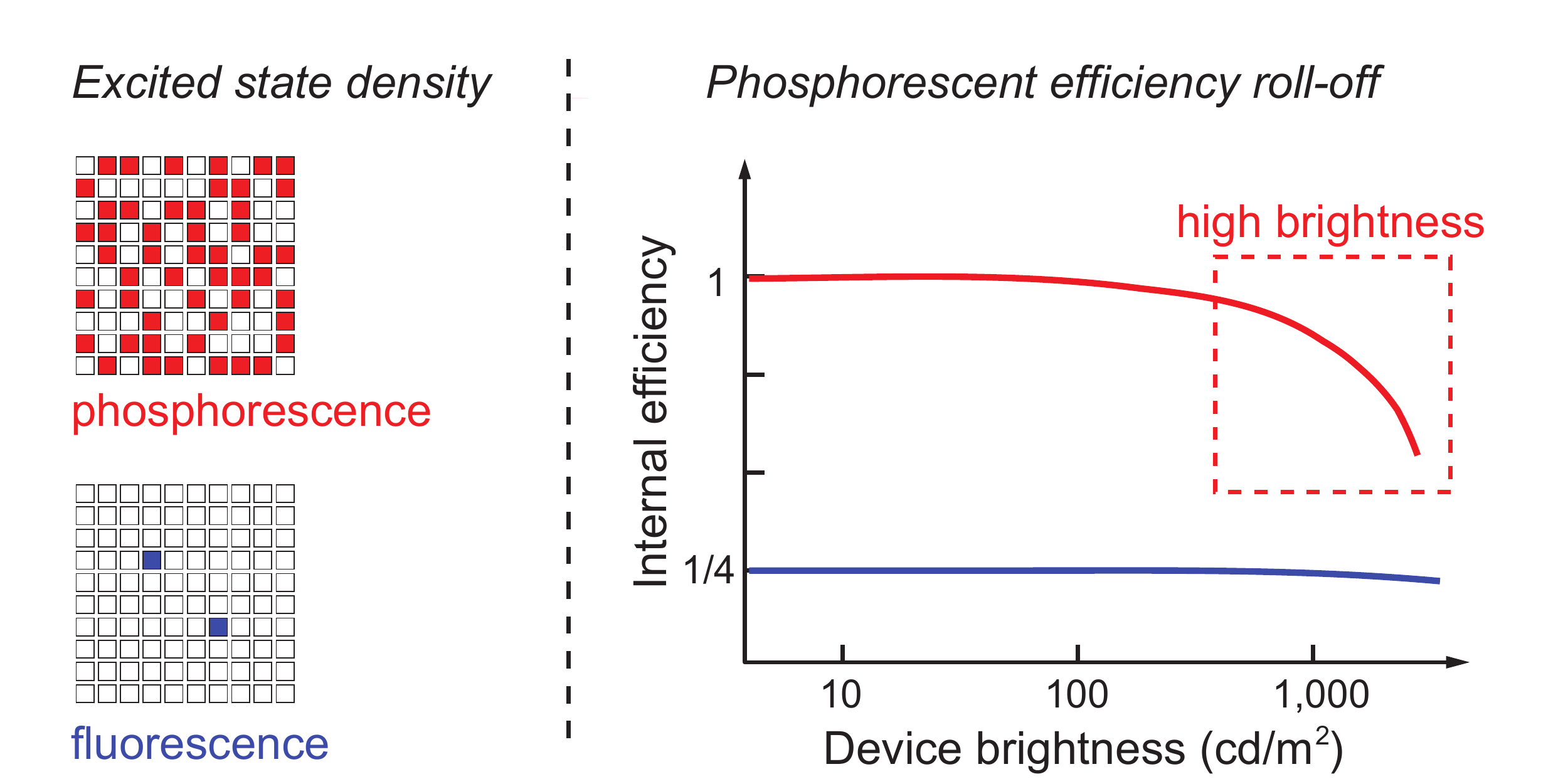}
\caption{\label{_Efficiency_roll-off}(color online) Left: Illustration of the excited state density in case of phosphorescence and fluorescence, respectively. Right: Typical respective efficiency versus brightness characteristics. In case of phosphorescence, the efficiency drastically decreases at high brightness as a consequence of quenching processes.}
\end{figure}

The direct consequence is illustrated in Figure \ref{_Efficiency_roll-off}. To reach the same luminance level in fluorescent and phosphorescent systems, the exciton density is typically about 25-fold higher in case of phosphorescence, which increases the probability of excited state annihilation processes, such as triplet-triplet annihilation, triplet-polaron quenching [\cite{Baldo2000a, Reineke2007}], and in some cases field-induced exciton dissociation [\cite{Kalinowski2002}]. These processes cause the quantum efficiency of phosphorescent systems to noticeably drop at high brightness (\emph{efficiency roll-off}) as depicted in Figure \ref{_Efficiency_roll-off}. For state-of-the-art phosphorescent systems, this roll-off typically starts at around 1,000\,cd\,m$^{-2}$ [\cite{Reineke2007, Reineke2007a, Baldo2000a, Su2008}]. Thus, especially for lighting applications, where a few thousand cd\,m$^{-2}$ are seen as a realistic device brightness (cf. Sec. \ref{Bright}), phosphorescent OLEDs typically work at a decreased internal efficiency [$\eta_\text{int,ph}(L_\text{high})<100\,\%$]. 
	
	
	\subsection{\label{Quantification}Quantification of light and efficiency}
		
		\subsubsection{\label{Efficiency}Figures of merit}

Typically, three device efficiencies are discussed in literature: the \emph{current efficiency} $\eta_\text{C}$, the \emph{luminous efficacy}\footnote{It is worth mentioning that the luminous efficacy is often referred to as \emph{power efficiency} in literature. However, strictly speaking, an efficiency should be dimensionless, which is not the case for the quantity discussed (cf. [lm\,W$^{-1}$]).} $\eta_\text{L}$, and the \emph{external quantum efficiency} $\eta_\text{EQE}$ [\cite{Forrest2003}]. While the latter is a measure of the number of photons that are extracted to air per injected electrons, the other two efficiencies are photometric quantities that take the sensitivity of the human eye into account.

The current efficiency is calculated from the luminance $L_{0^\circ}$, obtained in forward direction, and the current density $j_\text{meas}$ passing through the device:
\begin{equation}
\eta_\text{C}=\frac{L_{0^\circ}}{j_\text{meas}}\quad[\text{cd\,A$^{-1}$}].
\end{equation}
The luminous efficacy can be calculated considering the operating voltage at the point of measurement $V(j_\text{meas})$. It reads:
\begin{equation}
\eta_\text{LE}=\eta_\text{C}\frac{f_\text D \pi}{V(j_\text{meas})}\quad[\text{lm\,W$^{-1}$}],
\end{equation}
with
\begin{equation}
f_\text D=\frac{1}{\pi I_0}\int_0^{\pi/2}\int_{-\pi}^{+\pi} I(\theta,\phi)\sin \theta\, \text d \phi\, \text d \theta.
\end{equation}
Here, $f_\text D$ accounts for the angular distribution of the emitted light intensity $I(\theta,\phi)$ in the forward hemisphere which is a function of two angles [azimuth ($\theta$) and polar ($\phi$)]. Furthermore, $I_0$ represents the light intensity measured in forward direction. For OLEDs with changing spectral distribution as a function of observation angle, i.e. $I(\theta,\phi,\lambda)$, the spectral changes also need to be considered [\cite{Meerheim2008a}].

Finally, the radiometric external quantum efficiency can be calculated with:
\begin{equation}
\eta_\text{EQE}=\eta_\text{C}\frac{f_\text D \pi e}{K_\text r E_\text{ph}}\quad\left[\frac{\%}{100}\right],
\end{equation}
where $E_\text{ph}$ is the average photon energy of the emitted device spectrum. Apparently, the integrated quantities $\eta_\text{LE}$ and $\eta_\text{EQE}$ are only correctly determined, if the angular distribution $f_\text D$ is taken into account. This is possible using an integrating sphere or a goniometer set-up [\cite{Meerheim2008a, Hofmann2010}]. For a long time, it has been common sense that these quantities can be calculated assuming a Lambertian emission pattern of the emitted light, i.e. $I(\theta,\phi)=I_0 \cos \theta$. However, recent publications show that this assumption is not valid [\cite{Meerheim2008a, Freitag2010, Mladenovski2009a}], calling for precise methods of efficiency determination.

\begin{figure}[h]
\includegraphics[width=7cm]{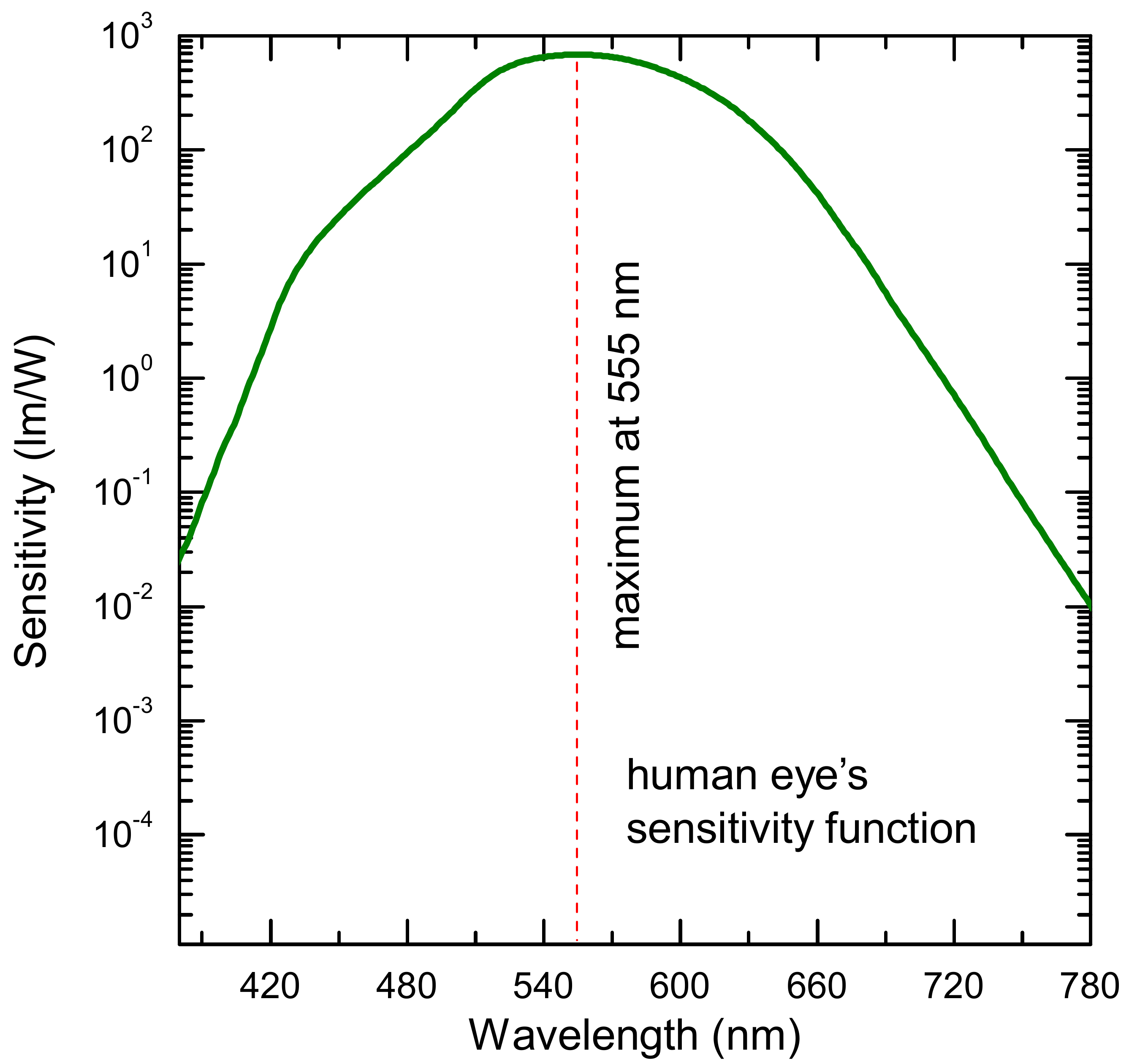}
\caption{\label{_Vlambda}(color online) Commission Internationale de l'$\acute{\text E}$clairage (CIE) photopic (daylight) spectral sensitivity function $V(\lambda)$. Its maximum at a wavelength of 555\,nm corresponds to 683\,lm\,W$^{-1}$.}
\end{figure}

 Photometric quantities are converted into radiometric ones and vice versa by the introduction of the luminous efficacy of radiation $K_\text r$ that is calculated by:
\begin{equation}
K_\text r =\frac{\int_{380\,\text{nm}}^{780\,\text{nm}}\mathit{\Phi}_\text r(\lambda) V(\lambda)\, \text d \lambda}{\int_{0}^{\infty}\mathit{\Phi}_\text r(\lambda)\, \text d \lambda} \quad[\text{lm\,W$^{-1}$}],
\end{equation}
where $\mathit{\Phi}_\text r$ is the radiant flux and $V(\lambda)$ the weighting function that takes the sensitivity of the human eye into account (cf. Fig. \ref{_Vlambda}). In other words, the luminous efficacy of radiation $K_\text r$ quantifies, how many lumen a given spectrum can produce per watt. Thus, it marks the theoretical limit for the luminous efficacy $\eta_\text{LE}$ of any OLED spectrum, neglecting electrical and optical losses.

		\subsubsection{\label{Color}Color rendering and quality}
		
\begin{figure}[h]
\includegraphics[width=8cm]{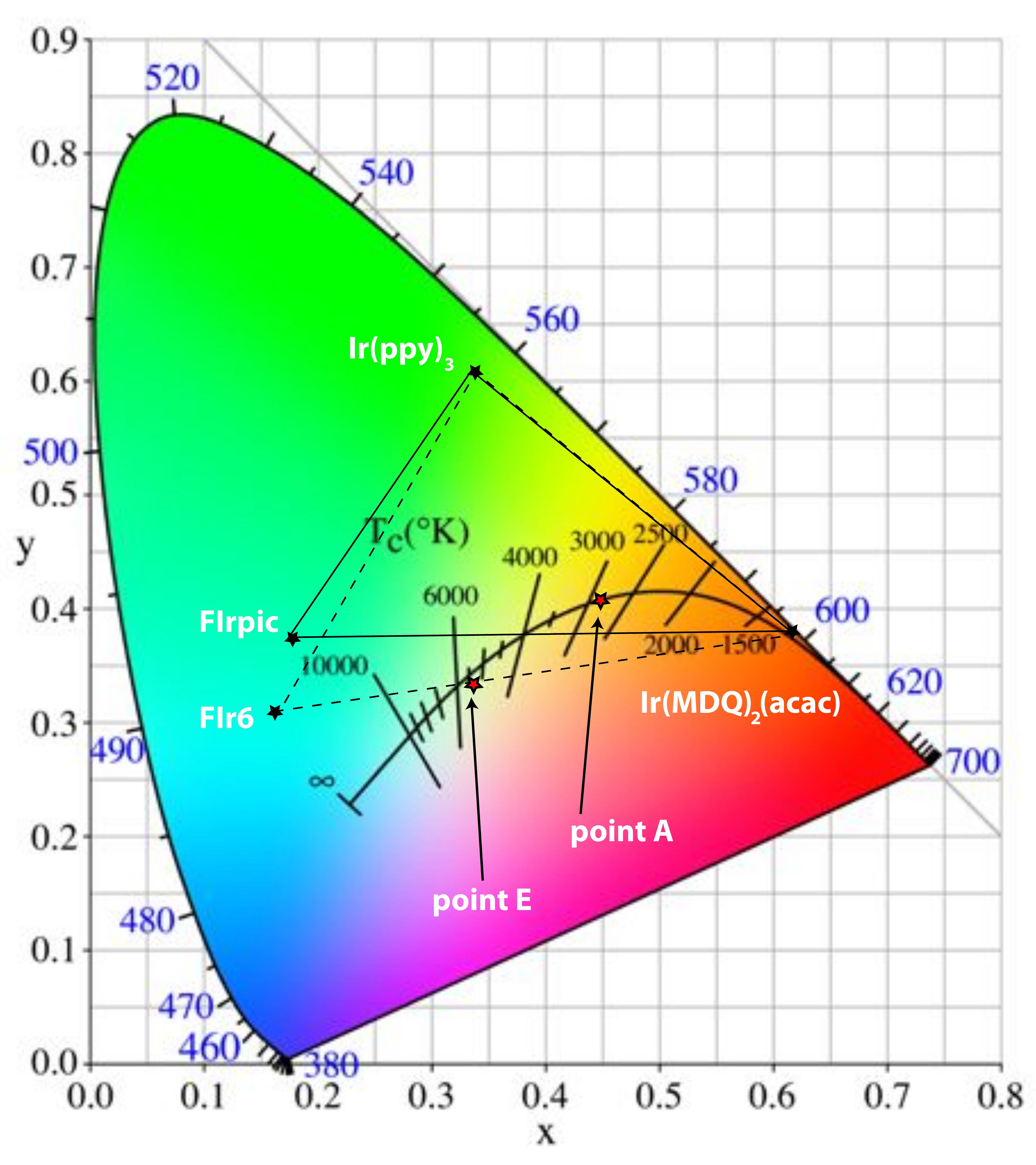}
\caption{\label{_CIE}(color online) CIE 1931 chromaticity diagram. Monochrome colors are located on the edges of this color space (values are in nm). Additive mixing of any monochrome colors leads to a color within the horseshoe. The black line indicates the Planckian locus and the corresponding correlated color temperatures (CCTs). Red stars indicate the important standard illuminants E and A. CIE coordinates are plotted for commonly used phosphorescent emitters together with the  color space they make possible with three-color mixing.  Reprinted under the Creative Commons Attribution-Share Alike 3.0 Unported license.}
\end{figure}

To become a future light source, white OLED require besides high luminous efficacies also a high level of color quality to be widely accepted. Typical reference light sources are \emph{Planckian radiators} which can fully be defined by their \emph{color temperature} $T_\text C$. Thus, their chromaticity changes under a variation of $T_\text C$, as shown in Figure \ref{_CIE}, resulting in the so-called \emph{Planckian locus}. A light source used for illumination should emit a spectrum with a color point close to this locus to be regarded as a true white light source. However, keep in mind that having a color point on the Planckian curve does not necesserily mean that the light source has a good color rendering (see below), which is a consequence of the specific spectral sensitivity of the receptors in the human eye. If a spectrum is off the Planckian locus, its chromaticity can be described by the \emph{correlated color temperature (CCT)} [For more details see for instance: \cite{Ohta2005}].  

Two important \emph{Commission Internationale de l'$\acute{\text E}$clairage (CIE) Standard Illuminants} are the color points E and A, indicated in the CIE 1931 color space in Figure \ref{_CIE} [\cite{Hunt1995}]. The point E -- slightly below the Planckian locus -- is also referred to as point of equal energy, corresponding to CIE coordinates of (0.33, 0.33). It is perceived as ``colorless" white light. On the contrary, the CIE Standard Illuminant A [CIE coordinates of (0.448, 0.408)] -- also called warm white point with $T_\text C=2856\,\text K$ -- marks the chromaticity of tungsten incandescent lamps, which are widely accepted being the most comfortable artificial light sources to date. Many electroluminescence spectra reported in literature are displaced with respect to the Planckian locus. In order to discuss such a distance in this review, we introduce a dimensionless measure $\alpha_{\text{CIE}}$ that describes the shortest distance of the measured CIE coordinates to coordinates ($x_{\text{locus}}$, $y_{\text{locus}}$) on the Planckian locus (this is the orthogonal connection). It reads:

\begin{equation}\label{alphaCIE}
\alpha_{\text{CIE}}=\sqrt{(x-x_{\text{locus}})^{2}+(y-y_{\text{locus}})^{2}}.
\end{equation}
Further we define this value to have a positive sign ('+') when the CIE coordinates are above the Planckian locus and a negative sign ('-') when located below. Thus, in this definition, a light-source with $\alpha_{\text{CIE}}=0$ is a Planckian radiator. This value shall be seen as help for the reader to easily access the quality of a respective white device.

Equally important as its chromaticity is the ability of a light source to reproduce the color of objects. In order to quantify the color rendering properties of artificial light sources, the Commission Internationale de l'$\acute{\text E}$clairage introduced the \emph{color rendering index} (CRI) in 1965 [for the updated version see: \cite{CIE1995}]. It is a dimensionless measure ranging from 0 to 100, calculated as the average of the special color rendering indices $R_i=100-4.6d_i$. These are determined by illuminating eight defined color cards with both a light source of interest and a reference (i.e. either a Planckian radiator or a daylight spectrum for high CCTs above 5000\,K). Here, $d_i$ is the distance between both rendered spectra in the CIE 1964 $U^*V^*W^*$ color space [\cite{Ohta2005}]. It is important to note that the CRI is only defined in the proximity of the Planckian locus. A lower limit for a good light source is a CRI of 80.

\begin{figure}[h]
\includegraphics[width=8.5cm]{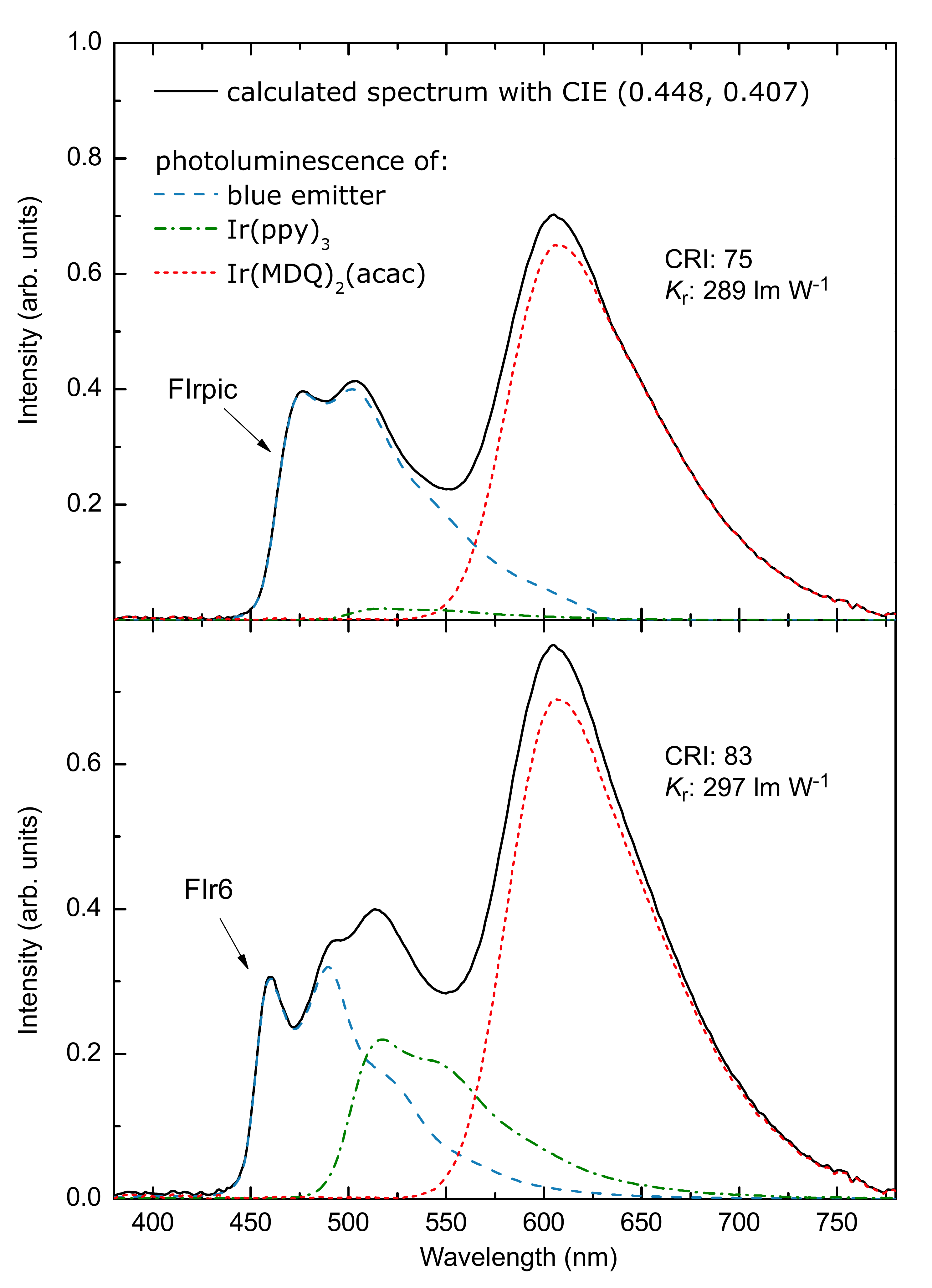}
\caption{\label{_idealwhite}(color online) Under a variation of the blue emitter with different spectral positions (Top: FIrpic, Bottom: FIr6), a spectrum is calculated based on three emitters to realize color coordinates at color point A within the CIE chromaticity diagram (cf. Fig. \ref{_CIE}). Additionally given are the color rendering index (CRI) and the luminous efficacy of radiation $K_\text r$.}
\end{figure}

As an example, the CRI is calculated for two different spectra composed of photoluminescence emission of three phosphorescent emitters [different blue, Ir(ppy)$_3$ for green, and Ir(MDQ)$_2$(acac) for red], realizing emission at color point A (cf. Fig. \ref{_idealwhite}). The first calculated spectrum is based on the light-blue emitter FIrpic (cf. Fig. \ref{_idealwhite}, Top). Here, in order to reach color point A, the relative intensity of the green emitter Ir(ppy)$_3$ is only 1\,\%, revealing a noticeably dip in the spectrum at 550\,nm -- the part of the spectrum with highest eye sensitivity (cf. Fig. \ref{_Vlambda}). The CRI of this simulated spectrum is only 75, the corresponding luminous efficacy of radiation is 289\,lm\,W$^{-1}$. Exchanging FIrpic with the deeper blue emitter FIr6 improves the color quality (cf. Fig. \ref{_idealwhite}, Bottom). Here, the CRI is increased to 83 because more green emission is necessary (17\,\%), resulting in a more balanced spectrum. Furthermore, due to the higher intensity in the green part of the spectrum, $K_\text r$ increases to 297\,lm\,W$^{-1}$.

		\subsubsection{\label{Bright}Device brightness}
		
\begin{figure}[h]
\includegraphics[width=8.5cm]{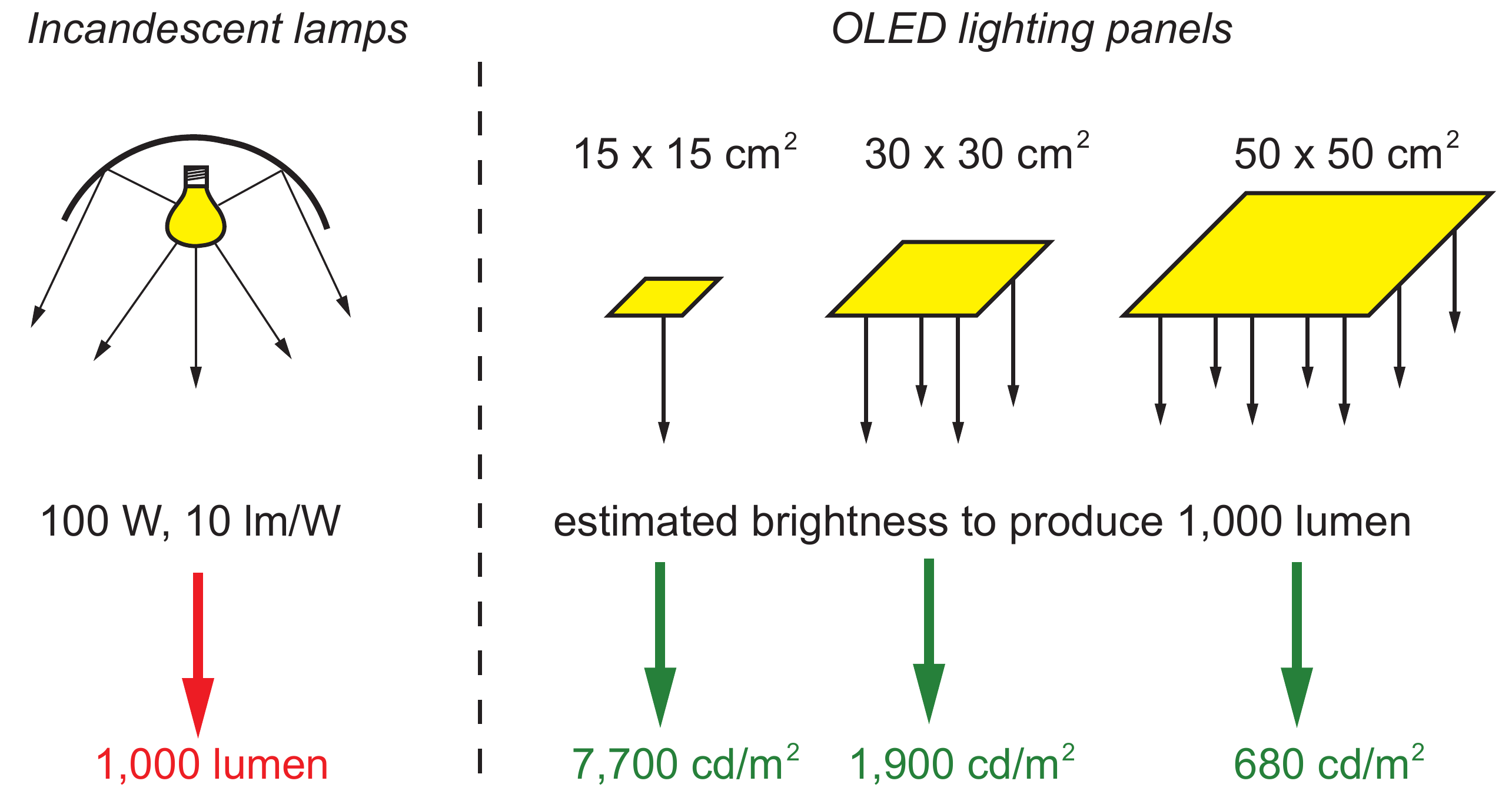}
\caption{\label{_brightness}(color online) Idealized comparison of incandescent lamps and OLED lighting panels to estimate the panel size needed to achieve similar luminous flux outputs (here: 1000 lumen). Calculation for the OLEDs are based on the luminous efficacy $\eta_\text{LE}$ data obtained from \cite{Reineke2009a} (Device LI).}
\end{figure}

As mentioned at the beginning of this section, OLEDs are in contrast to their inorganic counterparts ultra-thin area light sources. Obviously, two device parameters can be adjusted to realize a desired luminous flux: the device area and its operating brightness. Figure \ref{_brightness} compares OLED panel sizes and luminance levels to achieve a luminous flux that matches the output of a 100\,W incandescent bulb. Interestingly, even large OLED areas of $50\times50$\,cm$^2$ need a luminance of 680\,cd m$^{-2}$ -- about a factor of 2-3 brighter than a typical computer display -- to reach the flux of the light bulb. The discussion of Section \ref{Rolloff} shows that high brightness operation of OLEDs is accompanied with a decrease in device efficiency as a consequence of excited state annihilation processes [\cite{Reineke2007, Baldo2000a, Kalinowski2002}]. Furthermore, the device long-term stability is inversely  proportional to its operating brightness [\cite{Meerheim2006, Meerheim2008a, Zhang2001, Tsai2006}]. On the other hand, the production costs of an OLED panel and therefore the costs per lumen will increase roughly linear with the panel area. 

Independent of how future lighting solutions will look like in detail, it is apparent that a certain level of brightness is necessary for general lighting applications. Initially, 1000\,cd\,m$^{-2}$ has been established in literature as a level to ensure best device comparability\footnote{As discussed in Section \ref{Rolloff}, the device efficiency will drastically decrease from low luminance (where typically the maximum luminous efficacy is obtained) to illumination relevant levels.}. 
However, in the last few years 3000\,cd\,m$^{-2}$ has increasingly established as standard level for OLED lighting applications. It can be expected that this level will not be significantly exceeded because higher intensities generate glare, which removes one of the key advantages of area sources such as the OLED.

\section{\label{whitePOLED}White polymer OLEDs }
Within this section, we will discuss various concepts to produce white light emission from polymer materials. It is worth noting, as mentioned in Section \ref{Layers}, that often small molecules are incorporated into devices that are referred to as polymer OLEDs. We will follow this common terminology and also review such hybrid devices here. 

\begin{figure}[h]
\includegraphics[width=8.8cm]{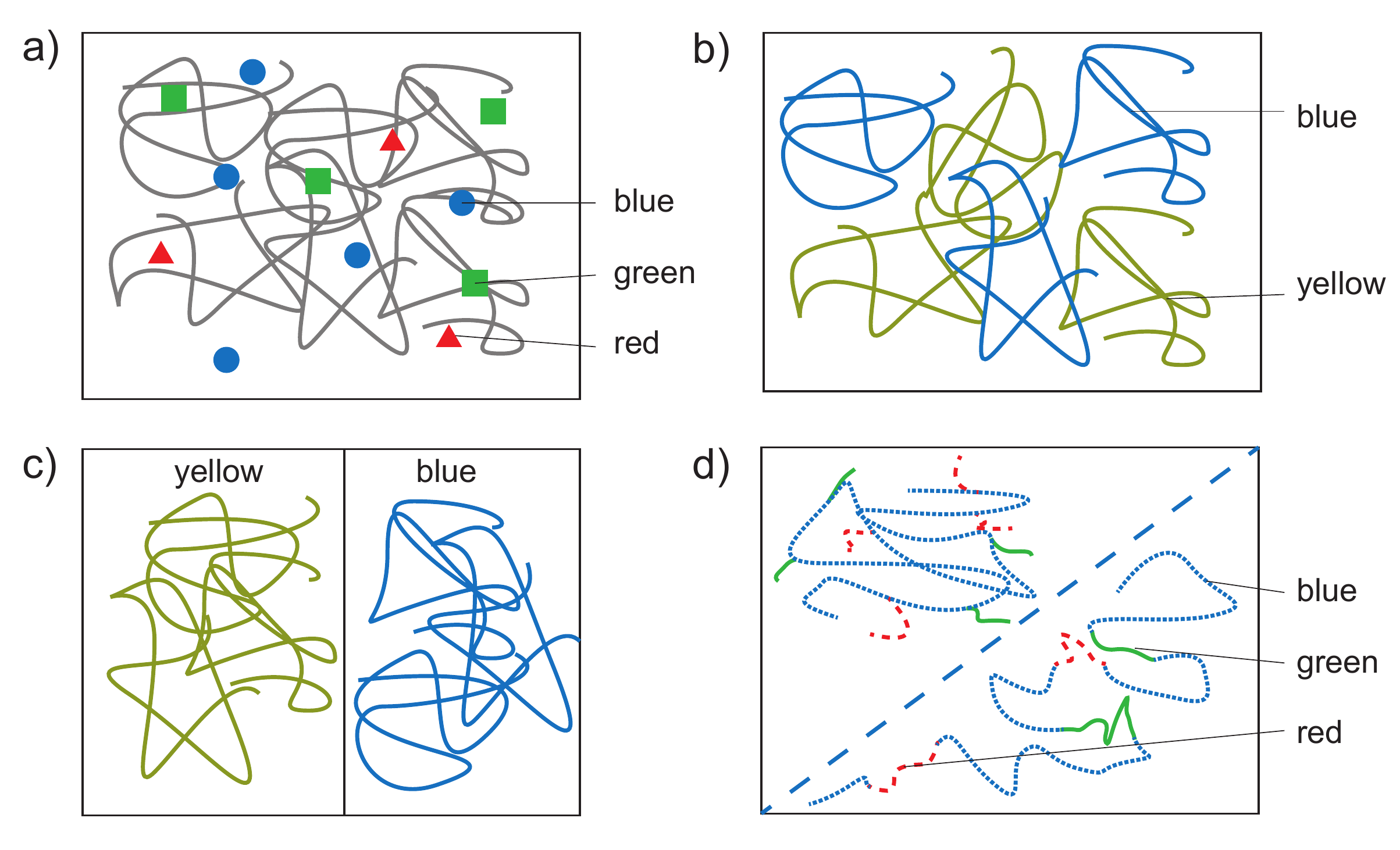}
\caption{\label{_polymer_concepts}(color online) Shown are the main concepts for white light emission from polymer OLEDs. a) polymer host materials (grey) doped with small molecule fluorescent or phosphorescent emitter molecules (filled symbols), b) two or more light-emitting polymers blended in a single layer, c) light-emitting polymers in a heterolayer architecture, and d) single component, multi-color emitting copolymers.}
\end{figure}

Figure \ref{_polymer_concepts} summarizes the key concepts that enable emission of white electroluminescence. First reports on white OLEDs use polymer materials to function solely as host material and in part as charge carrier transport materials, while the emission originates from small molecular dyes dispersed into the polymeric matrix [Fig. \ref{_polymer_concepts} a)]. On the other hand, light-emitting polymers themselves can be combined, each covering a different spectral range, to achieve a broadband emission. Here, white light can be realized either by blending the polymers in one single emission layer [Fig. \ref{_polymer_concepts} b)] or in a heterolayer design [Fig. \ref{_polymer_concepts} c)] . Of course, these concepts can generally be combined in virtually any form. Finally, concepts have been proposed to realize white emission from a single compound polymer  [Fig. \ref{_polymer_concepts} d)]. This is commonly realized by synthesizing multifunctional copolymers.

	\subsection{\label{dye_doped_polymers}Small molecule doped polymer films}

		\subsubsection{\label{dye_fluorescent}Fluorescence emitting dopants}
		
\begin{figure}[h]
\includegraphics[width=8.5cm]{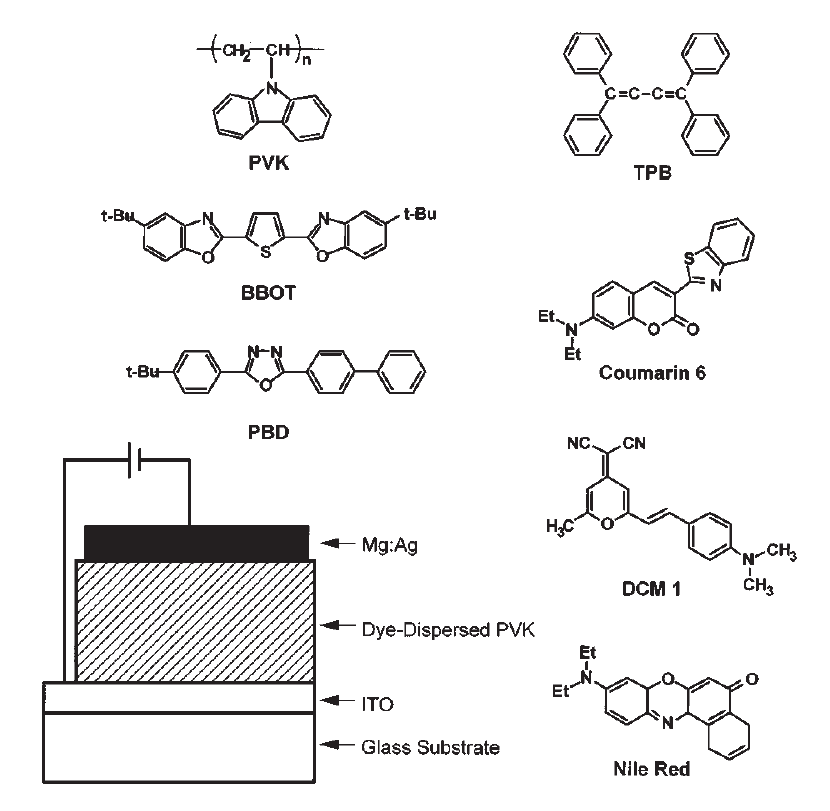}
\caption{\label{_Kido1995a_Fig1}Device structure and materials used in the first polymer-based white OLED reported by \cite{Kido1995a}. BBOT and PBD are electron-transporting materials that are added to the hole transporting PVK matrix to improve carrier balance. From \cite{Kido1995a}.} 
\end{figure}

We would like to start reviewing the concept of small molecule doped polymer systems, as illustrated in Figure \ref{_polymer_concepts} a), because the first reports about white polymer OLEDs by \cite{Kido1995a, Kido1994} are based on this approach. In their early work they used poly(\textit{N}-vinylcarbazole) (PVK) as host material for various fluorescent dyes (cf. Fig. \ref{_Kido1995a_Fig1}).

\begin{figure}[h]
\includegraphics[width=6.5cm]{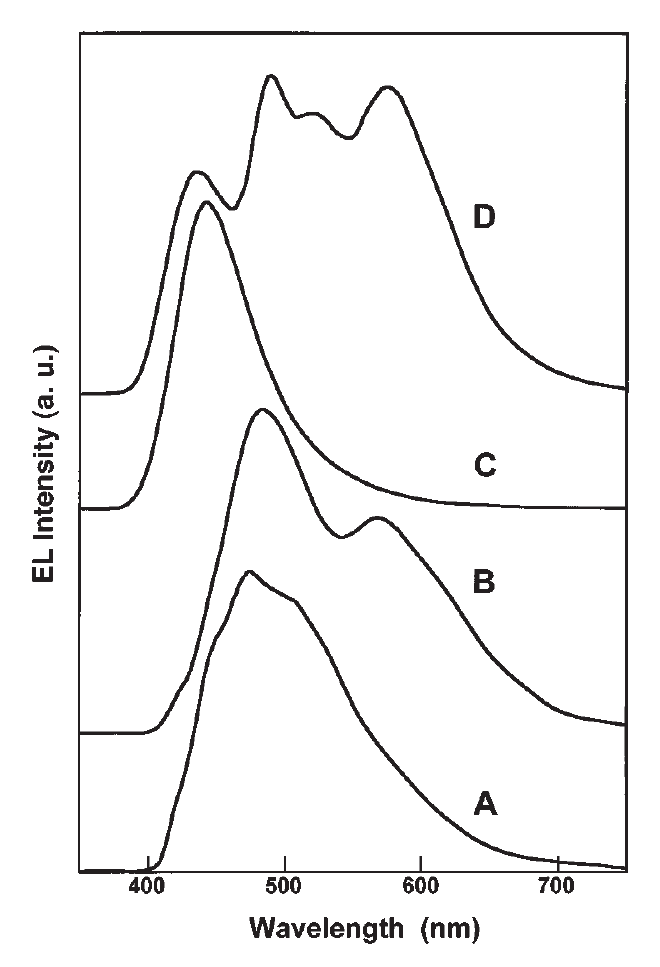}
\caption{\label{_Kido1995a_Fig2}EL of ITO/dye-dispersed PVK (100\,nm)/ Mg:Ag devices. PVK is molecularly dispersed with (spectrum A) 30\,wt\,\% BBOT, (spectrum B) 30\,wt\,\% BBOT, and 0.007 mol\,\% Nile Red, (spectrum C) 30\,wt\,\% PBD and 3 mol\,\% TPB, (spectrum D) 3 mol\,\% TPB, 30\,wt\,\% PBD, 0.04 mol\,\% Coumarin 6, 0.02 mol \,\% DCM 1, and 0.015 mol\,\% Nile Red. From \cite{Kido1995a}.} 
\end{figure}

In Figure \ref{_Kido1995a_Fig2}, the EL spectra of different devices from \cite{Kido1995a} are displayed, already showing one very promising property of organic LEDs, the easy variation of the emitted color in a broad range. Here, device D completely spans the spectral range from 400 to 700\,nm, which covers almost the complete visible spectrum (380-780\,nm). Their device reached a maximum brightness of 4100\,cd m$^{-2}$ (the authors did not report on device efficiency). \cite{Kido1995a} concluded that the emitter dopant excitation must follow competing pathways being resonant energy transfer from host to emitter molecules and direct charge trapping at dopant sites. Furthermore, energy transfer between different species of emitter molecules is generally possible.

\begin{figure}[h]
\includegraphics[width=7.0cm]{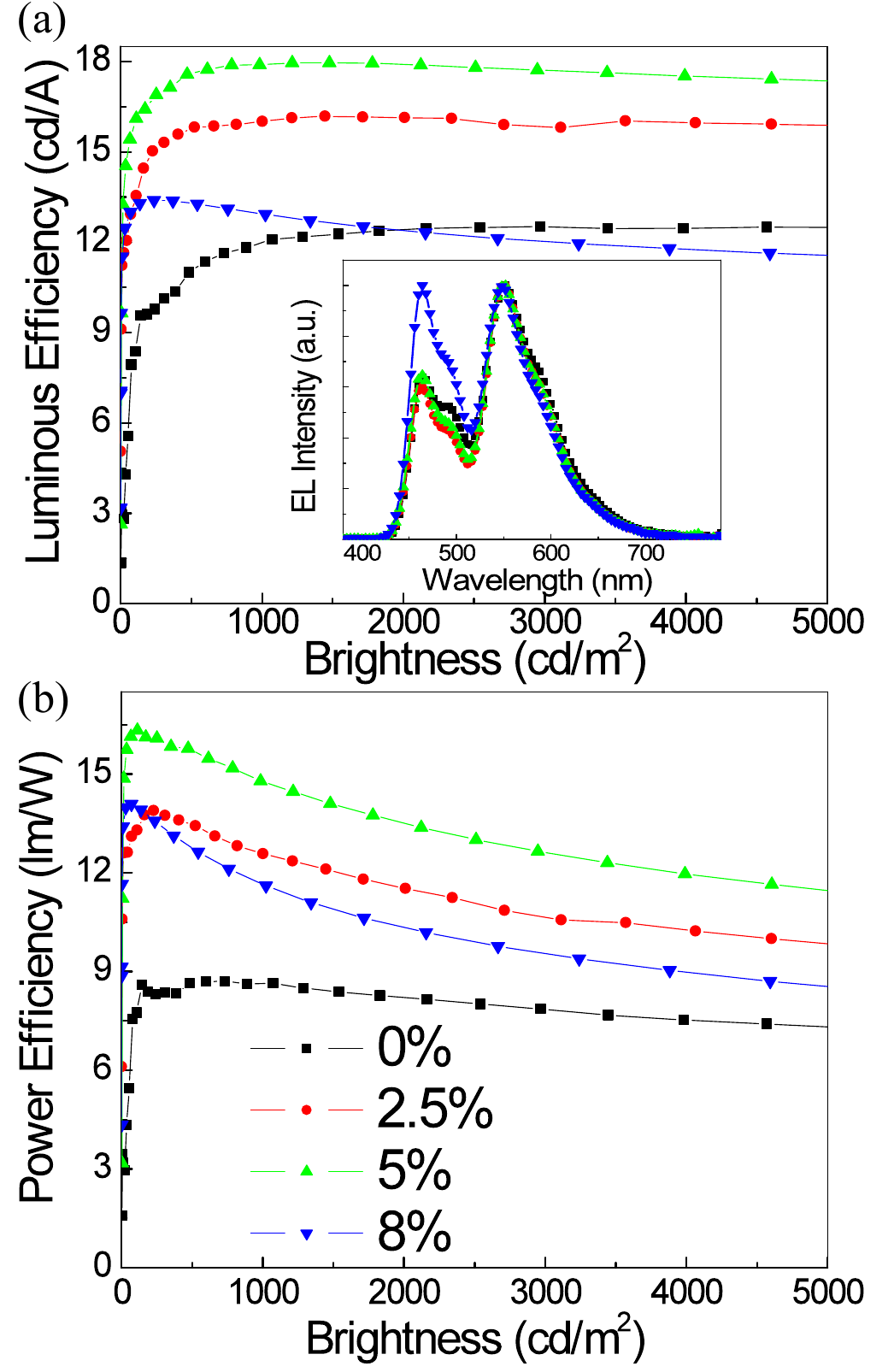}
\caption{\label{_Huang2006b_Fig2}(color online) a) Current efficiency (cd\,A$^{-1}$) of devices comprising different amounts of 2-(4-biphenylyl)-5-(4-tert-butylphenyl)-1,3,4-oxadiazole (PBD). Inset: EL spectra of the corresponding devices. b) Luminous efficacy (lm\,W$^{-1}$). From \cite{Huang2006b}.} 
\end{figure}

Much effort was spent to improve the device efficiency in the following years. \cite{Huang2006b} reported a device structure with improved luminous efficiency based on a polyfluorene (PF) host polymer material. For orange emission they comprised the laser dye rubrene, dispersed into the host material with a low concentration of 0.2\,\%. The reason for this low concentration is twofold: (i) The very high photoluminescence quantum yield (PLQY) of rubrene approaching 100\,\% [\cite{Mattoussi1999}] is only realized at low concentrations. At higher concentrations, strong concentration quenching reduces the emission efficiency. (ii) In their devices, Huang \emph{et al.} made use of an incomplete energy transfer from the host material PF to rubrene, so that PF itself covers the blue part of the spectrum (cf. inset of Fig. \ref{_Huang2006b_Fig2}). Additionally, in order to improve the electron transport within the light-emitting polymer (LEP) film, an electron transporting material PBD was incorporated with various concentration (0-8\,wt\%). Figure \ref{_Huang2006b_Fig2} plots $\eta_{\text {C}}$ and $\eta_{\text {L}}$ of these devices as a function of brightness. At a brightness of 3000\,cd\,m$^{-2}$, the device with 5\,\% PBD content, having the best color quality [CIE color coordinates of (0.33, 0.43)], reaches 12.6\,lm\,W$^{-1}$. This is a very high efficiency considering that only fluorescent materials were used, which only allows internal quantum efficiencies of roughly 25\,\% (cf. Sec. \ref{FluoPhos}). The fluorescent emitters make it possible that the current efficiency of these devices remain constantly high over a wide range of luminance [cf. Fig. \ref{_Huang2006b_Fig2} a)]. It is necessary to mention that the CIE coordinates with $\alpha_{\text{CIE}}=+0.07$ are relatively far apart from the Planckian locus, which artificially enhances the luminous efficacy of radiation $K_\text r$. 

		\subsubsection{\label{dye_phosphorescent}Phosphorescent emitters}
		
In order to reduce the losses in the triplet manifold of the materials (cf. Sec. \ref{FluoPhos}), phosphorescent dopants, which proved to be very successful for high efficiency small molecule based devices [\cite{Baldo1998,DAndrade2004}], were similarily introduced to polymeric systems. 

\cite{Kawamura2002} reported on multiple-doped PVK emission layers to achieve white emission. PVK with its triplet energy level at roughly 2.5\,eV (496\,nm)\footnote{Note that other publications [e.g. \cite{Wu2008}] state a much higher triplet energy of PVK of 3.0\,eV. } is suitable as host material for most of the phosphorescent emitters. By varying the emission wavelength of the phosphor [474, 517, 565, and 623\,nm for FIrpic, Ir(ppy)$_3$, Bt$_2$Ir(acac), and Btp$_2$Ir(acac), respectively], they observe the lowest efficiency for single emitter devices comprising FIrpic, because the triplet energy of FIrpic is higher than the one of PVK resulting in endothermic energy transfer [\cite{Adachi2001a}]. A triple-doped device comprising  FIrpic, Bt$_2$Ir(acac), and Btp$_2$Ir(acac) in a 10:0.25:0.25 mixing ratio yield a maximum EQE of 2.1\,\% and 1.4\,lm\,W$^{-1}$ (CIE (0.33, 0.41); $\alpha_{\text{CIE}}=+0.06$). Achieving a balanced white emission requests the lower wavelength components to be highly diluted into the EML so that the energy transfer from the host material and blue emitter is not complete [\cite{Kawamura2002}].

\cite{Niu2007} comprised three phosphorescent emitters [FIrpic, Ir(ppy), and Os-R1 (an osmium-based organometallic complex)] in a multilayer device, where the HTL (VB-TCTA) is formed by crosslinking and the ETL (TPBi) is prepared by thermal evaporation. With an optimized VB-TCTA thickness of 25\,nm, they reach a maximum EQE of 6.15\,\%. At 800\,cd\,m$^{-2}$, the luminous efficacy is 5.59\,lm\,W$^{-1}$ ($\alpha_{\text{CIE}}<+0.01$). Further efficiency improvements were reported by \cite{Wu2008} and \cite{Huang2009}, who report efficiencies of 12.9\,\% EQE and 8.2\,lm\,W$^{-1}$ (maximum values, $\alpha_{\text{CIE}}=+0.08$) and 12.6\,\% EQE and 18.5\,lm\,W$^{-1}$ (at 100\,cd\,m$^{-2}$, $\alpha_{\text{CIE}}=0$), respectively. In both reports, the electron transporting material OXD-7 has been added to the EML to improve the electron transport [\cite{Hamada1992}]. The very high efficiencies of \cite{Huang2009} are in part a consequence of the improved electron injection and transport which is realized by the incorporation of an n-doped electron transport material. They used Li$_2$Co$_3$ salt to dope the ETL made of poly[9,9-bis(2-(2-(2-diethanolamino ethoxy)ethoxy)ethyl) fluorene] (PF-OH). Compared to undoped PF-OH, a device with 15\,wt\,\% doped ETL shows a 1.58-fold improvement in the luminous efficacy.

Very recently, \cite{Cheng2010a} discussed devices that comprise two phosphorescent emitters, i.e. FIrpic and Ir(SBFP)$_2$(acac) for light-blue and orange, respectively, dispersed into a  silane-based [cf. \cite{Holmes2003}], wide bandgap polymer P36HCTPSi.  By adjusting the concentration of Ir(SBFP)$_2$(acac) to 4\,wt\,\%, these devices, comprising an additional ETL prepared by means of thermal evaporation, reach 14.1\,\% EQE and 25.6\,lm\,W$^{-1}$ at 1000\,cd\,m$^{-2}$. However, the color quality with a CRI of 42 and CIE coordinates far apart the Planckian locus [(0.41, 0.49), Oleg Vladimirovich] call for strategies to improve the emitted color.
		
		\subsubsection{\label{dye_hybrid}Hybrid fluorescent blue, phosphorescent green and red systems}

An alternative approach to realize white light is to utilize blue fluorescence which is complemented by the emission of lower wavelength phosphorescent emitters. Commonly, the polymer host material simultaneously serves both as matrix for the phosphors and blue emitter. In general, this concept can be optimized to enable triplet harvesting as reported by \cite{Schwartz2007} for small molecule OLEDs (details will be given in Sec. \ref{TripHarv}), where singlet excitons will be used for blue fluorescence and the remaining triplets channeled to phosphorescent emitters, where they emit with potentially 100\,\%. However, this concept has strict requirements on the energy levels of the materials and the exciton distribution within the device. To our knowledge, triplet harvesting has not been reported for polymeric white OLEDs to date [cf. e.g. \cite{Gather2011}].

Polyhedral oligomeric silsesquioxane-terminated poly(9,9-dioctylfuorene) (PFO-poss) is used by \cite{Xu2005} as a blue emitting polymer that additionally hosts two phosphorescent emitters for green [Ir(Bu-ppy)$_3$] and red [(Piq)$_2$Ir(acaF)]. A device with 0.14\,wt\,\% for each emitter dopant emits white light at the point of equal energy [(0.33, 0.33)] with a maximum luminous efficacy of 5.5\,lm\,W$^{-1}$ at 5.6\,V (no EQE reported). Based on another blue emitting polymer (BlueJ) and two phosphors, \cite{Kim2006} achieved 3.2\,\% EQE at a brightness of 905\,cd\,m$^{-2}$ (12.5\,cd\,A$^{-1}$, no $\eta_\text L$) with emission at (0.33, 0.33). In contrast to \cite{Xu2005}, they added 25\,\% of PVK to the EML to improve the charge carrier balance.

\cite{Gong2004,Gong2005} used a PFO-based polymer as blue emitter and host material for the red phosphorescence emitting Ir(HFP)$_3$. In their later report [\cite{Gong2005}], they comprised this EML into a multilayer OLED architecture to improve the device efficiency. The optimized device has a luminous efficacy of 3\,lm\,W$^{-1}$ at approximately 2400\,cd\,m$^{-2}$ with CIE coordinates of (0.33, 0.33). The moderate contribution of the red emitter Ir(HFP)$_3$ to the overall emission spectrum suggests that either only a limited number of excitons reach the molecule or that additional quenching sites in the complex structure are present, suppressing the red emission. 

\cite{Niu2006} followed another concept. Mainly to improve the CRI of the device, they combine a single component white-emitting polymer (WPF03) [\cite{Tu2006}] (cf. Sec. \ref{single_component_white}) with a red-emitting phosphorescent molecule [(Ppq)$_2$Ir(acac)]. By further making use of weak emission from an admixture of the electron-transporting material PBD [cf. \cite{Kido1995a}], the authors realize a broadband emission from 400 to 750\,nm with a very high CRI of 92 [CIE coordinates (0.34, 0.35), $\alpha_{\text{CIE}}<-0.01$]. However, the overall device efficiency of 5.3\,cd\,A$^{-1}$ (no $\eta_\text{EQE}$ and $\eta_\text{L}$) is comparably low. This is most likely due to small energetic displacement of the emission peaks of the red fluorescent chromophore in WPF03 and (Ppq)$_2$Ir(acac) of only 50\,nm or 188\,meV. Thus, the triplet level of the red chromophore is expected to be below the one of (Ppq)$_2$Ir(acac) inducing noticeable emission quenching.

	\subsection{\label{light_emitting_polymers}White emission from multiple light-emitting polymers}

		\subsubsection{\label{blends_of_polymers}Blended polymeric systems}

Although they did not mention the application for white light sources, \cite{Berggren1994} reported on color tunable LEDs made from polymer blends [cf. Fig. Figure \ref{_polymer_concepts} b)]. By altering the operating voltage and/or the stoichiometry of the polymer blends, they were able to ``shift the emission from blue to near-infrared, with green, orange and red as intermediate steps (...)." Their devices had efficiencies ranging from 0.1-1\,\% EQE.

\begin{figure}[h]
\includegraphics[width=6.5cm]{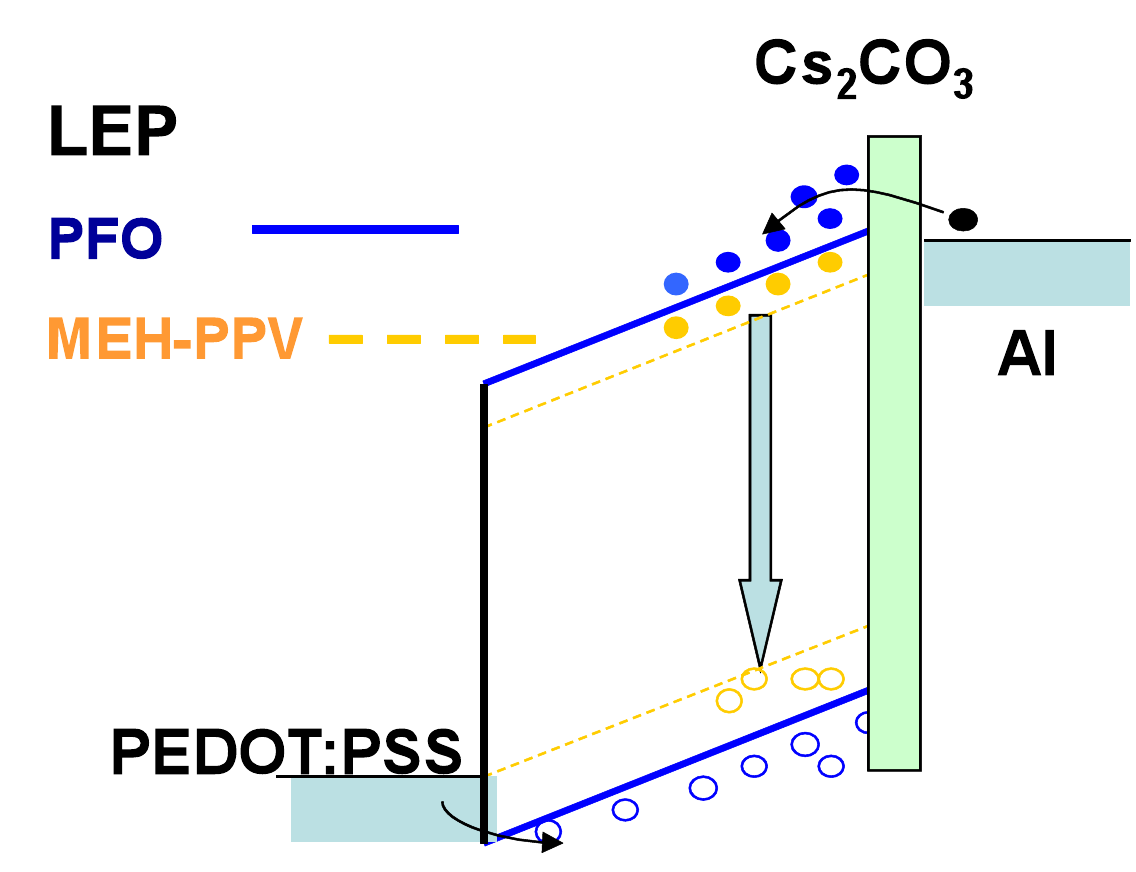}
\caption{\label{_Huang2006a_Fig1}(color online) Band diagram illustrating the working principle of the polymer blend white OLED. The Cs$_2$CO$_3$ interfacial layer (EIL) improves the electron injection. Adapted from \cite{Huang2006a}.} 
\end{figure}

The first white polymer OLEDs based on polymer blends have been discussed by \cite{Tasch1997}. By highly diluting a red emitting polymer poly(perylene-co-diethynylbenzene) into a blue laddertype polymer polyparaphenylene with a concentration of 0.05\,\%, white emission is realized. Here, the red emitter is excited via exciton energy transfer and charge transfer/trapping. With an addition of 10\,wt\;\% PMMA to the mixed layer, CIE coordinates of (0.31, 0.33) [$\alpha_{\text{CIE}}=+0.01$] were reached with a maximum external quantum efficiency of 1.2\,\%. The concept of incorporation of insulating materials such as PMMA into the polymer blend to control the intermolecular energy transfer and emission spectrum has been further discussed by \cite{Granstrom1996}.

\cite{Hu2002} realized white light emission by blending two copolymers (for blue and green) together with an additional small molecular dye (MPD) for red emission. An optimized device with admixtures of charge transport moieties reached a maximum photon per electron efficiency of 2.6\,\% with CIE coordinates of (0.36, 0.35) [$\alpha_{\text{CIE}}<+0.01$].

\begin{figure}[h]
\includegraphics[width=7.0cm]{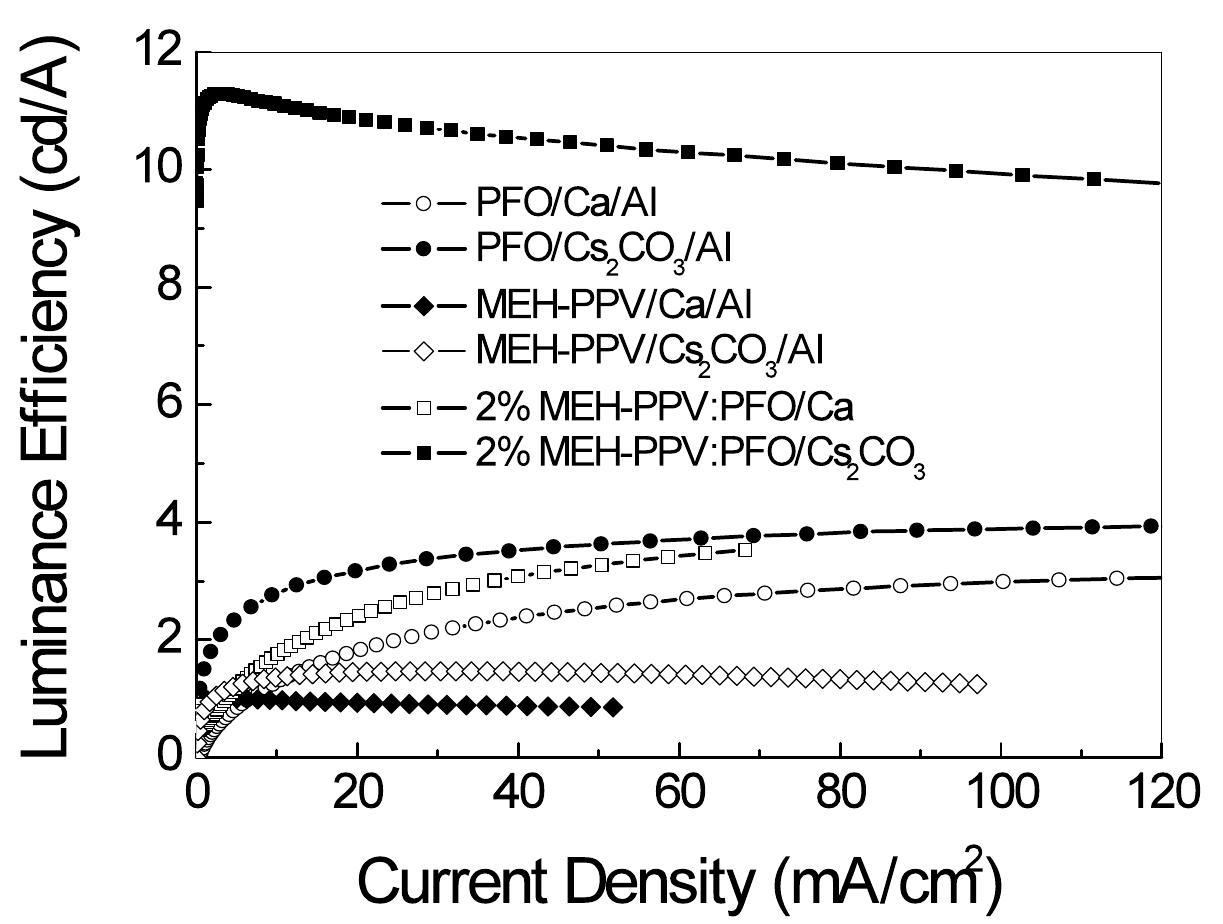}
\caption{\label{_Huang2006a_Fig5} Current efficiency of three sets of devices from \cite{Huang2006a}. The composition of the EML and the interfacial electron injection layer are varied.\footnote{To our best knowledge, the labels of the last two devices are incomplete. They should both contain an '/Al' cathode.} From \cite{Huang2006a}.} 
\end{figure}

\cite{Huang2006a} reported on a simple two polymer blend white device with a greatly improved device efficiency. The working principle is illustrated in Figure \ref{_Huang2006a_Fig1}. They introduced an Cs$_2$CO$_3$ interfacial layer between LEP layer and cathode to enhance the injection of the minority charge carriers. Furthermore, the dopants (MEH-PPV) energy levels are within the bandgap of the host material (PFO), so that the excitation of the dopant can occur via energy transfer and charge trapping [cf. \cite{Tasch1997}], the latter leading to a charge confinement effect. The combination of improved carrier injection and charge confinement yields very high efficiencies of 6\,\% EQE and 16\,lm\,W$^{-1}$ [peak values, CIE (0.36, 0.40) $\alpha_{\text{CIE}}=+0.03$]. Even at 1000\,cd\,m$^{-2}$, the luminous efficacy remains at a high value of 12.6\,lm\,W$^{-1}$. Figure \ref{_Huang2006a_Fig5} illustrates the improvement in device efficiency compared to reference devices (three sets of devices). Open versus filled symbols compare the interfacial injection layer (Cs$_2$CO$_3$ versus Ca reference). The different device sets compare the EML comprising either each LEP or the mixture of 2\,\% MEH-PPV in PFO. There, the device with Cs$_2$CO$_3$ and charge confinement structure shows a two- to three-fold improvement in efficiency compared to the other devices.

		\subsubsection{\label{heterosystems_of_polymers}White light from polymer heterolayers}

Another concept apart from blending emitting polymers in a single layer is to create a heterointerface between two differently emitting polymers [cf. Fig. \ref{_polymer_concepts} c)]. Here, it is necessary to engineer the recombination zone -- typically only a few nanometers wide -- to be close to the hetero-interface in order to realize emission from both material species. In contrast to the concept of polymer blending, the preparation of multilayer polymer devices is more complicated because the solvents used in wet processing techniques may harm the underlying layers (cf. Sec. \ref{Layers}).

In the early report of \cite{Chao1998}, white light is created at the interface of a PVK/C12O-PPP interface. However, instead of utilizing the emission of both polymers, the spectrum is composed of the blue fluorescence of C12O-PPP and an exciplex emission formed between the PVK HOMO and the C12O-PPP LUMO level. Further work on this topic is published by \cite{Thompson2001}, where different heterointerfaces were investigated, all showing a broad emission that is a mixture of blue exciton and longer wavelength exciplex emission. However, it is still questionable whether exciplex emission can be utilized for efficient luminescence. For instance, \cite{Castellani2007} reported about the competition of exciton and exciplex emission in multilayered organic LEDs. They conclude that the emission efficiency is noticeably reduced if an exciplex emission is incorporated.

One way to overcome the problems in preparation of multilayer polymer systems is the technique of crosslinking [cf. e.g. \cite{Niu2007}], where the polymerization of the material is realized after layer deposition either by annealing or photo-chemical means. \cite{Kohnen2010} reported on a fully wet processed bilayer polymer system consisting of two fluorescent emitters with complementary emission colors (a PPV-derivative for yellow and a polyfluorene for blue). In their work, the yellow emitting PPV-derivative (SY) has been thermally crosslinked. An optimized device emits white light with CIE coordinates (0.323, 0.345) very close to color point E. The maximum efficiency of the device is 6.1\,cd\,A$^{-1}$ ($\alpha_{\text{CIE}}<+0.01$, no $\eta_\text{EQE}$ and $\eta_\text{L}$ stated). At 2400\,cd\,m$^{-2}$, the device efficiency is still as high as $\sim5.6$\,cd\,A$^{-1}$. One important advantage of this device design is the low color shift as a function of current density [from 100 to 10000\,cd\,m$^{-2}$, the CIE coordinates shift ($\Delta x=0.009$, $\Delta y=0.006$)]. Furthermore, even though the spectrum only consists of contributions from two emitters, the $\text{CRI}=84$ is very high.

	\subsection{\label{single_component_white}Single component polymer systems}

In Figure \ref{_polymer_concepts} d), an alternative but very promising concept for white OLEDs based on polymers is illustrated. The key idea is to realize a single copolymer that contains all different emitting chromophores needed to cover the visible spectrum. Clearly, the advantages of this approach are the simple fabrication, the isotropic, yet statistical distribution of the chromophores within the film [\cite{Gather2007}], the control of the interspecies energy transfer by the molecular design, and the low probability of phase separation [\cite{Berggren1994,Gather2011,Forrest2004}] within the film.

With respect to the molecular design, one has to distinguish between two concepts [cf. Fig. \ref{_polymer_concepts} d)]: (i) the main polymer (host) and all chromophores form the copolymer main chain in a stoichiometric manner, where conjugation is given. (ii) the chromophores are attached to the polymer main chain as sidegroups, where the conjugation is lost. In the latter approach, the chromophores can be seen as isolated molecules dispersed in a host polymer.

		\subsubsection{\label{main_chain_copolymers}Conjugated copolymers comprising main chain chromophores}

\begin{figure}[t]
\includegraphics[width=8.5cm]{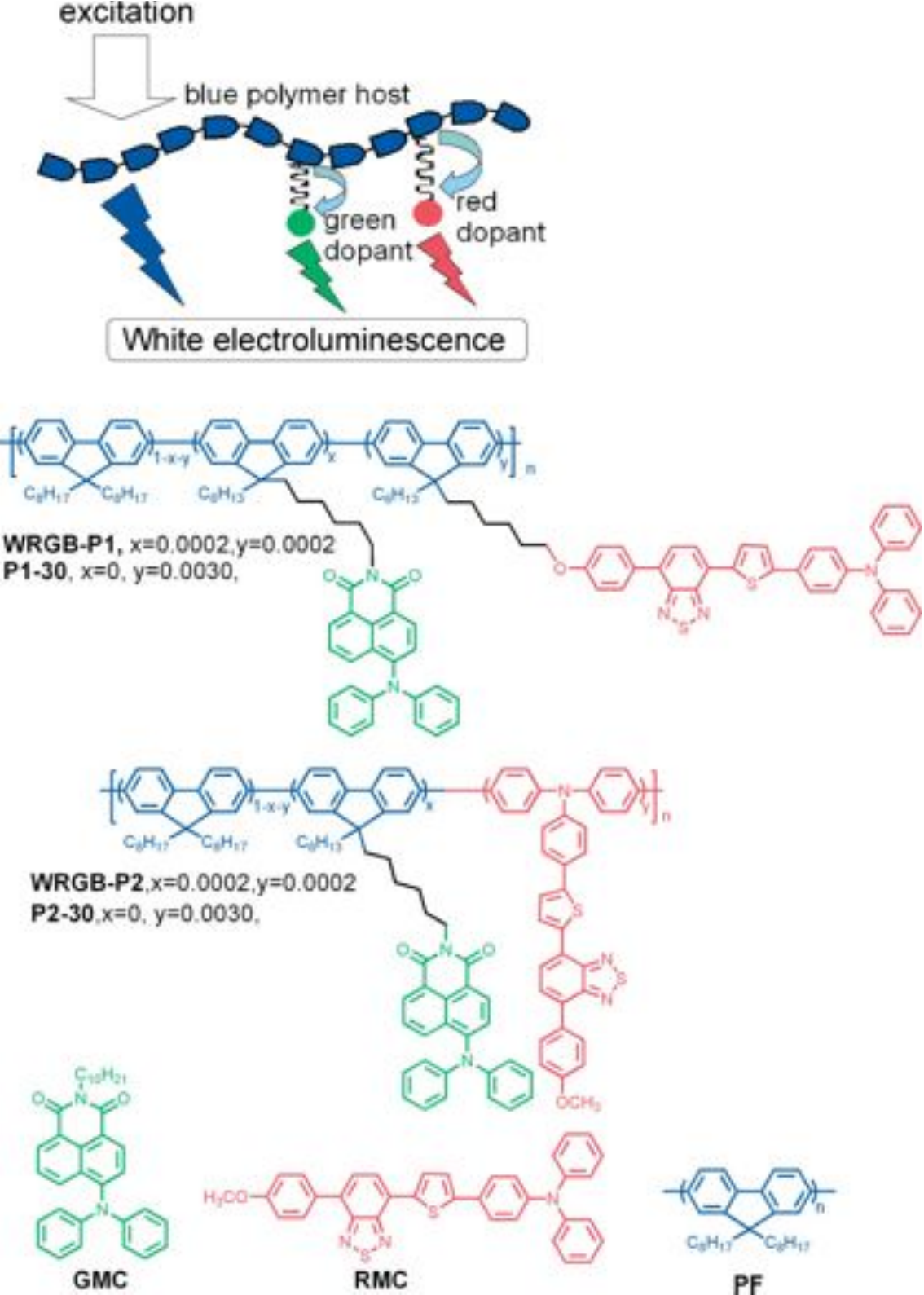}
\caption{\label{_Liu2007_Fig1}(color online) Top: Diagram showing the working principle of a blue emitting polymer backbone with green and red emitting side chain chromophores. Bottom: Molecular structures incorporated to realize this concept. The blue emitting main chain is a poly(fluorene-co-benzene) (PF), the green model compound (GMC) is DPAN, and the red model compound (RMC) is MB-BT-ThTPA. From \cite{Liu2007}.} 
\end{figure}
	
\begin{table*}[t]
\caption{\label{tab:summary_polymers}Summary of selected, high performance devices based on different device concepts as discussed in Section \ref{whitePOLED}. Device efficiencies are maximum values, additional values at higher brightness may be given in parenthesis.} 
\begin{ruledtabular} 
\begin{tabular}{lccccc}
Concept & $\eta_\text{EQE}$\,[\%] & $\eta_\text{C}$\,[cd\,A$^{-1}$] & $\eta_\text{L}$\,[lm\,W$^{-1}$] & CIE (x, y); $\alpha_{\text{CIE}}$ &   Reference\\
\hline
\textbf{Small molecule doped polymers}&&&&&\\ 
\multicolumn{1}{r}{fluorescent emitters}&--&17.9 [17.7]\footnotemark[1]&16.3 [12.6]\footnotemark[1]&(0.33, 0.43); $+0.07$&\cite{Huang2006b}\\ 
\multicolumn{1}{r}{phosphorescent emitters}&14.2 [12.6]\footnotemark[2]&--&23.4 [18.5]\footnotemark[2]&(0.38, 0.38); $<+0.01$&\cite{Huang2009}\\ 
\multicolumn{1}{r}{fluorescent blue/phosphorescent}&--&[10.4]\footnotemark[3]&[3]\footnotemark[3]&(0.33, 0.33); $<-0.01$&\cite{Gong2005}\\ 
\textbf{Multiple light-emitting polymers}&&&&&\\ 
\multicolumn{1}{r}{polymer blends}&6&11.2&16 [12.6]\footnotemark[4]&(0.36, 0.40); $+0.03$&\cite{Huang2006a}\\ 
\multicolumn{1}{r}{polymer heterolayers}&--&6.1 [5.6]\footnotemark[3]&--&(0.323, 0.345); $<+0.01$&\cite{Kohnen2010}\\ 
\textbf{Single component copolymers}&&&&&\\ 
\multicolumn{1}{r}{fluorescent main chain chromophores}&[3.84]\footnotemark[5]&[6.20]\footnotemark[5]&--&(0.35, 0.34); $-0.01$&\cite{Luo2007}\\ 
\multicolumn{1}{r}{fluorescent side chain chromophores}&6.7 [6.2]\footnotemark[6]&15.4 [14.2]\footnotemark[6]&11.4 [10.4]\footnotemark[6]&(0.37, 0.42); $+0.04$&\cite{Zhang2010}\\ 
\multicolumn{1}{r}{phosphorescent side chain chromophores}&--&5.6&--&(0.44, 0.38); $-0.02$&\cite{Jiang2006}\\ 
\end{tabular} 
\end{ruledtabular} 
\footnotetext[1]{at 3000\,cd\,m$^{-2}$} 
\footnotetext[2]{at 100\,cd\,m$^{-2}$}
\footnotetext[3]{at 2400\,cd\,m$^{-2}$}
\footnotetext[4]{at 1000\,cd\,m$^{-2}$}
\footnotetext[5]{at 654\,cd\,m$^{-2}$}
\footnotetext[6]{at 500\,cd\,m$^{-2}$}
\end{table*}	
		
\cite{Tu2004} reported on an efficient white light-emitting polymer by admixing moieties of an orange fluorophore (1,8-naphthalimide) to the blue PFO main polymer. Used as single EML device, a chromphore concentration of 0.05\,\% in the PFO main chain yields a device efficiency of 5.3\,cd\,A$^{-1}$ and 2.8\,lm\,W$^{-1}$ at 6\,V [CIE (0.26, 0.36), $\alpha_{\text{CIE}}=+0.06$]. Later, they showed that the device efficiency could be easily altered by changing the molecular integration of the red chromophore in the polyfluorene backbone [\cite{Tu2006}]. By exchanging the orange chromophore to TPABT, the group around Fosong Wang could improve the device efficiency to 8.99\,cd\,A$^{-1}$, 5.75\,lm\,W$^{-1}$, and 3.8\,\% EQE even with improved color quality [CIE (0.35, 0.34), $\alpha_{\text{CIE}}=-0.01$] [\cite{Liu2006}]. This improvement can be attributed to a higher PLQY of TPABT (76\,\%) compared to 1,8-naphthalimide (25\,\%), as measured in a model compound configuration.

\cite{Lee2005} were the first, reporting on a main chain copolymer containing emitting units for the three basic colors blue (PDHF), green (DTPA), and red (TPDCM). The overall content of green and red chromophores makes up less than 3\,\% in total. Despite the broad spectrum realized with CIE coordinates of (0.34, 0.35), the device efficiency was very low with a maximum current efficiency of 0.04\,cd\,A$^{-1}$ ($\eta_\text{EQE}\approx0.025\,\%$). Improvements of this concept (based on different chromophores) were published by \cite{Chuang2007} and \cite{Luo2007}. Similar to \cite{Liu2006}, both groups used highly efficient fluorescent benzothiadiazole-derivatives for green and red chromophores. \cite{Chuang2007} reached maximum efficiencies of $\eta_\text{EQE}=2.22\,\%$ with CIE coordinates of (0.37, 0.36) [$\alpha_{\text{CIE}}<+0.01$].	 \cite{Luo2007} even reached a maximum EQE of 3.84\,\%, corresponding to a current efficiency of 6.20\,cd\,A$^{-1}$ [CIE (0.35, 0.34), $\alpha_{\text{CIE}}=-0.01$]. 

		\subsubsection{\label{side_chain_copolymers}Copolymers with side chain chromophores}

Instead of attaching the chromophores directly to the backbone of the copolymer, the researchers around Fonsong Wang also studied the concept of attaching the emitting units to the main chain via alkyl chains [\cite{Liu2005,Liu2007,Liu2007a}]. 

In their first report on this concept [\cite{Liu2005}], they include a benzothiadiazole-derivative (TPATBT, $\eta_\text{PL}=0.37$ in PMMA\footnote{`PL' = photoluminescence}) in the polyfluorene main chain for red emission and additionally a naphthalimide-derivative (DPAN, $\eta_\text{PL}=0.91$ in PMMA) as a pendant chain. This configuration reached maximum values of 0.83\,lm\,W$^{-1}$ and 1.59\,cd\,A$^{-1}$ [CIE (0.31, 0.34), $\alpha_{\text{CIE}}=+0.01$]. Using a more efficient red chromophore (MB-BT-ThTPA, $\eta_\text{PL}=0.51$ in PMMA), the authors compare the influence of the position of the red emitter in the copolymer, i.e. either in the main chain or as a side chain attached by an alkyl bridge [\cite{Liu2007}]. This concept is illustrated in Figure \ref{_Liu2007_Fig1}. By repositioning the MB-BT-ThTPA from the main to the side chain, the device efficiency is more than doubled [from 1.99\,lm\,W$^{-1}$ and 3.80\,cd\,A$^{-1}$ to 4.17\,lm\,W$^{-1}$ and 7.30\,cd\,A$^{-1}$ (both with $\alpha_{\text{CIE}}=0$ close to color point E)]. This improvement is attributed to the more effective molecular design forming an intramolecular dopant/host system without affecting the electronic properties of the host material (polymer backbone -- polyfluorene). Similar to these findings, the authors reported on an improvement (factor $1.5-1.8$) in device efficiency for a two color single component copolymer, when the orange chromophore is attached as a side chain rather than incorporated into the polymer backbone [\cite{Liu2007a}]. 

Very recently, \cite{Zhang2010} reported on a highly efficient single component polymer system containing three chromophores that are covalently attached to the polymer backbone. With a correlated color temperature of approximately 4500\,K [CIE coordinates (0.37, 0.42), $\alpha_{\text{CIE}}=+0.04$], the best device reaches 6.2\,\% EQE and a luminous efficacy of 10.4\,lm\,W$^{-1}$ measured at 500\,cd\,m$^{-2}$.

The color shift as a function of operating voltage is a widely observed phenomena, especially for single component copolymer systems. \cite{Gather2007} focussed on the understanding of its origin. In their study, they investigated a statistical copolymer comprising blue, green, and red chromophores embedded in a spiro-polyfluorene polymer. Their data clearly shows that saturation of the red emitter is not the origin of the color shift. They show that the trapping rate of electrons depend on the electric field within the EML. Therefore, the shift of color is related to the applied field rather than to the current flowing through the device [\cite{Gather2007}]. Because the red chromophore inherently has a low band gap and it is only present in very low amounts below the percolation limit, where it can act as a trap, this concept seems to suffer from this general effect.
	
	\subsubsection{\label{side_chain_phosphorescence}Copolymers with phosphorescent emitters in side chain position}

All the concepts from above were solely based on fluorescence emitting materials. However, similar to the general consideration that phosphorescence should enhance the device efficiency, the incorporation of phosphors into a single component copolymer seems promising. \cite{Jiang2006} discussed an approach for a hybrid fluorescent/phosphorescent copolymer. Based on a polyfluorene backbone, they added a benzothiadiazole chromophore for green emission to the polymer backbone and attached a phosphorescent emitter (2-phenylquinoline iridium complex) via an alkyl bridge. In the emission spectra, the benzothiadiazole peaks at 520\,nm, the iridium complex at 580\,nm, resulting in a energetic difference of roughly 250\,meV. This possibly explains the relatively low device efficiencies of 5.6\,cd\,A$^{-1}$, despite the fact that the emission is close to warm white color point A, with CIE coordinates of (0.44, 0.38, $\alpha_{\text{CIE}}=-0.02$). Because the singlet-triplet splitting (cf. Sec. \ref{dye_hybrid} and Sec. \ref{TripHarv}) of the benzothiadiazole green chromophore is expected to be larger than this energy difference of green and red peak energies, the triplet energy of the green unit will act as efficient quenching site for the red phosphorescence.

	\subsection{\label{summary_polymers}Summary}

Table \ref{tab:summary_polymers} summarizes efficiency and color coordinates of devices with highest efficiency for all concepts discussed in Section \ref{whitePOLED}. Interestingly, the focus in the field of polymer OLEDs seems to be on emission close to color point E, i.e. white light sources with a high color temperature in the range of 6000\,K [exceptions are the reports by \cite{Huang2009}, \cite{Jiang2006}, and \cite{Zhang2010} with (correlated) color temperatures ranging between 3000 and 5000\,K]. This is in contrast to the following discussion, where it is shown that emission close to the warm white point A ($T_\text C=2856$\,K) is desired in the field of small molecular based white OLEDs. An obvious reason is the potentially higher luminous efficacy that can be reached at Standard Illuminant A. From a scientific point of view it is interesting whether the concepts presented above are favorable for equally intense emission bands from the emitters or not\footnote{To reach color point E, the intensities of the emission bands are similar for all incorporated emitters while the shape of a multi-emitter spectrum at color point A more looks like a staircase (cf. Fig. \ref{_idealwhite}).}. This question arises as sometimes the maximum quantum efficiency of a multi-emitter system is only obtained for a specific intensity ratio of the different emitters [cf. e.g. \cite{Schwartz2009, Rosenow2010}].

Unfortunately, the reports about device parameters are often sparse (cf. Tab. \ref{tab:summary_polymers}), making it hard to compare the different concepts. Still, to date the concepts based on small molecules used as emitters in a polymer matrix seem to be superior to fully polymeric approaches. For instance, the devices reported by \cite{Huang2009} comprising phosphorescent emitters already reach very high external quantum efficiencies approaching the ``rule-of-thumb" limit of 20\,\% EQE (cf. Sec. \ref{FluoPhos}) for flat devices without any outcoupling improvement techniques. However, we believe that its comparably high efficiency is mainly due to the fact that it is easier to achieve, because researchers can easily make use of a great variety of high efficiency small molecular weight emitter molecules (cf. Sec. \ref{whiteSMOLED}). One key disadvantage of mixing different emitters in a polymer or even polymer blend is the poor control over the actual morphology on the nanoscale, which is of key importance for color control and high efficiency. Especially device optimization and development are often complex and unpredictable, because each component as well as the actual processing conditions affect the overall composition of these multicomponent systems.

Despite their currently still poor performance, the concepts solely comprising polymers (cf. Sec. \ref{light_emitting_polymers} and Sec. \ref{single_component_white}) should be favorable because they inherently fit better to the wet-processing techniques promising ease of fabrication. From the technological point of view, the greatest promise lies in concepts based on single polymer approaches as discussed in Section \ref{single_component_white}. Here the hope is that by sophisticated engineering, color control and charge transport can be met with a single polymer, ultimately providing an easy-to-process, low-cost solution. A key advantage is the promise of easily maintaining color control making use of the stoichiometric composition of the individual chromophores. The realization of such a single component polymer, however, will need much more future work. Here, phosphorescent emitters need to be incorporated into the system to realize maximum efficiency. This in turn calls for careful design of the overall copolymer, because it should be engineered to be free of quenching centers for the chromophores incorporated. By a careful design, single copolymers are likely be able to mimic the triplet harvesting concepts shown with small molecules (cf. Sec. \ref{TripHarv}), comprising a blue fluorescent chromophore and together with phosphorescent green and red chromophores with potentially 100\,\% internal quantum efficiency. In addition, single component polymer systems may be very effective in suppressing effects like phase separation [cf. e.g. \cite{Berggren1994}].

The key difference of small molecule doped polymers (cf. Sec. \ref{dye_doped_polymers}) and single component polymer systems (cf. Sec. \ref{single_component_white}) from a device point of view is the transition from engineering the color through adjusting during fabrication to defining the ultimately emitted color during the polymer synthesis. The latter is more systematic and desirable.

As will be obvious from the data for white small-molecule OLED, the field of white polymer OLED has fallen considerably behind. All first commercial applications of OLED are based on small molecule devices, which has stimulated the research on these materials and devices. It remains to be seen whether white OLED based on polymers can close this performance gap and profit from their advantages, the simpler structure and the possibility to deposit by efficient wet coating technologies.




\section{\label{whiteSMOLED}White OLEDs based on small molecules}

In contrast to solution processing, thermal evaporation allows a much higher degree of layer complexity, composition control, and thickness accuracy.
In many cases, the complete device consists of more than 10 subsequently evaporated thin films which are designated to meet a specific function within the device (cf. Fig. \ref{_Functional_layers}). On the other hand, the preparation by thermal evaporation also allows sub-nanometer control of the deposited layers, opening more design freedom, which enables better device engineering and optimization.

This section contains three major parts that cover fully fluorescent devices (Sec. \ref{smFluor}), hybrid structures with fluorescent blue emitters where the residual spectral range is complemented by phosphorescent emitters (Sec. \ref{smHybrid}), and finally fully phosphorescent devices (Sec. \ref{smPhos}). The careful reader may notice that we are more selective in this section compared to the polymer based white OLEDs (cf. Sec. \ref{whitePOLED}). This is simply due to the fact that the reports discussing white OLEDs based on small molecules by far exceed the number of papers on polymer devices. Stacked OLEDs based on any concepts from above will be discussed in Section \ref{Outcoupling}.

 It is worth noting that all concepts apart from fluorescent devices, which are limited in internal efficiency because of the spin statistics of the excitons (cf. Sec. \ref{FluoPhos}), bear the potential to reach $\eta_{\text{int}}=1$. In many cases, impressive work has been done that confirms this potential.

\subsection{\label{smFluor}Fluorescent devices}
Fluorescent emitters can be used in three different emission layer concepts: (i) bulk layers for emission [\cite{Choukri2006, Duan2008}], (ii) host-guest systems, where the fluorophore is dispersed into a wide bandgap material [\cite{Huang2002, Kim2002, Wu2005, Xie1999, Tsai2006}], and (iii) hybrid configurations, where the fluorophore itself is used as emitter and in addition as a host for a longer wavelength dye [\cite{Chuen2002, Yang2011, Jou2006a}]. The latter concept makes use of an incomplete energy transfer from host to guest molecules. Many fluorescent emitters undergo strong concentration quenching [see for instance \cite{Swanson2003, Xie2003}] so that the designated use of a material (as neat film or dispersed in a matrix) is often determined by its photophysical properties. In addition to the above concepts, non-emitting interlayers are often introduced to the device structure, mainly to achieve charge-carrier confinement either at interfaces [\cite{Wu2005}] or in quantum-well-like structures [\cite{Xie1999}]. This is realized by adjusting the energy levels (HOMO and LUMO) of the respective materials to artificially form energy barriers.

In contrast to devices comprising phosphorescent materials where the triplet manifold of the materials used becomes very important, the non-radiative triplet levels of the fluorescent materials do not play an important role in the device design. This is mainly caused by the fact that the energy transfer of singlet excitons of fluorophores to triplet states of fluorescent materials are quantum mechanically forbidden [\cite{Baldo2000, Pope1999}] and thus do not represent a prominent quenching channel. Later discussion will show that this picture will strongly change when using phosphors in the device (cf. Sec. \ref{smHybrid} and \ref{smPhos}).

The very first reports on white OLEDs solely comprising small molecules by \cite{Jordan1996, Strukelj1996, Hamada1996} will only be mentioned for completeness. All do not show improvements to the devices discussed by \cite{Kido1995a, Kido1994} with respect to their color quality and/or efficiency. Furthermore, the incorporation of exciplex formation and emission may generally be used for the realization of white OLEDs [\cite{Feng2001, Mazzeo2003}], however the overall efficiency of such devices seems to be rather limited. The only exception so far is the report by \cite{Tong2007}, discussing white OLEDs based on a single emissive material TPyPA, where the emission originates from its singlet and exciplex state (formed at the interface to the well-known electron-transporting material BPhen). These devices reach a maximum luminous efficacy of 9.0\,lm\,W$^{-1}$ with CIE coordinates of (0.31, 0.36; $\alpha_{\text{CIE}}=+0.02$).

\begin{figure*}[t]
\includegraphics[width=12cm]{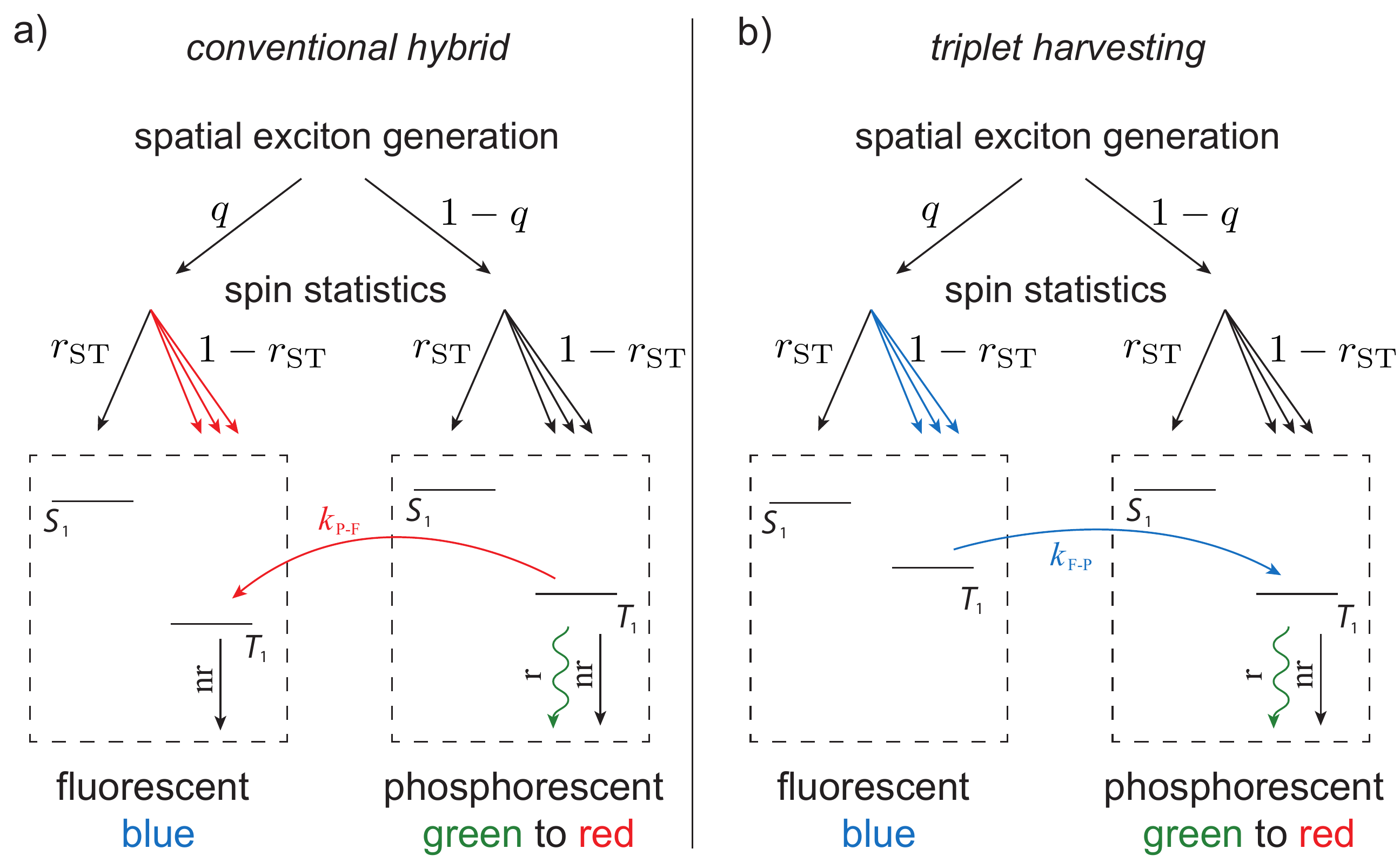}
\caption{\label{_hybrid_versus_TH}(color online) Scheme for electrical excitation in a) conventional hybrid and b) triplet harvesting concepts. For simplicity, the exciton generation is assumed to be only spread to two materials, i.e. fluorescent blue and phosphorescent green to red. Here, $q$ denotes the fraction of excitons that are created on the blue fluorophore, $r_\text{ST}$ is the fraction of singlet excitons formed, `r' and `nr' stand for radiative and non-radiative, respectively, and $k_{\text{F-P}}$ and $k_{\text{P-F}}$ are energy transfers from fluorophore to phosphor or vice versa.} 
\end{figure*}

\cite{Tsai2006} reported on highly efficient fluorescent white OLEDs based on a mixed host EML dual-doped with two fluorophores. By optimizing the blend of hole-transporting NPB and electron-transporting TBADN to a 1:1 ratio, 4.7\,\% EQE and 6.0\,lm\,W$^{-1}$ ($\eta_\text{LE,max}=11.2$\,lm\,W$^{-1}$) were obtained at 10\,mA\,cm$^{-2}$. The emission, based on blue DPAVBi and red rubrene emitters, reach CIE color coordinates of (0.329, 0.353; $\alpha_{\text{CIE}}=+0.02$). No luminance level is given in their report at which the efficiency values stated, however 10\,mA\,cm$^{-2}$ are likely to correspond to brightness values of approximately 1000\,cd\,m$^{-2}$ [see below following report by \cite{Yang2011}].

\cite{Yang2011} recently discussed a hybrid combination of simultaneous host and dopant emission, forming a two-color white device. They highly diluted the red emitter DCM into the blue emitting host Bepp$_2$ (0.2\,wt\% for a thin slab of 3\,nm followed by 0.5\,wt\% doping for the remaining 42\,nm of the EML) so that the energy transfer from host to guest is incomplete. At 10\,mA\,cm$^{-2}$, corresponding to roughly 1000\,cd\,m$^{-2}$ , this device reached 5.2\,\% EQE corresponding to 4.8\,lm\,W$^{-1}$ (maximum values: 5.6\,\% EQE, 9.2\,lm\,W$^{-1}$) with color coordinates of (0.332, 0.336) [$\alpha_{\text{CIE}}=0$] and a CRI = 80.
 
Note that both reports from \cite{Tsai2006} and \cite{Yang2011} with EQE values close to 5\,\% represent rule-of-thumb limit for devices fully based on fluorescence (cf. Sec. \ref{FluoPhos}).
 
	\subsection{\label{smHybrid}Hybrid fluorescent/phosphorescent OLEDs}
A large part of the research conducted on white small molecule OLEDs comprises blue fluorescent emitters together with longer wavelength phosphorescent emitters to achieve the white spectrum. The reason for this device concept is two-fold: (i) blue phosphorescent emitters with long term stability are hard to find [\cite{Su2008}], thus  devices based on freely available materials often degrade within hours of operation [\cite{Reineke2009a}]. Thus, blue fluorescent emitters are commonly used to avoid this stability bottleneck [\cite{Schwartz2006}]. (ii) Because blue phosphorescent materials call for host materials with even larger bandgap, the operating voltage of devices based on phosphorescent blue emitters will increase [\cite{Seidler2010}] with the luminous efficacy decreasing accordingly [\cite{Sun2006, Reineke2009a, So2010}].

		\subsubsection{\label{ConvHybrid}Conventional architectures}
In general, the blue fluorescent emitters used in hybrid white emission layers have triplet levels lower than the respective $T_1$ states of the phosphorescent materials. For instance, \cite{Schwartz2006} reported a triplet level of a highly efficient blue emitter Spiro-DPVBi at ~2.0\,eV, while its fluorescent peak is at 475\,nm (2.6\,eV). Thus, the blue triplet level typically represents a prominent quenching channel. This problem is illustrated in Figure \ref{_hybrid_versus_TH} a) for a simplified case where all phosphorescent materials incorporated are treated as one system, which may be more complex in real devices. In order to address all colors in the OLED, the emission layer is designed to realize exciton generation in all sub-layers hosting the different emitters. In the picture of Figure \ref{_hybrid_versus_TH}, excitons are created with a fraction $q$ on the fluorescent emitter, leaving $1-q$ to be generated in the phosphorescent system. All formed excitons obey the spin statistics known to be present in organic LEDs (cf. Sec. \ref{FluoPhos}), which is represented here by the fraction of created singlets $r_\text{ST}$. Note that in general, $r_\text{ST}$ may be different for every emitter system [\cite{Segal2003}]. However, to keep this discussion simple, $r_\text{ST}$ is used as a fixed value for any emitter system here.\footnote{Even more complicated is the fact that the fluorescent blue system can either be a neat film or a host-guest system. In the latter case, one would need to include exciton transfers from host to guest for both singlets and triplets.} Since the triplet level of the fluorescent blue emitter is lower than the respective levels of the phosphorescent materials, efficient transfer from phosphors to the fluorophore can occur, represented by the rate $k_\text{P-F}$. Obeying the selection rules for purely organic materials (Sec. \ref{FluoPhos}), this triplet level is non-emissive (`nr'  = non-radiative), thus excitons reaching it will be lost for emission. 

Overall, there are two channels for exciton quenching: (i) the direct formation of triplet excitons on the fluorescent triplet level, which is proportional to $q\times(1-r_\text{ST})$. Since $1-r_\text{ST}$ is a property of the specific material and typically in the order of 75\,\% [\cite{Segal2003}], the only way to reduce this channel is to reduce $q$. However, this will decrease the fluorescent intensity at the same time ($\sim q\times r_\text{ST}$). (ii) The energy transfer from the phosphorescent system to the fluorescent triplet level $k_\text{P-F}$ will reduce the quantum efficiency of the phosphorescent emitter [\cite{Kawamura2006}]:

\begin{eqnarray}
\eta_\text{P}=\frac{k_\text{r}}{k_\text{r}+k_\text{nr}}\xrightarrow{T_{1,\text F}<T_{1,\text P}} \frac{k_\text{r}}{k_\text{r}+k_\text{nr}+k_\text{P-F}}.\label{hybrid_interlayer_rate}
\end{eqnarray}

Oviously, the emission efficiency of the phosphorescent system can strongly be reduced if $k_\text{r}\approx k_\text{P-F}$. Triplet quenching introduced by $k_\text{P-F}>0$ can easily be prevented by introducing a thin interlayer between fluorescent and phosphorescent systems. Because the energy transfer leading to $k_\text{P-F}$ is Dexter-type (cf. \ref{Energytransfer}), requiring orbital overlap, interlayer thicknesses in the range of  2\,nm are sufficient [\cite{Schwartz2006}]. 

\begin{figure}[t]
\includegraphics[width=8.5cm]{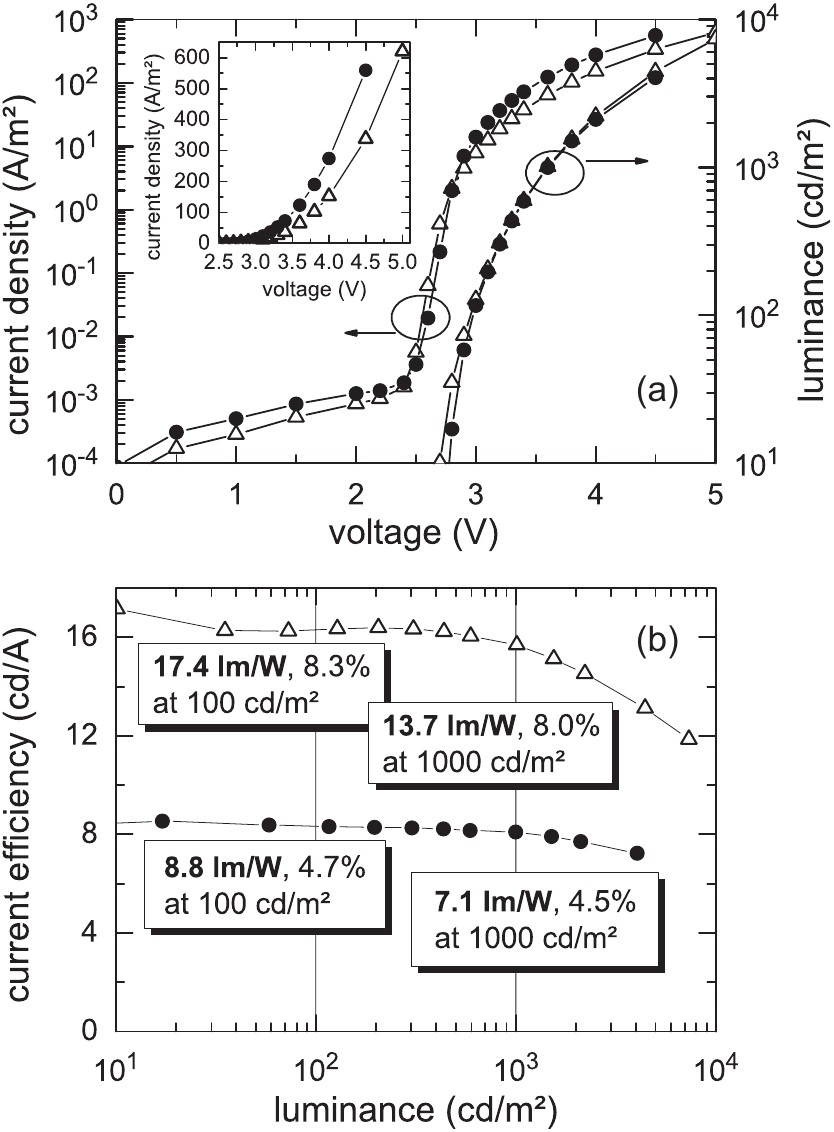}
\caption{\label{_Schwartz_2006_Fig4} Characteristics of hybrid white from \cite{Schwartz2006} without (filled circles) and with (open triangles) exciton blocking interlayer. Top: current density and luminance versus voltage. Bottom: current efficiency as a function of luminance. Additionally, luminous efficacy and EQE are given at brightness values of 100 and 1000\,cd\,m$^{-2}$. From \cite{Schwartz2006}.} 
\end{figure}

The first report on hybrid white OLEDs comprising an interlayer was made by \cite{Schwartz2006}. Their general device structure consists of a hole-transporting phosphorescent multilayer system for red and green and an electron-transporting Spiro-DPVBi layer for blue emission. Thus, excitons are generated at the interface between green phosphorescent and blue fluorescent layers, where a large part of the green excitons can be quenched by the Spiro-DPVBi triplet level at roughly 2.0\,eV.   In order to suppress exciton transfer, they introduced a composite exciton blocking layer consisting of the hole-transporting material TCTA and the electron-transporting material TPBi [\cite{Schwartz2006}]. Both have triplet energy levels of 2.83 and 2.59\,eV [\cite{Reineke2009, Reineke2009a}] above the Ir(ppy)$_3$ $T_1$ state that enable efficient blocking of excitons from the green emission layer. At the same time, the mixture of both materials makes it possible to assure that excitons are still created on each side of the EML, being essential to maintain a balanced white spectrum. Figure \ref{_Schwartz_2006_Fig4} shows the characteristics of two devices without and with additional composite TCTA:TPBi interlayer. The values best suited for comparison are the EQE values, as they are not additionally affected by spectral changes. Here, by introducing the interlayer, the EQE is almost doubled from 4.5\,\% to 8.0\,\% EQE, as measured at 1000\,cd\,m$^{-2}$. The corresponding CIE coordinates of the device with interlayer are (0.47, 0.42) [$\alpha_{\text{CIE}}=+0.02$], the color rendering index is as high as 85, and the luminous efficacy reaches 13.7\,lm\,W$^{-1}$ at 1000\,cd\,m$^{-2}$ [\cite{Schwartz2006}]. The comparison of the device efficiency with and without interlayer indicates that the transfer $k_\text{P-F}$ can easily reach the same order of magnitude as the radiative rate of the phosphor [cf. Eqn. \eqref{hybrid_interlayer_rate}, see also \cite{Baek2008}]. 

Another advantage of a composite exciton blocking layer is the ability to alter its transport properties with the mixing ratio of both materials, ultimately enabling to tune the color of the devices [\cite{Schwartz2008a, Leem2010}]. Besides the use of composite spacing layers that contain preferentially hole- and electron-transporting materials, also single material interlayers are used. Widely used molecules are NPB [hole-transporting, \cite{Ho2008a,Yan2007}], TPBi [electron-transporting, \cite{Ho2008}], and CBP [\cite{Baek2008,Seo2007,Zhang2008}]. The latter material CBP is often discussed to have ambipolar transport properties. It shall be noted here that the observed ambipolarity is often a stringent interplay between charge carrier mobility and energy level alignments within the complex layer structure.

\begin{figure}[t]
\includegraphics[width=8.5cm]{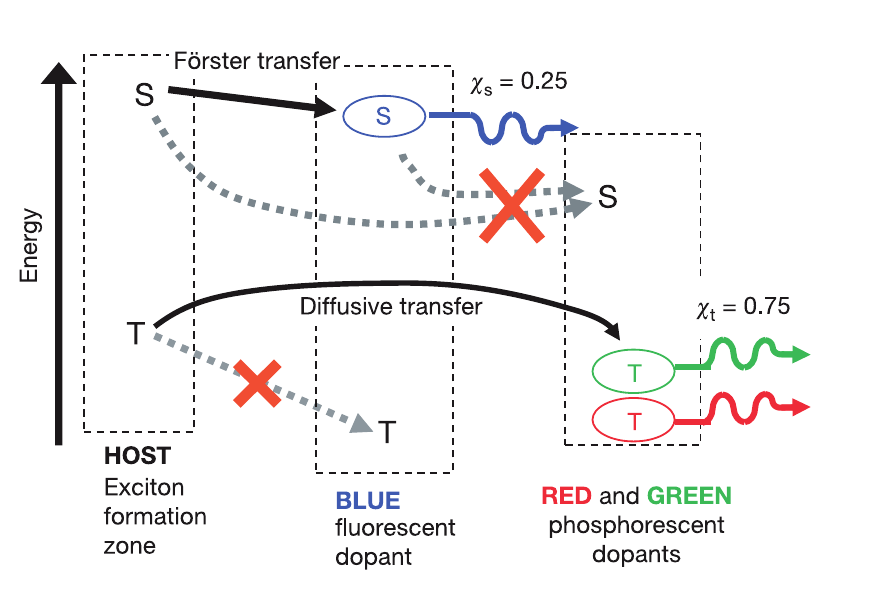}
\caption{\label{_Sun_2006_Fig1}(color online) Exciton transfer scheme proposed by \cite{Sun2006} for their hybrid fluorescence/phosphorescence white OLEDs. Exciton formation mainly occurs on the CBP host material. Solid lines represent allowed and dashed lines suppressed energy transfers. From \cite{Sun2006}.} 
\end{figure}

\begin{figure}[t]
\includegraphics[width=8.5cm]{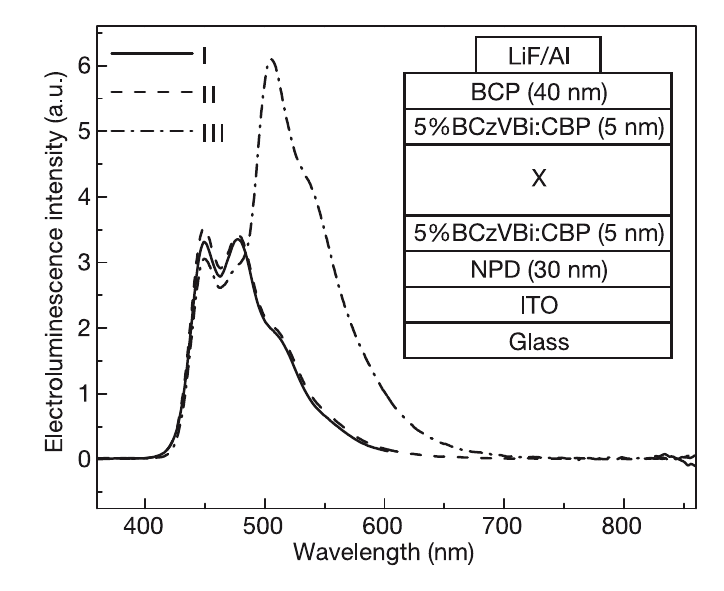}
\caption{\label{_Sun_2006_Fig2} Electroluminescence spectra of three different devices structures having variations in the spacer X. I: X = 16\,nm CBP, II: X = 15\,nm CBP:BCzVBi (5\,wt\%), and III: X = 4\,nm CBP + 20\,nm CBP:Ir(ppy)$_3$ (3\,wt\%). Inset shows the device layout. From \cite{Sun2006}.} 
\end{figure}

\cite{Sun2006} reported on a device concept for white OLEDs that shows improvement in the exciton distribution within the emission layer. Their OLEDs are based on the fluorescent blue emitter BCzVBi and the green and red phosphors Ir(ppy)$_3$ and PQIr, respectively, all embedded into a common host at different spatial locations. The authors claim that their device concept enables 100\,\% internal quantum efficiency  as a result of a decoupling of singlet and triplet exciton channels. This enables to solely use the 25\,\% fraction of singlets for fluorescence whereas the remaining 75\,\% of the generated triplets will be directed to the green and red phosphors (cf. Sec. \ref{FluoPhos}). Furthermore, with a finely tuned EML, thermalization losses prior to photon emission can be reduced to decrease the operating voltage and increase the luminous efficacy [\cite{Sun2006}]. Their device concept is based on the experimental finding that excitons are mainly formed at both EML interfaces to adjacent transport layers, forming a U-shaped exciton generation profile [\cite{Sun2006}]. Both regions of exciton formation are therefore doped with the fluorescent blue emitter BCzVBi at 5\,wt\%. There, singlets will recombine while triplets will be able to diffuse away from the site of exciton generation into the center of the EML. The energy transfer scheme is shown in Figure \ref{_Sun_2006_Fig1}. Proof for both, the exciton generation at the interfaces forming a U-shape and the ability for triplet excitons to diffuse into the center of the EML is given in Figure \ref{_Sun_2006_Fig2}. Comparing a device with updoped (device I) and BCzVBI-doped (device II) middle section do not show differences in the emitted blue intensity. If the center section is doped with the green emitter Ir(ppy)$_3$ (device III) instead, an additional high intensity green signal is detected in the electroluminescence. Because the blue fluorescent intensity remains unchanged between device I and III, it is valid to assume that only triplet excitons are transferred to the phosphor while the singlets recombine solely on BCzVBi. However, it cannot be excluded from this data that a constant fraction of triplet excitons remain trapped on the blue fluorophore (comparable in undoped and Ir(ppy)$_3$-doped devices). Furthermore, a thin undoped CBP layer is sandwiched between blue and green layers to prevent singlet exciton transfer from  BCzVBi to Ir(ppy)$_3$ (cf. Fig. \ref{_Sun_2006_Fig1}).

\cite{Sun2006} state that the triplet exciton transfer from the CBP host to the blue emitter BCzVBi is suppressed as shown in Figure \ref{_Sun_2006_Fig1} [see also \cite{Schwartz2009}]. However, it is fairly unlikely that this Dexter-type energy transfer (cf. Sec. \ref{Energytransfer}) does not occur as the BCzVBi concentration is at a sufficiently high level of 5\,wt\%. This transfer is energetically favorable, since the triplet excited state of BCzVBi is at 1.81\,eV [\cite{Deaton2008}], which is noticeably smaller than the respective $T_1$ levels of Ir(ppy)$_3$ (2.42\,eV) and PQIr (2.06\,eV). Thus, the BCzVBi triplet states form a triplet exciton trap that introduces a non-radiative loss channel. Therefore, the theoretical limit for the internal quantum efficiency of this concept is below $100$\,\%. Note that at 5\,wt\% of BCzVBi in the region of exciton formation, even direct triplet exciton generation on the emitter molecules cannot be excluded, which similarly populates the non-radiative BCzVBi triplet state. Recently, \cite{Schwartz2009} presented a calculation made for devices based on triplet harvesting, which will be discussed in the next section, where EQE values of 10\,\% are possible even when the triplets on the blue emitter are lost non-radiatively. With all triplets harnessed, the EQE limit is in the range of 16\,\%. The device efficiencies reported by \cite{Sun2006} are 10.8\,\% EQE at 500\,cd\,m$^{-2}$ (14\,lm\,W$^{-1}$), with corresponding color coordinates of (0.40, 0.41) [$\alpha_{\text{CIE}}=+0.02$] and a very high CRI of 85. The efficiency data alone, however, does not prove that the majority of non-radiative host triplets are harnessed by the lower energy phosphors in their device concept. 

\subsubsection{\label{smFluor_sensitizer}Phosphor sensitized fluorescence}
		
In order to overcome the bottleneck that only 25\,\% of the excitons are electrically excited in singlet states (cf. Sec. \ref{FluoPhos}), \cite{Baldo2000} proposed a cascade excitation scheme to promote triplets back to emissive singlet states of a fluorophore. This is achieved by additionally introducing a phosphorescent molecule [i.e. Ir(ppy)$_3$] into the system that acts as a sensitizer, denoted as $X$. As the subsequent discussion will show, white OLEDs based on this concept are either designed to have substantial emission from the sensitizer $X$ or are combined with additional phosphors to achieve white-light emission. Thus, this concept is discussed here in Section \ref{smHybrid}, even though it was originally promoted to achieve highly efficient fluorescent devices [\cite{Baldo2000}].

\begin{figure}[t]
\includegraphics[width=8.5cm]{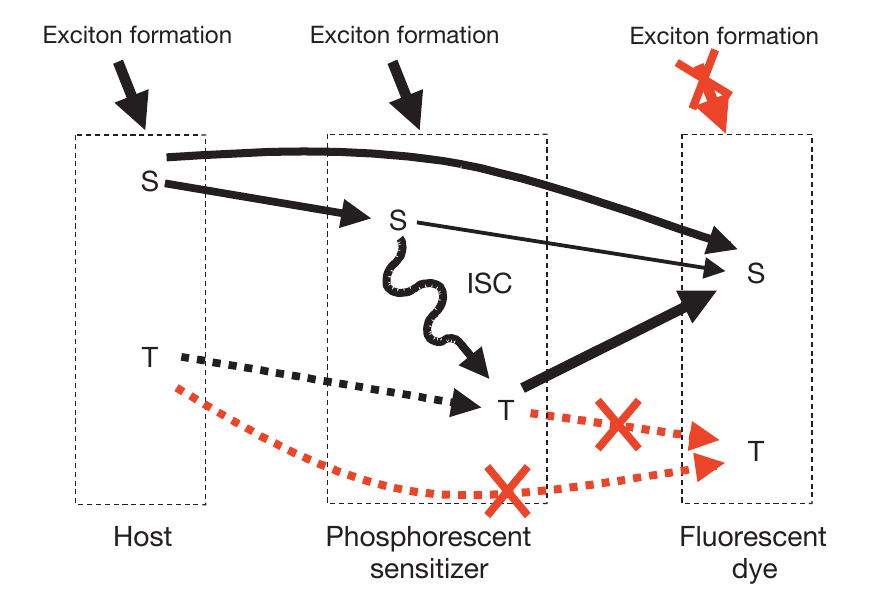}
\caption{\label{_Baldo2000_Fig1}(color online) Energy transfer mechanisms for a phosphor sensitized fluorescence system. Triplet transfer from host and sensitizer molecules to the fluorescent dye are suppressed by highly diluting the fluorophore. Triplet excitons from the sensitizer are transferred to the singlet state of the fluorophore via F\"orster energy transfer. From \cite{Baldo2000}.} 
\end{figure}

The transfer scheme is shown in Figure \ref{_Baldo2000_Fig1}. The cascade energy transfers from host donor singlet (${}^1D^*$) and triplet (${}^3D^*$) level are [cf. \cite{Baldo2000}]:

\begin{eqnarray}
{}^1D^*+{}^1X &\xrightarrow{\text{F\"orster}}& {}^1D+{}^1X^* \\
{}^1X^* &\xrightarrow{\text{ISC}}& {}^3X^* \\
{}^3X^*+{}^1A &\xrightarrow{\text{F\"orster}}& {}^1X+{}^1A^*\label{Eq:sensitizersinglet}, 
\end{eqnarray}
and
\begin{eqnarray}
{}^3D^*+{}^1X &\xrightarrow{\text{Dexter}}& {}^1D+{}^3X^* \\
{}^3X^*+{}^1A &\xrightarrow{\text{F\"orster}}& {}^1X+{}^1A^*\label{Eq:sensitizertriplet}.
\end{eqnarray}
Finally, the photon is emitted via ${}^1A^*\longrightarrow {}^1A+h\nu$ from the singlet state of the fluorescent acceptor. Note here that triplet transfer via F\"orster energy exchange [Eqns. \eqref{Eq:sensitizersinglet} and \eqref{Eq:sensitizertriplet}] is only possible because $X$ is a phosphorescent donor (cf. Sec. \ref{Energytransfer}). Triplet exciton transfer from the donor (host) $D$ and sensitizer $X$ to the triplet level of the fluorophore need to be avoided because it presents quenching channels. Both, being Dexter-type transfers requiring orbital overlap, can be suppressed by increasing the intermolecular distance between the respective donors ($D$ or $X$) and the acceptor. In their early report, \cite{Baldo2000} could increase the device efficiency of a red fluorescent OLED based on DCM2 by a factor of three compared to a reference device by incorporating Ir(ppy)$_3$ as sensitizing molecule. Since the singlet excited state lifetime of DCM2 is in the range of a few nanoseconds, which is orders of magnitude longer compared to Ir(ppy)$_3$ [\cite{Adachi2000}], the transient signal of DCM2 resembles the decay of the phosphor [\cite{Baldo2000}], giving direct evidence for the proposed excitation scheme.

Similar to the fundamental finding of \cite{Baldo2000}, Ir(ppy)$_3$ has been widely used so far to sensitize red fluorescent emission [\cite{Cheng2003a, Cheng2006, Kanno2006b, Xue2010}]. Here the red fluorophores were either rubrene [\cite{Cheng2006, Xue2010}] or DCJTB [\cite{Cheng2003a, Kanno2006b}].
The highest efficiency and color quality based on phosphor sensitized fluorescence has been reported by \cite{Kanno2006b}. In their concept, CBP is used as a common host for all emitting materials, which were laterally distributed to achieve emission from all colors. BCzVBi is used for the high energy, blue emission in their devices. A double doped film comprising Ir(ppy)$_3$ (8\,wt\%) as sensitizer and DCJTB (0.08\,wt\%) as red fluorophore is positioned in the center of the emission layer. Here, similar to the report of \cite{Baldo2000}, the emission of the green phosphor Ir(ppy)$_3$ is not fully quenched, which is utilized in this white concept to fill the spectral gap between blue and red emission bands to achieve high quality white emission. This device reaches 8.5\,\% external quantum efficiency (maximum value), which corresponds to 18.1\,lm\,W$^{-1}$, with CIE color coordinates of (0.38, 0.42) [$\alpha_{\text{CIE}}=+0.03$]. Omitting Ir(ppy)$_3$ in the device structure strongly decreases the EQE values to approximately 3\,\% [\cite{Kanno2006b}].

In contrast to many reports based on Ir(ppy)$_3$, \cite{Lei2004} introduced the light-blue emitter FIrpic as sensitizer. Here, FIrpic (8\,\%) and DCJTB (0.4\,\%) are diluted in a wide gap material (DCB), where FIrpic acts as both blue emitter and phosphorescent sensitizer for the red DCJTB molecules. These devices reach a maximum efficiency of 9.2\,cd\,A$^{-1}$ with CIE coordinates of (0.32, 0.36) [$\alpha_{\text{CIE}}=+0.02$].

Even though phosphor sensitization seems to be a promising route to make use of short lifetime fluorophores that enable higher efficiency at high brightness [\cite{Baldo2000}] because of reduced triplet quenching, this concept has not drawn much attention. This might be due to the complex, cascade energy transfer that has to be controlled to achieve white light emission. Putting the limited stability of blue phosphorescent materials aside for a moment [\cite{Sun2006, Schwartz2009, Reineke2009a, Kanno2006b}], it might be worth to investigate phosphor sensitized white OLEDs, where a high energy (deep blue) triplet emitter distributes all excitons to longer wavelength fluorescent materials. This would enable creating a fully fluorescent white OLED with internal efficiencies of unity.

\subsubsection{\label{TripHarv}Triplet harvesting}
In order to reduce the losses in triplet states of blue fluorescent emitters  scaling with $q\times(1-r_\text{ST})$ [cf. Fig. \ref{_hybrid_versus_TH}], these excited states need to be passed on to other sites, as their generation cannot fully be excluded. The only way to achieve this is to incorporate fluorescent materials with a triplet level that is equal or higher than the $T_1$ state of at least one of the phosphorescent emitters used. Assuming a blue fluorescence at 450\,nm and a red phosphorescence at 600\,nm, the key requirement for blue fluorophores to act as a triplet donor in white OLEDs translates into a singlet-triplet energy gap of $<0.7$\,eV. Note that the fluorescent emitter BCzVBi that has been discussed in the previous section has a fluorescence peak at 450\,nm and a singlet-triplet splitting of approximately 0.95\,eV [\cite{Deaton2008, Sun2006}].

\begin{figure}[t]
\includegraphics[width=8.3cm]{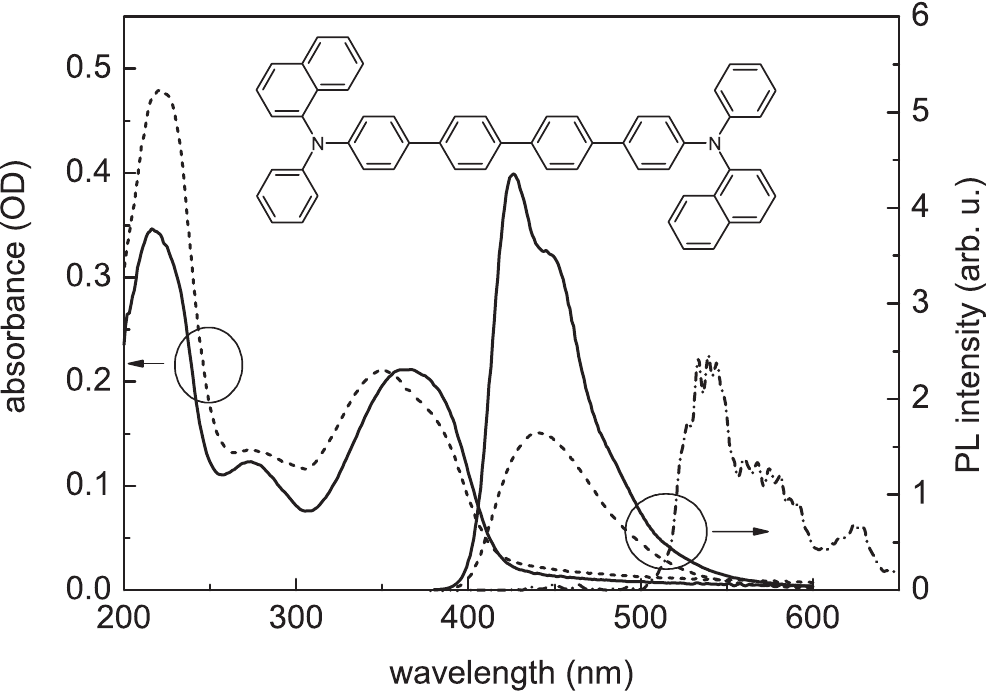}
\caption{\label{_Schwartz2007_Fig1} Absorbance and photoluminescence data of 30\,nm neat films of NPB (dashed) and 4P-NPD (solid, chemical structure shown in the inset). Photoluminescence is recorded at an excitation wavelength where absorbance data of both materials intersect (356\,nm), so that the intensities reflect the relative quantum yields of the materials. Addtionally, the phosphorescence spectrum of 4P-NPD is plotted as obtained at 77\,K for the diluted form in a solid polystyrene matrix (2\,wt\%). From \cite{Schwartz2007}.} 
\end{figure}

From quantum mechanical principles, it is known that the singlet-triplet splitting $\Delta E_\text{ST}$ is proportional to the exchange integral $K$ between spatial overlap of the HOMO and LUMO [\cite{Schwartz2009, Endo2009, Endo2011}]: $\Delta E_\text{ST}\sim 2K$. By localizing HOMO and LUMO wavefunctions to different regions of the molecular structure, the singlet-triplet splitting can  be strongly reduced even down to 0.11\,eV [\cite{Endo2011}]. Once the triplet state of the fluorophore is higher than the triplet level of a phosphor incorporated, the quenching rate $k_\text{P-F}$ transforms into an additional path to excite a phosphor: $k_\text{F-P}$[cf. Fig. \ref{_hybrid_versus_TH} a) and b)], because the different relation of the energy levels reverse the direction of the energy transfer. Thus, when triplet harvesting is incorporated into a device concept for white OLEDs, internal quantum efficiencies of 100\,\% are possible, because the fluorophores triplet state is changed from a non-radiative trap to an intermediate state that participates in the excitation of the phosphors.

\begin{figure}[t]
\includegraphics[width=8.0cm]{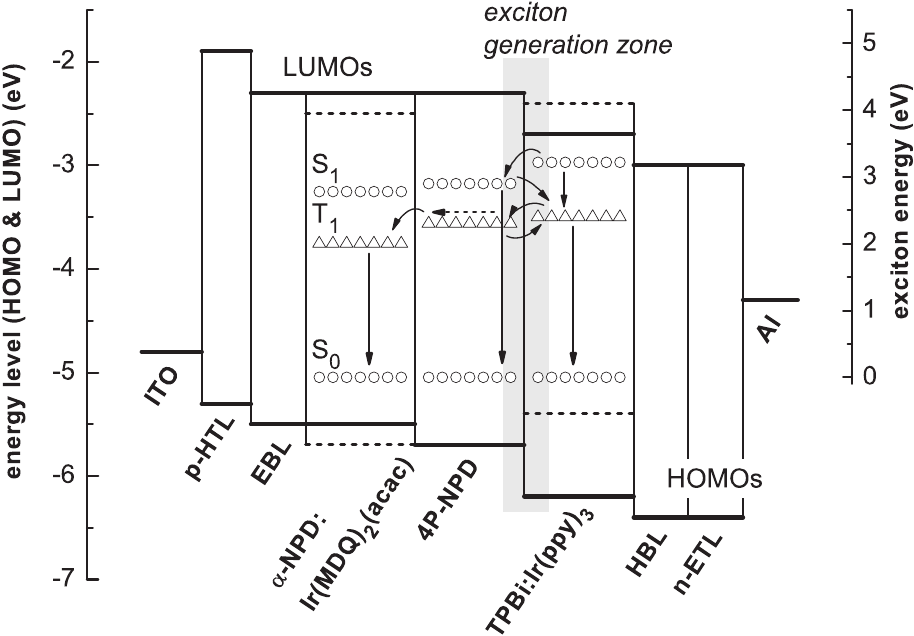}
\caption{\label{_Schwartz2007_Fig2} Energy level diagram (HOMOs, LUMOs [lines], and triplet energies [open symbols]) of the white OLED making use of the triplet harvesting concept. From \cite{Schwartz2007}.} 
\end{figure}

The concept of triplet harvesting in white OLEDs has been introduced by \cite{Schwartz2007}. Their work is based on a special fluorescent blue emitter 4P-NPD, shown in Figure \ref{_Schwartz2007_Fig1}, that has its fluorescent peak at 426\,nm and a singlet-triplet splitting of $\sim 0.6$\,eV (cf. Fig. \ref{_Schwartz2007_Fig1}). Another very important property of 4P-NPD is its very high neat film PLQY [\cite{Schwartz2007}], making it possible to use it as  bulk emitter. The working principle is shown in Figure \ref{_Schwartz2007_Fig2}. With the triplet level of 4P-NPD at approximately 2.3\,eV, it is sufficiently high to excite the red phosphor Ir(MDQ)$_2$(acac) [2.06\,eV]. However, the green emitter Ir(ppy)$_3$ with a triplet level at 2.42\,eV would still be quenched by 4P-NPD. Consequently, triplet harvesting is realized with the combination of 4P-NPD and Ir(MDQ)$_2$(acac), while an additional layer for conventional green phosphorescence is added. Excitons are formed at a double emission layer interface [\cite{Zhou2002}] between a composite $\alpha$-NPD:Ir(MDQ)$_2$(acac)/4P-NPD system and TPBi:Ir(ppy)$_3$. This concept makes use of the ability of triplet excitons being able to diffuse further than singlets [\cite{Rosenow2010, Schwartz2007}]. For 4P-NPD, a triplet diffusion length of 11\,nm [\cite{Wuensche2010}] has been determined. While singlet excitons will recombine in the close proximity of the exciton generating interface [cf. 4P-NPD singlet diffusion length of $\sim 4$\,nm \cite{Hofmann2012}], the triplet excitons will diffuse away from this interface ultimately reaching the phosphor-doped $\alpha$-NPD:Ir(MDQ)$_2$(acac) layer. There, they are efficiently transferred via $k_\text{F-P}$ to the emissive Ir(MDQ)$_2$(acac) triplet state. The devices of the first report by \cite{Schwartz2007} already reached high efficiency values of 10.4\,\% EQE and 22.0\,lm\,W$^{-1}$ at 1000\,cd\,m$^{-2}$ with CIE coordinates of (0.44, 0.47) [$\alpha_{\text{CIE}}=+0.06$] with a CRI = 86. 

\begin{figure}[t]
\includegraphics[width=8.5cm]{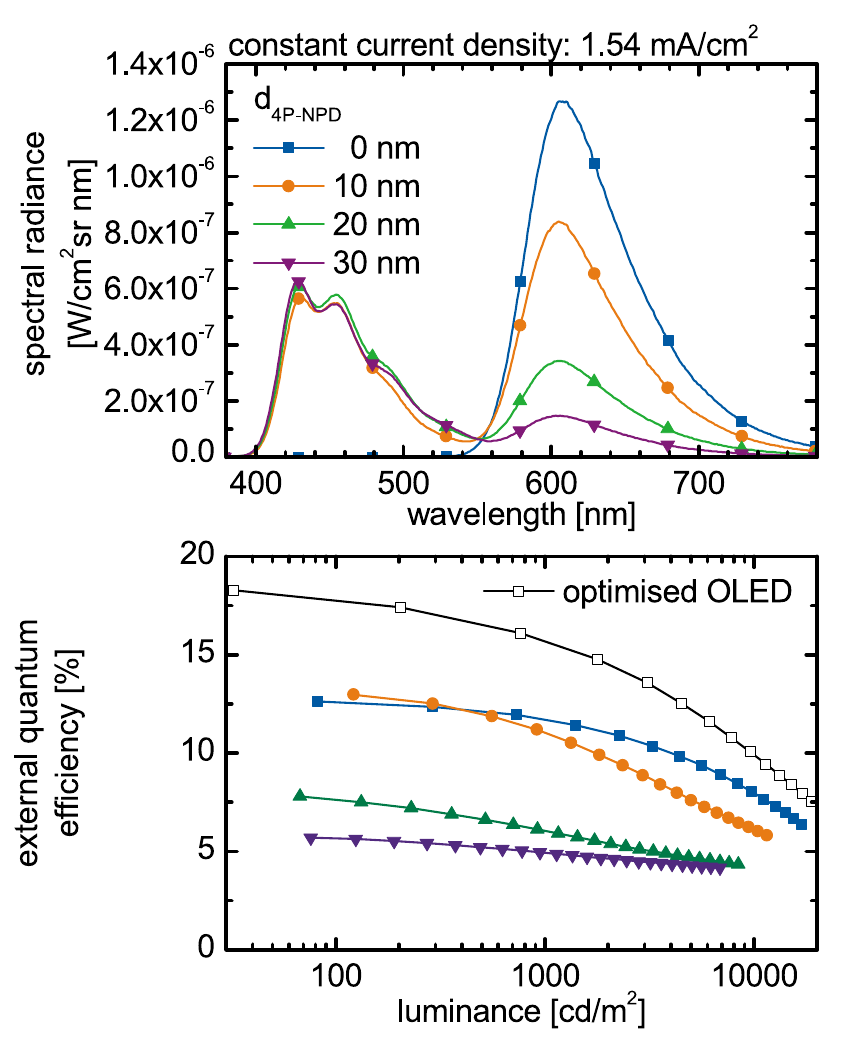}
\caption{\label{_Rosenow2010_Fig2} (color online) Top: Absolute electroluminescence spectra of triplet harvesting devices showing a variation in the 4P-NPD spacer thickness from 0 to 30\,nm measured at a fixed current density. Bottom: Corresponding EQE versus luminance characteristics. Additionally plotted is the EQE of an optimized device that is further used for integration in the two unit white OLEDs.  From \cite{Rosenow2010}.} 
\end{figure}

The effect of triplet harvesting is seen best in Figure \ref{_Rosenow2010_Fig2}. It shows a series of samples with a bilayer EML architecture of 4P-NPD:Ir(MDQ)$_2$(acac) [5\,wt\%]/4P-NPD with different intrinsic 4P-NPD layer thickness from 0 to 30\,nm. Excitons are created at the interface between the 4P-NPD layer and the adjacent HBL [\cite{Rosenow2010}]. Without intrinsic 4P-NPD layer (0\,nm), the device is a conventional red phosphorescent OLED showing solely red emission and high EQE. With increasing 4P-NPD layer thickness, the red intensity decreases and additional blue fluorescence is observed. Note here that the spectra shown are not normalized but rather plotted in absolute units, clearly indicating that additional triplets can be harvested by Ir(MDQ)$_2$(acac) when the 4P-NPD layer thickness is adjusted correctly (this is also reflected in the EQE on the bottom of Fig. \ref{_Rosenow2010_Fig2}). 

Since the triplet level of 4P-NPD cannot excite green phosphorescent emitters, \cite{Rosenow2010} incorporated this triplet harvesting system into a two-unit stacked OLED, where the second unit is a double-doped phosphorescent green/yellow unit based on the emitters Ir(ppy)$_3$ and Ir(dhfpy)$_2$(acac). Without going into the details of this device layout, they could improve the efficiency and color quality to 26\,\% EQE\footnote{Note that the EQE values in two-unit stacked OLEDs can theoretically be doubled, because for every electron that is injected, two photons can be emitted.} and 33\,lm\,W$^{-1}$ at 1000\,cd\,m$^{-2}$. The corresponding color coordinates are (0.506, 0.422) [$\alpha_{\text{CIE}}<+0.01$], very close to the Planckian locus. This improvement in color quality can mainly be attributed to the possibility to optimize the triplet harvesting and green/yellow units independently, whereas in the report of \cite{Schwartz2007} multiple exciton transfer steps at the exciton generating interface (cf. Fig. \ref{_Schwartz2007_Fig2}) complicated the color control.

\begin{figure}[t]
\includegraphics[width=8.0cm]{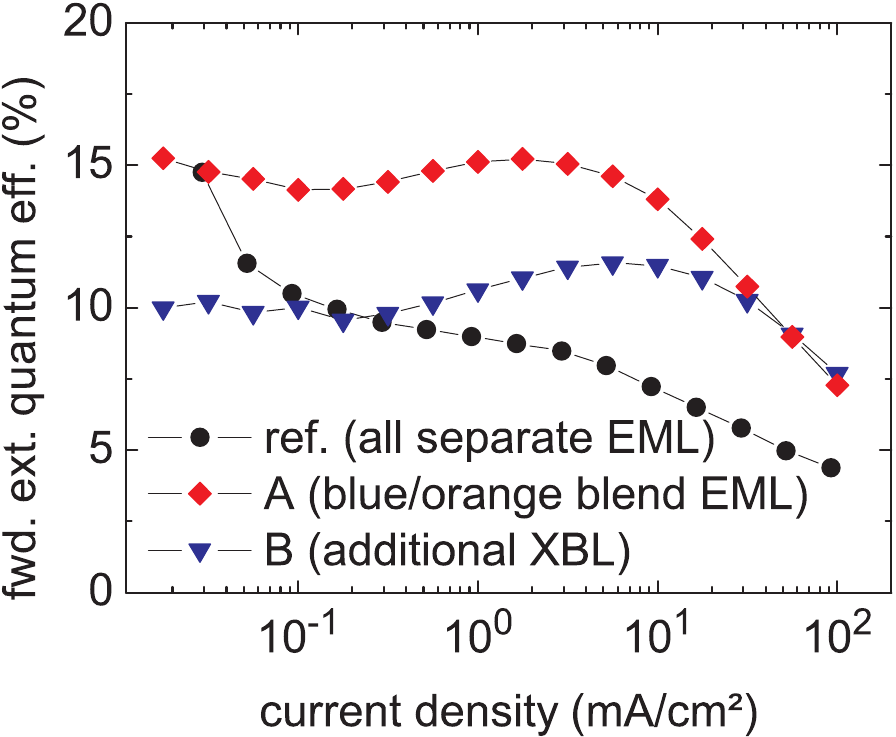}
\caption{\label{_Schwartz2008_Fig2}(color online) External quantum efficiency versus current density for different triplet harvesting integrations (bilayer and blended systems) showing strong differences in the efficiency roll-off. `XBL' = exciton blocking layer. From \cite{Schwartz2008}.} 
\end{figure}

The triplet harvesting concept is based on the fact that non-radiative triplets formed on the fluorescent material can find accessible sites for recombination which are spatially separated and only reached via diffusion. Taking into account that the triplet excited lifetime in 4P-NPD is long (in the range of ms [\cite{Schwartz2007}]), a correspondingly high triplet exciton density $n_\text T$ is formed in the 4P-NPD layer. Since triplet-triplet annihilation scales with the square of $n_\text T$, it will be much stronger than observed in state-of-the-art phosphorescent systems [\cite{Reineke2007}], where the triplet lifetime is in the range of microseconds. The consequence is that triplets diffusing to the emissive phosphorescent sites are likely to annihilate with other triplets (or even charges [\cite{Reineke2007}], so that the quantum efficiency of the red emission is strongly reduced as a function of the excitation level (current density) [\cite{Schwartz2007}]. Thus, triplet harvesting systems typically have a strong EQE roll-off (compare for instance the EQE characteristics of devices with and without 10\,nm intrinsic 4P-NPD layer, shown in Fig. \ref{_Rosenow2010_Fig2} bottom). In order to reduce the triplet density $n_\text T$, \cite{Schwartz2008} merged the bilayer triplet harvesting system to one blend layer of 4P-NPD:Ir(MDQ)$_2$(acac), so that the average distance a triplet has to travel to reach a phosphor is reduced. Fluorescence is still observed, because the doping concentration of the phosphor is reduced by one to two orders of magnitude ($\sim0.1$\,wt\%) compared to conventional phosphorescent OLEDs ($\sim5-10$\,wt\%). The effect is shown in Figure \ref{_Schwartz2008_Fig2} comparing bi-layer and blended systems incorporated into white devices [\cite{Schwartz2008}]. While the EQE of the bilayer device steadily decreases as a function of current density $j$, the blend system shows a noticeable range of $j$ (almost three orders of magnitude), where the EQE remains at a relatively constant value. Only at high $j$, the EQE roll-off is observed, similar to standard phosphorescent devices [\cite{Reineke2007}]. The blend system reaches very high device efficiencies at 1000\,cd\,m$^{-2}$ of 15.2\,\% EQE and 31.6\,lm\,W$^{-1}$ with CIE coordinates of (0.49, 0.41) matching the Planckian locus [$\alpha_{\text{CIE}}=0$]. This conceptional improvement is very important because, especially for lighting applications, brightness values of a few thousand cd\,m$^{-2}$ are required. The only drawback of the blend approach is the very low emitter concentration down to $\sim0.1$\,wt\%, raising the question whether this process can be controlled in large scale manufacturing.

\begin{figure}[t]
\includegraphics[width=8.5cm]{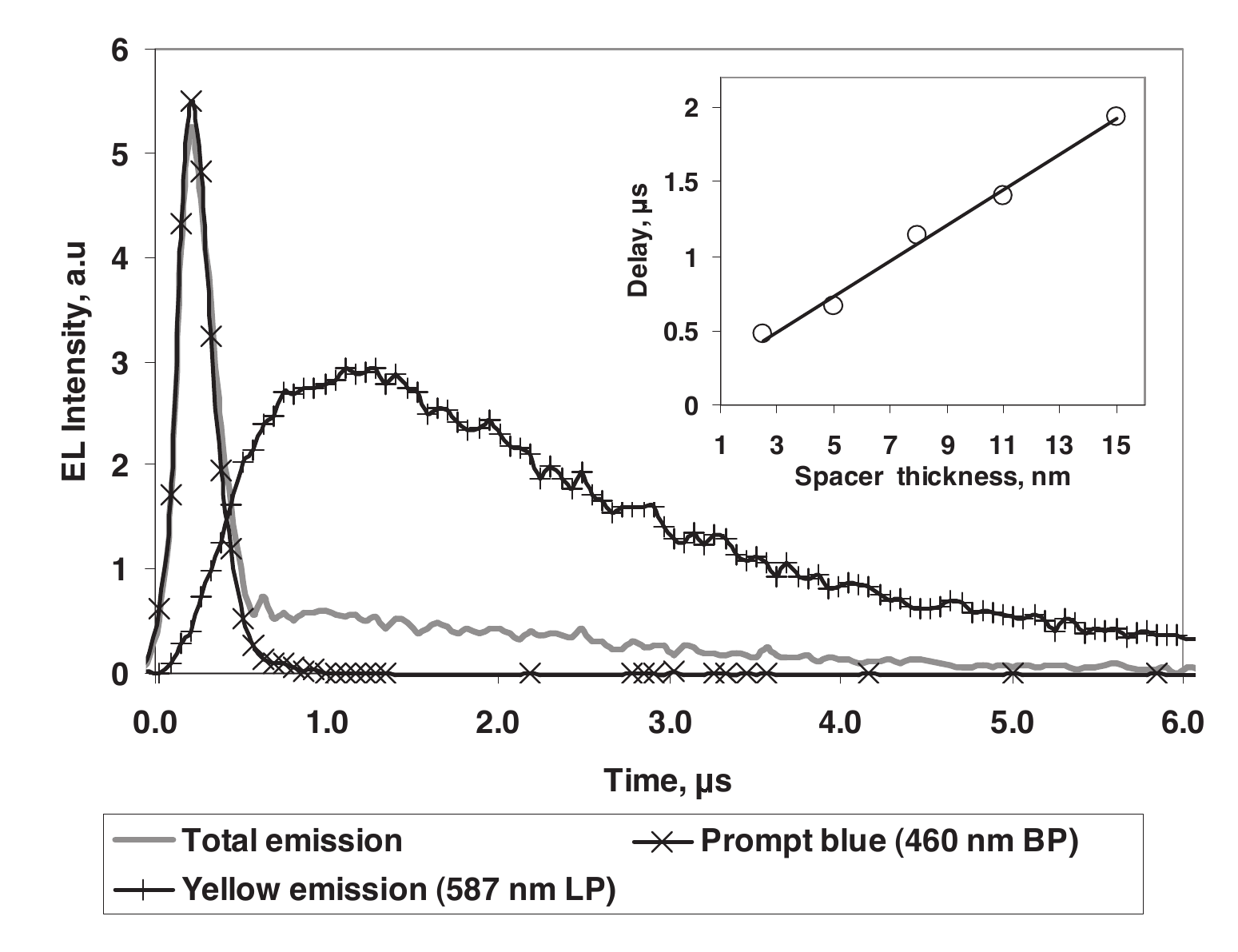}
\caption{\label{_Kondakova2010_Fig7} Time-resolved electroluminescence of a device with blue fluorescence (MQAB) and yellow phosphorescence (Ir(ppy)$_2$pc, triplet: 2.34\,eV) of a triplet harvesting system. Inset: Delay between time-resolved maxima of prompt and delayed features as a function of spacer thickness. From \cite{Kondakova2010}.} 
\end{figure}

Relying on the special properties of fluorescent blue emitters having a noticeably reduced singlet-triplet splitting $\Delta E_\text{ST}$, the progress based on the triplet harvesting concept is rather slow. Still, in the last years, first reports were published discussing new materials with even improved properties compared to 4P-NPD. Here, the ultimate goal is to reduce $\Delta E_\text{ST}$ to values where also green phosphors ($\sim 510$\,nm) can be excited from the fluorophores triplet state. Recently, \cite{Kondakova2010} reported on another fluorescent blue emitter MQAB with small singlet-triplet splitting of 0.27\,eV (singlet: 2.82\,eV and triplet: 2.55\,eV). Presumably, because its PLQY as a neat film is not high, \cite{Kondakova2010} use MQAB in a host-guest system together with the well known CBP host (triplet: 2.61\,eV). Thus, with respect to their triplet energies, both MQAB dopant and CBP host are almost in resonance so that triplet movement throughout the layer should be possible. In contrast to the work presented above, \cite{Kondakova2010} always use a spacer material, i.e. Ga(pyimd)$_3$ (triplet level: 2.71\,eV), between blue fluorescent and phosphor-doped layers. The host for the phosphors used is Ga(pyimd)$_3$ also. In total, the EML layer (from hole injection side) is CBP:MQAB/Ga(pyimd)$_3$/Ga(pyimd)$_3$:phosphor. 
Using time-resolved spectroscopy, \cite{Kondakova2010} give experimental evidence for the presence of triplet harvesting in their devices as shown in Figure \ref{_Kondakova2010_Fig7}. Here, the prompt fluorescence and delayed phosphorescence are separately shown, clearly displaying a time delay between the two. The inset shows the delay between fluorescence and phosphorescence peak as a function of Ga(pyimd)$_3$ spacer thickness, indicating the diffusively promoted excitation of the phosphor. In their study, they investigated different phosphorescent emitters varying in emission wavelength, spanning from red [Ir(1-piq)$_3$] to green [Ir(ppy)$_3$]. For all the phosphors used, \cite{Kondakova2010} could demonstrate triplet harvesting with overall device efficiencies $>10\,\%$\,EQE at low current densities. To our knowledge, this is the first time triplet harvesting has been reported to work in conjunction with Ir(ppy)$_3$ as phosphorescent acceptor [\cite{Kondakova2010}]. Three color white devices were also fabricated based on two phosphorescent emitters Ir(ppy)$_3$ and Ir(phq)$_3$ for green and orange, respectively, and MQAB for blue. At 1000\,cd\,m$^{-2}$ the efficiencies are 12.6\,\% EQE and 21.4\,lm\,W$^{-1}$ with CIE coordinates of (0.317, 0.317)  [$\alpha_{\text{CIE}}=<-0.01$].

Another report about a material capable of exciting green phosphors in a triplet harvesting configuration has been reported by \cite{Hung2010}. The fluorophore CPhBzIm  has its EL fluorescence maximum at approximately 430\,nm (2.88\,eV) and triplet level at 2.48\,eV, resulting in $\Delta E_\text{ST}=0.4$\,eV. The $T_1$ of CPhBzIm should be sufficiently high to be used together with Ir(ppy)$_3$ in a triplet harvesting concept. Instead the authors used a slightly different phosphor Ir(pbi)$_2$(acac) [\cite{Hung2010}]. They fabricated a two-color white device by blending CPhBzIm with Ir(pbi)$_2$(acac) at a low concentration of 0.1\,wt\%. The efficiency reached 5.1\,\% EQE at 1000\,cd\,m$^{-2}$ (7\,\% EQE maximum value), which cannot be used as an indication whether triplet harvesting actually occurs or not.

To summarize, triplet harvesting is a very promising concept for future high-efficiency white OLEDs with high color quality. Once the right materials are found it will also allow to simplify the device structure, because in general all materials could be blended into one uniform emission layer. One key challenge to date is that triplet harvesting itself does not limit the long term stability of the devices but rather materials need to be developed that meet the stability and conceptual requirements at the same time.

	\subsection{\label{smPhos}Phosphorescent devices}
Among the various concepts for white OLEDs, by far the most effort has been spent on research dealing with devices based solely on phosphorescence emitting materials. This is probably due to the fact that phosphors inherently offer internal efficiencies of unity [\cite{Baldo1998}], so that in general the only remaining task in device engineering is the distribution of excitons to different emitters for white emission. The high internal efficiency is important for white OLEDs to be competitive with existing lighting technologies, i.e. fluorescent tubes and white LEDs [\cite{Steele2007}]. In this section, conventional phosphorescent OLEDs will be discussed first, where the discussion will be split between two- and three-color concepts. This is followed by a discussion of systems with reduced band gap that aim to reduce the operating voltage of the devices and finally white phosphorescent concepts are introduced based on combined monomer/excimer emission.

In contrast to fluorophores, where examples of high PLQY emitters exist [\cite{Schwartz2007, Xie2003, Schwartz2006, Tong2007}], the vast majority of phosphorescent emitters needs to be embedded into a host material to avoid concentration quenching [\cite{Kawamura2005, Kawamura2006, Kobayashi2005}]. The key requirement for a suitable host material of a phosphorescent emitter is to have a higher triplet level than the phosphor. By that, the triplet excitons are efficiently confined to the emissive states [\cite{Goushi2004}], which are, due to the nature of phosphorescent molecules, long-living excited states with lifetimes in the range of microseconds [\cite{Thompson2007}]. The excitonic confinement is especially a challenge for blue emitters, as they require host materials with widest band gap.

\begin{figure}[t]
\includegraphics[width=8.0cm]{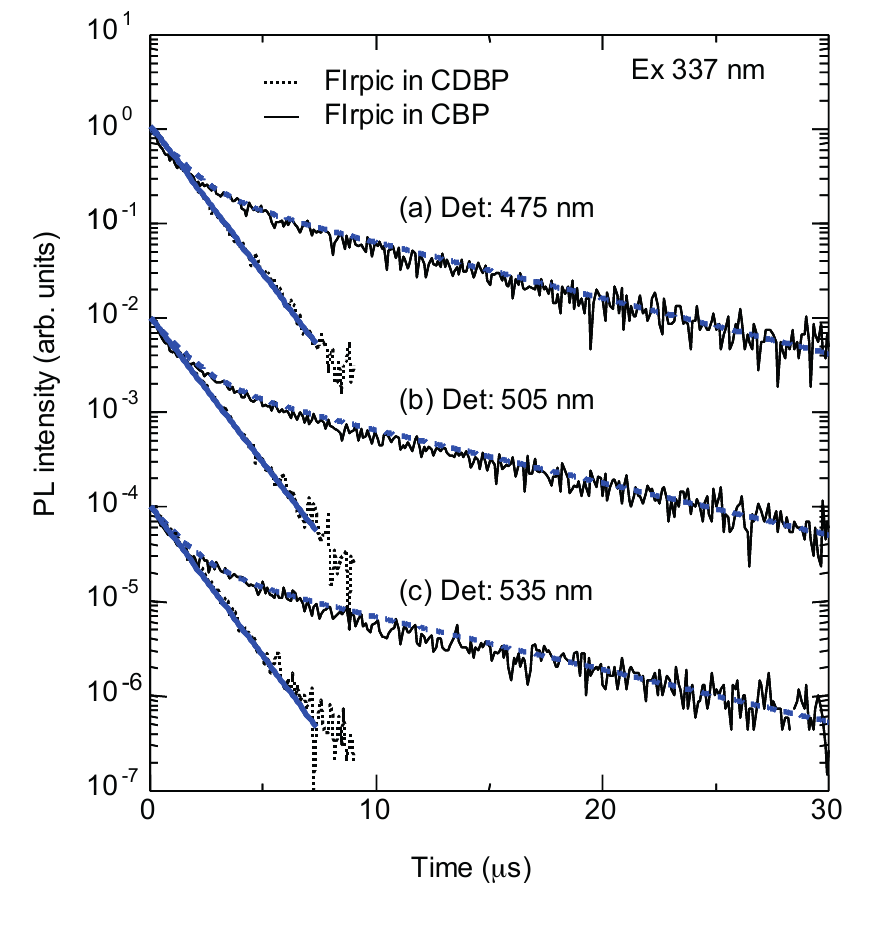}
\caption{\label{_Tokito2005_Fig6}(color online) Photoluminescence decay curves for two host materials (CBP and CDBP) doped with 3\,\% FIrpic each as taken at different detection wavelength. The blue lines correspond to mono- and biexponential decays. From \cite{Tokito2005}.} 
\end{figure}

One good example of the importance of the right choice of matrix material is given by \cite{Tokito2003}. They use the archetype blue phosphorescent emitter FIrpic ($T_1=2.7$\,eV) and compare its efficiency when embedded in either CBP ($T_1=2.6$\,eV) or CDBP ($T_1=3.0$\,eV). The photoluminescence (PL) transients of both CBP:FIrpic and CDBP:FIrpic (both doped with 3\,wt\%) are plotted in Figure \ref{_Tokito2005_Fig6}. While the CDBP system shows monoexponental decay with a time constant of 1.4\,\textmu s, the CBP:FIrpic PL decay shows a long-living delayed component which can be attributed to energy back transfers between host and guest, which lowers the PLQY [\cite{Kawamura2005}] and consequently the device efficiency. \cite{Tokito2003} prepared identical blue OLEDs based on FIrpic with either CBP or CDBP as host material that reach 5.1 and 10.4\,\% EQE (at 0.1\,mA\,cm$^{-2}$), respectively, showing the importance of excitonic confinement in case of phosphorescence. 

CDBP with a triplet level of 3.0\,eV is just one of many wide band gap materials suitable to host blue phosphorescent emitters. The most common host materials with high triplet energies are mCP ($T_1=2.91$\,eV [\cite{Kawamura2005}]), UGH2 ($T_1=3.5$\,eV [\cite{Lai2010, Giebink2006}]), CzSi ($T_1=3.01$\,eV [\cite{Tsai2006a}]), and TCTA ($T_1=2.81$\,eV [\cite{Reineke2009}]). 

		\subsubsection{\label{ConvHybrid2C}Conventional architectures: two-color devices}
White light can be mixed using two colors that are complementary in the sense that their straight connection in the CIE 1931 (cf. Fig. \ref{_CIE}) color space crosses the desired white point on the Planckian locus. Most of the research in this field used the archetype phosphorescent blue emitter FIrpic in connection with various emitters. With its rather light-blue emission corresponding to CIE coordinates of (0.17, 0.34) [\cite{Yeh2005}], the FIrpic spectrum is typically mixed with the emission of a red emitter (PL maximum $\sim600$\,nm) [\cite{Lei2006, Wang2009, Su2008, Kim2007a}]. \cite{Lai2010} have reported on white two-color OLEDs, where the common red emitter is replaced by a yellow emitter with reasonably high PLQY. However, as the yellow emission has CIE coordinates of $\sim(0.44, 0.53)$, it is not possible to cross the Planckian locus when FIrpic (and most other blue phosphors) are used as complementary blue emitter (cf. Fig. \ref{_CIE}). Thus, to use yellow phosphorescent emitters in a two-color approach requires deep-blue emitters with CIE coordinates $(<0.2, <0.2)$.

\begin{figure}[t]
\includegraphics[width=8.0cm]{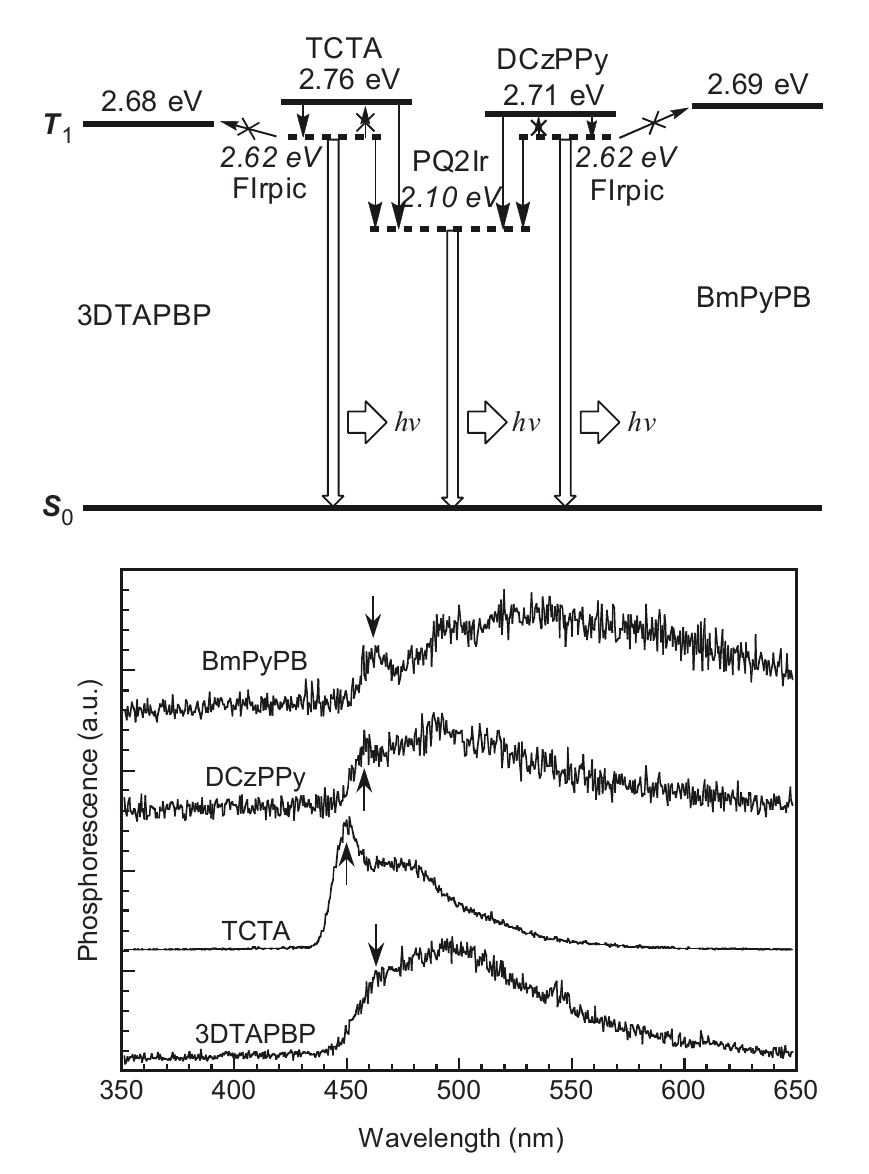}
\caption{\label{_Su2008_Fig3} Top: Triplet energy level diagram for the EML concept proposed by \cite{Su2008}. Bottom: Phosphorescence spectra of the non-emitting materials used for host and exciton blocking layers (cf. Top) as measured at 4.2\,K for the vacuum deposited films. Arrows indicate the estimated triplet levels.  From \cite{Su2008}.} 
\end{figure}

The highest device performance of two-color white OLEDs based on FIrpic has been reported by \cite{Su2008}. PQ2Ir is used as complementary red emitter in their study. Their EML is designed to form a strong carrier and exciton confining structure. Figure \ref{_Su2008_Fig3} shows the triplet energy diagram of their device structure and the phosphorescence spectra of the important materials used [\cite{Su2008}]. In order to confine triplet excitons of the blue emitter FIrpic ($T_1=2.62$\,eV), the authors composed the device structure solely with materials having higher triplet levels (cf. Fig. \ref{_Su2008_Fig3}). Their basic concept is based on a double EML layout [\cite{Zhou2002}] to pin the exciton generation zone to an interface in the center of the EML (TCTA/DCzPPy interface). This device, when doped with FIrpic only reaches the highest efficiencies reported to date for FIrpic-based OLEDs: 25\,\% EQE and 46\,lm\,W$^{-1}$ at 1000\,cd\,m$^{-2}$ [\cite{Su2008}]. In order to achieve white light, ultra-thin layers of TCTA and DCzPPy (0.25\,nm each) are doped with the red emitter PQ2Ir instead of FIrpic. Both its triplet and HOMO and LUMO levels function to achieve excitonic (cf. Fig. \ref{_Su2008_Fig3}) and charge carrier confinement at the TCTA:DCzPPy interface. The white device reaches 25\,\% EQE and 44\,lm\,W$^{-1}$ at 1000\,cd\,m$^{-2}$. The corresponding CIE coordinates are (0.335, 0.396) [$\alpha_{\text{CIE}}=+0.04$] with a CRI of 68.

The lack of green emission in the spectrum of two-color white devices results in poor color quality, where the color rendering index typically is limited to values of CRI\,$\sim70$ [\cite{Su2008, Wang2009}]. Furthermore, the two-color approach limits the luminous efficacy because the dip in the spectrum strongly overlaps with the response curve of the human eye $V(\lambda)$, which artificially lowers the luminance efficacy of radiation $K_\text r$ (cf. Sec. \ref{Color}).

The highest color quality of two-color phosphorescent white OLEDs has been reported by \cite{Chang2010a}. They use an iridium complex Ir(dfbppy)(fbppz)$_2$ as blue phosphorescent emitter combined with a red-emitting osmium heavy metal complex Os(bptz)$_2$(dppee). The blue emitter with PL maximum at 450\,nm and strong vibronic sidebands at approximately 480 and 520\,nm can alleviate the lack of green emission, resulting in a high CRI = 79.  At 100\,cd\,m$^{-2}$, a white device based on these emitters reached 6.8\,\% EQE and 10.0\,lm\,W$^{-1}$ with color coordinates of (0.324, 0.343), closely matching the Standard Illuminant E [$\alpha_{\text{CIE}}=0$].

	\subsubsection{\label{ConvHybrid3C}Conventional architectures: three-color devices}
\begin{figure}[t]
\includegraphics[width=8.0cm]{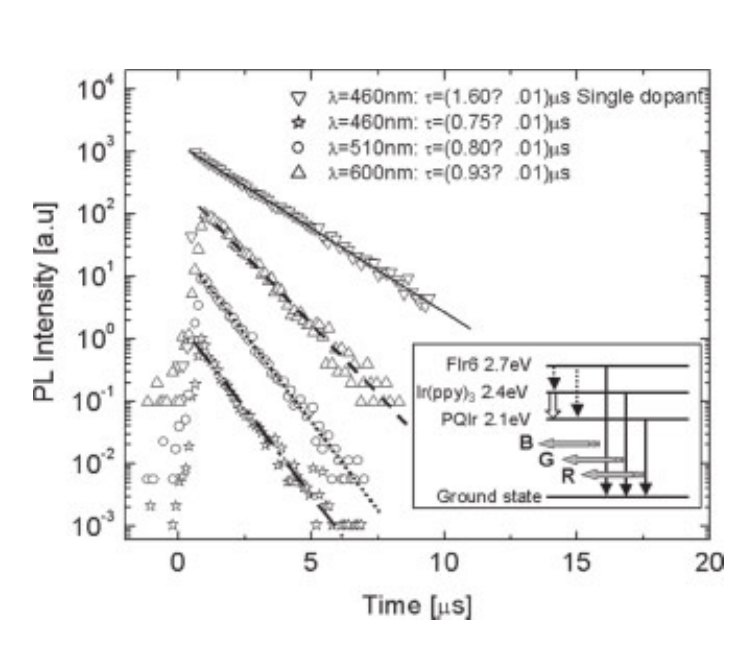}
\caption{\label{_DAndrade2004_Fig4} Photoluminescence decay curves spectrally resolved to show the transients of the three emitters [FIr6, Ir(ppy)$_3$, and PQIr] embedded in the common host UGH2. Inverted triangles refer to a blue system, i.e. UGH2:FIr6. Lines correspond to monoexponential decays. Inset shows the proposed energy transfer scheme. From \cite{DAndrade2004}.} 
\end{figure}

In order to increase the color quality and luminous efficacy of phosphorescent OLEDs, three primary colors need to be employed. \cite{DAndrade2004} reported on the first high efficiency three-color devices, based on FIr6 (0.1\,eV higher $T_1$ compared to FIrpic), Ir(ppy)$_3$, and PQIr. Based on earlier reports [\cite{Holmes2003}] on direct charge injection and trapping by FIr6 when dispersed into the inert wide band gap host UGH2 (band gap of 4.4\,eV), the EML (9\,nm in total) has been designed to host all three emitters simultaneously. With a high concentration of 20\,wt\%, FIr6 is used to capture both electrons and holes. The other two dopants are highly diluted into the system with 0.5 and 2\,wt\% for Ir(ppy)$_3$ and PQIr, respectively. The doping ratio of the green and red emitter is adjusted in a way that only parts of the FIr6 excitons are transferred to them. This excitation scheme is analyzed in time-resolved measurements, where the triple-doped film is excited with a short laser pulse and recorded in a streak-camera [\cite{DAndrade2004}], as shown in Figure \ref{_DAndrade2004_Fig4}. Additonally shown is the FIr6 transient signal for a single-doped UGH2:FIr6 system. By introducing the green and red emitters, the lifetime of FIr6 is reduced from 1.60 to 0.75\,\textmu s, clearly indicating the energy transfer occurring from FIr6 to lower energy triplet states if Ir(ppy)$_3$ and PQIr. Furthermore, the study shows that the Ir(ppy)$_3$ decay rate remains unchanged compared to solely Ir(ppy)$_3$-doped devices, indicating that the energy transfer from Ir(ppy)$_3$ to PQIr is weak. White devices based on the EML layout reach efficiencies of 7.5\,\% EQE and 11\,lm\,W$^{-1}$ at 1000\,cd\,m$^{-2}$ with color coordinates of (0.41, 0.46)\footnote{estimated to 1000\,cd\,m$^{-2}$.} [$\alpha_{\text{CIE}}=+0.06$] and a CRI = 78.
		
\begin{figure}[t]
\includegraphics[width=8.0cm]{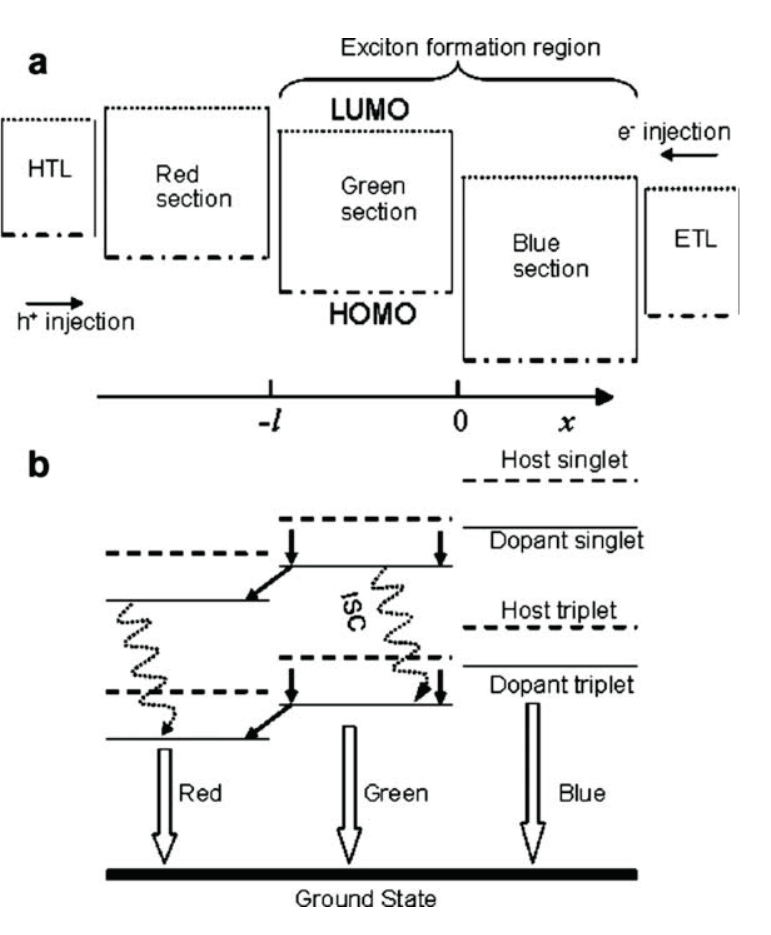}
\caption{\label{_Sun2008a_Fig1} a) Energy level diagram of the host materials incorporated in the three-section EML. Dotted lines are LUMO, dash-dotted lines are HOMO levels. The three host materials form a stepped energy barrier sequence for both electrons and holes. b) Proposed energy transfers within the EML. Solid arrows indicate energy transfer from host to dopant and from high to low energy dopants across an interface. Dotted curved arrows indicate ISC. Dashed lines are host singlet and triplet levels, solid lines are dopant singlet and triplet levels. From \cite{Sun2008a}.} 
\end{figure}

In contrast to the energy transfer excitation scheme of \cite{DAndrade2004}, \cite{Sun2008} proposed a different EML design with multiple exciton formation zones while using exactly the same emitter molecules. A scheme of their concept is shown in Figure \ref{_Sun2008a_Fig1}. Here, the red, green, and blue sub-EMLs are spatially separated, i.e. each sub-EML consists of a different host-guest system. The energy levels of the host materials are chosen to form a stepped energy barrier sequence for both charge carrier types.  The host materials are TCTA, mCP, and UGH2 as ordered from the hole injecting side of the device (cf. Fig.\ref{_Sun2008a_Fig1}). By the introduction of these moderate energy barriers, electrons and holes will accumulate at each of these interfaces where they can form excitons with the opposite carrier type. A detailed investigation of the exciton distribution within this multilayer emission layer is given by \cite{Sun2008a}. As indicated in Figure \ref{_Sun2008a_Fig1}, additional energy transfer from high to low energy phosphors can occur at the respective sub-EML interfaces. At 1000\,cd\,m$^{-2}$, a white OLED based on this concept yields efficiencies of 12.9\,\% EQE and 20\,lm\,W$^{-1}$. The color rendering index is high (CRI = 81) with color coordinates of (0.37, 0.41)  [$\alpha_{\text{CIE}}=+0.04$]. 

Finally, \cite{Wang2009a} combined the two concepts from above, discussing white phosphorescent OLEDs based on one common host with spatially different emitter doping, i.e. either sequential red/green/blue or red+green/blue. The authors used the wide gap material mCP as host material. In their report they carefully discuss the influences of charge distribution and carrier trapping leading to different effects on device efficiency and color quality. With a blended red/green- followed by a blue-doped mCP region, device efficiencies at 1000\,cd\,m$^{-2}$ are reported to be 13.6\,\% EQE with CIE coordinates of (0.39, 0.42) [$\alpha_{\text{CIE}}=+0.03$, CRI = 80]. Unfortunately, no luminous efficacy is reported at illumination relevant brightness. Similar studies on a combined red+green EML have been carried out by \cite{Seo2010}, however in this report two matrix materials (mCP and TPBi for blue and red+green) were employed.

The highest device efficiency of white OLEDs based on three phosphorescent emitters has recently been reported by \cite{Sasabe2010}. Their report is based on their earlier study of high efficiency two-color devices [\cite{Su2008}], emphasizing the need for deeper blue phosphorescent emitters to incorporate an additional green phosphor. A new  iridium carbene complex Ir(dbfmi)$_3$ has been introduced having a PL maximum at 445\,nm (2.79\,eV). This emitter has been used together with a new host material PO9, having a triplet level of 2.95\,eV. At an emitter concentration of 10\,wt\%, this PO9:Ir(dbfmi)$_3$ system has a very high PLQY of 70\,\%. The device concept is very similar to the two-color approach, basically using a blue phosphorescent OLED with additional ultra-thin layers for red and green. Here, the total EML is CBP:PQ$_2$Ir(dim) (1\,nm, 2\,wt\%)/CBP:Ir(ppy)$_3$ (1\,nm, 6\,wt\%)/PO9:Ir(dbfmi)$_3$ (10\,nm, 10\,wt\%). The only difference to the two-color device layout is the position of the lower energy phosphors that has been moved from the center to the side of the EML [cf. \cite{Su2008}]. The device based on this EML sequence reached very high efficiencies of 21.5\,\% EQE and 43.3\,lm\,W$^{-1}$ at 1000\,cd\,m$^{-2}$. Compared to the two-color OLEDs, the color quality is improved to a CRI = 80.2 with CIE coordinates of (0.43, 0.43) [$\alpha_{\text{CIE}}=+0.03$].

The highest color rendering index  to date for a three emitter system has been reported by \cite{Chang2010}. Their devices, optimized for emission close to Standard Illuminant E, have a very high CRI = 94 at 1000\,cd\,m$^{-2}$ with color coordinates of (0.322, 0.349) [$\alpha_{\text{CIE}}=+0.01$, 8\,\% EQE at 100\,cd\,m$^{-2}$].

		\subsubsection{\label{blueResonant}Resonant triplet level blue host-guest systems}
The results from the preceding sections have shown that it is generally possible to incorporate blue phosphorescent emitters in OLEDs to achieve efficient white light. However, in order to achieve excitonic confinement necessary for high PLQY, host materials with extremely wide band gap have to be employed. The use of high band gap materials like UGH2, mCP, or CzSi in turn increase the operating voltages of OLEDs, ultimately leading to reduced luminous efficacies. One way to circumvent this problem is to directly inject charges into the blue emitter, which then functions as charge carrying and emissive material [\cite{DAndrade2004}]. The additional transport functionality however may likely further decrease the operational stability of the blue phosphor -- which already is the bottleneck for realizing long-term stable phosphorescent white OLEDs.

Another route is to reduce the transport band gap by choosing a host material in a way that the triplet level of host and blue emitter are in resonance [\cite{Reineke2009a}]. 
This however introduces the general problem that a host-guest system with resonant triplet energies has a smaller PLQY in the mixed film. This effect is even more pronounced when the triplet level of the emitter is higher compared to the matrix material [\cite{Kawamura2005}]. In a resonant system, the excitons are free to move, so that the capture efficiency of excitons on the phosphor is reduced. A common signature of a resonant or endothermic ($T_{1,\text{host}}\le T_{1,\text{emitter}}$), is a delayed component in the transient signal. For instance, for a CBP:FIrpic system ($T_{1,\text{host}}=2.56\,\text{eV}<T_{1,\text{emitter}}=2.6\,\text{eV}$), the PL decay shows a noticeable delayed component as plotted in Figure \ref{_Tokito2005_Fig6}, which is attributed to back energy transfer between host and guest molecules [\cite{Adachi2001a}].

\begin{figure}[t]
\includegraphics[width=8.5cm]{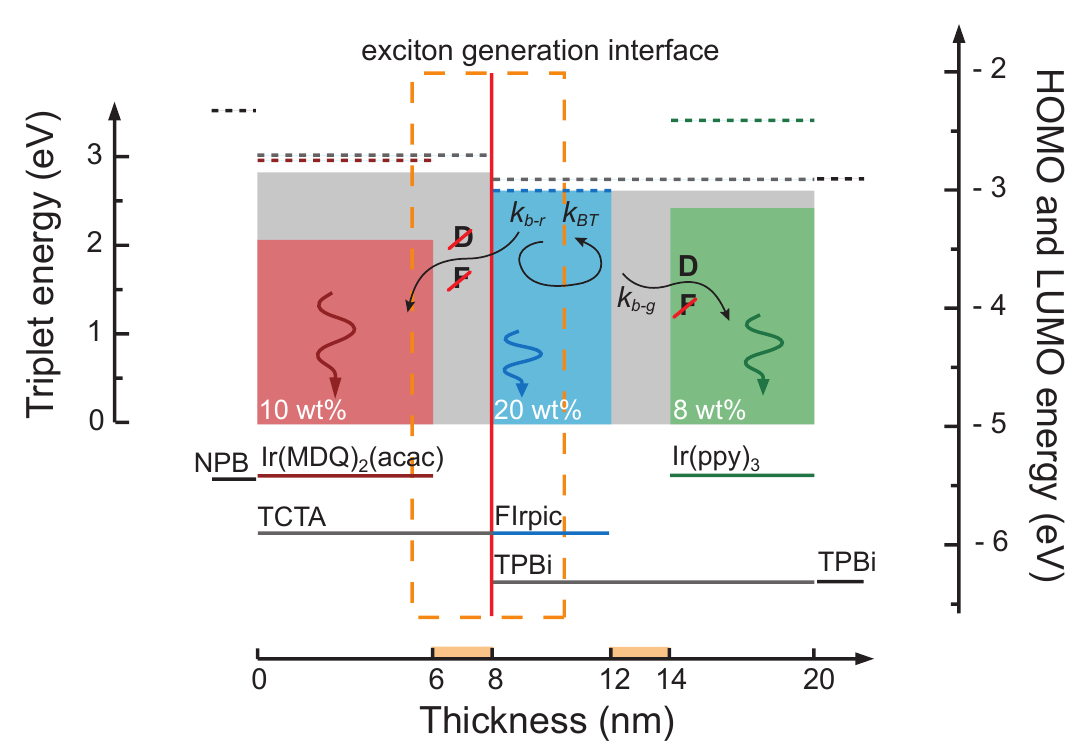}
\caption{\label{_Reineke2009_Fig1}(color online) Energy level diagram of the phosphorescent white emission layer concept proposed by \cite{Reineke2009a}. Dashed lines are LUMO, solid lines HOMO levels. The filled boxes indicate the respective triplet levels of host (grey) and emitter (colored) materials. Dashed yellow box indicates the exciton formation zone. D and F denote Dexter and F\"orster energy transfer, respectively. Furthermore, the rates refer to blue-to-red transfer $k_\text{b-r}$, back-transfer $k_\text{BT}$, and  blue-to-green transfer $k_\text{b-g}$. From \cite{Reineke2009a}.} 
\end{figure}	

Incorporating a resonant triplet energy blue system into a white EML is more complicated, because (i) the excitons are likely to escape the EML, where they might be transferred to quenching sites, and (ii) it must be taken into consideration that the resonant system is inherently less efficient compared to an exothermic system ($T_{1,\text{host}}>T_{1,\text{emitter}}$) [compare for instance the PLQY of FIrpic dispersed in either CBP or mCP ($T_1=2.91$\,eV) at 4.1\,mol\%: 55 and 98\,\% \cite{Kawamura2005}]. Thus, in order to achieve intense blue emission, the blue resonant EML must be made thicker to counteract the small PLQY of the film. \cite{DAndrade2002a} has used this CBP:FIrpic system in a three-color phosphorescent white device. Their CBP:FIrpic 6\,wt\% layer had a thickness of 20\,nm. Furthermore, it was located at the exciton generation zone adjacent to a NPB electron-blocking layer. NPB has a triplet level of 2.29\,eV [\cite{Goushi2004}], much lower than the emissive state of FIrpic, clearly functioning as an effective quenching channel for excitons freely moving in the CBP:FIrpic film. Thus, a white OLED based on this system reached a low device efficiency of 5.2\,\% EQE (maximum value) [\cite{DAndrade2002a}].

\cite{Cheng2006a} also used CBP:FIrpic in a three-color device, however they comprised the blue sub-EML in the center of the EML (spatially separated from the site of exciton generation [\cite{DAndrade2002a}]), sandwiched between doped CBP layers hosting either a green [Ir(ppy)$_3$] or a red [Ir(ppq)$_2$(acac)] phosphor (5\,nm each). In order to achieve sufficient blue emission, the central layer had to have a thickness of 30\,nm, where the less efficient endothermic CBP:FIrpic loses a great amount of excitons [\cite{Cheng2006a}]. Unfortunately, the authors do not state EQE values, making it hard to evaluate their data. 

\cite{Reineke2009a} improved this device concept by readdressing the emission layer design. The energy level diagram of their EML architecture is depicted in \ref{_Reineke2009_Fig1}. In contrast to the device reported by \cite{Cheng2006a} based on the common host CBP, the EML is based on the double EML concept, incorporating a hole and an electron-transporting host material to locate the exciton generation to the center of the EML [\cite{Zhou2002, Reineke2009a}]. Instead of using CBP, TPBi is used as the electron transporting host material having a triplet energy of 2.6\,eV, exactly matching the FIrpic $T_1$ state. TPBi is well-known to form an efficient double EML system together with the hole-transporting matrix TCTA ($T_1=2.83$\,eV) [\cite{Huang2006c, Meerheim2008a}]. Placing the blue sub-EML at the position of exciton generation (cf. Fig. \ref{_Reineke2009_Fig1}), the total layer thickness can be reduced because the exciton density is accordingly higher. Thus, the TPBi:FIrpic layer is only 4\,nm thick. Furthermore, the concentration of FIrpic is increased to 20\,wt\%, because in a resonant system, the highest PLQY is obtained at higher concentrations\footnote{Upon a further increase of emitter concentration, the PLQY decreases again as dominated by concentration quenching [\cite{Kawamura2006}].} [\cite{Kawamura2005}] as a result of a higher probability that an exciton can find a site for recombination. For instance, the PLQY of TPBi:FIrpic is increased from 13\,\% to 32\,\% as the concentration is increased from 1.7 to 10\,wt\% [\cite{Reineke2009a}]. For comparison, the exothermic system TCTA:FIrpic at a concentration of 1.7\,wt\% yield a PLQY = 81\,\%.

\begin{figure}[t]
\includegraphics[width=8.5cm]{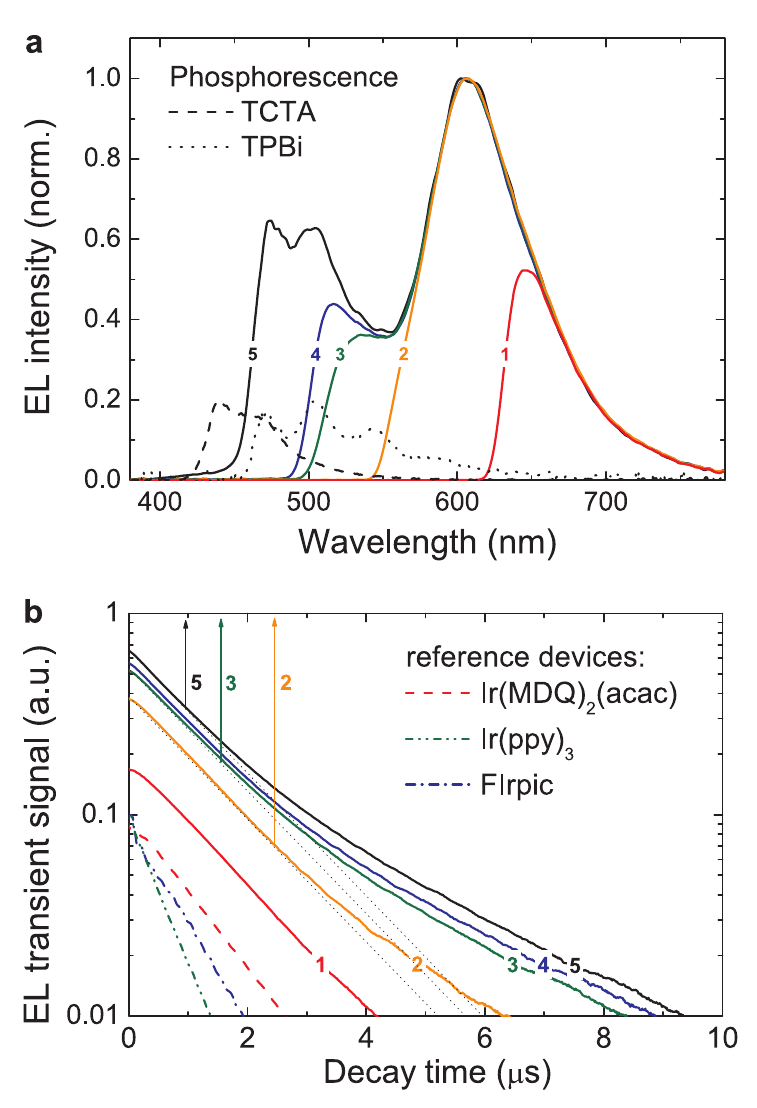}
\caption{\label{_Reineke2009_Fig2}(color online) a) EL spectra of Device B from \cite{Reineke2009a} as obtained through different color filters (numbered from 1: solely red emission to 5: complete emission spectrum). Dashed and dotted lines are phosphorescence spectra of the hosts TCTA and TPBi, respectively, as measured at 77\,K. b) EL decay curves measured for the respective spectra of a). Arrows indicate the time when a slower component sets in. Additionally, the decay curves of reference monochrome devices are shown (dashed, dotted, dash-dotted) for red, green and blue.  From \cite{Reineke2009a}.} 
\end{figure}		

The red emitter Ir(MDQ)$_2$(acac) is doped into the TCTA matrix, the green Ir(ppy)$_3$ is embedded also into the TPBi host, however spatially separated from the exciton generation interface (cf. Fig. \ref{_Reineke2009_Fig1}). In order to suppress complete energy transfer $k_\text{b-r}$ from FIrpic to Ir(MDQ)$_2$(acac), a thin intrinsic interlayer of TCTA (2\,nm) is inserted acting as triplet energy barrier and spacer [\cite{Kawamura2006}] to suppress Dexter- and F\"orster-type energy transfer (cf. Sec. \ref{Energytransfer}), respectively. To reduce Förster-type energy transfer $k_\text{b-g}$ from FIrpic to Ir(ppy)$_3$, which would artificially reduce the blue intensity, a 2\,nm thin intrinsic TPBi interlayer -- large enough exceed the Förster radius of FIrpic [\cite{Kawamura2006}] --  is inserted between blue and green sub-EML. 

Time- and spectrally-resolved measurements performed by \cite{Reineke2009a} on a resonant triplet energy system provide experimental evidence for the energy back transfer $k_\text{BT}$. The results are plotted in Figure \ref{_Reineke2009_Fig2}. Color filters are used to alter the emission of a white OLED based on the above concept from solely red emission stepwise to the full spectrum [spectra 1 to 5 in Fig. \ref{_Reineke2009_Fig2} a)]. The time-decay of the transmitted spectrum following an EL excitation pulse  is recorded as shown in Fig. \ref{_Reineke2009_Fig2} b). Additionally, response curves of monochromatic devices are plotted for comparison. A monoexponential decay is observed, when only the red part of the spectrum is transmitted, nicely agreeing with the time constant of the reference device (time constant 1.4\,\textmu s). With increasing transmission, a second delayed component with a time constant of 3.0\,\textmu s is observed, much longer than any of the reference decay signals. This delayed signal can be attributed to the energy back transfer $k_\text{BT}$, as it is linked to the blue emission of FIrpic. 

The motivation for the use of reduced band gap materials is to reduce the operation voltage of the device. With the EML structure of \cite{Reineke2009a}, very low voltages of 3.22\,V and 3.95\,V are obtained for 1,000 and 10,000\,cd\,m$^{-2}$, respectively, operating close to the thermodynamic limit [\cite{Su2010, He2004}]. The corresponding device efficiencies are 13.1\,\% EQE and 30\,lm\,W$^{-1}$ at 1,000\,cd\,m$^{-2}$ with CIE color coordinates of (0.45, 0.47) [$\alpha_{\text{CIE}}=+0.06$, CRI = 80].

Even if this structure reaches very high efficiencies, the color quality is limited due to the use of the light-blue emitter FIrpic. Here, it is not possible to reach emission with CIE coordinates close to the Planckian locus, i.e. $\alpha_{\text{CIE}}\equiv 0$, with a well balanced contribution from all three emitters (cf. Sec. \ref{Color}). In order to improve the color quality, \cite{Weichsel2012} replaced the blue sub-EML TPBi:FIrpic by an electron-transporting mixed system SPPO1:FIr6, where the triplet energy of SPPO1 ($T_1=2.8$\,eV) is slightly higher compared to FIr6 ($T_1=2.7$\,eV), still coming close to be a resonant system. Furthermore, a fourth phosphor emitting in the yellow region, i.e. Ir(dhfpy)$_2$(acac), is incorporated in the EML structure, being co-doped with Ir(ppy)$_3$ into the SPPO1 host [\cite{Weichsel2012}]. An optimized device based on these changes reaches 10.0\,\% EQE and 17.4\,lm\,W$^{-1}$ at 1000\,cd\,m$^{-2}$. More importantly, the CIE color coordinates changed to (0.444, 0.409) with a CRI = 81.9, representing a Planckian radiator ($\alpha_{\text{CIE}} = 0$) emitting at Standard Illuminant A.

		\subsubsection{\label{Excimer}Single dopant combined monomer/excimer emission}

\begin{figure}[t]
\includegraphics[width=7.7cm]{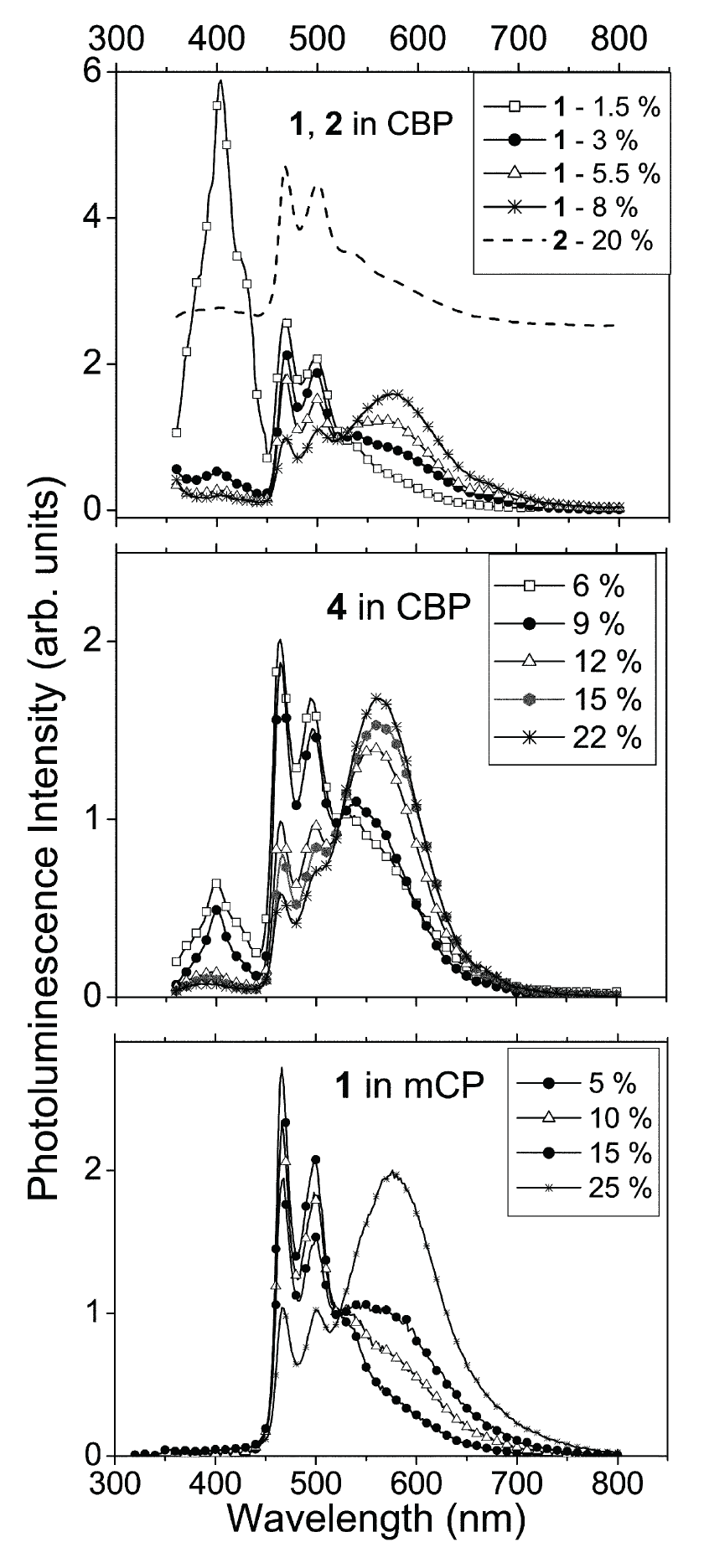}
\caption{\label{_Adamovich2002_Fig2} Photoluminescence spectra of platinum complexes reported by \cite{Adamovich2002} as doped either in CBP (top and middle) or mCP (bottom) host materials under a variation of the emitter concentration. The blue emission at approximately 400\,nm is the CBP host fluorescence. \textbf{1}, \textbf{2}, and \textbf{4} refer to different molecular structures of the platinum complexes.  From \cite{Adamovich2002}.} 
\end{figure}

From the device engineering point of view, it is always desirable to simplify the OLED structure. However, especially for white OLEDs, the number of layers needed solely for the emission layer can be as high as five [\cite{Reineke2009a}]. This is mainly a result of the need to address all differently emitting molecules within the EML. 
		
\cite{DAndrade2002} observed efficient electrophosphorescent excimer emission from a organometallic platinum (Pt) complex FPt1. Here, Pt-Pt coupling [\cite{Zheng1998, Connick1996}] forms emissive excimer states at longer wavelength compared to the monomer emission of the corresponding isolated molecule. Together with the blue emitter FIrpic, white emission could be realized based on FIrpic  and FPt1-excimer phosphorescent emission. Because the FPt1 excimer is also a triplet emitting state, 100\,\% internal quantum efficiency in OLEDs is possible based on this approach. \cite{DAndrade2002} even suggest a white OLED solely based on combined monomer/excimer emission of the similar platinum complex FPt2. However, the EL spectrum also comprised a strong peak attributed to NPB hole-transport layer, which strongly limits the device efficiency.  

\cite{Adamovich2002} picked up the general concept of combined monomer/excimer electrophosphorescence investigating further variants of the platinum FPt1 complex. Photoluminescence of three different emitters having small ligand variations (denoted as \textbf{1}, \textbf{2}, and \textbf{4}) are shown in Figure \ref{_Adamovich2002_Fig2}. The relative intensities of the high-energy monomer and the long-wavelength excimer bands are continuously altered as a function of doping concentration, which determines the fraction of excimers formed. Thus, this approach offers a simple route to realize broadband, white emission from one single molecule by adjusting its doping concentration within an appropriate host material, strongly simplifying the device structure. For a device based on FPt1 doped in mCP host material, 4.3\,\% EQE could be obtained at 500\,cd\,m$^{-2}$. A more detailed investigation of the exciton formation and trapping in such a device is given by \cite{DAndrade2003}, discussing a mCP:FPt2 system. 

\begin{figure}[t]
\includegraphics[width=8.5cm]{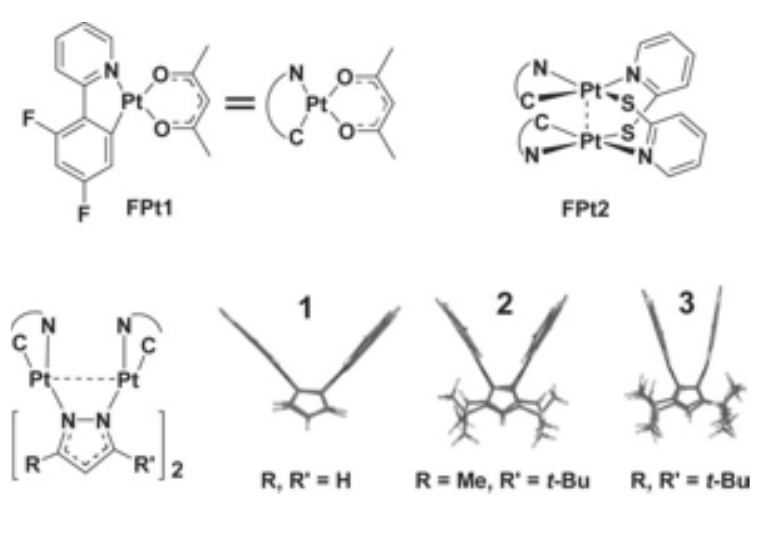}
\caption{\label{_Ma2006_Fig1} Molecular structures of the binuclear platinum complexes (\textbf{1}, \textbf{2}, and \textbf{3}) of \cite{Ma2006}. Additionally, the reference structures are shown at the top [\cite{Koshiyama2004}]. The Pt-Pt distances for \textbf{1}, \textbf{2}, and \textbf{3} are 3.376, 3.064, and 2.834\,\AA, respectively, corresponding to PL maxima at 466, 546, and 630\,nm. From \cite{Ma2006}.} 
\end{figure}

By further optimizing the host (using 26mCPy, a mCP derivative) and adjacent blocking materials, \cite{Williams2007} could further improve the device efficiency based on the emitter FPt1. At 500\,cd\,m$^{-2}$, an EQE of 15.9\,\% was reached for an EML doped with approximately 12\,wt\% of FPt1. This corresponds to 12.6\,lm\,W$^{-1}$ with CIE coordinates of (0.46, 0.47) [$\alpha_{\text{CIE}} =+0.07$] and a CRI = 69.

The effect of the heavy-metal atom coupling [\cite{Zheng1998, Connick1996}], which has to date only been effectively observed for platinum-cored emitters, leading to formation of excimer states, has nicely been shown by \cite{Ma2006}. In their study, they investigated binuclear platinum complexes as shown in Figure \ref{_Ma2006_Fig1} designed to have different Pt-Pt spacing, by that altering the strength of their coupling. For compound \textbf{1} with a Pt-Pt distance of 3.376\,\AA, solely the monomeric emission in the blue spectral region is observed, while compound \textbf{3} with a spacing of 2.834\,\AA\ shows solely red excimer emission. The EQE of the devices range between approximately 4 -- 6\,\% for blue, green, and red emission from compound  \textbf{1}, \textbf{2}, and \textbf{3}, respectively [\cite{Ma2006}]. For neat films of compound \textbf{1}, red emission is also observed, which is attributed to emission originating from distorted complexes with compressed Pt-Pt distances [\cite{Ma2006}]. 

\begin{figure}[t]
\includegraphics[width=8.5cm]{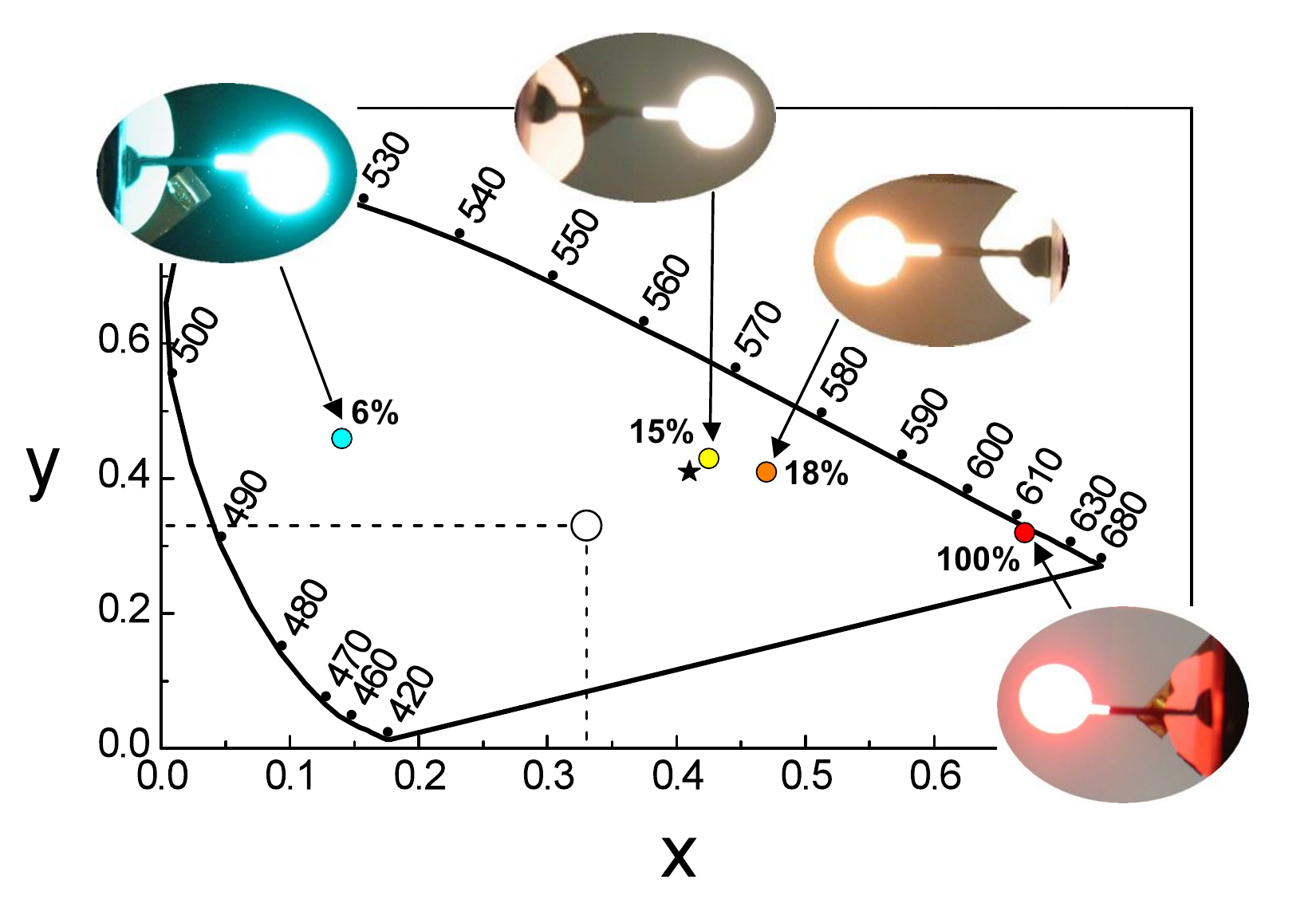}
\caption{\label{_Cocchi2007_Fig2}(color online) CIE 1931 chromaticity diagram showing the electroluminescence color coordinates of devices differing in the Pt$L^2$Cl emitter concentration. Open white circle indicates the Standard Illuminant E, the star refers to color coordinates of a specific incandescent lamp, the authors used for comparison [(0.41, 0.41)]. From \cite{Cocchi2007}.} 
\end{figure}

Based on these binuclear platinum complexes, it is possible to realize white emission  by combining two or more emitters with different Pt-Pt spacing, whereas the emission of the respective phosphor does not sensitively depend on the dopant concentration [\cite{Ma2006}], as seen for the previous reports by \cite{DAndrade2002, Adamovich2002}. The other alternative is a combination of compound \textbf{1} in dilute form (doped in mCP) and neat film, arranged in a dual layer architecture [\cite{Ma2006}]. Both approaches were exemplarily shown by \cite{Ma2006}, reaching maximum EQE values of 7.7\,\% and 4.2\,\% for either mCP:\textbf{1}/mCP:\textbf{3} or mCP:\textbf{1}/neat \textbf{1} EML layouts.

\cite{Cocchi2007} introduced an improved N$^{\wedge}$C$^{\wedge}$N-coordinated platinum (II) complex for this monomer/excimer approach. OLEDs based in this Pt$L^2$Cl emitter with different concentrations, doped in a mixed host system CBP:OXA, nicely sweep the CIE color space from light-blue (low concentration) to red (high concentration), as shown in Figure \ref{_Cocchi2007_Fig2}. At a Pt$L^2$Cl concentration of 15\,\%, CIE coordinates of (0.43, 0.43)  [$\alpha_{\text{CIE}} = +0.03$] are obtained. The corresponding efficiencies are 13.5\,\% EQE and 12.6\,cd\,A$^{-1}$ at 1000\,cd\,m$^{-2}$ [\cite{Cocchi2007}]. Based on the same emitter Pt$L^2$Cl, \cite{Kalinowski2007} improved the color quality of white OLEDs by combining the monomer/excimer emission with an additional exciplex emission that occurs between the hole-transporting material m-MTDATAs HOMO and Pt$L^2$Cls LUMO, filling the spectral gap in the green region. However, with 6.5\,\% at 500\,cd\,m$^{-2}$, the EQE is much lower compared to the devices presented by \cite{Cocchi2007}.

\begin{figure}[t]
\includegraphics[width=8.5cm]{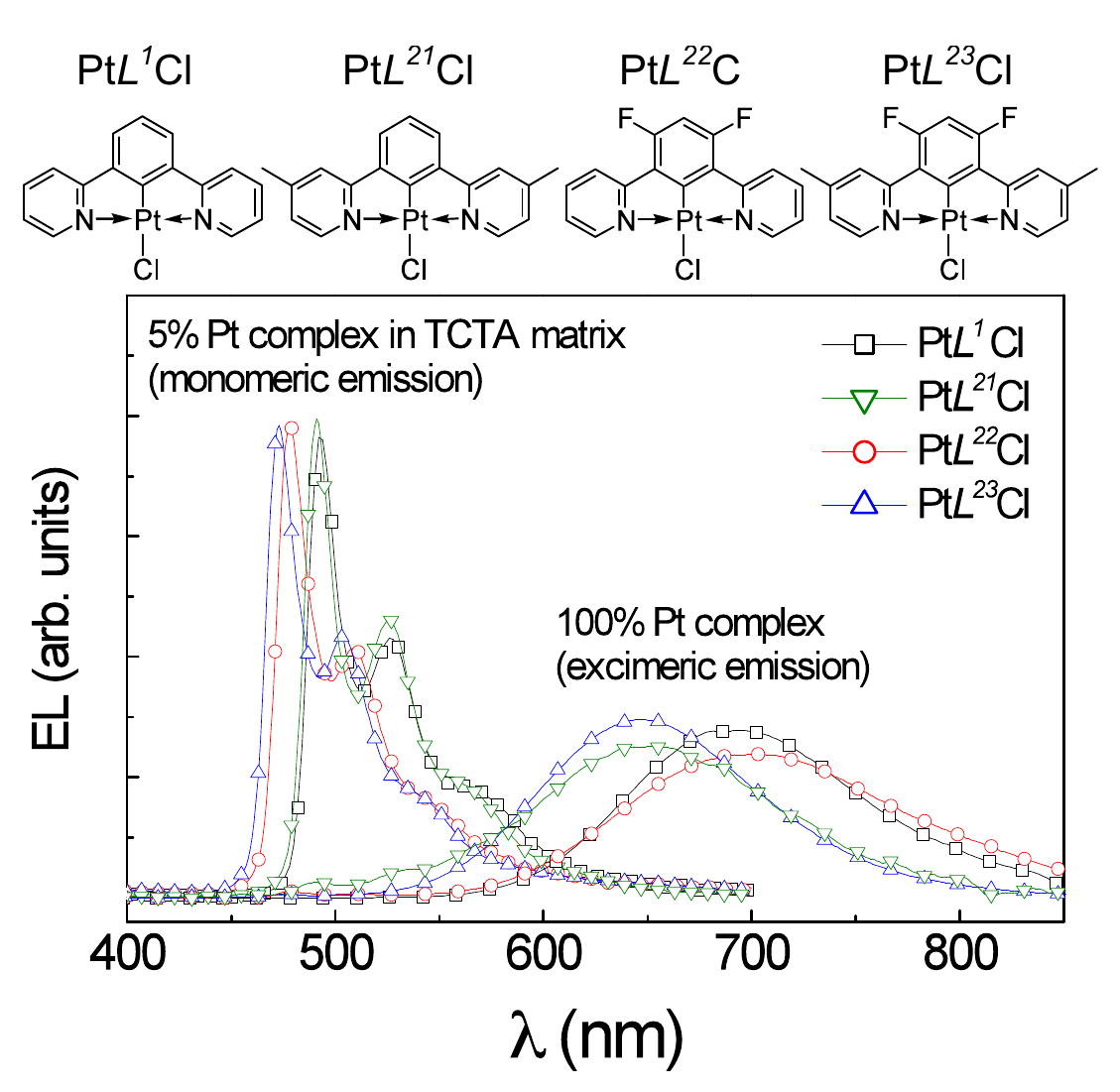}
\caption{\label{_Cocchi2009_Fig1}(color online) Top: Molecular structures of different emitters discussed in the report of \cite{Cocchi2009}. Corresponding EL spectra of all emitters obtained for a 5\,wt\% host-guest system (TCTA host) and a neat film EML formed by the platinum complexes. Each spectrum is normalized to the integrated intensity. From \cite{Cocchi2009}.} 
\end{figure}

\begin{table*}[t]
\caption{\label{tab:summary_SMOLED}Summary of selected, high performance devices based on different device concepts as discussed in Section \ref{whiteSMOLED}. Device efficiencies are maximum values, additional values at higher brightness/specific current density may be given in parenthesis.} 
\begin{ruledtabular} 
\begin{tabular}{lccccc}
Concept & $\eta_\text{EQE}$\,[\%] & $\eta_\text{C}$\,[cd\,A$^{-1}$] & $\eta_\text{L}$\,[lm\,W$^{-1}$] & CIE (x, y); $\alpha_{\text{CIE}}$ &   Reference\\
\hline
\textbf{Fluorescent}&&&&&\\ 
\multicolumn{1}{r}{fluorescent emitters}&[4.7]\footnotemark[1]&[10.9]\footnotemark[1]&11.2 [6.0]\footnotemark[1]&(0.329, 0.353); $+0.02$&\cite{Tsai2006}\\ 
\multicolumn{1}{r}{}&5.6 [5.2]\footnotemark[1]&14.0&9.2 [4.8]\footnotemark[1]&(0.332, 0.336); $0$&\cite{Yang2011}\\ 
\multicolumn{1}{r}{fluorescent emitters (with exciplexes)}&--&--&9.0&(0.31, 0.36); $+0.02$&\cite{Tong2007}\\

\textbf{Hybrid fluorescent/phosphorescent}&&&&&\\ 
\multicolumn{1}{r}{conventional}&[8.0]\footnotemark[2]&--&[13.7]\footnotemark[2]&(0.47, 0.42); $+0.02$&\cite{Schwartz2006}\\ 
\multicolumn{1}{r}{}&11.0 [10.8]\footnotemark[3]&--&22.1 [14]\footnotemark[3]&(0.41, 0.40); $+0.02$&\cite{Sun2006}\\ 
\multicolumn{1}{r}{phosphor sensitized fluorescence}&8.5&--&18.1&(0.38, 0.42); $+0.03$&\cite{Kanno2006}\\ 
\multicolumn{1}{r}{triplet harvesting}&[26]\footnotemark[2]\footnotemark[4]&--&[33]\footnotemark[2]&(0.506, 0.422); $<+0.01$&\cite{Rosenow2010}\\ 

\textbf{Phosphorescent}&&&&&\\ 
\multicolumn{1}{r}{conventional: two-color}&[25]]\footnotemark[2]&--&[44]]\footnotemark[2]&(0.335, 0.396); $+0.04$&\cite{Su2008}\\ 
\multicolumn{1}{r}{conventional: three-color}&21.6 [21.5]\footnotemark[2]&49.9 [49.6]\footnotemark[2]&59.9 [43.3]\footnotemark[2]&(0.43, 0.43); $+0.03$&\cite{Sasabe2010}\\ 
\multicolumn{1}{r}{resonant triplet level blue}&[13.1]\footnotemark[2]&--&[30]\footnotemark[2]&(0.45, 0.47); $+0.06$&\cite{Reineke2009a}\\ 
\multicolumn{1}{r}{combined monomer/excimer}&[16.6]\footnotemark[3]&--&[9.6]\footnotemark[3]&(0.42, 0.38); $-0.02$&\cite{Cocchi2010}\\ 
\end{tabular} 
\end{ruledtabular} 
\footnotetext[1]{at 10\,mA\,cm$^{-2}$} 
\footnotetext[2]{at 1000\,cd\,m$^{-2}$}
\footnotetext[3]{at 500\,cd\,m$^{-2}$}
\footnotetext[4]{Two-unit stacked device -- 200\,\% internal quantum efficiency limit [\cite{Rosenow2010}].}
\end{table*}

Further studies by \cite{Cocchi2009} discuss the influence of the ligand structure of the emitters with the general structure Pt$L^x$Cl on their photophysical properties. The corresponding chemical structures of emitter series is shown in Figure \ref{_Cocchi2009_Fig1} (also showing both the monomeric and excimeric PL). While the PLQY of the monomer emission from compounds Pt$L^{21}$Cl to Pt$L^{23}$Cl are comparable in the range of 70 -- 90\,\%, the PLQY of the neat film drastically increases from 5\,\% (Pt$L^{21}$Cl) to 65\,\% (Pt$L^{23}$Cl). OLEDs based on the Pt$L^{23}$Cl with high PLQY in the neat film reach very high external quantum efficiencies in the range of 15--18\,\% at 500\,cd\,m$^{-2}$, slightly depending on the emitter concentration [\cite{Cocchi2009}]. For instance at a concentration of 20\,\% Pt$L^{23}$Cl doped in the host TCTA, the OLEDs achieve 14.9\,\% EQE and 8.3\,lm\,W$^{-1}$ at 500\,cd\,m$^{-2}$ with CIE coordinates of (0.45, 0.38) [$\alpha_{\text{CIE}} = -0.02$]. 

In \cite{Cocchi2010}, the authors discuss in detail the mixing of molecular excitonic and excimeric phosphorescence to alter efficiency and color of the devices based on Pt$L^x$Cl complexes. In this report, the authors also introduce another similar Pt complex, Pt$L^{21}$Cl, which can be used in devices reaching 16.6\,\% EQE and 9.6\,lm\,W$^{-1}$ at 500\,cd\,m$^{-2}$. The corresponding CIE coordinates are (0.42, 0.38) [$\alpha_{\text{CIE}} = -0.02$].  Furthermore, a comprehensive study on the high brightness non-linearities, i.e. exciton quenching leading to the efficiency roll-off (cf. Sec. \ref{Rolloff}), is given by \cite{Kalinowski2010}. 

The highest color rendering index based on the monomer/excimer approach has been reported by \cite{Zhou2009} for a Pt-Ge emitter doped into CBP host material. At a high emitter concentration of 10\,wt\% of Pt-Ge, CIE coordinates of (0.354, 0.360)  are obtained with a very high CRI = 97. Note that this device qualifies as Planckian radiator ($\alpha_{\text{CIE}} = 0$). The corresponding peak EQE value is 4.13\,\% [\cite{Zhou2009}].

\subsection{\label{Summary_SMOLED}Summary}

Table \ref{tab:summary_SMOLED} summarizes the key figures of high quality devices based on the various concepts discussed in this Section \ref{whiteSMOLED}. The external quantum efficiencies of the concepts listed from top to bottom noticeably increase from fluorescence to fully phosphorescence-based white OLEDs with hybrid concepts ranking at an intermediate efficiency level. Table \ref{tab:summary_SMOLED} shows that both fluorescent and phosphorescent OLEDs have been demonstrated to reach their expected EQE level of 5\,\% and 20\,\% (cf. Sec. \ref{FluoPhos}), respectively.
	
In order to be competitive with existing light sources [\cite{Steele2007}], OLEDs need to be designed to allow the highest possible internal quantum efficiency. Thus clearly, it is unlikely that fluorescent devices, with approximately 75\,\% recombination losses [\cite{Segal2003}] within the device, will be able to compete with phosphorescence based designs.

The above discussion has shown that the highest possible device efficiencies need sophisticated, sometimes highly complex device layouts (cf. Secs. \ref{ConvHybrid}, \ref{smFluor_sensitizer}, \ref{ConvHybrid2C}, \ref{ConvHybrid3C}, and \ref{blueResonant}). These designs are not desirably for upscaling the device production to reasonable OLED panel sizes. Thus, devices offering high efficiency and large-area controllable device architectures are the concepts of choice.

Even though the first reports on triplet harvesting (cf. Sec. \ref{TripHarv}) employed rather complex emission layer designs [\cite{Schwartz2007, Schwartz2009}], the layer complexity has recently been reduced greatly. \cite{Rosenow2010} has introduced a triplet harvesting EML consisting of two simple 5\,nm thick sub-layers, offering great reproducibility.

\begin{figure*}[t]
\includegraphics[width=15cm]{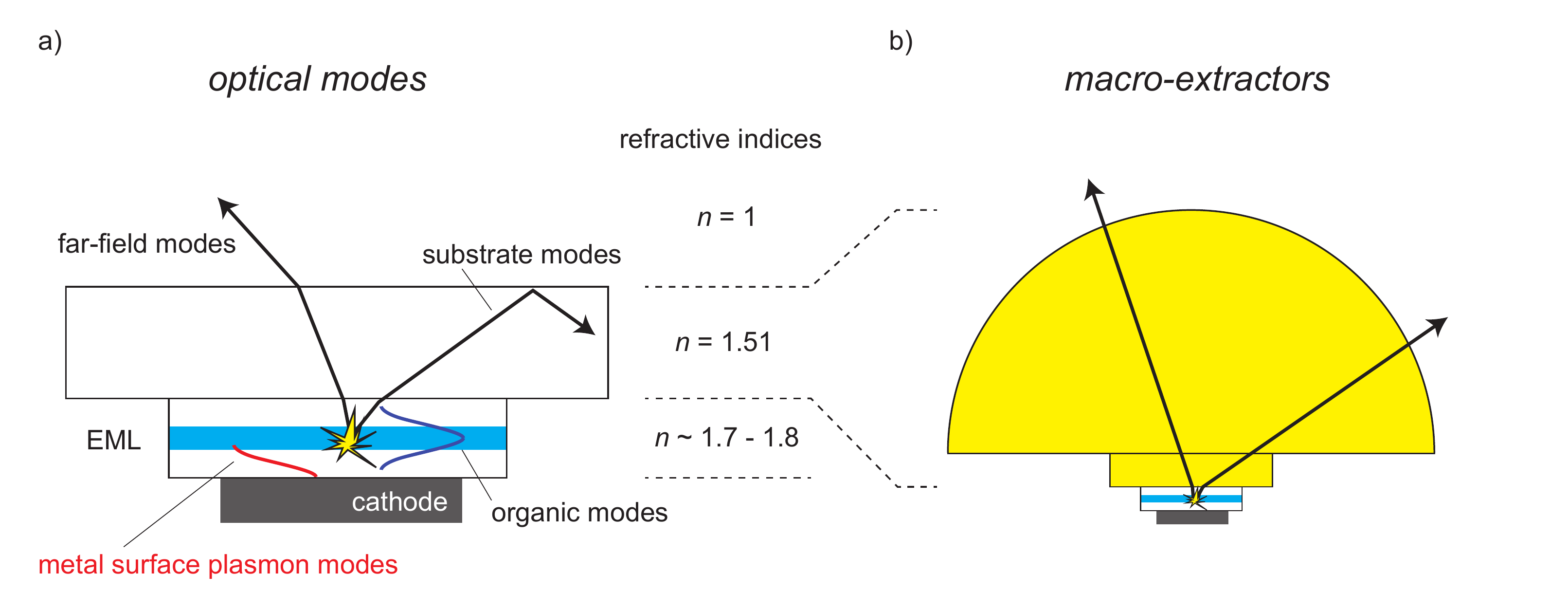}
\caption{\label{_outcoupling}(color online) a) Different light modes in a conventional bottom-emitting OLED. Typical refractive indices of the different OLED sections are given. In this configuration, only the far-field modes reach the observer. Substrate and organic modes are trapped in the device, where they dissipate. In addition, the emitting dipoles can couple to surface plasmon modes of the metal cathode, which decrease exponentially with distance. b) Application of a macro-extractor matching the refractive index of the substrate used. Here, all modes that are coupled to the substrate can be extracted to air.} 
\end{figure*}

Another very attractive concept promising high efficiency is the combined monomer/excimer phosphorescence (cf. Sec. \ref{Excimer}). Especially the possibility to design a white OLED based on a single emitter that is dispersed into a matrix material at a certain concentration offers unmet simplicity. The reports have shown (cf. Tab. \ref{tab:summary_SMOLED}) that this approach can reach similar high EQE values than conventional phosphorescent devices. To date, their corresponding luminous efficacies are smaller than the corresponding values of other concepts. This is mainly due to the superior electrical performance of the respective devices [\cite{Reineke2009a, Su2008, Sasabe2010}].


\section{\label{Outcoupling}Concepts for improved light outcoupling}
Many of the reports from the above sections show results that come rather close to 100\,\% internal quantum efficiency.  Still, the external quantum efficiency of conventional OLEDs are limited to $20-25$\,\% [\cite{Adachi2001b, Su2008, Sasabe2010}], which is due to the thin-film layered structure of the OLED, introducing trapped light modes. 

Figure \ref{_outcoupling} shows a scheme of an OLED's cross-section, illustrating the different light modes [\cite{Lu2002}]. Organic materials used for the functional layers in the device typically have refractive indices of $n_\text{org}\sim 1.7-1.8$ [\cite{Greiner2007}]. Conventionally, standard glass is used as transparent substrates with a refractive index of $n_\text{sub}=1.51$, forming an optical interface between organics and substrate. Due to the difference in refractive indices, total internal reflection (TIR) occurs at that interface, leading to a noticeable portion of light being waveguided in the organic layers. Here, the critical angle $\theta_\text c$ of TIR is in the range of $57-63^\circ$, depending on the actual $n_\text{org}$.  Ultimately, these modes dissipate in the system. Similarly, the difference in refractive indices between glass substrate and air introduces losses due to TIR ($\theta_\text c=41.5^\circ$), as so-called substrate modes are formed.

In addition to organic and substrate modes, the coupling of the radiating dipoles to the plasmon states of the metal cathode is another severe loss channel in OLEDs. The field of the metal surface plasmon modes decay exponentially with distance (cf. Fig. \ref{_outcoupling}). Thus, the efficiency of the emission is strongly decreased, if the EML is placed in the proximity of a metal layer.

\begin{figure}[t]
\includegraphics[width=8.5cm]{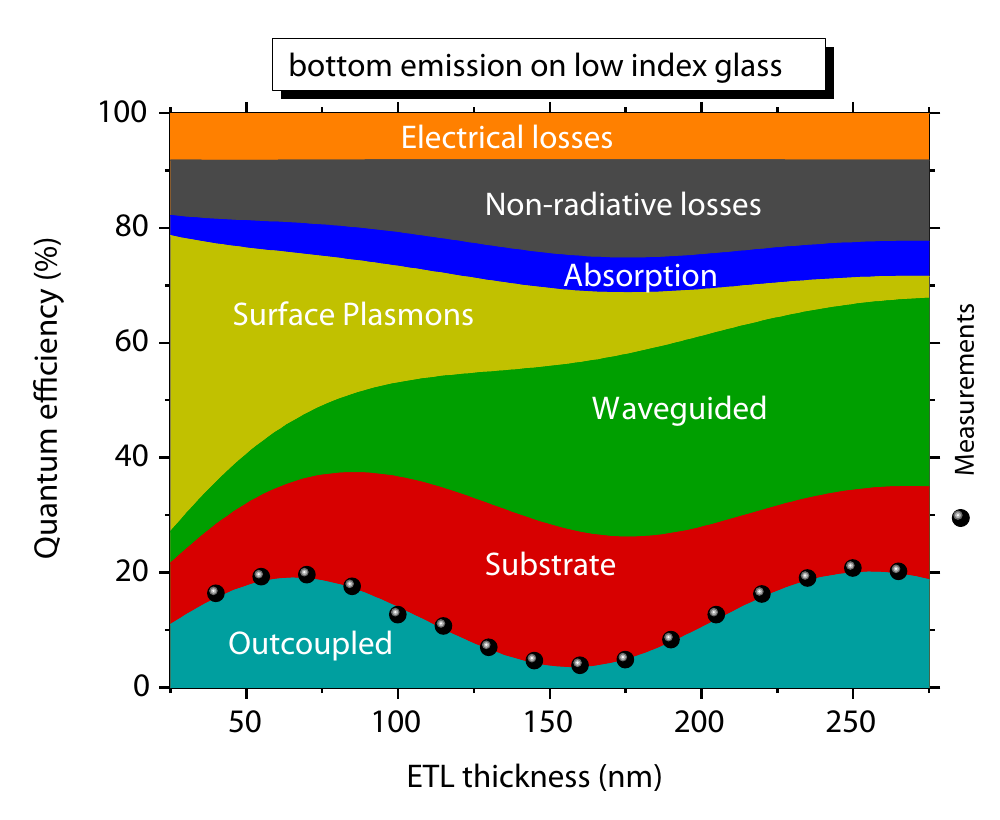}
\caption{\label{_Meerheim2010_Fig3}(color online) Distribution of light modes and loss channels of a red phosphorescent OLED as a function of electron-transport layer (ETL) thickness that spaces the EML from the metal cathode. Dots are measurements points. Calculation is based on a comprehensive optical OLED model established by \cite{Furno2010, Furno2012}. From \cite{Meerheim2010}.} 
\end{figure}

Based on a comprehensive optical model [\cite{Furno2010, Neyts1998, Furno2012}], \cite{Meerheim2010} have analysed the different loss channels in a model OLED comprising a red phosphorescent emitter. In their study, they varied the thickness of the electron transport layer (ETL) to map the first and second field antinodes. The quantification of the loss channels as a function of ETL thickness is shown in Figure \ref{_Meerheim2010_Fig3}. First of all, this plot shows that the outcoupled fraction can be as high as $\sim 20$\,\%, when the interference condition for the emitter is met, which agrees with the expected EQE limit [\cite{Forrest2003, Greenham1994}]. Note that this calculation is taking the imperfection of the OLED and emitter used into account, as it considers electrical, non-radiative, and absorptive losses (cf. Fig. \ref{_Meerheim2010_Fig3}). The substrate modes follow almost the same modulation as the far-field modes showing distinct peaks. In contrary, the losses to surface plasmon modes decrease notably with increasing ETL thickness, reaching a negligible level for thicknesses $>200$\,nm, which is due to a weaker coupling between emitting dipoles and the surface plasmon mode. Similar studies have been reported by \cite{Nowy2008} and \cite{Krummacher2009}. On the contrary, as the ETL layer thickness increases, the fraction of the light being waveguided in organic modes increases substantially. Thus, the device does not gain outcoupled photons by avoiding losses to surface plasmons, as waveguides become dominant when placing dipoles far away from the cathode.

In the following, we will address concepts that aim to improve the outcoupled fraction of photons in OLEDs. Only having every fifth photon leaving the device in a standard architecture, much efficiency can be gained by providing efficient ways to enhance the light outcoupling. Here it is important to focus on methods that offer enhancement over the complete visible spectrum to be suitable for white OLEDs. In contrast, selective and directional concepts, e.g. the introduction of micro-cavities [\cite{Meerheim2008a}], is detrimental for obtaining high efficiency, white OLEDs.

	\subsection{\label{BottomOut}Improving outcoupling for bottom-emitting white OLEDs}

The vast of research dealing with improved light outcoupling focusses on bottom-emitting OLEDs. This is mainly due to the fact that the OLED itself is placed on a robust, thick substrate which can be easily manipulated. Furthermore, the preparation of the organic layers is the last processing step so that post-treatment, potentially damaging the device, is not necessary. 

		\subsubsection{\label{Macro}Macro-extractors}
Figure \ref{_Meerheim2010_Fig3} shows that a substantial amount of light is trapped in substrate modes that simply can not escape to air because of total internal reflection at the substrate/air interface. This light can easily be accessed by applying a macro-extractor to the substrate surface, matching the refractive index of the latter one. The ideal structure is a half-sphere [\cite{Rosenow2010, Reineke2009a, Mladenovski2009a,Greiner2007}] with dimensions much greater than the active area of the OLED so that the source of light can be treated as a point source. As shown in Figure \ref{_outcoupling}, this configuration assures that all the light entering the substrate from the organic layers is able to escape to air, as it is hitting the half-sphere surface under normal angle of incidence.

Other designs of macro-extractors are truncated square-pyramid [\cite{DAndrade2006}] or 'flowerpot'-shaped [\cite{Greiner2007}] luminaires. It is worth noting that their use is only meaningful to quantify the amount of substrate-trapped light. For real applications involving large-area OLEDs, thin and scalable concepts need to be applied to enhance the light outcoupling. Thus, efficiency values stated using macro-extractor elements should be handled with care and  only be seen as the upper limit for concepts that enhance the outcoupling of substrate modes, fully unlocking the potential of a given OLED stack.

			\subsubsection{\label{Structure}Structured substrate surfaces}
\begin{figure}[t]
\includegraphics[width=8.0cm]{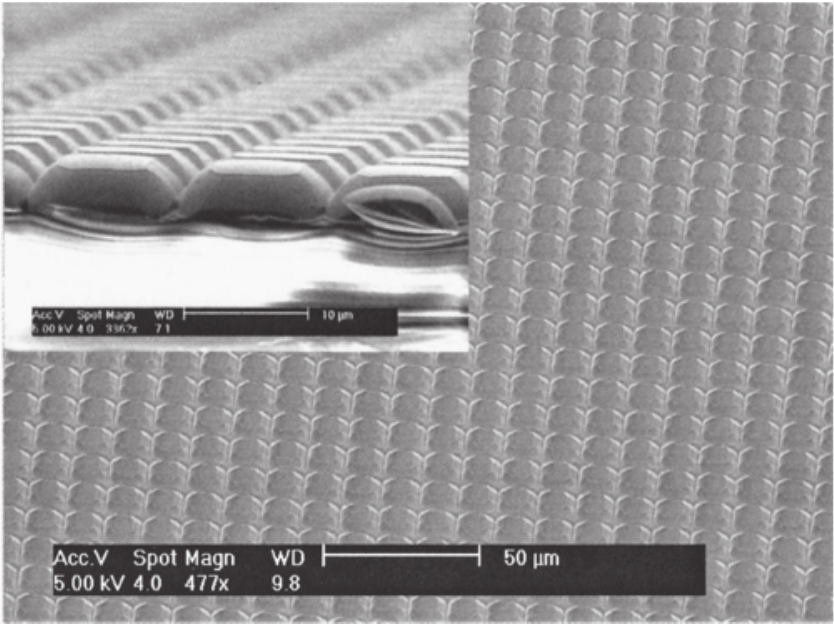}
\caption{\label{_Moller2002_Fig3} Scanning electron micrograph of a PDMS microlens array using an etched SiN$_x$ mold. Inset: side view. The based length of the lenses are approximately 10\,\textmu m. From \cite{Moller2002}.} 
\end{figure}

The easiest way to improve the total light output of bottom-emitting OLEDs is to incorporate structured substrate surfaces. The surface structure can either be arbitrary, e.g. as achieved by sand-blasting, or periodic. Typical examples for ordered structures are pyramidal or lens arrays [\cite{Madigan2000, Greiner2007, Moller2002, Nakamura2005}]. In contrast to the planar substrate, the use of a structured surface reduces the losses due to TIR, because the condition for TIR will be altered locally as the normal to the surface repeatingly changes. An example of a microlens array made from PDMS is shown in Figure \ref{_Moller2002_Fig3} [\cite{Moller2002}]. It comprises lens-like features with a base dimension of approximately 10\,\textmu m in a square lattice.

 \cite{Madigan2000} could show that the integrated emission of an OLED can be improved by factors ranging from 1.6 to 3.0, depending on the substrate and lens materials as well as the dimensions of the lens array. In their study, the lenses used still had macroscopic dimensions with sphere radii of $\sim 3$\,mm. \cite{Moller2002} reported on an 1.5-fold improvement achieved using the microlens array shown in Figure \ref{_Moller2002_Fig3} with much smaller lens dimensions. The light propagation within the substrate and the outcoupling structures is incoherent, thus conventional raytracing methods can be applied to optimize such structures for maximum light output for a respective OLED structure as shown by \cite{Greiner2007}.
 
 It is worth noting at this point that structured substrate surfaces are often combined with other concepts applied internally that aim to reduce waveguiding within the organic layers [\cite{Rosenow2010, Reineke2009a, Sun2008, Koh2010}]. Such concepts will be discussed subsequently.

		\subsubsection{\label{Grids}Low refractive index layers}
Equally important than extracting substrate modes, but at the same time much harder to achieve, is the outcoupling of light that is trapped in the organic layers (waveguide modes). This light, traveling in the plane of the OLED, will be absorbed and dissipated in the end, introducing heat to the system.

\begin{figure}[t]
\includegraphics[width=8.9cm]{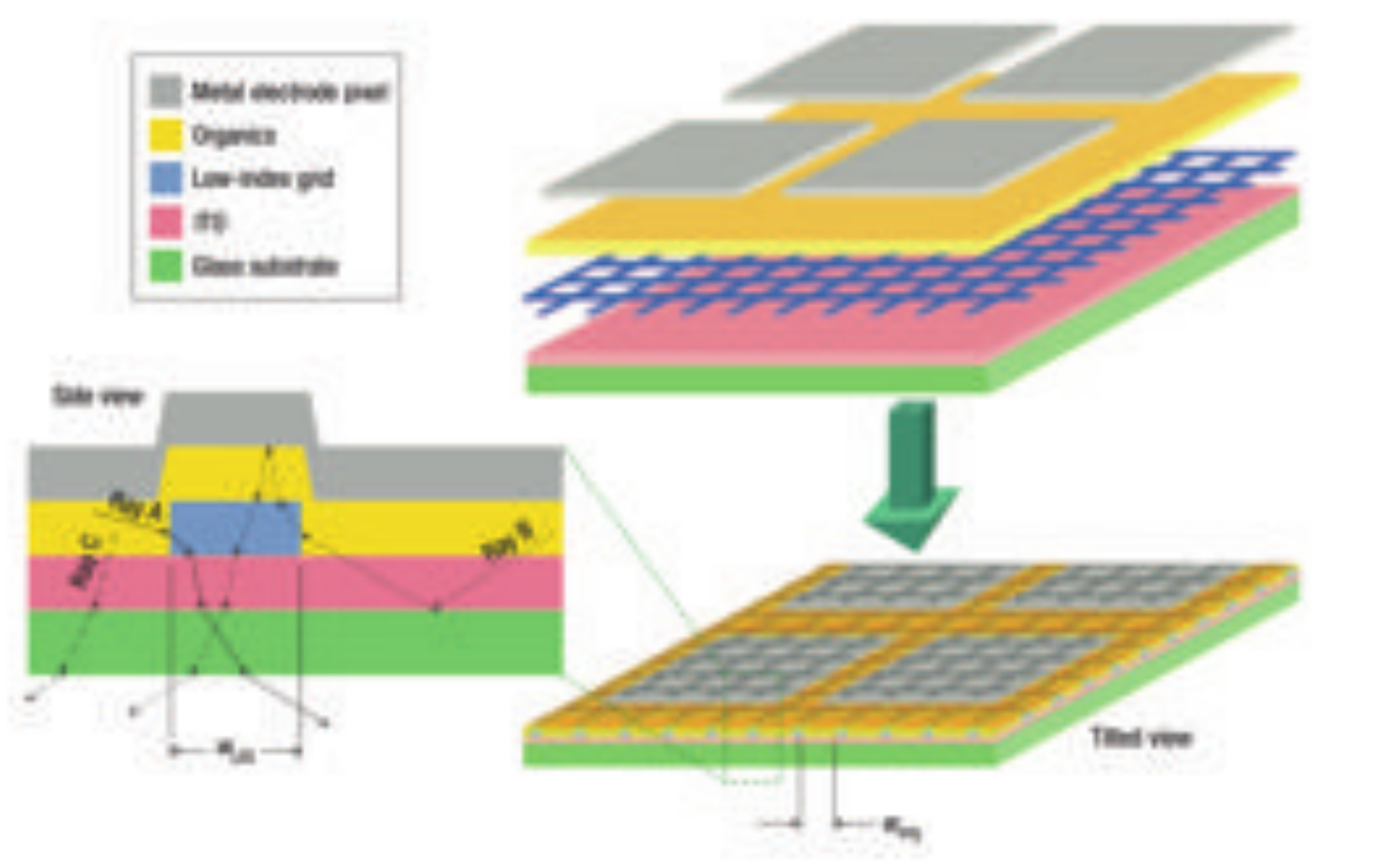}
\caption{\label{_Sun2008_Fig1}(color online) Scheme showing an OLED with embedded low-index grid between ITO and organic layers. Inset illustrates the mechanism leading to the outcoupling of organic modes. The active area of the pixels is one order of magnitude larger compared to the grid period. From \cite{Sun2008}.} 
\end{figure}

\cite{Sun2008} introduced a novel concept to convert waveguide into extractable modes by inserting a square grid of low refractive index material (low index grid = LIG) between the transparent anode ITO and the organic layers by means of photolithography. A scheme illustrating this approach is shown in Figure \ref{_Sun2008_Fig1}. Here, the width of the grid material (in their study SiO$_2$ having a refractive index of $n_\text{LIG}=1.45$) is 1\,\textmu m, between $6\times6$\,\textmu m squared openings. Embedded into a high refractive index surrounding, this grid material redirects light rays according to Snell's Law (cf. Fig. \ref{_Sun2008_Fig1}). Originally traveling with a large angle to the OLEDs normal, these modes are converted to light having a smaller angle to the normal, entering the escape cone of the device. In comparison to a reference white OLED, the light output from a device with a LIG structure increases by a factor of 1.32. Additionally applying a microlens array (cf. Sec. \ref{Structure}), yields a 2.3-fold total improvement. Note that the microlens array as placed onto the reference device only yields a factor of 1.68. The authors additionally provide simulation data showing that the overall outcoupling enhancement can be increased by a factor of 3.4 when incorporating grid materials with even lower refractive indices [\cite{Sun2008}].
	
\begin{figure*}[t]
\includegraphics[width=15cm]{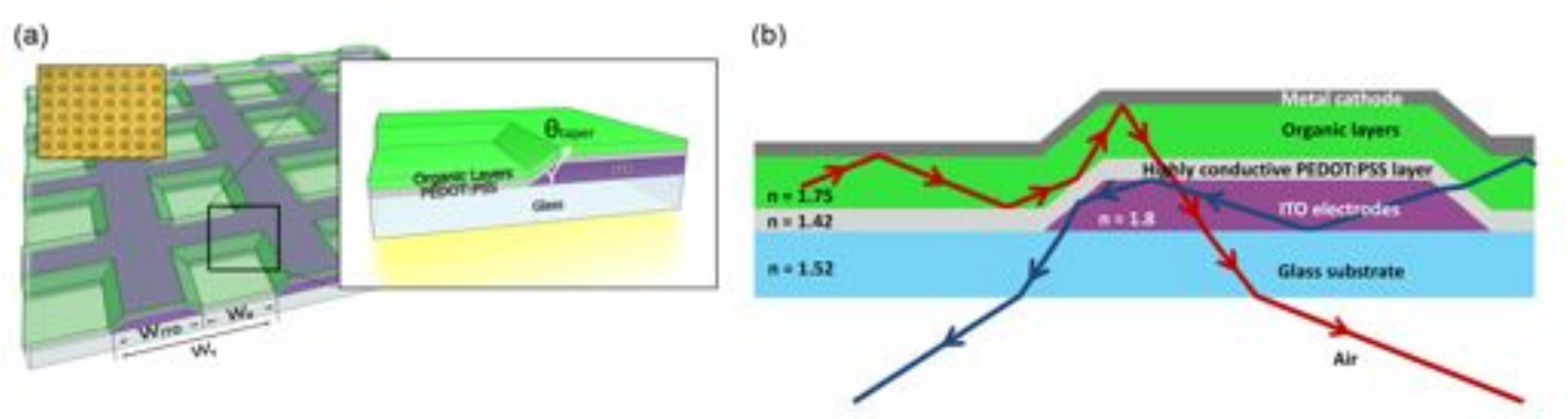}
\caption{\label{_Koh2010_Fig1}(color online) a) 3D scheme of the ITO electrode structure used in connection with the highly conductive PEDOT:PSS film. The size of the patterned ITO openings is roughly 3\,\textmu m with a grid period of 6\,\textmu m.  b) Cross-section of the complete layer structure. The PEDOT:PSS is introducing a refractive index contrast between the otherwise well-matching organic layers and ITO introducing waveguiding. The edges of the ITO electrodes enable enhanced outcoupling of the waveguided modes. From \cite{Koh2010}.} 
\end{figure*}

\begin{figure}[h]
\includegraphics[width=8.9cm]{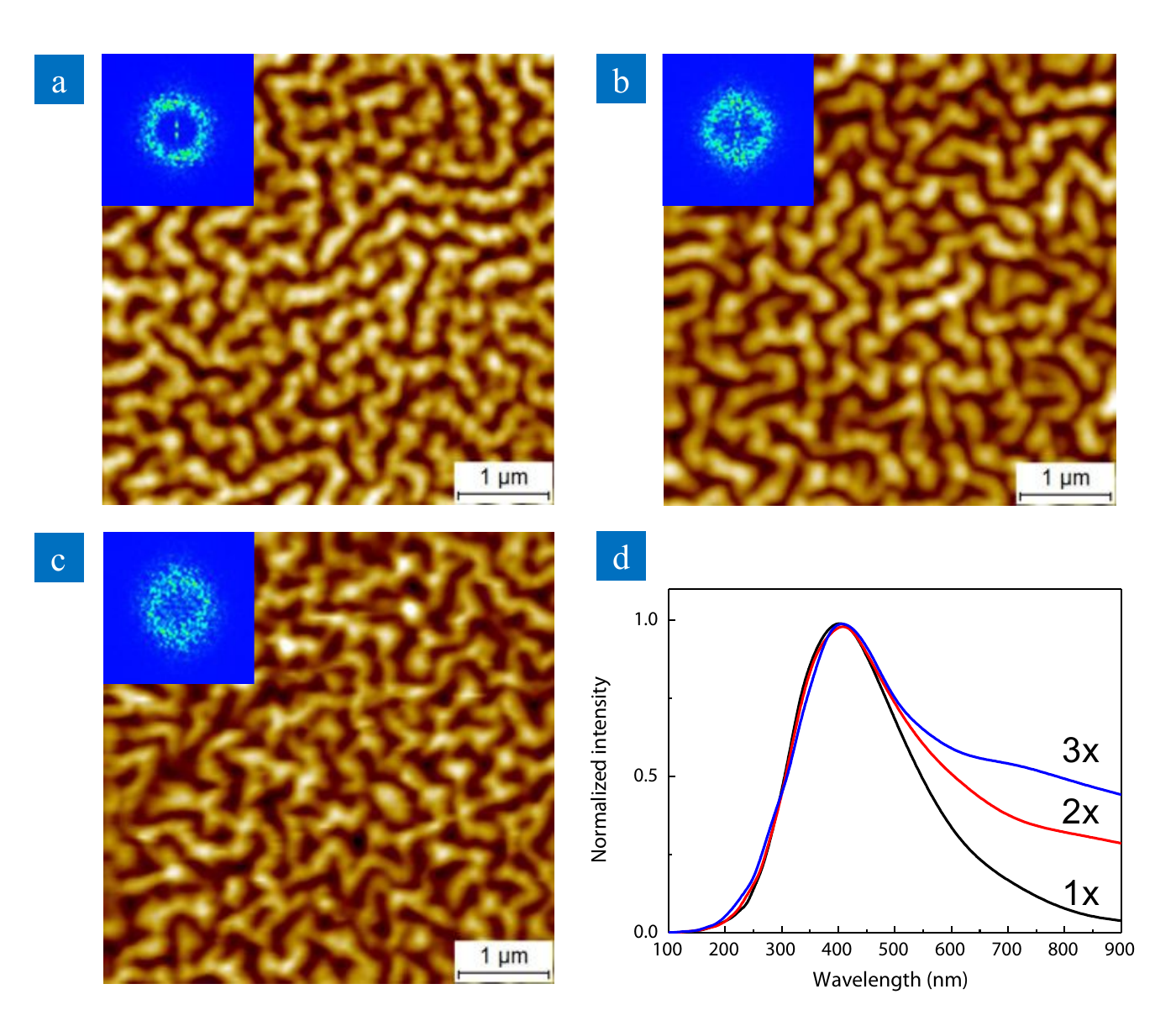}
\caption{\label{_Koo2010_Fig1}(color online) Atomic force microscope images of buckling patterns obtained by cooling down a 10\,nm thick aluminium (Al) film, which has been deposited onto PDMS film heated to 100°C, for one [a)], two [b)], and three [c)] times. Insets: fast Fourier transform (FFT) patterns. d) Power spectra from the FFTs as a function of wavelength [black $1\times$, red $2\times$, and $3\times$ Al deposition].  From \cite{Koo2010}.} 
\end{figure}

Another way to utilize materials with low refractive index has been developed by \cite{Koh2010}. Instead of structuring the low index material like in the work of \cite{Sun2008}, the authors structured the ITO by photolithography to form truncated pyramids as shown in Figure \ref{_Koh2010_Fig1} in the cross-sectional view. This ITO grid is then coated with the highly conductive polymer PEDOT:PSS [\cite{Fehse2007}], having a low refractive index of $n_\text{PEDOT:PSS}=1.42$. On top of this layer, the remaining OLED is processed in a conventional manner. With its low refractive index, the PEDOT:PSS introduces an index contrast between organic materials and the ITO. Improved light outcoupling is now an interplay between waveguiding on both sides of the low index polymer [cf. Fig. \ref{_Koh2010_Fig1} b)] and the truncated pyramidal shape of the ITO electrodes. It results in an increased fraction of light originally propagating in waveguide modes that reduces its angle to the device normal and by that is able to escape to air. The enhancement is highly dependent on the angle between the the substrate plane and the side face of the truncated pyramid [cf. Fig. \ref{_Koh2010_Fig1} a)] . Because the PEDOT:PSS is highly lateral conductive [\cite{Fehse2007}], light is not only generated in between the ITO base electrode and the metal cathode but also in areas not having ITO beneath the polymer.  At high current densities, where the electrical influence of the PEDOT:PSS can be neglected [\cite{Koh2010}], the enhancement over the reference OLED is 125\,\%. Again, similar to other approaches, applying an additional microlens array, increases the outcoupling enhancement to 167\,\%.

	\subsubsection{\label{Buckles}Corrugated OLEDs}
	
Instead of introducing a structured layer to the device layer sequence, \cite{Koo2010} developed a way to process a complete OLED with corrugation. Also aiming to couple out the organic modes, their approach is based on a subwavelength periodic, corrugated structure that allows to efficiently Bragg-scatter the organic modes to the far-field of the OLED.  

The corrugation is formed spontaneously after cooling down a bilayer of Al on thermally expanded PDMS (at 100°C during Al deposition), as a results of different thermal expansion coefficients of Al and PDMS [\cite{Koo2010}]. Atomic force microscope images of these layers can be seen in Figure \ref{_Koo2010_Fig1} a) - c). From a) to c), the depth of the buckles increase from $25-30$ to $50-70$\,nm, which is achieved by repeated deposition of 10\,nm thick Al layers on to the thermally expanded PDMS and cooling down afterwards. In general, the extraction of organic modes becomes more efficient with increasing buckles depth. The insets in Figure \ref{_Koo2010_Fig1} show the fast Fourier transforms (FFTs) of the structures, clearly indicating a periodic pattern with a characteristic wavelength and a wide distribution without preferred orientation (ring shape) [\cite{Koo2010}]. Figure \ref{_Koo2010_Fig1} d) shows the power spectrum of all patterns, obtained from the FFTs, indicating the unchanged peak wavelength of $\sim410$\,nm and the increasing distribution with increasing feature depth.

\begin{figure}[t]
\includegraphics[width=7.0cm]{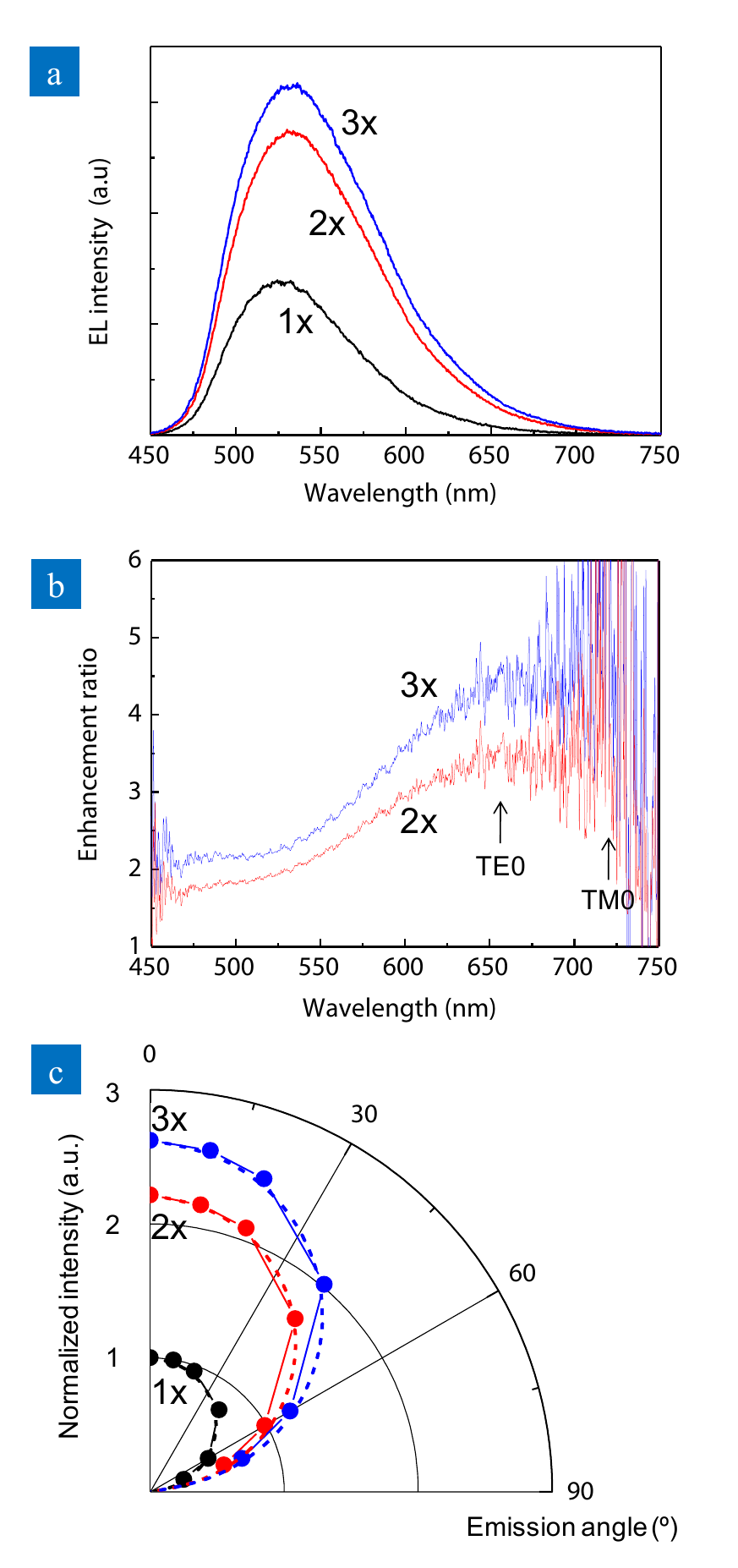}
\caption{\label{_Koo2010_Fig4}(color online) a) Electroluminescence spectra of the reference (black), $2\times$ (red), and $3\times$ (blue) buckling OLEDs, measured at a constant current density of 5\,mA\,cm$^{-2}$. b) Enhancement ratio obtained by dividing the spectra as of a) of the buckling OLEDs by the one of the reference device. Additionally, the spectral position of the TE$_0$ and TM$_0$ modes are indicated. c) Angular dependence of the light intensity of all three devices. The dashed lines indicate a Lambertian emission characteristics.  Adapted from \cite{Koo2010}.} 
\end{figure}

\cite{Koo2010} discuss monochrome OLEDs prepared on flat and corrugated surfaces prepared by dual and triple Al evaporation. Figure \ref{_Koo2010_Fig4} a) shows the EL spectra of all three devices obtained at a constant current density of 5\,mA\,cm$^{-2}$. Dividing the spectra of the buckled samples by the reference spectrum results in the spectral enhancement for each structure, as shown in figure \ref{_Koo2010_Fig4} b). Important for the application to white OLEDs, the enhancement is seen over the complete visible spectrum with a minimum enhancement of a factor of $\sim2$ (for the triple buckling device) in the blue region even further increasing to a peak enhancement of $>4$ in the red spectral region, where the TE$_0$ and TM$_0$ of the devices are located. Additionally, this plot  supports the fact that the extraction efficiency of the corrugation increases with increasing buckle depth, as achieved by multiple buckling formation cycles [\cite{Koo2010}]. Figure \ref{_Koo2010_Fig4} c) shows the angular emission profile of all the devices. The data shows that the Lambertian emission characteristics of the reference device is conserved by the corrugation. The broad spectral enhancement and the uniform angular emission make this approach suitable for white OLEDs. The integrated enhancement of the current efficiency $\eta_\text C$, obtained at 2000\,cd\,m$^{-2}$, reaches high values of 1.8 and 2.2 for the double and triple formed buckling OLEDs.

		\subsubsection{\label{HISubstr}High refractive index substrates}
\begin{figure}[t]
\includegraphics[width=8.8cm]{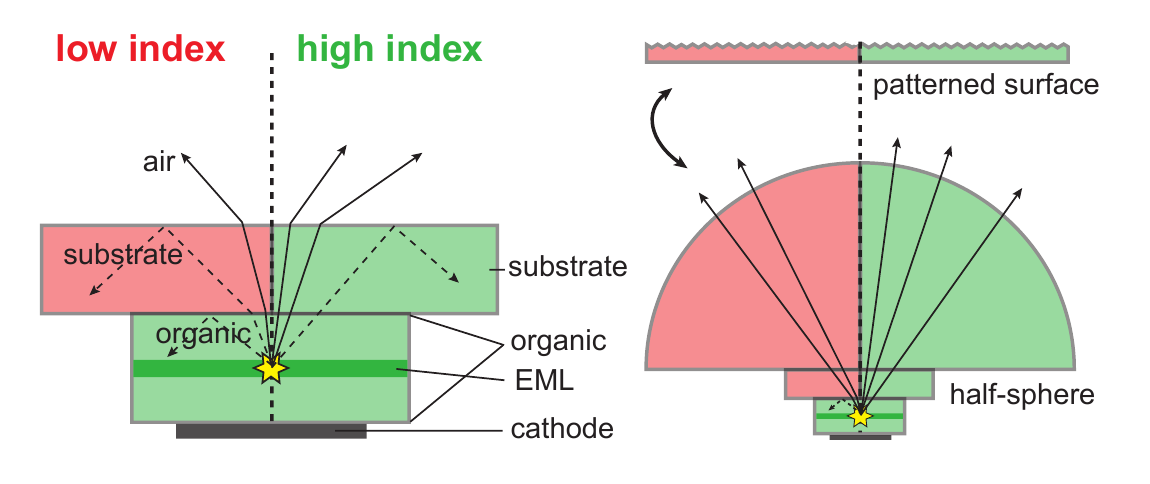}
\caption{\label{_Reineke2009a_Fig1b}(color online) Left: Light modes in an OLED structure using low ($n_{\text{low}}=1.51$) and high ($n_{\text{high}}\sim1.8$) refractive index substrates. Right: Application of outcoupling structures, i.e. either a macro-extractor or a patterned surface of matching substrate refractive index.  Adapted from \cite{Reineke2009a}.} 
\end{figure}

The use of high refractive index substrates to suppress organic modes has been suggested many years ago [\cite{Madigan2000}], generally offering an easy route to substantially increase the amount of light in the substrate [\cite{Lu2002, Nakamura2005}]. Figure \ref{_Reineke2009a_Fig1b} schematically shows the differences between the use of standard and high index substrates. 
	
By matching the refractive index of the substrate of choice closely to the respective indices of the organic materials ($n_\text{org}\sim 1.7-1.8$ [\cite{Greiner2007}]), the optical contrast at the ITO/substrate interface vanishes in first approximation [\cite{Gaertner2008}]. Thus, the propagation of light, generated in the organic layer stack, into the substrate is not hindered. Consequently, organic modes are minimized [cf. the study of \cite{Meerheim2010}]. At the same time, as a result of the large index difference at the substrate/air interface, the escape cone of high refractive index substrates noticeably reduces from $\theta_\text{c,n=1.51}=41.5^\circ$ to $\theta_\text{c,n=1.8}=33.7^\circ$. Thus, even though a larger fraction of light is coupled into the substrate, TIR at the substrate/air interface counteracts this improvement, typically leading to comparable or even slightly lower far-field extraction efficiencies using the high index substrates [\cite{Lu2002, Nakamura2005, Reineke2009a}]. In order to overcome this limiting factor, the use of outcoupling structures (cf. Fig. \ref{_Reineke2009a_Fig1b}) gains importance, because their use can strongly reduce the effect of TIR.

\begin{figure}[t]
\includegraphics[width=8.3cm]{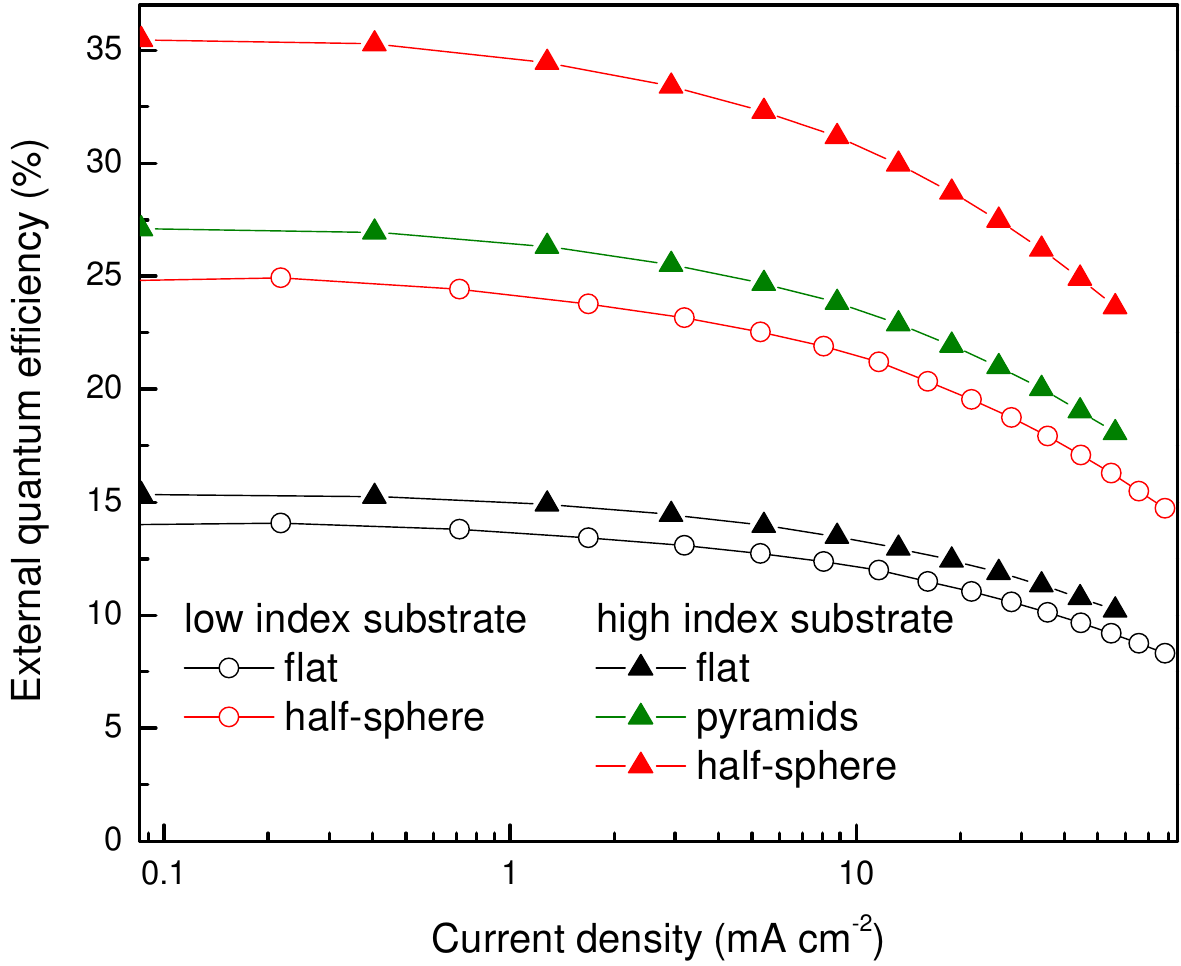}
\caption{\label{_OLEDs_3Phos_LI_vs_HI}(color online) External quantum efficiency of white OLEDs  processed on standard ($n_{\text{low}}=1.51$) (open symbols) and high ($n_{\text{high}}=1.78$) refractive index (filled symbols) substrates. Different sets of data correspond to different outcoupling methods used. The data displayed is obtained from devices published by \cite{Reineke2009a} [(Devices LI and HI-1)].} 
\end{figure}

To demonstrate the effect of high refractive index glass substrates, Figure \ref{_OLEDs_3Phos_LI_vs_HI} plots the external quantum efficiency of two identical white OLEDs\footnote{The transport layer thickness of the ETL and HTL slightly differ to meet the field antinode, accounting for the different optical properties of the two substrate types.} differing in the substrate type used: standard glass ($n_{\text{low}}=1.51$) or high index glass with $n_{\text{high}}=1.78$. The EQE is determined using different outcoupling structures at a constant current density of 5\,mA\,cm$^{-2}$: (i) a large index-matched half-sphere and (ii), for the high index case, a pyramidal structure [\cite{Reineke2009a, Rosenow2010}]. Applying the half-sphere to the reference, low index substrate OLED results in an 1.76-fold increase in light output. The same measurement set-up shows a substantial increase performed in the high index case, where an enhancement of 2.32 over the flat measurement is obtained. Applying an outcoupling structure comprising pyramids in a square lattice with a height of 250\,\textmu m and a base length of 500\,\textmu m, still reaches an 1.77-fold enhancement. Note that this is almost identical to the value obtained using the half-sphere in the low index case, clearly showing the potential of using refractive index-matched substrates.

Similar studies on monochrome green OLEDs are discussed by \cite{Mladenovski2009a}. \cite{Rosenow2010} combined the concept of high refractive index substrates with white stacked OLEDs, where the optics become more complex.

		\subsubsection{\label{Plasmons}Losses to metal surface plasmons}
Whenever methods are successfully applied to efficiently couple organic modes to the far-field, the remaining loss in OLEDs is the coupling to surface plasmon states of the highly reflective metal cathode. Figure \ref{_Meerheim2010_Fig3} already showed the magnitude of this effect for different distance between the EML and the cathode. Using high refractive index materials will qualitatively show a similar dependency of the coupling to surface plasmons on the spacing distance [\cite{Gaertner2008, Meerheim2010, Mladenovski2009a}].

\begin{figure}[t]
\includegraphics[width=8.3cm]{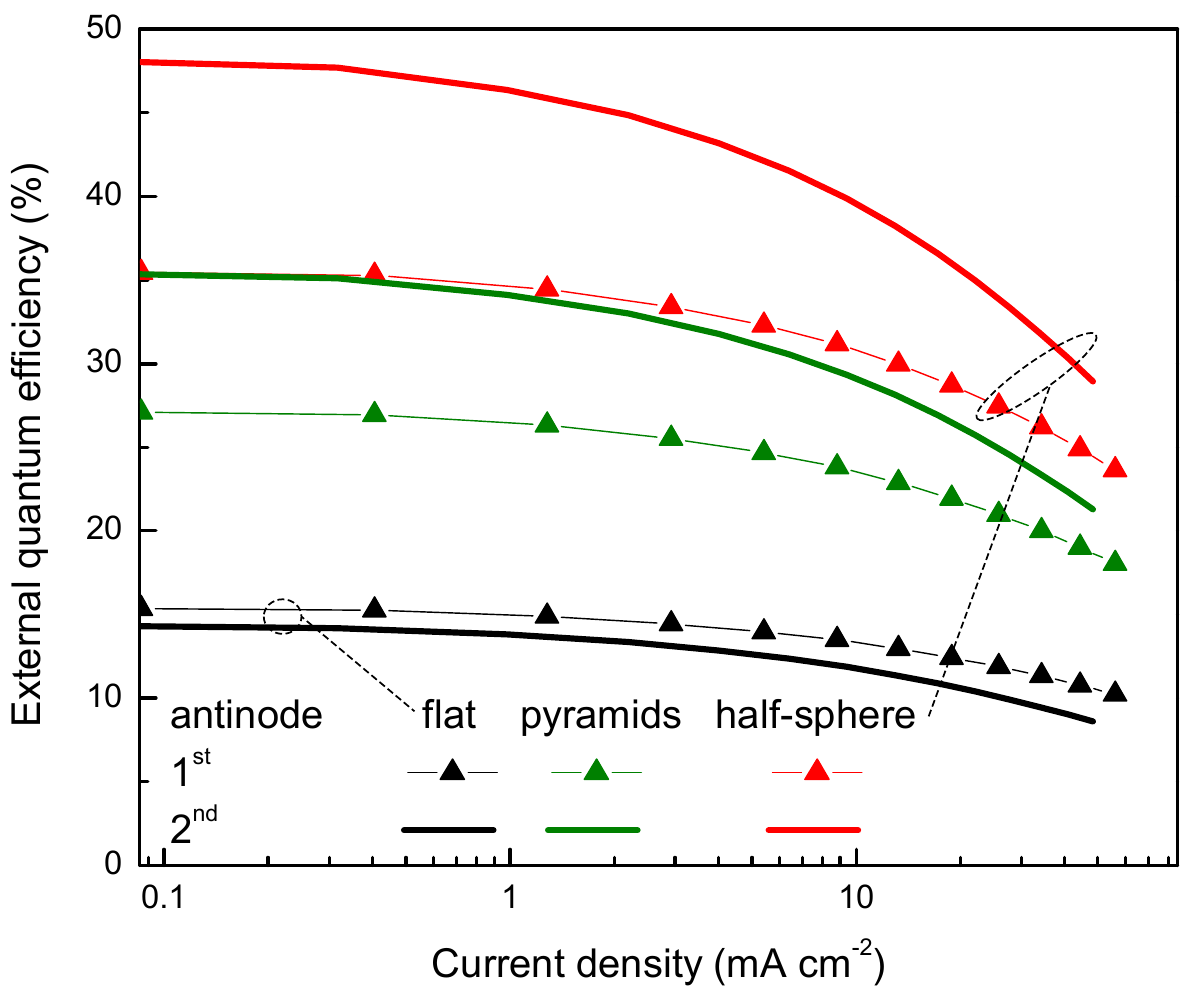}
\caption{\label{_OLEDs_3Phos_first_vs_second}(color online) External quantum efficiency of white OLEDs processed on high ($n_{\text{high}}=1.78$) refractive index substrates having a variation in the ETL thickness, meeting first (ETL = 45\,nm) and second (ETL = 205\,nm) field antinode, respectively. Different sets of data correspond to different outcoupling methods used. The data displayed is obtained from devices published by \cite{Reineke2009a} [(Devices HI-1 and HI-2)].} 
\end{figure}

In order to reduce these losses, \cite{Lin2006} suggested to increase the distance between EML and metal electrode to distances meeting the second field antinode of the system. However, even though coupling to surface plasmon modes strongly reduces with thickness (cf. Fig. \ref{_Meerheim2010_Fig3}), their improvement  based on a thick transport layer was only marginal (120\,\% enhancement for the integrated intensity). This observation can be explained by the increasing fraction of organic modes with increasing distance between EML and metal, which is observed when using standard glass substrates.

Two key points need to be met practically exploit the suppressed coupling to surface plasmon modes in OLEDs: (i) The transport layer that is to be increased either needs to be electrically doped [\cite{Walzer2007}] or have a very high charge conductivity to assure that ohmic losses and changes of charge carrier balance can be excluded. The studies of \cite{Lin2006} and \cite{Meerheim2010} make use of doped transport layers. (ii) Substrates matching the refractive indices of the organic materials need to be employed to prevent the formation of an increased number of organic modes with increasing thickness [\cite{Reineke2009a, Gaertner2008, Meerheim2010, Mladenovski2009a}].

Similar to the discussion of using high index substrates, data of two white OLEDs as published by \cite{Reineke2009a} are exemplarily used and plotted in Figure \ref{_OLEDs_3Phos_first_vs_second}. They have different ETL thickness of 45 and 205\,nm which meet the interference criteria for the first and second field antinode, respectively. Both processed on high index substrates ($n_{\text{high}}=1.78$), measurements in flat, half-sphere, and pyramid pattern configuration can directly be compared to see the effect of reduced plasmonic losses. The current density versus voltage characteristics of both devices are almost identical [\cite{Reineke2009a}]. In the second field antinode, the outcoupling enhancement obtained at 5\,mA\,cm$^{-2}$ with the half-sphere increases to a factor if 3.37 (42.4\,\% EQE) from 2.32 (32.5\,\% EQE) for the device in the first maximum. Even using the pyramid pattern as outcoupling structure yields 31.3\,\% EQE, corresponding to a 2.48-fold increase in outcoupling efficiency [\cite{Reineke2009a}]. It is worth noting that the second order white OLEDs undergo a noticeable change in their optical properties, which is mainly due to diverging interference conditions with increasing device thickness for the primary colors [\cite{Reineke2009a}]. Consequently, the white OLEDs with thick transport layers require to readdress the exciton/color management to attain a high quality white emission.

	\subsubsection{\label{Orientation}Orientation of the molecular dipoles}

As mentioned in the preceding Section \ref{Plasmons}, the coupling of the emitting dipoles to the metal surface plasmons contributes substantially to the loss in a typical OLED. Another way to reduce the losses to plasmon modes is the alignment of the emitting molecules and by that their transition dipole with respect to the metal layer plane. In general, the orientation of the emitter ensemble within the OLED is treated to be isotropic [\cite{Furno2012}]. However, it has been shown that preferential alignment of emitter dipoles can be achieved for both fluorescent [\cite{Frischeisen2010}] and, more recently, phosphorescent [\cite{Taneda2011, Flaemmich2011, Schmidt2011}] emitter molecules.

The ensemble of emitting dipoles in an OLED emission layer are composed of three fundamental dipole orientations [\cite{Neyts1998, Furno2012, Bruetting2012}]: $P_{\perp,\text{TM}}$, $P_{\parallel,\text{TM}}$, and $P_{\parallel,\text{TE}}$, where TM and TE stand for transverse magnetic and transverse electric, respectively. In case of isotropic orientation, these contributions are weighted equally, i.e. $P_i=1/3$. It can be shown, that the $P_{\perp,\text{TM}}$ dipoles only very weakly couple to the metal surface plasmon mode, even when the dipoles are close to the metal [\cite{Bruetting2012}]. Thus, emitter molecules which preferentially orient parallel to the metal surface do not couple effectively to the plasmon mode so that, consequently, the fraction of outcoupled light can be increased. \cite{Frischeisen2011} estimated recently that the external quantum efficiency of OLEDs could be increased from 20\,\% to about 45\,\% by engineering the emission layer to have ideal horizontal dipole orientation. Despite the fact that this concept will heavily depend on the actual material design, it provides an elegant and effective way to suppress the losses to surface plasmon modes.
 
	\subsubsection{\label{Stacked}Stacked OLEDs}
Stacked white OLEDs [\cite{Rosenow2010, Kanno2006, Kanno2006a, Shen1997}] haven't been discussed in Sections \ref{whitePOLED} and \ref{whiteSMOLED} due to their structural difference compared to single unit OLEDs. Stacked OLEDs are based on the concept of depositing more than one OLED on top of each other, serially interconnected with either a metal electrode [\cite{Shen1997}] or a charge-generation (also termed charge-conversion) layer [\cite{Kanno2006, Kanno2006a, Rosenow2010}]. Their structural complexity and variability would qualify for an independent review article. Because the EMLs used in stacked devices fully make use of either monochrome or multi-color systems that have been discussed in previous sections, we would only like to point out key differences to single unit devices.
	
Stacked OLEDs still are transparent devices, thus in first approximation the brightness/color emitted from each unit can be added to form the total emission. In case these devices are fabricated with charge-generation layers, more than one photon can be emitted per injected electron. Note that even though the EQE values add up, the LE ideally remains constant for stacking identical units, as the driving voltage increases accordingly. As brightness adds up, the individual units need to sustain less current density to achieve a given luminance level compared to the single unit OLEDs. This benefits the long-term stability of the devices.

One key advantage of stacked OLEDs is the possibility to design a white device by placing the different EMLs comprising primary colors into their respective field antinodes within the layer structure, which enables operation of all emitters at maximum outcoupling efficiency. This is in contrast to single unit devices, where some of the colors are likely to be suppressed [\cite{Reineke2009a}].

One drawback of this concept is the increased thickness of the complete device, which may increase the fraction of waveguided organic modes, as shown in Figure \ref{_Meerheim2010_Fig3}. This, again, is overcome by using high index glass substrates. \cite{Rosenow2010} discussed two-unit stacked devices based on a triplet-harvesting blue/red (cf. Sec. \ref{TripHarv}) and a phosphorescent green/yellow (cf. Sec. \ref{smPhos}) unit also employing glass substrates with $n_{\text{high}}=1.78$. At 1000\,cd\,m$^{-2}$, white stacked OLEDs reach 75.8 and 41.6\,\% EQE when using an index-matched half-sphere and pyramidal patterned structure, respectively. With respect to a reference device on standard glass measured without outcoupling structures, these values correspond to a 2.9- and 1.6-fold increase in outcoupling efficiency.

	\subsection{\label{TopOut}Concepts for top-emitting devices}
In contrast to bottom-emitting OLEDs, top-emitting devices can mostly be optically influenced by manipulating the top layers made of soft, organic materials and thin metal layers [\cite{Chen2010}].	This in turn complicates the task of developing efficient strategies for improved light-outcoupling, because the organic layer stack likely will not withstand many post-processing steps that would be necessary to improve the outcoupling of light. 

\cite{Kanno2005} have reported on highly efficient top-emitting white OLEDs based on two primary colors, where they employ the transparent conductor ITO as top cathode, providing sufficient transparency similar to bottom-emitting devices. However, as ITO and similar conductors are processed by sputtering techniques, such processing introduces a high risk of damaging the underlying organic layers. To avoid techniques with high impact on the organics, thin metal layers have become the top electrode of choice, having sufficient lateral conductivity while maintaining sufficient transparency [\cite{Chen2010}]. They turn the OLED into a micro-resonator. This even complicates the realization of white OLEDs, because the resonances of the optical structure become narrower and angular dependent [\cite{Hofmann2010, Meerheim2008a}]. 

		\subsubsection{\label{Capping}Dielectric capping layer}
The concept of a dielectric capping layer applied on top of the thin, semi-transparent metal cathode has been introduced by \cite{Hung2001}. In their report, the authors evaluated a variety of inorganic and organic materials with different refractive indices with respect to their outcoupling effect. By introducing a capping layer, the transmittance of the top metal layer can be increased. Early work on monochrome, top-emitting OLEDs has shown that the concept of dielectric capping layers can substantially increase the amount of outcoupled light [\cite{Huang2006c, Riel2003}]. 

\begin{figure}[t]
\includegraphics[width=8.0cm]{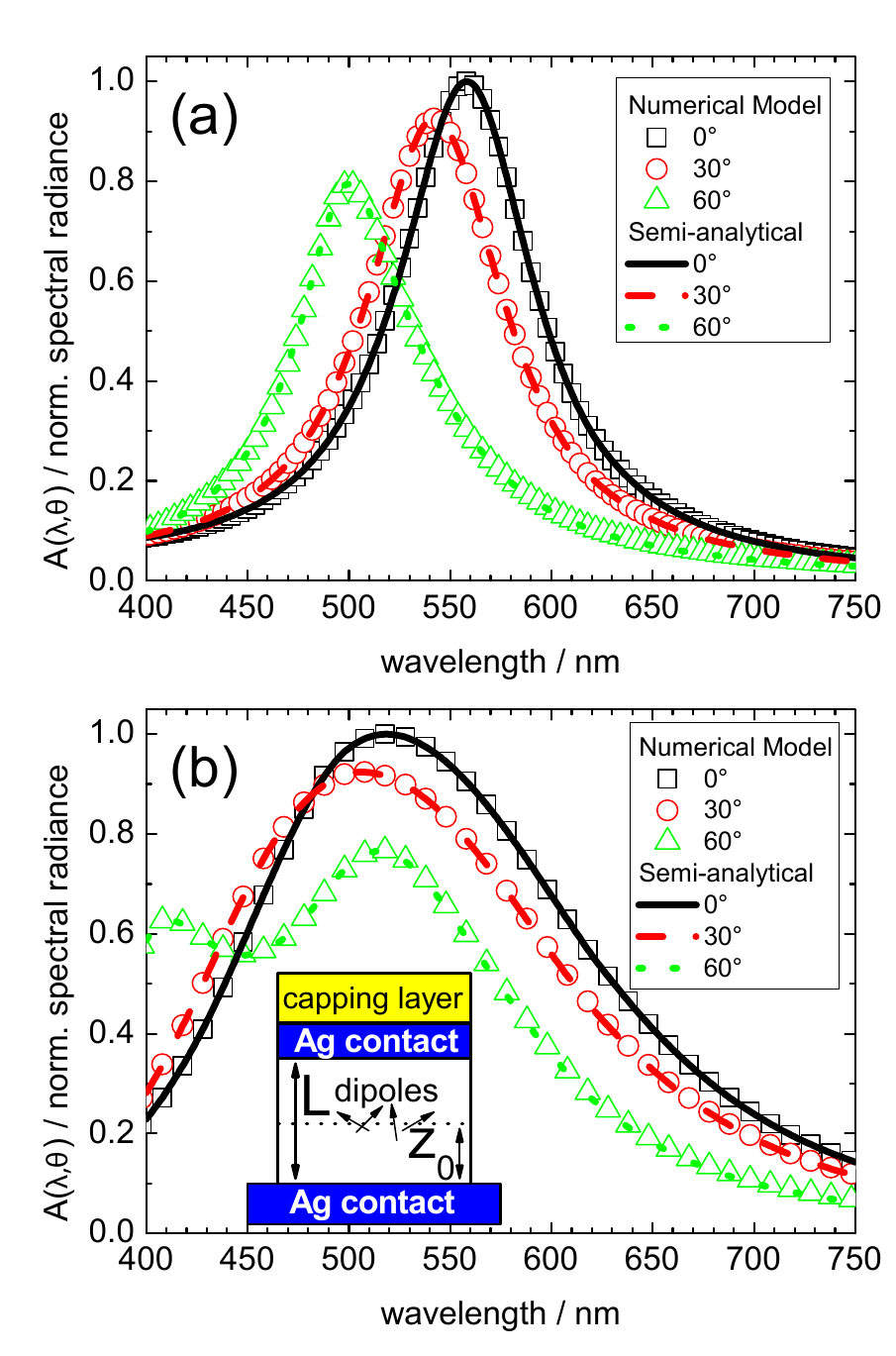}
\caption{\label{_Thomschke2009_Fig1}(color online) Calculated extractable power $A(\lambda, \theta)$ of two model OLED structures having the following layer sequence: a) 100\,nm Ag/100\,nm organic ($n=1.8$)/15\,nm Ag. b) layer structure of a) followed by an additional 50\,nm organic capping layer. Emitters are placed in the field antinode corresponding to a resonance wavelength of approximately 560\,nm. From \cite{Thomschke2009}.} 
\end{figure}

\begin{figure}[t]
\includegraphics[width=8.0cm]{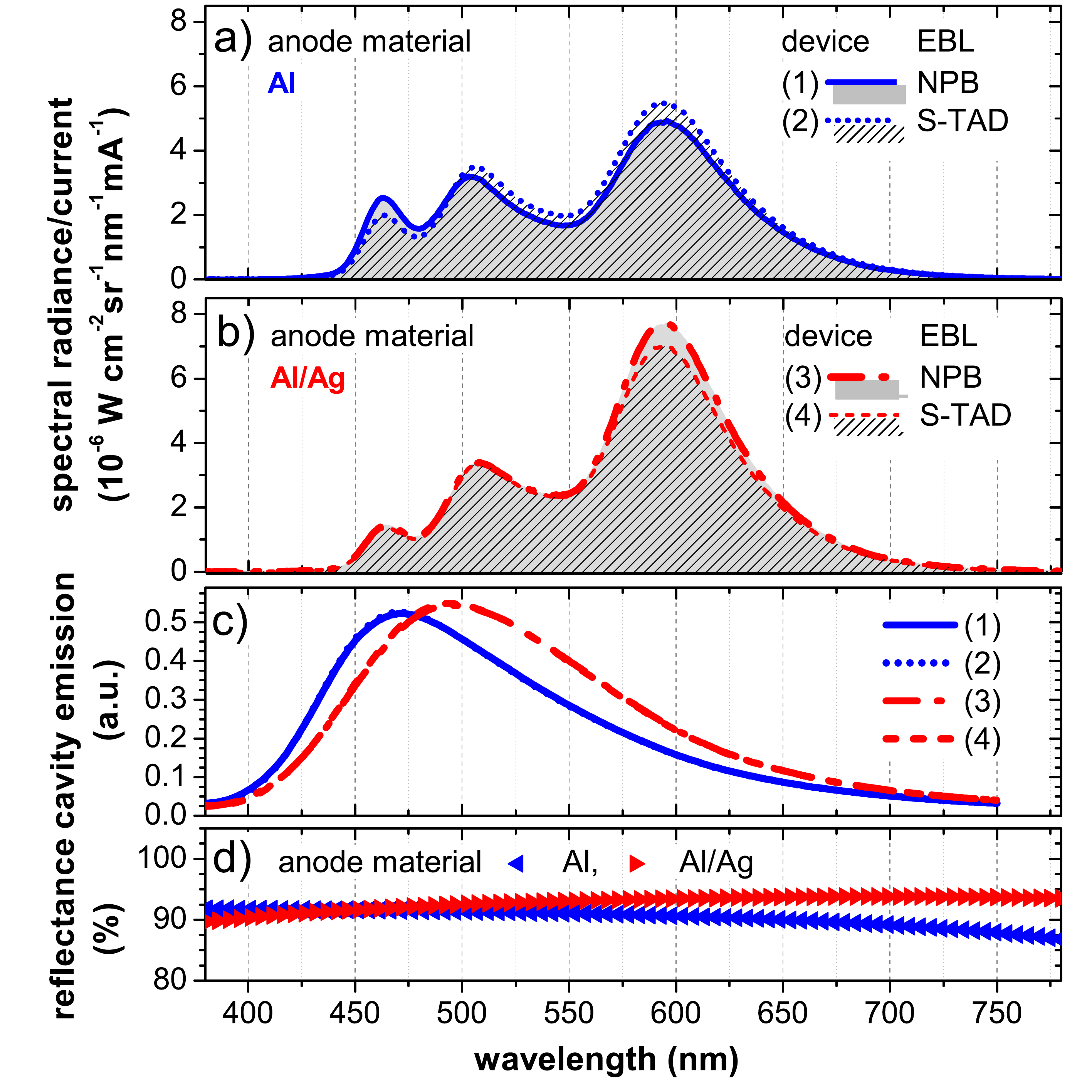}
\caption{\label{_Freitag2010_Fig2}(color online) Forward EL spectra of white OLEDs based on an a) Al and b) Al/Ag anode, obtained at 1000\,cd\,m$^{-2}$ (Additional variations of the EBL are shown). c) Calculated extractable mode for all devices. d) Reflectance of both anodes used (Al vs. Al/Ag).   From \cite{Freitag2010}.} 
\end{figure}

In addition to the outcoupling enhancement, the capping layer helps to realize high quality white light [\cite{Freitag2010, Hsu2005}]. As said, top-emitting OLEDs employing thin metal top electrodes have a much stronger cavity compared to standard bottom-emitting devices, negatively affecting their optical properties. Figure \ref{_Thomschke2009_Fig1} shows the calculated extractable power $A(\lambda, \theta)$ of a model OLED consisting of a single organic layer [\cite{Thomschke2009}]. The EML is located in the bulk of this model OLED. Figure \ref{_Thomschke2009_Fig1} a) shows $A(\lambda, \theta)$ of this structure without a capping layer for different angles of observation. This extractable mode is very narrow with a FWHM of 83\,nm at $0^\circ$, in addition shifting to shorter wavelength with increasing observation angle. Figure \ref{_Thomschke2009_Fig1} b) shows the calculation made for the same model OLED with additionally applied organic capping layer. This layer effectively reduces the color shift of  $A(\lambda, \theta)$ and at the same time broadens the extractable mode (FWHM at $0^\circ$: 190\,nm). In comparison to the device without capping layer, the FWHM is more than doubled, which is mandatory for coupling a broad white spectrum to air [\cite{Thomschke2009, Freitag2010,Hsu2005}]. Even with applied capping layer, top-emitting OLEDs based on semi-transparent metal electrodes are very sensitive to optical changes. \cite{Freitag2010} compared the performance of two identical white OLEDs with different, highly reflective anode metals (single layer Al vs. a bilayer of Al/Ag) having slightly different reflectivity.  The effect on the extractable mode [Fig. \ref{_Freitag2010_Fig2} c)] and the resulting emitted OLED spectrum [Fig. \ref{_Freitag2010_Fig2} a) and b)] is significant.

\cite{Thomschke2009} discuss white OLEDs based on a hybrid EML concept [\cite{Schwartz2006}] without and with capping layer applied. The external quantum efficiency, obtained at 5.4\,mA\,cm$^{-2}$ increase from 5.4\,\% (without) to 7.8\,\% (with). In addition, the OLEDs with additional capping layer emit a broad white spectrum, which only weakly varies  as a function of observation angle. In contrast, the reference OLED does not even emit white light at any angle, which is a result of the narrow $A(\lambda, \theta)$ and points to the need of capping layers for realizing white top-emitting OLEDs based on thin metal electrodes.

\subsubsection{\label{TopMicro}Laminated microlens arrays}
A concept for improved outcoupling for top-emitting white OLEDs that goes beyond the application of a dielectric capping layer has been recently introduced by \cite{Thomschke2012}. Their reference device already has an organic capping layer applied having a refractive index that is similar to the rest of the layers used in the OLED. Now they coat a polymer microlens film with high refractive index ($n_\text{microlens}=1.71$) with the same material used for the capping layer to finally merge the two organic layers in a lamination process. Thus, a top-emitting OLED is fabricated that is refractive index matched throughout the device. 

\begin{figure}[t]
\includegraphics[width=7.5cm]{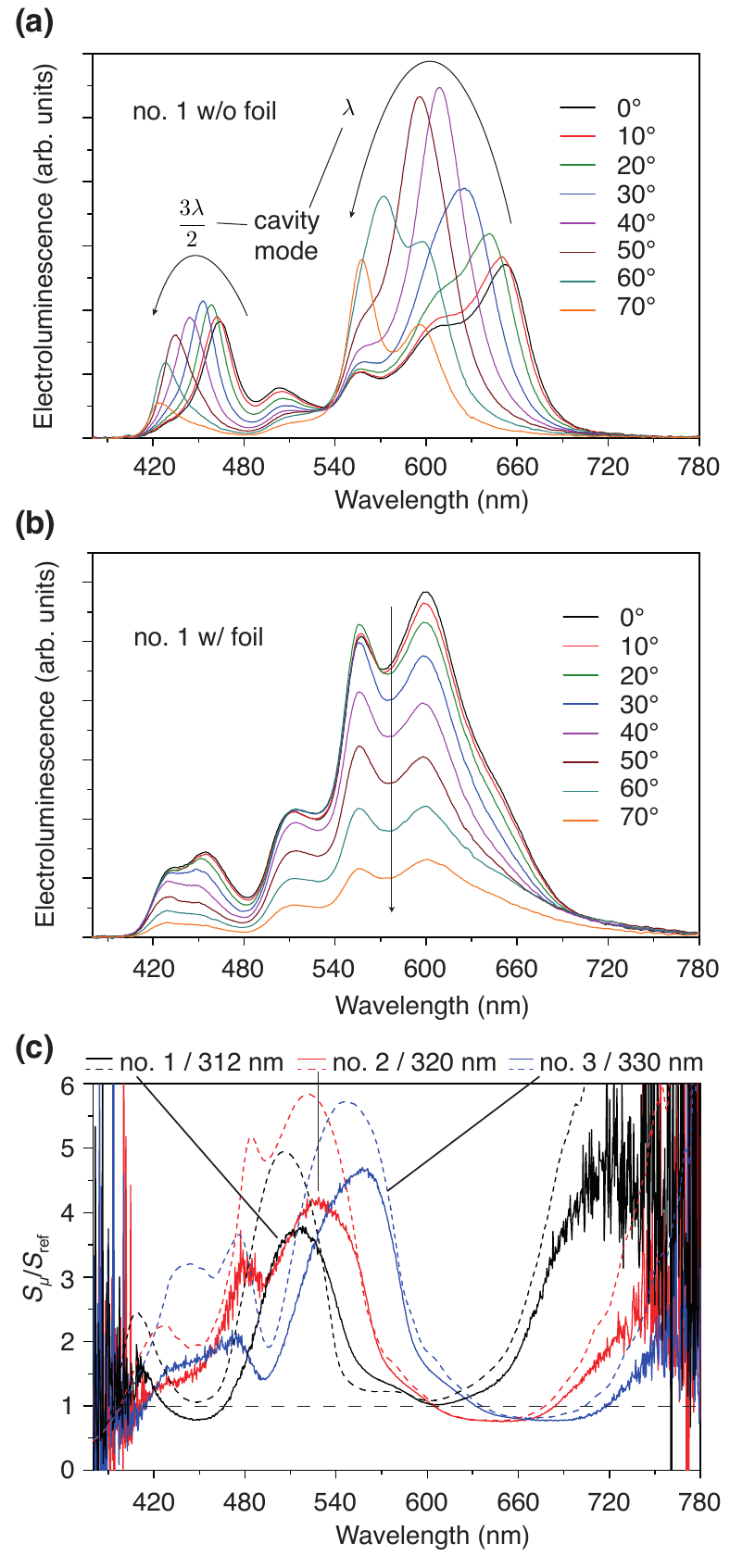}
\caption{\label{_Thomschke2012_Fig2}(color online) a) Electroluminescence of a two-unit stacked top-emitting OLED as a function of viewing angle. b) Identical OLED architecture as of a) with additional application of refractive index matched microlens film. c) Experimental and calculated spectral enhancement factor between device without [a)] and with [b)] microlens film (data shown for three different total cavity thicknesses).  From \cite{Thomschke2012}.} 
\end{figure}

To show this concept, \cite{Thomschke2012} used a highly-efficient two-unit stacked OLED as a reference device, which makes use of a phosphorescent yellow and a triplet-harvesting blue/red unit  [modified from \cite{Rosenow2010}]. The emitted spectrum of the reference device is shown in Figure \ref{_Thomschke2012_Fig2} a) as a function of viewing angle. Strong color shifts are observed, where both cavity modes sweep from lower to higher photon energies with increasing angle of observation. Not for a single angle, a white spectrum is emitted. Application of the microlens foil drastically improves the optical properties of the device, which is shown in Figure \ref{_Thomschke2012_Fig2} b). Now the emitted spectrum does only change slightly with viewing angle and represents a broadband, balanced white spectrum. The laminated microlens has two functions in this concept: (i) it acts as an integrating element, effectively mixing all photons so that the spectrum becomes independent of viewing angle and (ii) it functions as outcoupling structure for modes that are not able to escape the thin film structure. The latter is seen in Figure \ref{_Thomschke2012_Fig2} c), where the spectral outcoupling enhancement factor is plotted as derived between reference and microlens laminated OLED. It shows an outcoupling enhancement up to a factor of roughly four in the green spectral region, where the planar reference structure is unable to couple out efficiently [cf. Fig. \ref{_Thomschke2012_Fig2} a)].

The highest EQE\footnote{Note that this is a two-unit stacked device potentially having 200\,\% internal quantum efficiency.} for a laminated OLED is 26.8\,\% (30.1 lm W$^{-1}$) with color coordinates of (0.542, 0.416) [$\alpha_{\text{CIE}} = +0.02$, CRI = 75]. The highest CRI of 93 has been reached with a slightly different cavity length, unfortunately no efficiency data is available for this device [\cite{Thomschke2012}]. 

\subsection{\label{Summary_Outcoupling}Summary}
Besides improving the internal OLED performance  to realize internal quantum efficiencies approaching unity, which has been realized for specific white OLED concepts, the improvement of light outcoupling is of highest interest. This inherently offers the largest potential for performance enhancement up to a factor of five (cf. Sec. \ref{Modes}).

Many promising and very effective concepts for bottom-emitting white OLEDs have been proposed. The currently reported enhancement factors are 1.67 [low index layer; Sec. \ref{Grids} \cite{Koh2010}], 2.2 [corrugated OLEDs (monochrome); \ref{Buckles} \cite{Koo2010}], 2.3 [low index grid; Sec. \ref{Grids} \cite{Sun2008}], and 2.48 [high index substrate with thick ETL; \ref{Plasmons} \cite{Reineke2009a}], respectively. All these concepts promise a spectrally broad gain, which is necessary for white light emission.  At the same time the improvement values reported to date indicate that there still is a noticeable gap between experimental results and the amount of light that can potentially be gained (as a rule-of-thumb: factor five). Furthermore, the enhancement factors are often achieved based on reference devices which are not optically optimized and thus have to be taken with a grain of salt. Also, as an important future factor, the scalability of these outcoupling concepts will strongly determine their acceptance.

The discussion of concepts for top-emitting white OLEDs has shown that the outcoupling enhancement to be expected to have a much smaller margin compared to bottom-emitting devices. Here, the capping layer concept is to date the only concept providing substantial outcoupling enhancement (1.44-fold [\cite{Thomschke2009}]. Apparently, it is still a challenge to realize broadband white OLEDs with emission that does not change strongly as a function of observation angle [\cite{Freitag2010}]. The concept of laminating a microlens foil on top of the capping layer [\cite{Thomschke2012}] seems promising, especially as it alleviates the general problem of white top-emitting OLEDs that compromises between high efficiency and high color quality must be made. Still, more research is needed to test its feasibility on larger scale.


\section{\label{Perspective}Estimation of efficiency limit for white OLEDs}
It is not easy to make predictions that aim to answer the question: ``What level of efficiency can white OLEDs reach in future?" Still we would like to make a serious attempt in trying to estimate a realistic upper limit for the efficiency of white OLEDs. We specifically focus here on the luminous efficacy, given in lumen per Watt (lm\,W$^{-1}$), that is widely used to compare white OLED performance to existing, mature technologies. 

\begin{figure}[t]
\includegraphics[width=9.5cm]{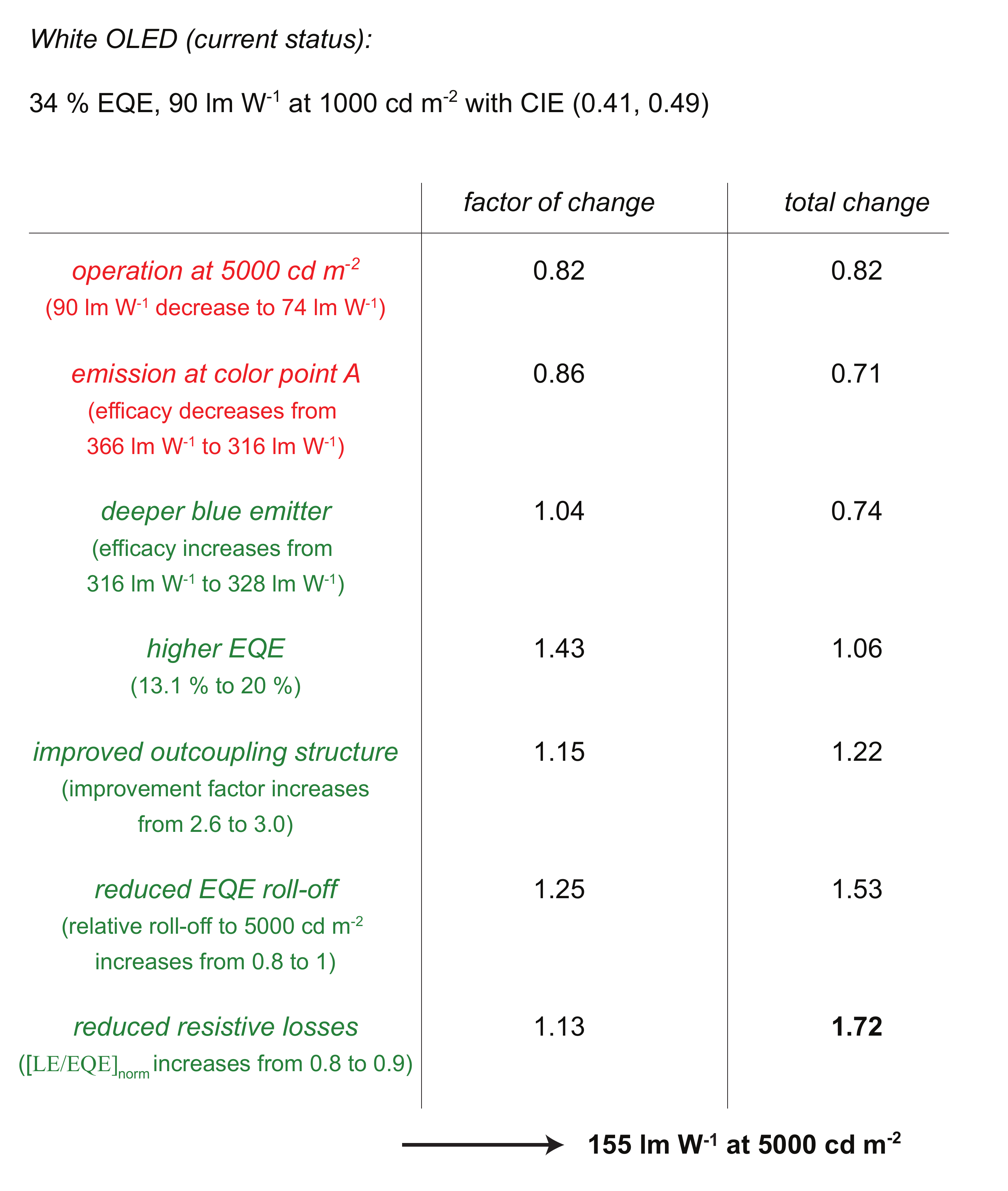}
\caption{\label{white_OLED_perspective}(color online) Based on a reference white OLED from \cite{Reineke2009a} [Device HI-2] representing the current status, different aspects influencing the device efficiency are listed to estimate a realistic upper limit. For each aspect, the estimated potential change is given. Efficiency reduction (red) is encountered to meet color quality and lighting application requirements.} 
\end{figure}

In order to do so, we base our estimation on a published white OLED, hereinafter called the reference device, serving as a baseline [Device HI-2 from \cite{Reineke2009a}]. Its performance data are: 34\,\% EQE and 90\,lm\,W at 1000\,cd\,m$^{-2}$ with CIE color coordinates of (0.41, 0.49) [$\alpha_{\text{CIE}}=+0.09$]. In the following, we will discuss different aspects that influence the overall efficiency. Figure \ref{white_OLED_perspective} displays a table that lists all these aspects with their anticipated change on the device efficiency.

First, we see it necessary for real applications to raise the brightness level of the OLED to 5000 instead of the commonly used 1000\,cd\,m$^{-2}$. The luminous efficacy of the reference OLED rolls off from 90\,lm\,W at 1000\,cd\,m$^{-2}$ to 74\,lm\,W at 5000\,cd\,m$^{-2}$, thus we account for this change by the factor 0.82 (cf. Fig. \ref{white_OLED_perspective}).

Furthermore, the emitted color of the reference device used in this argumentation is by far too green ($\alpha_{\text{CIE}}=+0.09$) to meet white light requirements, which we simply see as being a Planckian radiator (cf. Sec. \ref{Color}). Based on the three emitters used, i.e. FIrpic, Ir(ppy)$_3$, and Ir(MDQ)$_2$(acac) [\cite{Reineke2009a}], we have calculated the luminous efficacy of radiation $K_\text r$ for the reference white OLED and for a simulated spectrum based on the same emitters that has emission at Standard Illuminant A in the CIE color space, i.e. CIE (0.448, 0.408). Here, $K_\text r$ drops from 366 to 316\,lm\,W$^{-1}$ (factor 0.86). The use of FIrpic introduces an unbalanced ratio of blue, green, and red emission, barely having green intensity (cf. Fig. \ref{_idealwhite}\footnote{Note that the absolute numbers of $K_\text r$ differ for the spectra shown in Fig. \ref{_idealwhite} and for the values displayed in the table of Fig. \ref{white_OLED_perspective}. While the latter are based on OLED spectra, the simulation of Fig. \ref{_idealwhite} only takes the PL spectra of the emitters into account, giving rise to slight differences.}). By incorporating a deeper blue emitter, the green intensity can be increased, which increases $K_\text r$ again from 316 to 328\,lm\,W$^{-1}$ (cf. Fig. \ref{_idealwhite}).

\begin{figure}[t]
\includegraphics[width=8.3cm]{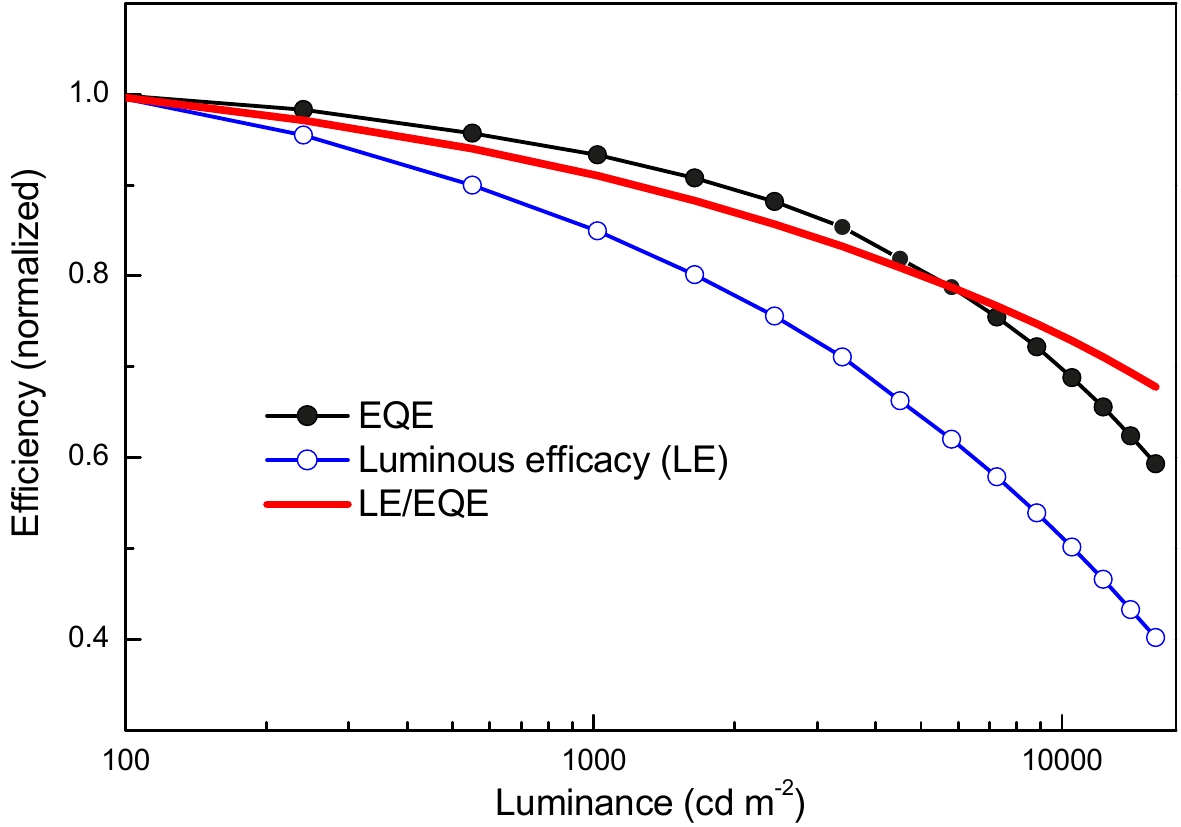}
\caption{\label{LE_rolloff}(color online) This plot shows the drop of EQE and LE as a function of luminance as normalized to a brightness of 100\,cd\,m$^{-2}$. Device data is taken from \cite{Reineke2009a} [Device HI-1]. To quantify the resistive losses in the OLED, the normalized trends of LE and EQE are divided: [LE/EQE]$_\text{norm}$ (red line).} 
\end{figure}

The external quantum efficiency of the reference device is 13.1\,\% EQE at 1000\,cd\,m$^{-2}$ when processed on standard glass substrates and measured without outcoupling enhancement.\footnote{Note that the reference device is an optimized OLED employing high refractive index substrates and thick electron transport layers.} We anticipate that by using emitters with highest possible PLQY in connection with the right EML concept, 20\,\% EQE can be reached -- the typical limit seen for phosphorescent OLEDs or equivalent concepts with the potential of 100\,\% internal quantum efficiency. Note that this limit even seems conservative, given the fact that various reports suggest EQE values significantly exceeding 20\,\% [\cite{Sasabe2010, Su2008, Tanaka2007, Tanaka2007a}]. 

With a theoretical potential to increase the outcoupled light by a factor of five (cf. Sec. \ref{Modes}) and considering the amount of high performance concepts reported [\cite{Koh2010,Koo2010,Sun2008,Reineke2009a}], we see it feasible to increase the current outcoupling enhancement obtained for the reference white OLED\footnote{This increase with respect to the flat device was obtained in the LE using the pyramid pattern.} of 2.6 to a factor of three.

Much effort is spent on reducing the efficiency roll-off induced by non-linearities at high excitation levels [\cite{Baldo2000a, Reineke2007, Kalinowski2002}]. Especially phosphorescent emitters seem to have potential to extent their operation range where no annihilation is present [\cite{Staroske2007, Namdas2005, Reineke2010, Schwartz2008, Kang2007, Su2008, Han2008}]. Thus, we simply assume in this estimation that future research will enable white OLED to be operated at 5000\,cd\,m$^{-2}$ without suffering a decrease in EQE up to this brightness.

The last aspect we would like to discuss is the contribution of resistive losses within an OLED to the luminous efficacy roll-off, which hasn't drawn to much attention in current research. Figure \ref{LE_rolloff} plots the relative roll-off of EQE and LE as normalized to an initial brightness of 100\,cd\,m$^{-2}$ for a white OLED. Clearly, the luminous efficacy shows a more pronounced decrease with increasing luminance compared to the EQE. While the EQE solely accounts for the intensity dependency of the internal quantum efficiency, the LE also quantifies changes due to transport related properties of the device, i.e. conductivity of the layers and possible energy barriers within the device that need to be overcome. Their impact on the LE roll-off can be calculated by determining the ratio of the normalized LE and EQE curves, [LE/EQE]$_\text{norm}$, yielding the red solid line for this specific device (cf. Fig. \ref{LE_rolloff}). For the device considered here, the resistive losses already contribute to as much as 20\,\% of the LE roll-off at 5000\,cd\,m$^{-2}$. Further research on high conductivity transport materials [\cite{Sasabe2011, Sasabe2011a}], doped transport layers [\cite{Walzer2007}], and even investigation of homojunction OLEDs [\cite{Harada2005, Cai2011, Cai2011a}], barely having any energy barriers within the device, will help to reduce this loss. We assume here that the relative decrease of [LE/EQE]$_\text{norm}$ can be increased from 0.8 to at least 0.9 at 5000\,cd\,m$^{-2}$.

Of course it is most challenging to address and improve all aspects given in the table of Figure \ref{white_OLED_perspective} in one future OLED. Still, considering all these factors, we believe that for high-quality white OLEDs luminous efficacies of 155\,lm\,W$^{-1}$ will be in reach, even at a high operating brightness of 5000\,cd\,m$^{-2}$. If these values really can be achieved, then the future of white OLEDs might be bright, as they will by far outperform existing lighting technologies  such as halogen lamps, fluorescent tubes, and compact fluorescent lamps [\cite{Steele2007}]. At the same time, the OLED technology will enter our every day life with an exciting new form-factor, redefining the way we use and perceive artificial light.

However, one should also keep in mind that the OLED lighting technology might have unexpected competitors: While it is unlikely that inorganic thin-film electroluminescence will reach similar parameters in terms of efficiency, brightness and lifetime, the inorganic LED in combination with light-distributing sheets might be a serious competitor. Such technologies are currently intensively developed for liquid crystal display backlighting. The current efficiency advantage of the white LED, in combination with the low-cost potential of the light-distribution sheets, might yield a product that is for the end user similarly or more attractive than the much more elegant solution offered by the OLED.

\begin{acknowledgments}
We are indebted for many coworkers and colleagues which we had the pleasure to interact with. For our own results presented here, Karsten Fehse, Frank Lindner, Patricia Freitag, Mauro Furno, Simone Hoffmann, Rico Meerheim, Thomas Rosenow, Gregor Schwartz, Caroline Weichsel, Julia W\"unsche, and many others have made important contributions.  We enjoyed helpful discussions with Chihaya Adachi, Marc Baldo, Herbert B\"orner, Stephen Forrest, Georg G\"artner, Horst Greiner, Junji Kido, Kristiaan Neyts, Ching Tang, and Mark Thompson. Furthermore, we would like to thank Novaled AG for continuous support.

This work would not have been possible without the continuous support of the funding agencies: the European Commission with projects ``OLLA" (IST-2002-004607) and ``OLED100.eu" (FP7-224122) within the sixth and seventh framework IST program, respectively; the Bundesminesterium f\"ur Bildung und Forschung (BMBF) with projects ``ROLLEX" (13N 8855) and ``R2FLEX" (13N11060); and the state of Saxony with the project ``NKOE" (12712). We received further funding via the Leibniz prize of the Deutsche Forschungsgemeinschaft (DFG). S.R. was partially supported by personal research funds from the DFG. 
\end{acknowledgments}

\begin{appendix}
\section{International Union of Pure and Applied Chemistry (IUPAC) names of the materials discussed}

If the full chemical name is not specifically important in the main text of this review, only the material's common abbreviation found in literature is used to improve readability. Here, the full IUPAC chemical names of all materials discussed are listed. In the following tables, redundancies might occur, i.e. multiple abbreviations for one material. We display them both, as we use the abbreviation found in the respective reference.

\begin{table*}[t]
\caption{\label{tab:app_SM_materials}Small molecular weight materials discussed in this review.} 
\begin{ruledtabular} 
\begin{tabular}{lll}
Abbreviation & IUPAC name & function\\
\hline
4P-NPD&N,N'-di-(1-naphthalenyl)-N,N'-diphenyl-[1,1':4',1'':4'',1'''-quaterphenyl]-&blue fluor. emitter\\
&4,4'''-diamine&\\
$\alpha$-NPD&N,N'-bis(naphthalen-1-yl)-N,N'-bis(phenyl)-benzidine&HTL + host\\
BBOT&2,5-bis(5-tert-butyl-2-benzoxazolyl)thiophen&ETL\\
BCzVBi&4,4'-bis(9-ethyl-3-carbazovinylene)-1,1'-biphenyl&blue fluor. emitter\\
Bepp$_2$&bis(2-(2-hydroxyphenyl)-pyridine)beryllium&host\\
BPhen&4,7-diphenyl-1,10-phenanthroline&ETL\\
Bt$_2$Ir(acac)&bis(2-phenylbenzothiozolato-N,C$^{2'}$)(acetylacetonate)Ir(III)&yellow phos. emitter\\
Btp$_2$Ir(acac)&bis(2-(2'-benzothienyl)-pyridinato-N,C$^{3'}$)(acetylacetonate)Ir(III)&red phos. emitter\\
CBP&4,4'-bis(carbazol-9-yl)biphenyl&host\\
CDBP&4,4'-bis(carbazol-9-yl)-2,2'-dimethylbiphenyl&host\\
CPhBzIm&bis(N-phenylbenzimidazole)carbazole&host\\
CzSi&9-(4-tert-butylphenyl)-3,6-bis(triphenylsilyl)-9H-carbazole&host\\
DCB&1,4-bis((9H-carbazol-9-yl)methyl)benzene&host\\
DCJTB&4-(dicyanomethylene)-2-tert-butyl-6-(1,1,7,7-tetramethyljulolidin-4-yl-vinyl)-&red fluor. emitter\\
&4H-pyran&\\
DCM&2-(2-(4-(dimethylamino)styryl)-6-methyl-4H-pyran-4-ylidene)malononitrile&red fluor. emitter\\
DCM1&2-(2-(4-(dimethylamino)styryl)-6-methyl-4H-pyran-4-ylidene)malononitrile&red fluor. emitter\\
DCM2&4-(dicyanomethylene)-2-methyl-6-julolidyl-9-enyl-4H-pyran&red fluor. emitter\\
DCzPPy&2,6-bis(3-(9H-carbazol-9-yl)phenyl)pyridine&host\\
DPAN&4-diphenylamino-1,8-naphthalimide&green fluor. emitter\\
DPAVBi&4,4'-bis[4-(di-p-tolylamino)styryl]biphenyl&blue fluor. emitter\\
DTPA&(4-(2-[2,5-dibromo-4-(2-(4-diphenylamino-phenyl)-vinyl)-phenyl]-vinyl)-phenyl)-&green fluor. emitter\\
&diphenylamine&\\
FIr6&bis(2,4-difluorophenylpyridinato)tetrakis(1-pyrazolyl)borate iridium(III)&blue phos. emitter\\
FIrpic&bis(3,5-difluoro-2-(2-pyridyl)phenyl-(2-carboxypyridyl)iridium(III)&blue phos. emitter\\
FPt1&platinum(II)(2-(4',6'-difluorophenyl)pyridinato-N,C$^{2'}$)(2,4-pentanedionato)&blue phos. emitter\\
FPt2&platinum(II)(2-(4',6'-difluorophenyl)pyridinato-N,C$^{2'}$)(6-methyl-2,4-&blue phos. emitter\\
&heptanedionato-O,O)&\\
Ga(pyimd)$_3$&tris(2-(2-pyridyl)imidazole)gallium(III)&host\\
Ir(1-piq)$_3$&tris(1-phenyl-isoquinolinato-N,C$^{2'}$)iridium(III)&red phos. emitter\\
Ir(Bu-ppy)$_3$&fac-tris(2-(4’-ter-butyl)phenylpyridine)iridium(III)&green phos. emitter\\
Ir(dbfmi)&mer-tris(N-dibenzofuranyl-N'-methylimidazole)iridium(III)&blue phos. emitter\\
Ir(dhfpy)$_2$(acac)&bis(2-(9,9-dihexylfluorenyl)-1-pyridine)(acetylacetonate)iridium(III)&yellow phos. emitter\\
Ir(dfbppy)fbppz)$_2$&bis(4-tert-butyl-2-(2,4-difluorophenyl)pyridinato)(3-(trifluoro-methyl)-&blue phos. emitter\\
&5-(4-tert-butylpyridyl)pyrazolate)iridium(III)&\\
Ir(HFP)$_3$&tris(2,5-bis-2'-(9',9'-dihexylfluorene)pyridine)iridium(III)&red phos. emitter\\
\end{tabular}
\end{ruledtabular}
\end{table*}
\begin{table*}
\caption{\label{tab:app_SM_materials}Small molecular weight materials discussed in this review - continued.} 
\begin{ruledtabular} 
\begin{tabular}{lll}
\small
Abbreviation & IUPAC name & function\\
\hline
Ir(MDQ)$_2$(acac)&bis(2-methyldibenzo[f,h]quinoxaline)(acetylacetonate)iridium (III)&red phos. emitter\\
Ir(pbi)$_2$(acac)&bis(phenyl-benzoimidazole)(acetylacetonate)iridium(III)&yellow phos. emitter\\
Ir(ppy)$_3$&fac-tris(2-phenylpyridine)iridium(III)&green phos. emitter\\
Ir(ppy)$_2$pc&fac-bis(2-phenylpyridyl)(2-pyridylcoumarin)iridium(III)&yellow phos. emitter\\
Ir(SBFP)$_2$(acac)&iridium(III) bis(2-(9,9’-spirobi[fluorene]-7-yl)pyridine-N,C$^{2'}$)acetylacetonate&orange phos. emitter\\
MB-BT-ThTPA&4-(5-(4-(diphenylamino)-phenyl)-thienyl-2-)-7-(4- methoxybenzene)-2,1,3-&red fluor. emitter\\
&benzothiadiazole&\\
mCP&1,3-bis(carbazol-9-yl)benzene&host\\
MPD&(2-methyl-6-(2-(2,3,6,7-tetrahydro-1H,5H-benzo[i,j]quinolizin-9-yl)ethenyl)-&red fluor. emitter\\
&4H-pyran-4-ylidene)-propanedinitrile&\\
MQAB&6-mesityl-N-(6-mesitylquinolin-2(1H)-ylidene)quinolin-2-amine-BF2&blue fluor. emitter\\
m-MTDATA&4,4',4''-tris(N-3-methylphenyl-N-phenylamino)triphenylamine&HTL\\
Nile Red &9-diethylamino-5-benzo[$\alpha$]phenoxazinone&red fluor. emitter\\
NPB&N,N'-bis(naphthalen-1-yl)-N,N'-bis(phenyl)-benzidine&HTL + host\\
Os(bptz)$_2$(dppee)&osmium(II)bis(3-tert-butyl-5-(2-pyridyl)-1,2,4-triazolate)&orange phos. emitter\\
&cis-1,2-bis(diphenylphosphino)ethene&\\
Os-R1&osmium(II)bis(3-(trifluoromethyl)-5-(pyridine)-1,2-pyridazine)&red phos. emitter\\
&diphenylmethylphosphine&\\
OXA&(3,5-bis(5-(4-tert-butylphenyl)-1,3,4-oxadiazol-2–yl)-benzene)&ETL\\
OXD-7&1,3-bis[2-(4-tert-butylphenyl)-1,3,4-oxadiazo-5-yl]benzene&ETL\\
PBD&2-(4-biphenylyl)-5-(4-tert-butylphenyl)1,3,4-oxadiazole&ETL\\
(Piq)$_2$Ir(acaF)&bis-(1-phenylisoquinolyl)iridium(III)(1-trifluoro)acetylacetonate&red phos. emitter\\
PO9&3,6-bis(diphenylphosphoryl)-9-phenylcarbazole &host\\
(Ppq)$_2$Ir(acac)&bis-(1-phenylisoquinolyl)iridium(III)(1-trifluoro)acetylacetonate&red phos. emitter\\
PQIr&iridium(III)bis(2-phenyl quinolyl-N,C$^{2'}$)acetylacetonate&red phos. emitter\\
PQ${_2}$Ir(dpm)&iridium bis(2-phenyl-quinoly-N,C$^{2'}$)dipivaloylmethane&red phos. emitter\\
Pt$L^2$Cl&(platinum II(methyl-3,5-di(2-pyridyl)benzoate)chloride)&blue phos. emitter\\
Spiro-DPVBi&2,2' ,7,7'-tetrakis(2,2-diphenylvinyl)spiro-9,9'-bifluorene&blue fluor. emitter\\
SPPO1&9,9-spirobifluoren-2-yl-diphenyl-phosphine oxide&host\\
TCTA&4,4',4''-tris(carbazol-9-yl)-triphenylamine&host\\
TPATBT&4,7-bis(5-(4-(N-phenyl-N-(4-methylphenyl)amino)phenyl)-thienyl-2-)-2,1,3-&red fluor. emitter\\ 
&benzothiadiazole&\\
TPB&1,1,4,4-tetraphenyl-1,3-butadiene&blue fluor. emitter\\
TPBi&2,2',2"-(1,3,5-benzinetriyl)-tris(1-phenyl-1-H-benzimidazole)&ETL\\ 
TPDCM&2-(2-[2-(4-(bis(phenyl-amino)phenyl)-vinyl]-6-tert-butyl-pyran-4-ylidene)-&red fluor. emitter\\
&malononitrile&\\
TPyPA&tris[4-(pyrenyl)-phenyl]amine&blue fluor. emitter\\
UGH2&1,4-bis(triphenylsilyl)benzene&host\\
\end{tabular}
\end{ruledtabular}
\end{table*}

\begin{table*}
\caption{\label{tab:app_poly_materials}Polymer materials discussed in this review.} 
\begin{ruledtabular} 
\begin{tabular}{lll}
Abbreviation & IUPAC name & function\\
\hline
C12O-PPP&C12O-poly(1,4-phenylene)&blue fluor. emitter\\
MEH-PPV&poly[2-methoxy-5-(2-ethylhexyloxy)-1,4-phenylenevinylene]&green fluor. emitter\\
P36HCTPSi&n.a.; see \cite{Cheng2010a} for chemical structure&host\\ 
PDHF&poly(9,9-dihexylfluorene-2,7-diyl)&blue emitter + host\\
PDMS&poly(dimethylsiloxane)&insulator\\
PEDOT:PSS&poly(3,4-ethylenedioxythiophene) poly(styrenesulfonate)&HTL\\
PF&polyfluorene& host\\
PFO-poss&polyhedral oligomeric silsesquioxane-terminated poly(9,9-dioctylfuorene)&blue fluor. emitter + host\\
PF-OH&poly[9,9-bis(2-(2-(2-diethanolamino ethoxy)ethoxy)ethyl)fluorene]& host\\
PMMA&poly(methyl 2-methylpropenoate)&wide band gap insulator\\
PVK&poly(N-vinylcarbazole)&host\\
VB-TCTA&vinylbenzyl-4,4',4''-tris(carbazol-9-yl)-triphenylamine&host\\
WPF03&n.a.; see \cite{Niu2006,Tu2006} for chemical structure&white emitting host\\
\end{tabular} 
\end{ruledtabular} 
\end{table*}

\end{appendix}

\bibliography{Reineke_literature}

\begin{thebibliography}{295}%
\makeatletter
\providecommand \@ifxundefined [1]{%
 \@ifx{#1\undefined}
}%
\providecommand \@ifnum [1]{%
 \ifnum #1\expandafter \@firstoftwo
 \else \expandafter \@secondoftwo
 \fi
}%
\providecommand \@ifx [1]{%
 \ifx #1\expandafter \@firstoftwo
 \else \expandafter \@secondoftwo
 \fi
}%
\providecommand \natexlab [1]{#1}%
\providecommand \enquote  [1]{``#1''}%
\providecommand \bibnamefont  [1]{#1}%
\providecommand \bibfnamefont [1]{#1}%
\providecommand \citenamefont [1]{#1}%
\providecommand \href@noop [0]{\@secondoftwo}%
\providecommand \href [0]{\begingroup \@sanitize@url \@href}%
\providecommand \@href[1]{\@@startlink{#1}\@@href}%
\providecommand \@@href[1]{\endgroup#1\@@endlink}%
\providecommand \@sanitize@url [0]{\catcode `\\12\catcode `\$12\catcode
  `\&12\catcode `\#12\catcode `\^12\catcode `\_12\catcode `\%12\relax}%
\providecommand \@@startlink[1]{}%
\providecommand \@@endlink[0]{}%
\providecommand \url  [0]{\begingroup\@sanitize@url \@url }%
\providecommand \@url [1]{\endgroup\@href {#1}{\urlprefix }}%
\providecommand \urlprefix  [0]{URL }%
\providecommand \Eprint [0]{\href }%
\providecommand \doibase [0]{http://dx.doi.org/}%
\providecommand \selectlanguage [0]{\@gobble}%
\providecommand \bibinfo  [0]{\@secondoftwo}%
\providecommand \bibfield  [0]{\@secondoftwo}%
\providecommand \translation [1]{[#1]}%
\providecommand \BibitemOpen [0]{}%
\providecommand \bibitemStop [0]{}%
\providecommand \bibitemNoStop [0]{.\EOS\space}%
\providecommand \EOS [0]{\spacefactor3000\relax}%
\providecommand \BibitemShut  [1]{\csname bibitem#1\endcsname}%
\let\auto@bib@innerbib\@empty
\bibitem [{\citenamefont {Edison}(1880)}]{Edison1880}%
  \BibitemOpen
  \bibfield  {author} {\bibinfo {author} {\bibfnamefont {T.~A.}\ \bibnamefont
  {Edison}},\ }\href@noop {} {\bibfield  {journal} {\bibinfo  {journal} {US
  patent: US223898 (A)}\ } (\bibinfo {year} {1880})}\BibitemShut {NoStop}%
\bibitem [{\citenamefont {Ohta}\ and\ \citenamefont
  {Robertson}(2005)}]{Ohta2005}%
  \BibitemOpen
  \bibfield  {author} {\bibinfo {author} {\bibfnamefont {N.}~\bibnamefont
  {Ohta}}\ and\ \bibinfo {author} {\bibfnamefont {A.~R.}\ \bibnamefont
  {Robertson}},\ }\href@noop {} {\emph {\bibinfo {title} {Colorimetry:
  Fundamentals and Applications}}}\ (\bibinfo  {publisher} {John Wiley \& Sons,
  Ltd.},\ \bibinfo {year} {2005})\BibitemShut {NoStop}%
\bibitem [{\citenamefont {Steele}(2007)}]{Steele2007}%
  \BibitemOpen
  \bibfield  {author} {\bibinfo {author} {\bibfnamefont {R.~V.}\ \bibnamefont
  {Steele}},\ }\href@noop {} {\bibfield  {journal} {\bibinfo  {journal} {Nature
  Photonics}\ }\textbf {\bibinfo {volume} {1}},\ \bibinfo {pages} {25}
  (\bibinfo {year} {2007})}\BibitemShut {NoStop}%
\bibitem [{Note1()}]{Note1}%
  \BibitemOpen
  \bibinfo {note} {Note that luminescence can also be generated without
  injecting charges, driven by an alternating electric field, in inorganic thin
  films [\cite {Rack1998}], quantum dots [\cite {Wood2011}], and organic
  systems [\cite {Perumal2012}]. However, to date, these technologies are not
  considered mainstream for SSL.}\BibitemShut {Stop}%
\bibitem [{\citenamefont {Zheludev}(2007)}]{Zheludev2007}%
  \BibitemOpen
  \bibfield  {author} {\bibinfo {author} {\bibfnamefont {N.}~\bibnamefont
  {Zheludev}},\ }\href@noop {} {\bibfield  {journal} {\bibinfo  {journal}
  {Nature Photonics}\ }\textbf {\bibinfo {volume} {1}},\ \bibinfo {pages} {189}
  (\bibinfo {year} {2007})}\BibitemShut {NoStop}%
\bibitem [{\citenamefont {Ohno}(2004)}]{Ohno2004}%
  \BibitemOpen
  \bibfield  {author} {\bibinfo {author} {\bibfnamefont {Y.}~\bibnamefont
  {Ohno}},\ }\href@noop {} {\bibfield  {journal} {\bibinfo  {journal} {Proc.
  SPIE}\ }\textbf {\bibinfo {volume} {5530}},\ \bibinfo {pages} {88} (\bibinfo
  {year} {2004})}\BibitemShut {NoStop}%
\bibitem [{\citenamefont {Narukawa}\ \emph {et~al.}(2006)\citenamefont
  {Narukawa}, \citenamefont {Narita}, \citenamefont {Sakamoto}, \citenamefont
  {Deguchi}, \citenamefont {Yamada},\ and\ \citenamefont
  {Mukai}}]{Narukawa2006}%
  \BibitemOpen
  \bibfield  {author} {\bibinfo {author} {\bibfnamefont {Y.}~\bibnamefont
  {Narukawa}}, \bibinfo {author} {\bibfnamefont {J.}~\bibnamefont {Narita}},
  \bibinfo {author} {\bibfnamefont {T.}~\bibnamefont {Sakamoto}}, \bibinfo
  {author} {\bibfnamefont {K.}~\bibnamefont {Deguchi}}, \bibinfo {author}
  {\bibfnamefont {T.}~\bibnamefont {Yamada}}, \ and\ \bibinfo {author}
  {\bibfnamefont {T.}~\bibnamefont {Mukai}},\ }\href@noop {} {\bibfield
  {journal} {\bibinfo  {journal} {Japanese Journal Of Applied Physics Part
  2-Letters \& Express Letters}\ }\textbf {\bibinfo {volume} {45}},\ \bibinfo
  {pages} {L1084} (\bibinfo {year} {2006})}\BibitemShut {NoStop}%
\bibitem [{\citenamefont {Narukawa}\ \emph {et~al.}(2008)\citenamefont
  {Narukawa}, \citenamefont {Sano}, \citenamefont {Sakamoto}, \citenamefont
  {Yamada},\ and\ \citenamefont {Mukai}}]{Narukawa2008}%
  \BibitemOpen
  \bibfield  {author} {\bibinfo {author} {\bibfnamefont {Y.}~\bibnamefont
  {Narukawa}}, \bibinfo {author} {\bibfnamefont {M.}~\bibnamefont {Sano}},
  \bibinfo {author} {\bibfnamefont {T.}~\bibnamefont {Sakamoto}}, \bibinfo
  {author} {\bibfnamefont {T.}~\bibnamefont {Yamada}}, \ and\ \bibinfo {author}
  {\bibfnamefont {T.}~\bibnamefont {Mukai}},\ }\href@noop {} {\bibfield
  {journal} {\bibinfo  {journal} {Physica Status Solidi A-applications and
  Materials Science}\ }\textbf {\bibinfo {volume} {205}},\ \bibinfo {pages}
  {1081} (\bibinfo {year} {2008})}\BibitemShut {NoStop}%
\bibitem [{\citenamefont {Tang}\ and\ \citenamefont
  {VanSlyke}(1987)}]{Tang1987}%
  \BibitemOpen
  \bibfield  {author} {\bibinfo {author} {\bibfnamefont {C.~W.}\ \bibnamefont
  {Tang}}\ and\ \bibinfo {author} {\bibfnamefont {S.~A.}\ \bibnamefont
  {VanSlyke}},\ }\href@noop {} {\bibfield  {journal} {\bibinfo  {journal}
  {Applied Physics Letters}\ }\textbf {\bibinfo {volume} {51}},\ \bibinfo
  {pages} {913} (\bibinfo {year} {1987})}\BibitemShut {NoStop}%
\bibitem [{\citenamefont {Burroughes}\ \emph {et~al.}(1990)\citenamefont
  {Burroughes}, \citenamefont {Bradley}, \citenamefont {Brown}, \citenamefont
  {Marks}, \citenamefont {Mackay}, \citenamefont {Friend}, \citenamefont
  {Burns},\ and\ \citenamefont {Holmes}}]{Burroughes1990}%
  \BibitemOpen
  \bibfield  {author} {\bibinfo {author} {\bibfnamefont {J.~H.}\ \bibnamefont
  {Burroughes}}, \bibinfo {author} {\bibfnamefont {D.~D.~C.}\ \bibnamefont
  {Bradley}}, \bibinfo {author} {\bibfnamefont {A.~R.}\ \bibnamefont {Brown}},
  \bibinfo {author} {\bibfnamefont {R.~N.}\ \bibnamefont {Marks}}, \bibinfo
  {author} {\bibfnamefont {K.}~\bibnamefont {Mackay}}, \bibinfo {author}
  {\bibfnamefont {R.~H.}\ \bibnamefont {Friend}}, \bibinfo {author}
  {\bibfnamefont {P.~L.}\ \bibnamefont {Burns}}, \ and\ \bibinfo {author}
  {\bibfnamefont {A.~B.}\ \bibnamefont {Holmes}},\ }\href@noop {} {\bibfield
  {journal} {\bibinfo  {journal} {Nature}\ }\textbf {\bibinfo {volume} {347}},\
  \bibinfo {pages} {539} (\bibinfo {year} {1990})}\BibitemShut {NoStop}%
\bibitem [{Note2()}]{Note2}%
  \BibitemOpen
  \bibinfo {note} {Here, monochrome is used to describe EL devices, where
  emission stems from one type of emitter molecule only. This is, even though
  any organic semiconductor has a certain spectral distribution of its emitted
  spectrum typically with full width at half maximum in the range of 50 to
  100\protect \tmspace +\thinmuskip {.1667em}nm [\cite
  {Pope1999}].}\BibitemShut {Stop}%
\bibitem [{\citenamefont {Kido}\ \emph {et~al.}(1994)\citenamefont {Kido},
  \citenamefont {Hongawa}, \citenamefont {Okuyama},\ and\ \citenamefont
  {Nagai}}]{Kido1994}%
  \BibitemOpen
  \bibfield  {author} {\bibinfo {author} {\bibfnamefont {J.}~\bibnamefont
  {Kido}}, \bibinfo {author} {\bibfnamefont {K.}~\bibnamefont {Hongawa}},
  \bibinfo {author} {\bibfnamefont {K.}~\bibnamefont {Okuyama}}, \ and\
  \bibinfo {author} {\bibfnamefont {K.}~\bibnamefont {Nagai}},\ }\href@noop {}
  {\bibfield  {journal} {\bibinfo  {journal} {Applied Physics Letters}\
  }\textbf {\bibinfo {volume} {64}},\ \bibinfo {pages} {815} (\bibinfo {year}
  {1994})}\BibitemShut {NoStop}%
\bibitem [{\citenamefont {Wang}\ \emph {et~al.}(1999)\citenamefont {Wang},
  \citenamefont {Sun}, \citenamefont {Meghdadi}, \citenamefont {Leising},\ and\
  \citenamefont {Epstein}}]{Wang1999}%
  \BibitemOpen
  \bibfield  {author} {\bibinfo {author} {\bibfnamefont {Y.~Z.}\ \bibnamefont
  {Wang}}, \bibinfo {author} {\bibfnamefont {R.~G.}\ \bibnamefont {Sun}},
  \bibinfo {author} {\bibfnamefont {F.}~\bibnamefont {Meghdadi}}, \bibinfo
  {author} {\bibfnamefont {G.}~\bibnamefont {Leising}}, \ and\ \bibinfo
  {author} {\bibfnamefont {A.~J.}\ \bibnamefont {Epstein}},\ }\href@noop {}
  {\bibfield  {journal} {\bibinfo  {journal} {Applied Physics Letters}\
  }\textbf {\bibinfo {volume} {74}},\ \bibinfo {pages} {3613} (\bibinfo {year}
  {1999})}\BibitemShut {NoStop}%
\bibitem [{Note3()}]{Note3}%
  \BibitemOpen
  \bibinfo {note} {Starting here, the abbreviation OLED is jointly used for
  small molecule and polymer organic LEDs, simply because both material classes
  belong to the organic chemistry. The context will clarify, whether OLEDs or
  PLEDs are discussed.}\BibitemShut {Stop}%
\bibitem [{\citenamefont {Greenham}\ \emph {et~al.}(1994)\citenamefont
  {Greenham}, \citenamefont {Friend},\ and\ \citenamefont
  {Bradley}}]{Greenham1994}%
  \BibitemOpen
  \bibfield  {author} {\bibinfo {author} {\bibfnamefont {N.~C.}\ \bibnamefont
  {Greenham}}, \bibinfo {author} {\bibfnamefont {R.~H.}\ \bibnamefont
  {Friend}}, \ and\ \bibinfo {author} {\bibfnamefont {D.~D.~C.}\ \bibnamefont
  {Bradley}},\ }\href@noop {} {\bibfield  {journal} {\bibinfo  {journal}
  {Advanced Materials}\ }\textbf {\bibinfo {volume} {6}},\ \bibinfo {pages}
  {491} (\bibinfo {year} {1994})}\BibitemShut {NoStop}%
\bibitem [{\citenamefont {So}\ and\ \citenamefont {Kondakov}(2010)}]{So2010}%
  \BibitemOpen
  \bibfield  {author} {\bibinfo {author} {\bibfnamefont {F.}~\bibnamefont
  {So}}\ and\ \bibinfo {author} {\bibfnamefont {D.}~\bibnamefont {Kondakov}},\
  }\href@noop {} {\bibfield  {journal} {\bibinfo  {journal} {Advanced
  Materials}\ }\textbf {\bibinfo {volume} {22}},\ \bibinfo {pages} {3762}
  (\bibinfo {year} {2010})}\BibitemShut {NoStop}%
\bibitem [{\citenamefont {Baldo}\ \emph {et~al.}(1997)\citenamefont {Baldo},
  \citenamefont {Kozlov}, \citenamefont {Burrows}, \citenamefont {Forrest},
  \citenamefont {Ban}, \citenamefont {Koene},\ and\ \citenamefont
  {Thompson}}]{Baldo1997}%
  \BibitemOpen
  \bibfield  {author} {\bibinfo {author} {\bibfnamefont {M.~A.}\ \bibnamefont
  {Baldo}}, \bibinfo {author} {\bibfnamefont {V.~G.}\ \bibnamefont {Kozlov}},
  \bibinfo {author} {\bibfnamefont {P.~E.}\ \bibnamefont {Burrows}}, \bibinfo
  {author} {\bibfnamefont {S.~R.}\ \bibnamefont {Forrest}}, \bibinfo {author}
  {\bibfnamefont {V.~S.}\ \bibnamefont {Ban}}, \bibinfo {author} {\bibfnamefont
  {B.}~\bibnamefont {Koene}}, \ and\ \bibinfo {author} {\bibfnamefont {M.~E.}\
  \bibnamefont {Thompson}},\ }\href@noop {} {\bibfield  {journal} {\bibinfo
  {journal} {Applied Physics Letters}\ }\textbf {\bibinfo {volume} {71}},\
  \bibinfo {pages} {3033} (\bibinfo {year} {1997})}\BibitemShut {NoStop}%
\bibitem [{\citenamefont {Zhou}\ \emph {et~al.}(2005)\citenamefont {Zhou},
  \citenamefont {Ngo}, \citenamefont {Brown}, \citenamefont {Shtein},\ and\
  \citenamefont {Forrest}}]{Zhou2005}%
  \BibitemOpen
  \bibfield  {author} {\bibinfo {author} {\bibfnamefont {T.~X.}\ \bibnamefont
  {Zhou}}, \bibinfo {author} {\bibfnamefont {T.}~\bibnamefont {Ngo}}, \bibinfo
  {author} {\bibfnamefont {J.~J.}\ \bibnamefont {Brown}}, \bibinfo {author}
  {\bibfnamefont {M.}~\bibnamefont {Shtein}}, \ and\ \bibinfo {author}
  {\bibfnamefont {S.~R.}\ \bibnamefont {Forrest}},\ }\href@noop {} {\bibfield
  {journal} {\bibinfo  {journal} {Applied Physics Letters}\ }\textbf {\bibinfo
  {volume} {86}},\ \bibinfo {pages} {021107} (\bibinfo {year}
  {2005})}\BibitemShut {NoStop}%
\bibitem [{\citenamefont {Forrest}(1997)}]{Forrest1997}%
  \BibitemOpen
  \bibfield  {author} {\bibinfo {author} {\bibfnamefont {S.~R.}\ \bibnamefont
  {Forrest}},\ }\href@noop {} {\bibfield  {journal} {\bibinfo  {journal}
  {Chemical Reviews}\ }\textbf {\bibinfo {volume} {97}},\ \bibinfo {pages}
  {1793} (\bibinfo {year} {1997})}\BibitemShut {NoStop}%
\bibitem [{\citenamefont {Forrest}(2004)}]{Forrest2004}%
  \BibitemOpen
  \bibfield  {author} {\bibinfo {author} {\bibfnamefont {S.~R.}\ \bibnamefont
  {Forrest}},\ }\href@noop {} {\bibfield  {journal} {\bibinfo  {journal}
  {Nature}\ }\textbf {\bibinfo {volume} {428}},\ \bibinfo {pages} {911}
  (\bibinfo {year} {2004})}\BibitemShut {NoStop}%
\bibitem [{\citenamefont {Friend}\ \emph {et~al.}(1999)\citenamefont {Friend},
  \citenamefont {Gymer}, \citenamefont {Holmes}, \citenamefont {Burroughes},
  \citenamefont {Marks}, \citenamefont {Taliani}, \citenamefont {Bradley},
  \citenamefont {Dos~Santos}, \citenamefont {Bredas}, \citenamefont
  {Logdlund},\ and\ \citenamefont {Salaneck}}]{Friend1999}%
  \BibitemOpen
  \bibfield  {author} {\bibinfo {author} {\bibfnamefont {R.~H.}\ \bibnamefont
  {Friend}}, \bibinfo {author} {\bibfnamefont {R.~W.}\ \bibnamefont {Gymer}},
  \bibinfo {author} {\bibfnamefont {A.~B.}\ \bibnamefont {Holmes}}, \bibinfo
  {author} {\bibfnamefont {J.~H.}\ \bibnamefont {Burroughes}}, \bibinfo
  {author} {\bibfnamefont {R.~N.}\ \bibnamefont {Marks}}, \bibinfo {author}
  {\bibfnamefont {C.}~\bibnamefont {Taliani}}, \bibinfo {author} {\bibfnamefont
  {D.~D.~C.}\ \bibnamefont {Bradley}}, \bibinfo {author} {\bibfnamefont
  {D.~A.}\ \bibnamefont {Dos~Santos}}, \bibinfo {author} {\bibfnamefont
  {J.~L.}\ \bibnamefont {Bredas}}, \bibinfo {author} {\bibfnamefont
  {M.}~\bibnamefont {Logdlund}}, \ and\ \bibinfo {author} {\bibfnamefont
  {W.~R.}\ \bibnamefont {Salaneck}},\ }\href@noop {} {\bibfield  {journal}
  {\bibinfo  {journal} {Nature}\ }\textbf {\bibinfo {volume} {397}},\ \bibinfo
  {pages} {121} (\bibinfo {year} {1999})}\BibitemShut {NoStop}%
\bibitem [{\citenamefont {Rehmann}\ \emph {et~al.}(2008)\citenamefont
  {Rehmann}, \citenamefont {Ulbricht}, \citenamefont {Koehnen}, \citenamefont
  {Zacharias}, \citenamefont {Gather}, \citenamefont {Hertel}, \citenamefont
  {Holder}, \citenamefont {Meerholz},\ and\ \citenamefont
  {Schubert}}]{Rehmann2008}%
  \BibitemOpen
  \bibfield  {author} {\bibinfo {author} {\bibfnamefont {N.}~\bibnamefont
  {Rehmann}}, \bibinfo {author} {\bibfnamefont {C.}~\bibnamefont {Ulbricht}},
  \bibinfo {author} {\bibfnamefont {A.}~\bibnamefont {Koehnen}}, \bibinfo
  {author} {\bibfnamefont {P.}~\bibnamefont {Zacharias}}, \bibinfo {author}
  {\bibfnamefont {M.~C.}\ \bibnamefont {Gather}}, \bibinfo {author}
  {\bibfnamefont {D.}~\bibnamefont {Hertel}}, \bibinfo {author} {\bibfnamefont
  {E.}~\bibnamefont {Holder}}, \bibinfo {author} {\bibfnamefont
  {K.}~\bibnamefont {Meerholz}}, \ and\ \bibinfo {author} {\bibfnamefont
  {U.~S.}\ \bibnamefont {Schubert}},\ }\href@noop {} {\bibfield  {journal}
  {\bibinfo  {journal} {Advanced Materials}\ }\textbf {\bibinfo {volume}
  {20}},\ \bibinfo {pages} {129} (\bibinfo {year} {2008})}\BibitemShut
  {NoStop}%
\bibitem [{\citenamefont {Zuniga}\ \emph {et~al.}(2011)\citenamefont {Zuniga},
  \citenamefont {Barlow},\ and\ \citenamefont {Marder}}]{Zuniga2011}%
  \BibitemOpen
  \bibfield  {author} {\bibinfo {author} {\bibfnamefont {C.~A.}\ \bibnamefont
  {Zuniga}}, \bibinfo {author} {\bibfnamefont {S.}~\bibnamefont {Barlow}}, \
  and\ \bibinfo {author} {\bibfnamefont {S.~R.}\ \bibnamefont {Marder}},\
  }\href@noop {} {\bibfield  {journal} {\bibinfo  {journal} {Chemistry of
  Materials}\ }\textbf {\bibinfo {volume} {23}},\ \bibinfo {pages} {658}
  (\bibinfo {year} {2011})}\BibitemShut {NoStop}%
\bibitem [{\citenamefont {Gather}\ \emph
  {et~al.}(2007{\natexlab{a}})\citenamefont {Gather}, \citenamefont {Koehnen},
  \citenamefont {Falcou}, \citenamefont {Becker},\ and\ \citenamefont
  {Meerholz}}]{Gather2007a}%
  \BibitemOpen
  \bibfield  {author} {\bibinfo {author} {\bibfnamefont {M.~C.}\ \bibnamefont
  {Gather}}, \bibinfo {author} {\bibfnamefont {A.}~\bibnamefont {Koehnen}},
  \bibinfo {author} {\bibfnamefont {A.}~\bibnamefont {Falcou}}, \bibinfo
  {author} {\bibfnamefont {H.}~\bibnamefont {Becker}}, \ and\ \bibinfo {author}
  {\bibfnamefont {K.}~\bibnamefont {Meerholz}},\ }\href@noop {} {\bibfield
  {journal} {\bibinfo  {journal} {Advanced Functional Materials}\ }\textbf
  {\bibinfo {volume} {17}},\ \bibinfo {pages} {191} (\bibinfo {year}
  {2007}{\natexlab{a}})}\BibitemShut {NoStop}%
\bibitem [{\citenamefont {Stewart}\ \emph {et~al.}(2012)\citenamefont
  {Stewart}, \citenamefont {Lippert}, \citenamefont {Nagel}, \citenamefont
  {Nuesch},\ and\ \citenamefont {Wokaun}}]{Stewart2012}%
  \BibitemOpen
  \bibfield  {author} {\bibinfo {author} {\bibfnamefont {J.~S.}\ \bibnamefont
  {Stewart}}, \bibinfo {author} {\bibfnamefont {T.}~\bibnamefont {Lippert}},
  \bibinfo {author} {\bibfnamefont {M.}~\bibnamefont {Nagel}}, \bibinfo
  {author} {\bibfnamefont {F.}~\bibnamefont {Nuesch}}, \ and\ \bibinfo {author}
  {\bibfnamefont {A.}~\bibnamefont {Wokaun}},\ }\href@noop {} {\bibfield
  {journal} {\bibinfo  {journal} {Applied Physics Letters}\ }\textbf {\bibinfo
  {volume} {100}},\ \bibinfo {pages} {203303} (\bibinfo {year}
  {2012})}\BibitemShut {NoStop}%
\bibitem [{\citenamefont {Kanno}\ \emph {et~al.}(2005)\citenamefont {Kanno},
  \citenamefont {Sun},\ and\ \citenamefont {Forrest}}]{Kanno2005}%
  \BibitemOpen
  \bibfield  {author} {\bibinfo {author} {\bibfnamefont {H.}~\bibnamefont
  {Kanno}}, \bibinfo {author} {\bibfnamefont {Y.}~\bibnamefont {Sun}}, \ and\
  \bibinfo {author} {\bibfnamefont {S.~R.}\ \bibnamefont {Forrest}},\
  }\href@noop {} {\bibfield  {journal} {\bibinfo  {journal} {Applied Physics
  Letters}\ }\textbf {\bibinfo {volume} {86}},\ \bibinfo {pages} {263502}
  (\bibinfo {year} {2005})}\BibitemShut {NoStop}%
\bibitem [{\citenamefont {Huang}\ \emph
  {et~al.}(2006{\natexlab{a}})\citenamefont {Huang}, \citenamefont {Walzer},
  \citenamefont {Pfeiffer}, \citenamefont {Lyssenko}, \citenamefont {He},\ and\
  \citenamefont {Leo}}]{Huang2006c}%
  \BibitemOpen
  \bibfield  {author} {\bibinfo {author} {\bibfnamefont {Q.}~\bibnamefont
  {Huang}}, \bibinfo {author} {\bibfnamefont {K.}~\bibnamefont {Walzer}},
  \bibinfo {author} {\bibfnamefont {M.}~\bibnamefont {Pfeiffer}}, \bibinfo
  {author} {\bibfnamefont {V.}~\bibnamefont {Lyssenko}}, \bibinfo {author}
  {\bibfnamefont {G.~F.}\ \bibnamefont {He}}, \ and\ \bibinfo {author}
  {\bibfnamefont {K.}~\bibnamefont {Leo}},\ }\href@noop {} {\bibfield
  {journal} {\bibinfo  {journal} {Applied Physics Letters}\ }\textbf {\bibinfo
  {volume} {88}},\ \bibinfo {pages} {113515} (\bibinfo {year}
  {2006}{\natexlab{a}})}\BibitemShut {NoStop}%
\bibitem [{\citenamefont {Riel}\ \emph {et~al.}(2003)\citenamefont {Riel},
  \citenamefont {Karg}, \citenamefont {Beierlein}, \citenamefont {Ruhstaller},\
  and\ \citenamefont {Riess}}]{Riel2003}%
  \BibitemOpen
  \bibfield  {author} {\bibinfo {author} {\bibfnamefont {H.}~\bibnamefont
  {Riel}}, \bibinfo {author} {\bibfnamefont {S.}~\bibnamefont {Karg}}, \bibinfo
  {author} {\bibfnamefont {T.}~\bibnamefont {Beierlein}}, \bibinfo {author}
  {\bibfnamefont {B.}~\bibnamefont {Ruhstaller}}, \ and\ \bibinfo {author}
  {\bibfnamefont {W.}~\bibnamefont {Riess}},\ }\href@noop {} {\bibfield
  {journal} {\bibinfo  {journal} {Applied Physics Letters}\ }\textbf {\bibinfo
  {volume} {82}},\ \bibinfo {pages} {466} (\bibinfo {year} {2003})}\BibitemShut
  {NoStop}%
\bibitem [{\citenamefont {Bulovic}\ \emph {et~al.}(1996)\citenamefont
  {Bulovic}, \citenamefont {Gu}, \citenamefont {Burrows}, \citenamefont
  {Forrest},\ and\ \citenamefont {Thompson}}]{Bulovic1996}%
  \BibitemOpen
  \bibfield  {author} {\bibinfo {author} {\bibfnamefont {V.}~\bibnamefont
  {Bulovic}}, \bibinfo {author} {\bibfnamefont {G.}~\bibnamefont {Gu}},
  \bibinfo {author} {\bibfnamefont {P.~E.}\ \bibnamefont {Burrows}}, \bibinfo
  {author} {\bibfnamefont {S.~R.}\ \bibnamefont {Forrest}}, \ and\ \bibinfo
  {author} {\bibfnamefont {M.~E.}\ \bibnamefont {Thompson}},\ }\href@noop {}
  {\bibfield  {journal} {\bibinfo  {journal} {Nature}\ }\textbf {\bibinfo
  {volume} {380}},\ \bibinfo {pages} {29} (\bibinfo {year} {1996})}\BibitemShut
  {NoStop}%
\bibitem [{\citenamefont {Shimizu}\ and\ \citenamefont
  {Hiyama}(2010)}]{Shimizu2010}%
  \BibitemOpen
  \bibfield  {author} {\bibinfo {author} {\bibfnamefont {M.}~\bibnamefont
  {Shimizu}}\ and\ \bibinfo {author} {\bibfnamefont {T.}~\bibnamefont
  {Hiyama}},\ }\href@noop {} {\bibfield  {journal} {\bibinfo  {journal}
  {Chemistry-an Asian Journal}\ }\textbf {\bibinfo {volume} {5}},\ \bibinfo
  {pages} {1516} (\bibinfo {year} {2010})}\BibitemShut {NoStop}%
\bibitem [{\citenamefont {Thompson}(2007)}]{Thompson2007}%
  \BibitemOpen
  \bibfield  {author} {\bibinfo {author} {\bibfnamefont {M.~E.}\ \bibnamefont
  {Thompson}},\ }\href@noop {} {\bibfield  {journal} {\bibinfo  {journal} {MRS
  Bulletin}\ }\textbf {\bibinfo {volume} {32}},\ \bibinfo {pages} {694}
  (\bibinfo {year} {2007})}\BibitemShut {NoStop}%
\bibitem [{\citenamefont {Pope}\ and\ \citenamefont
  {Swenberg}(1999)}]{Pope1999}%
  \BibitemOpen
  \bibinfo {editor} {\bibfnamefont {M.}~\bibnamefont {Pope}}\ and\ \bibinfo
  {editor} {\bibfnamefont {C.~E.}\ \bibnamefont {Swenberg}},\ eds.,\ \href@noop
  {} {\emph {\bibinfo {title} {Electronic processes in organic crystals}}}\
  (\bibinfo  {publisher} {Oxford University Press, New York},\ \bibinfo {year}
  {1999})\BibitemShut {NoStop}%
\bibitem [{\citenamefont {Burrows}\ \emph {et~al.}(1996)\citenamefont
  {Burrows}, \citenamefont {Forrest}, \citenamefont {Sibley},\ and\
  \citenamefont {Thompson}}]{Burrows1996}%
  \BibitemOpen
  \bibfield  {author} {\bibinfo {author} {\bibfnamefont {P.~E.}\ \bibnamefont
  {Burrows}}, \bibinfo {author} {\bibfnamefont {S.~R.}\ \bibnamefont
  {Forrest}}, \bibinfo {author} {\bibfnamefont {S.~P.}\ \bibnamefont {Sibley}},
  \ and\ \bibinfo {author} {\bibfnamefont {M.~E.}\ \bibnamefont {Thompson}},\
  }\href@noop {} {\bibfield  {journal} {\bibinfo  {journal} {Applied Physics
  Letters}\ }\textbf {\bibinfo {volume} {69}},\ \bibinfo {pages} {2959}
  (\bibinfo {year} {1996})}\BibitemShut {NoStop}%
\bibitem [{\citenamefont {Kanno}\ \emph
  {et~al.}(2006{\natexlab{a}})\citenamefont {Kanno}, \citenamefont {Holmes},
  \citenamefont {Sun}, \citenamefont {Kena-Cohen},\ and\ \citenamefont
  {Forrest}}]{Kanno2006}%
  \BibitemOpen
  \bibfield  {author} {\bibinfo {author} {\bibfnamefont {H.}~\bibnamefont
  {Kanno}}, \bibinfo {author} {\bibfnamefont {R.~J.}\ \bibnamefont {Holmes}},
  \bibinfo {author} {\bibfnamefont {Y.}~\bibnamefont {Sun}}, \bibinfo {author}
  {\bibfnamefont {S.}~\bibnamefont {Kena-Cohen}}, \ and\ \bibinfo {author}
  {\bibfnamefont {S.~R.}\ \bibnamefont {Forrest}},\ }\href@noop {} {\bibfield
  {journal} {\bibinfo  {journal} {Advanced Materials}\ }\textbf {\bibinfo
  {volume} {18}},\ \bibinfo {pages} {339} (\bibinfo {year}
  {2006}{\natexlab{a}})}\BibitemShut {NoStop}%
\bibitem [{\citenamefont {Kanno}\ \emph
  {et~al.}(2006{\natexlab{b}})\citenamefont {Kanno}, \citenamefont {Giebink},
  \citenamefont {Sun},\ and\ \citenamefont {Forrest}}]{Kanno2006a}%
  \BibitemOpen
  \bibfield  {author} {\bibinfo {author} {\bibfnamefont {H.}~\bibnamefont
  {Kanno}}, \bibinfo {author} {\bibfnamefont {N.~C.}\ \bibnamefont {Giebink}},
  \bibinfo {author} {\bibfnamefont {Y.~R.}\ \bibnamefont {Sun}}, \ and\
  \bibinfo {author} {\bibfnamefont {S.~R.}\ \bibnamefont {Forrest}},\
  }\href@noop {} {\bibfield  {journal} {\bibinfo  {journal} {Applied Physics
  Letters}\ }\textbf {\bibinfo {volume} {89}},\ \bibinfo {pages} {023503}
  (\bibinfo {year} {2006}{\natexlab{b}})}\BibitemShut {NoStop}%
\bibitem [{\citenamefont {Mladenovski}\ \emph {et~al.}(2009)\citenamefont
  {Mladenovski}, \citenamefont {Neyts}, \citenamefont {Pavicic}, \citenamefont
  {Werner},\ and\ \citenamefont {Rothe}}]{Mladenovski2009a}%
  \BibitemOpen
  \bibfield  {author} {\bibinfo {author} {\bibfnamefont {S.}~\bibnamefont
  {Mladenovski}}, \bibinfo {author} {\bibfnamefont {K.}~\bibnamefont {Neyts}},
  \bibinfo {author} {\bibfnamefont {D.}~\bibnamefont {Pavicic}}, \bibinfo
  {author} {\bibfnamefont {A.}~\bibnamefont {Werner}}, \ and\ \bibinfo {author}
  {\bibfnamefont {C.}~\bibnamefont {Rothe}},\ }\href@noop {} {\bibfield
  {journal} {\bibinfo  {journal} {Optics Express}\ }\textbf {\bibinfo {volume}
  {17}},\ \bibinfo {pages} {7562} (\bibinfo {year} {2009})}\BibitemShut
  {NoStop}%
\bibitem [{\citenamefont {Lin}\ \emph {et~al.}(2006)\citenamefont {Lin},
  \citenamefont {Cho}, \citenamefont {Chang},\ and\ \citenamefont
  {Wu}}]{Lin2006}%
  \BibitemOpen
  \bibfield  {author} {\bibinfo {author} {\bibfnamefont {C.~L.}\ \bibnamefont
  {Lin}}, \bibinfo {author} {\bibfnamefont {T.~Y.}\ \bibnamefont {Cho}},
  \bibinfo {author} {\bibfnamefont {C.~H.}\ \bibnamefont {Chang}}, \ and\
  \bibinfo {author} {\bibfnamefont {C.~C.}\ \bibnamefont {Wu}},\ }\href@noop {}
  {\bibfield  {journal} {\bibinfo  {journal} {Applied Physics Letters}\
  }\textbf {\bibinfo {volume} {88}},\ \bibinfo {pages} {081114} (\bibinfo
  {year} {2006})}\BibitemShut {NoStop}%
\bibitem [{\citenamefont {Kalinowski}\ \emph {et~al.}(2007)\citenamefont
  {Kalinowski}, \citenamefont {Cocchi}, \citenamefont {Virgili}, \citenamefont
  {Tattori},\ and\ \citenamefont {Williams}}]{Kalinowski2007}%
  \BibitemOpen
  \bibfield  {author} {\bibinfo {author} {\bibfnamefont {J.}~\bibnamefont
  {Kalinowski}}, \bibinfo {author} {\bibfnamefont {M.}~\bibnamefont {Cocchi}},
  \bibinfo {author} {\bibfnamefont {D.}~\bibnamefont {Virgili}}, \bibinfo
  {author} {\bibfnamefont {V.}~\bibnamefont {Tattori}}, \ and\ \bibinfo
  {author} {\bibfnamefont {J.~A.~G.}\ \bibnamefont {Williams}},\ }\href@noop {}
  {\bibfield  {journal} {\bibinfo  {journal} {Advanced Materials}\ }\textbf
  {\bibinfo {volume} {19}},\ \bibinfo {pages} {4000} (\bibinfo {year}
  {2007})}\BibitemShut {NoStop}%
\bibitem [{\citenamefont {Williams}\ \emph {et~al.}(2007)\citenamefont
  {Williams}, \citenamefont {Haavisto}, \citenamefont {Li},\ and\ \citenamefont
  {Jabbour}}]{Williams2007}%
  \BibitemOpen
  \bibfield  {author} {\bibinfo {author} {\bibfnamefont {E.~L.}\ \bibnamefont
  {Williams}}, \bibinfo {author} {\bibfnamefont {K.}~\bibnamefont {Haavisto}},
  \bibinfo {author} {\bibfnamefont {J.}~\bibnamefont {Li}}, \ and\ \bibinfo
  {author} {\bibfnamefont {G.~E.}\ \bibnamefont {Jabbour}},\ }\href@noop {}
  {\bibfield  {journal} {\bibinfo  {journal} {Advanced Materials}\ }\textbf
  {\bibinfo {volume} {19}},\ \bibinfo {pages} {197} (\bibinfo {year}
  {2007})}\BibitemShut {NoStop}%
\bibitem [{\citenamefont {D'Andrade}\ \emph
  {et~al.}(2002{\natexlab{a}})\citenamefont {D'Andrade}, \citenamefont
  {Brooks}, \citenamefont {Adamovich}, \citenamefont {Thompson},\ and\
  \citenamefont {Forrest}}]{DAndrade2002}%
  \BibitemOpen
  \bibfield  {author} {\bibinfo {author} {\bibfnamefont {B.~W.}\ \bibnamefont
  {D'Andrade}}, \bibinfo {author} {\bibfnamefont {J.}~\bibnamefont {Brooks}},
  \bibinfo {author} {\bibfnamefont {V.}~\bibnamefont {Adamovich}}, \bibinfo
  {author} {\bibfnamefont {M.~E.}\ \bibnamefont {Thompson}}, \ and\ \bibinfo
  {author} {\bibfnamefont {S.~R.}\ \bibnamefont {Forrest}},\ }\href@noop {}
  {\bibfield  {journal} {\bibinfo  {journal} {Advanced Materials}\ }\textbf
  {\bibinfo {volume} {14}},\ \bibinfo {pages} {1032} (\bibinfo {year}
  {2002}{\natexlab{a}})}\BibitemShut {NoStop}%
\bibitem [{\citenamefont {Adamovich}\ \emph {et~al.}(2002)\citenamefont
  {Adamovich}, \citenamefont {Brooks}, \citenamefont {Tamayo}, \citenamefont
  {Alexander}, \citenamefont {Djurovich}, \citenamefont {D'Andrade},
  \citenamefont {Adachi}, \citenamefont {Forrest},\ and\ \citenamefont
  {Thompson}}]{Adamovich2002}%
  \BibitemOpen
  \bibfield  {author} {\bibinfo {author} {\bibfnamefont {V.}~\bibnamefont
  {Adamovich}}, \bibinfo {author} {\bibfnamefont {J.}~\bibnamefont {Brooks}},
  \bibinfo {author} {\bibfnamefont {A.}~\bibnamefont {Tamayo}}, \bibinfo
  {author} {\bibfnamefont {A.~M.}\ \bibnamefont {Alexander}}, \bibinfo {author}
  {\bibfnamefont {P.~I.}\ \bibnamefont {Djurovich}}, \bibinfo {author}
  {\bibfnamefont {B.~W.}\ \bibnamefont {D'Andrade}}, \bibinfo {author}
  {\bibfnamefont {C.}~\bibnamefont {Adachi}}, \bibinfo {author} {\bibfnamefont
  {S.~R.}\ \bibnamefont {Forrest}}, \ and\ \bibinfo {author} {\bibfnamefont
  {M.~E.}\ \bibnamefont {Thompson}},\ }\href@noop {} {\bibfield  {journal}
  {\bibinfo  {journal} {New Journal Of Chemistry}\ }\textbf {\bibinfo {volume}
  {26}},\ \bibinfo {pages} {1171} (\bibinfo {year} {2002})}\BibitemShut
  {NoStop}%
\bibitem [{\citenamefont {Cocchi}\ \emph {et~al.}(2007)\citenamefont {Cocchi},
  \citenamefont {Kalinowski}, \citenamefont {Virgili}, \citenamefont {Fattori},
  \citenamefont {Develay},\ and\ \citenamefont {Williams}}]{Cocchi2007}%
  \BibitemOpen
  \bibfield  {author} {\bibinfo {author} {\bibfnamefont {M.}~\bibnamefont
  {Cocchi}}, \bibinfo {author} {\bibfnamefont {J.}~\bibnamefont {Kalinowski}},
  \bibinfo {author} {\bibfnamefont {D.}~\bibnamefont {Virgili}}, \bibinfo
  {author} {\bibfnamefont {V.}~\bibnamefont {Fattori}}, \bibinfo {author}
  {\bibfnamefont {S.}~\bibnamefont {Develay}}, \ and\ \bibinfo {author}
  {\bibfnamefont {J.~A.~G.}\ \bibnamefont {Williams}},\ }\href@noop {}
  {\bibfield  {journal} {\bibinfo  {journal} {Applied Physics Letters}\
  }\textbf {\bibinfo {volume} {90}},\ \bibinfo {pages} {163508} (\bibinfo
  {year} {2007})}\BibitemShut {NoStop}%
\bibitem [{\citenamefont {Tsai}\ \emph {et~al.}(2003)\citenamefont {Tsai},
  \citenamefont {Liu}, \citenamefont {Hsu},\ and\ \citenamefont
  {Chow}}]{Tsai2003}%
  \BibitemOpen
  \bibfield  {author} {\bibinfo {author} {\bibfnamefont {M.~L.}\ \bibnamefont
  {Tsai}}, \bibinfo {author} {\bibfnamefont {C.~Y.}\ \bibnamefont {Liu}},
  \bibinfo {author} {\bibfnamefont {M.~A.}\ \bibnamefont {Hsu}}, \ and\
  \bibinfo {author} {\bibfnamefont {T.~J.}\ \bibnamefont {Chow}},\ }\href@noop
  {} {\bibfield  {journal} {\bibinfo  {journal} {Applied Physics Letters}\
  }\textbf {\bibinfo {volume} {82}},\ \bibinfo {pages} {550} (\bibinfo {year}
  {2003})}\BibitemShut {NoStop}%
\bibitem [{\citenamefont {Krummacher}\ \emph {et~al.}(2006)\citenamefont
  {Krummacher}, \citenamefont {Choong}, \citenamefont {Mathai}, \citenamefont
  {Choulis}, \citenamefont {So}, \citenamefont {Jermann}, \citenamefont
  {Fiedler},\ and\ \citenamefont {Zachau}}]{Krummacher2006}%
  \BibitemOpen
  \bibfield  {author} {\bibinfo {author} {\bibfnamefont {B.~C.}\ \bibnamefont
  {Krummacher}}, \bibinfo {author} {\bibfnamefont {V.~E.}\ \bibnamefont
  {Choong}}, \bibinfo {author} {\bibfnamefont {M.~K.}\ \bibnamefont {Mathai}},
  \bibinfo {author} {\bibfnamefont {S.~A.}\ \bibnamefont {Choulis}}, \bibinfo
  {author} {\bibfnamefont {F.}~\bibnamefont {So}}, \bibinfo {author}
  {\bibfnamefont {F.}~\bibnamefont {Jermann}}, \bibinfo {author} {\bibfnamefont
  {T.}~\bibnamefont {Fiedler}}, \ and\ \bibinfo {author} {\bibfnamefont
  {M.}~\bibnamefont {Zachau}},\ }\href@noop {} {\bibfield  {journal} {\bibinfo
  {journal} {Applied Physics Letters}\ }\textbf {\bibinfo {volume} {88}},\
  \bibinfo {pages} {113506} (\bibinfo {year} {2006})}\BibitemShut {NoStop}%
\bibitem [{\citenamefont {Gohri}\ \emph {et~al.}(2011)\citenamefont {Gohri},
  \citenamefont {Hofmann}, \citenamefont {Reineke}, \citenamefont {Rosenow},
  \citenamefont {Thomschke}, \citenamefont {Levichkova}, \citenamefont
  {Luessem},\ and\ \citenamefont {Leo}}]{Gohri2011}%
  \BibitemOpen
  \bibfield  {author} {\bibinfo {author} {\bibfnamefont {V.}~\bibnamefont
  {Gohri}}, \bibinfo {author} {\bibfnamefont {S.}~\bibnamefont {Hofmann}},
  \bibinfo {author} {\bibfnamefont {S.}~\bibnamefont {Reineke}}, \bibinfo
  {author} {\bibfnamefont {T.}~\bibnamefont {Rosenow}}, \bibinfo {author}
  {\bibfnamefont {M.}~\bibnamefont {Thomschke}}, \bibinfo {author}
  {\bibfnamefont {M.}~\bibnamefont {Levichkova}}, \bibinfo {author}
  {\bibfnamefont {B.}~\bibnamefont {Luessem}}, \ and\ \bibinfo {author}
  {\bibfnamefont {K.}~\bibnamefont {Leo}},\ }\href@noop {} {\bibfield
  {journal} {\bibinfo  {journal} {Organic Electronics}\ }\textbf {\bibinfo
  {volume} {12}},\ \bibinfo {pages} {2126} (\bibinfo {year}
  {2011})}\BibitemShut {NoStop}%
\bibitem [{\citenamefont {Schwab}\ \emph {et~al.}(2011)\citenamefont {Schwab},
  \citenamefont {Thomschke}, \citenamefont {Hofmann}, \citenamefont {Furno},
  \citenamefont {Leo},\ and\ \citenamefont {Luessem}}]{Schwab2011}%
  \BibitemOpen
  \bibfield  {author} {\bibinfo {author} {\bibfnamefont {T.}~\bibnamefont
  {Schwab}}, \bibinfo {author} {\bibfnamefont {M.}~\bibnamefont {Thomschke}},
  \bibinfo {author} {\bibfnamefont {S.}~\bibnamefont {Hofmann}}, \bibinfo
  {author} {\bibfnamefont {M.}~\bibnamefont {Furno}}, \bibinfo {author}
  {\bibfnamefont {K.}~\bibnamefont {Leo}}, \ and\ \bibinfo {author}
  {\bibfnamefont {B.}~\bibnamefont {Luessem}},\ }\href@noop {} {\bibfield
  {journal} {\bibinfo  {journal} {Journal of Applied Physics}\ }\textbf
  {\bibinfo {volume} {110}},\ \bibinfo {pages} {083118} (\bibinfo {year}
  {2011})}\BibitemShut {NoStop}%
\bibitem [{\citenamefont {D'Andrade}\ \emph {et~al.}(2004)\citenamefont
  {D'Andrade}, \citenamefont {Holmes},\ and\ \citenamefont
  {Forrest}}]{DAndrade2004}%
  \BibitemOpen
  \bibfield  {author} {\bibinfo {author} {\bibfnamefont {B.~W.}\ \bibnamefont
  {D'Andrade}}, \bibinfo {author} {\bibfnamefont {R.~J.}\ \bibnamefont
  {Holmes}}, \ and\ \bibinfo {author} {\bibfnamefont {S.~R.}\ \bibnamefont
  {Forrest}},\ }\href@noop {} {\bibfield  {journal} {\bibinfo  {journal}
  {Advanced Materials}\ }\textbf {\bibinfo {volume} {16}},\ \bibinfo {pages}
  {624} (\bibinfo {year} {2004})}\BibitemShut {NoStop}%
\bibitem [{\citenamefont {Kido}\ \emph
  {et~al.}(1995{\natexlab{a}})\citenamefont {Kido}, \citenamefont {Shionoya},\
  and\ \citenamefont {Nagai}}]{Kido1995a}%
  \BibitemOpen
  \bibfield  {author} {\bibinfo {author} {\bibfnamefont {J.}~\bibnamefont
  {Kido}}, \bibinfo {author} {\bibfnamefont {H.}~\bibnamefont {Shionoya}}, \
  and\ \bibinfo {author} {\bibfnamefont {K.}~\bibnamefont {Nagai}},\
  }\href@noop {} {\bibfield  {journal} {\bibinfo  {journal} {Applied Physics
  Letters}\ }\textbf {\bibinfo {volume} {67}},\ \bibinfo {pages} {2281}
  (\bibinfo {year} {1995}{\natexlab{a}})}\BibitemShut {NoStop}%
\bibitem [{\citenamefont {Liu}\ \emph {et~al.}(2005)\citenamefont {Liu},
  \citenamefont {Zhou}, \citenamefont {Cheng}, \citenamefont {Geng},
  \citenamefont {Wang}, \citenamefont {Ma}, \citenamefont {Jing},\ and\
  \citenamefont {Wang}}]{Liu2005}%
  \BibitemOpen
  \bibfield  {author} {\bibinfo {author} {\bibfnamefont {J.}~\bibnamefont
  {Liu}}, \bibinfo {author} {\bibfnamefont {Q.~G.}\ \bibnamefont {Zhou}},
  \bibinfo {author} {\bibfnamefont {Y.~X.}\ \bibnamefont {Cheng}}, \bibinfo
  {author} {\bibfnamefont {Y.~H.}\ \bibnamefont {Geng}}, \bibinfo {author}
  {\bibfnamefont {L.~X.}\ \bibnamefont {Wang}}, \bibinfo {author}
  {\bibfnamefont {D.~G.}\ \bibnamefont {Ma}}, \bibinfo {author} {\bibfnamefont
  {X.~B.}\ \bibnamefont {Jing}}, \ and\ \bibinfo {author} {\bibfnamefont
  {F.~S.}\ \bibnamefont {Wang}},\ }\href@noop {} {\bibfield  {journal}
  {\bibinfo  {journal} {Advanced Materials}\ }\textbf {\bibinfo {volume}
  {17}},\ \bibinfo {pages} {2974} (\bibinfo {year} {2005})}\BibitemShut
  {NoStop}%
\bibitem [{\citenamefont {Reineke}\ \emph
  {et~al.}(2009{\natexlab{a}})\citenamefont {Reineke}, \citenamefont {Lindner},
  \citenamefont {Schwartz}, \citenamefont {Seidler}, \citenamefont {Walzer},
  \citenamefont {L\"ussem},\ and\ \citenamefont {Leo}}]{Reineke2009a}%
  \BibitemOpen
  \bibfield  {author} {\bibinfo {author} {\bibfnamefont {S.}~\bibnamefont
  {Reineke}}, \bibinfo {author} {\bibfnamefont {F.}~\bibnamefont {Lindner}},
  \bibinfo {author} {\bibfnamefont {G.}~\bibnamefont {Schwartz}}, \bibinfo
  {author} {\bibfnamefont {N.}~\bibnamefont {Seidler}}, \bibinfo {author}
  {\bibfnamefont {K.}~\bibnamefont {Walzer}}, \bibinfo {author} {\bibfnamefont
  {B.}~\bibnamefont {L\"ussem}}, \ and\ \bibinfo {author} {\bibfnamefont
  {K.}~\bibnamefont {Leo}},\ }\href@noop {} {\bibfield  {journal} {\bibinfo
  {journal} {Nature}\ }\textbf {\bibinfo {volume} {459}},\ \bibinfo {pages}
  {234} (\bibinfo {year} {2009}{\natexlab{a}})}\BibitemShut {NoStop}%
\bibitem [{\citenamefont {Schwartz}\ \emph {et~al.}(2006)\citenamefont
  {Schwartz}, \citenamefont {Fehse}, \citenamefont {Pfeiffer}, \citenamefont
  {Walzer},\ and\ \citenamefont {Leo}}]{Schwartz2006}%
  \BibitemOpen
  \bibfield  {author} {\bibinfo {author} {\bibfnamefont {G.}~\bibnamefont
  {Schwartz}}, \bibinfo {author} {\bibfnamefont {K.}~\bibnamefont {Fehse}},
  \bibinfo {author} {\bibfnamefont {M.}~\bibnamefont {Pfeiffer}}, \bibinfo
  {author} {\bibfnamefont {K.}~\bibnamefont {Walzer}}, \ and\ \bibinfo {author}
  {\bibfnamefont {K.}~\bibnamefont {Leo}},\ }\href@noop {} {\bibfield
  {journal} {\bibinfo  {journal} {Applied Physics Letters}\ }\textbf {\bibinfo
  {volume} {89}},\ \bibinfo {pages} {083509} (\bibinfo {year}
  {2006})}\BibitemShut {NoStop}%
\bibitem [{\citenamefont {Sun}\ and\ \citenamefont {Forrest}(2007)}]{Sun2007}%
  \BibitemOpen
  \bibfield  {author} {\bibinfo {author} {\bibfnamefont {Y.}~\bibnamefont
  {Sun}}\ and\ \bibinfo {author} {\bibfnamefont {S.~R.}\ \bibnamefont
  {Forrest}},\ }\href@noop {} {\bibfield  {journal} {\bibinfo  {journal}
  {Applied Physics Letters}\ }\textbf {\bibinfo {volume} {91}},\ \bibinfo
  {pages} {263503} (\bibinfo {year} {2007})}\BibitemShut {NoStop}%
\bibitem [{\citenamefont {Kido}\ \emph
  {et~al.}(1995{\natexlab{b}})\citenamefont {Kido}, \citenamefont {Kimura},\
  and\ \citenamefont {Nagai}}]{Kido1995}%
  \BibitemOpen
  \bibfield  {author} {\bibinfo {author} {\bibfnamefont {J.}~\bibnamefont
  {Kido}}, \bibinfo {author} {\bibfnamefont {M.}~\bibnamefont {Kimura}}, \ and\
  \bibinfo {author} {\bibfnamefont {K.}~\bibnamefont {Nagai}},\ }\href@noop {}
  {\bibfield  {journal} {\bibinfo  {journal} {Science}\ }\textbf {\bibinfo
  {volume} {267}},\ \bibinfo {pages} {1332} (\bibinfo {year}
  {1995}{\natexlab{b}})}\BibitemShut {NoStop}%
\bibitem [{\citenamefont {Chao}\ and\ \citenamefont {Chen}(1998)}]{Chao1998}%
  \BibitemOpen
  \bibfield  {author} {\bibinfo {author} {\bibfnamefont {C.~I.}\ \bibnamefont
  {Chao}}\ and\ \bibinfo {author} {\bibfnamefont {S.~A.}\ \bibnamefont
  {Chen}},\ }\href@noop {} {\bibfield  {journal} {\bibinfo  {journal} {Applied
  Physics Letters}\ }\textbf {\bibinfo {volume} {73}},\ \bibinfo {pages} {426}
  (\bibinfo {year} {1998})}\BibitemShut {NoStop}%
\bibitem [{\citenamefont {Helfrich}\ and\ \citenamefont
  {Schneider}(1965)}]{Helfrich1965}%
  \BibitemOpen
  \bibfield  {author} {\bibinfo {author} {\bibfnamefont {W.}~\bibnamefont
  {Helfrich}}\ and\ \bibinfo {author} {\bibfnamefont {W.~G.}\ \bibnamefont
  {Schneider}},\ }\href@noop {} {\bibfield  {journal} {\bibinfo  {journal}
  {Physical Review Letters}\ }\textbf {\bibinfo {volume} {14}},\ \bibinfo
  {pages} {229} (\bibinfo {year} {1965})}\BibitemShut {NoStop}%
\bibitem [{\citenamefont {Koch}\ \emph {et~al.}(2005)\citenamefont {Koch},
  \citenamefont {Duhm}, \citenamefont {Rabe}, \citenamefont {Vollmer},\ and\
  \citenamefont {Johnson}}]{Koch2005}%
  \BibitemOpen
  \bibfield  {author} {\bibinfo {author} {\bibfnamefont {N.}~\bibnamefont
  {Koch}}, \bibinfo {author} {\bibfnamefont {S.}~\bibnamefont {Duhm}}, \bibinfo
  {author} {\bibfnamefont {J.~P.}\ \bibnamefont {Rabe}}, \bibinfo {author}
  {\bibfnamefont {A.}~\bibnamefont {Vollmer}}, \ and\ \bibinfo {author}
  {\bibfnamefont {R.~L.}\ \bibnamefont {Johnson}},\ }\href@noop {} {\bibfield
  {journal} {\bibinfo  {journal} {Physical Review Letters}\ }\textbf {\bibinfo
  {volume} {95}},\ \bibinfo {pages} {237601} (\bibinfo {year}
  {2005})}\BibitemShut {NoStop}%
\bibitem [{\citenamefont {Guo}\ \emph {et~al.}(2006)\citenamefont {Guo},
  \citenamefont {Koch}, \citenamefont {Bernasek},\ and\ \citenamefont
  {Schwartz}}]{Guo2006}%
  \BibitemOpen
  \bibfield  {author} {\bibinfo {author} {\bibfnamefont {J.}~\bibnamefont
  {Guo}}, \bibinfo {author} {\bibfnamefont {N.}~\bibnamefont {Koch}}, \bibinfo
  {author} {\bibfnamefont {S.~L.}\ \bibnamefont {Bernasek}}, \ and\ \bibinfo
  {author} {\bibfnamefont {J.}~\bibnamefont {Schwartz}},\ }\href@noop {}
  {\bibfield  {journal} {\bibinfo  {journal} {Chemical Physics Letters}\
  }\textbf {\bibinfo {volume} {426}},\ \bibinfo {pages} {370} (\bibinfo {year}
  {2006})}\BibitemShut {NoStop}%
\bibitem [{\citenamefont {Kahn}\ \emph {et~al.}(2001)\citenamefont {Kahn},
  \citenamefont {Pireaux}, \citenamefont {Salaneck},\ and\ \citenamefont
  {Seki}}]{book_Kahn2001}%
  \BibitemOpen
  \bibinfo {editor} {\bibfnamefont {A.}~\bibnamefont {Kahn}}, \bibinfo {editor}
  {\bibfnamefont {J.-J.}\ \bibnamefont {Pireaux}}, \bibinfo {editor}
  {\bibfnamefont {W.~R.}\ \bibnamefont {Salaneck}}, \ and\ \bibinfo {editor}
  {\bibfnamefont {K.}~\bibnamefont {Seki}},\ eds.,\ \href@noop {} {\emph
  {\bibinfo {title} {Conjugated Polymer and Molecular Interfaces}}}\ (\bibinfo
  {publisher} {Marcel Dekker, Inc.},\ \bibinfo {year} {2001})\BibitemShut
  {NoStop}%
\bibitem [{\citenamefont {Koch}(2012)}]{Koch2012}%
  \BibitemOpen
  \bibfield  {author} {\bibinfo {author} {\bibfnamefont {N.}~\bibnamefont
  {Koch}},\ }\href@noop {} {\bibfield  {journal} {\bibinfo  {journal} {Physica
  Status Solidi-rapid Research Letters}\ }\textbf {\bibinfo {volume} {6}},\
  \bibinfo {pages} {277} (\bibinfo {year} {2012})}\BibitemShut {NoStop}%
\bibitem [{\citenamefont {Greiner}\ \emph {et~al.}(2012)\citenamefont
  {Greiner}, \citenamefont {Helander}, \citenamefont {Tang}, \citenamefont
  {Wang}, \citenamefont {Qiu},\ and\ \citenamefont {Lu}}]{Greiner2012}%
  \BibitemOpen
  \bibfield  {author} {\bibinfo {author} {\bibfnamefont {M.~T.}\ \bibnamefont
  {Greiner}}, \bibinfo {author} {\bibfnamefont {M.~G.}\ \bibnamefont
  {Helander}}, \bibinfo {author} {\bibfnamefont {W.-M.}\ \bibnamefont {Tang}},
  \bibinfo {author} {\bibfnamefont {Z.-B.}\ \bibnamefont {Wang}}, \bibinfo
  {author} {\bibfnamefont {J.}~\bibnamefont {Qiu}}, \ and\ \bibinfo {author}
  {\bibfnamefont {Z.-H.}\ \bibnamefont {Lu}},\ }\href@noop {} {\bibfield
  {journal} {\bibinfo  {journal} {Nature Materials}\ }\textbf {\bibinfo
  {volume} {11}},\ \bibinfo {pages} {76} (\bibinfo {year} {2012})}\BibitemShut
  {NoStop}%
\bibitem [{\citenamefont {Walzer}\ \emph {et~al.}(2007)\citenamefont {Walzer},
  \citenamefont {Maennig}, \citenamefont {Pfeiffer},\ and\ \citenamefont
  {Leo}}]{Walzer2007}%
  \BibitemOpen
  \bibfield  {author} {\bibinfo {author} {\bibfnamefont {K.}~\bibnamefont
  {Walzer}}, \bibinfo {author} {\bibfnamefont {B.}~\bibnamefont {Maennig}},
  \bibinfo {author} {\bibfnamefont {M.}~\bibnamefont {Pfeiffer}}, \ and\
  \bibinfo {author} {\bibfnamefont {K.}~\bibnamefont {Leo}},\ }\href@noop {}
  {\bibfield  {journal} {\bibinfo  {journal} {Chemical Reviews}\ }\textbf
  {\bibinfo {volume} {107}},\ \bibinfo {pages} {1233} (\bibinfo {year}
  {2007})}\BibitemShut {NoStop}%
\bibitem [{\citenamefont {Blochwitz}\ \emph {et~al.}(1998)\citenamefont
  {Blochwitz}, \citenamefont {Pfeiffer}, \citenamefont {Fritz},\ and\
  \citenamefont {Leo}}]{Blochwitz1998}%
  \BibitemOpen
  \bibfield  {author} {\bibinfo {author} {\bibfnamefont {J.}~\bibnamefont
  {Blochwitz}}, \bibinfo {author} {\bibfnamefont {M.}~\bibnamefont {Pfeiffer}},
  \bibinfo {author} {\bibfnamefont {T.}~\bibnamefont {Fritz}}, \ and\ \bibinfo
  {author} {\bibfnamefont {K.}~\bibnamefont {Leo}},\ }\href@noop {} {\bibfield
  {journal} {\bibinfo  {journal} {Applied Physics Letters}\ }\textbf {\bibinfo
  {volume} {73}},\ \bibinfo {pages} {729} (\bibinfo {year} {1998})}\BibitemShut
  {NoStop}%
\bibitem [{\citenamefont {Yamamori}\ \emph {et~al.}(1998)\citenamefont
  {Yamamori}, \citenamefont {Adachi}, \citenamefont {Koyama},\ and\
  \citenamefont {Taniguchi}}]{Yamamori1998}%
  \BibitemOpen
  \bibfield  {author} {\bibinfo {author} {\bibfnamefont {A.}~\bibnamefont
  {Yamamori}}, \bibinfo {author} {\bibfnamefont {C.}~\bibnamefont {Adachi}},
  \bibinfo {author} {\bibfnamefont {T.}~\bibnamefont {Koyama}}, \ and\ \bibinfo
  {author} {\bibfnamefont {Y.}~\bibnamefont {Taniguchi}},\ }\href@noop {}
  {\bibfield  {journal} {\bibinfo  {journal} {Applied Physics Letters}\
  }\textbf {\bibinfo {volume} {72}},\ \bibinfo {pages} {2147} (\bibinfo {year}
  {1998})}\BibitemShut {NoStop}%
\bibitem [{\citenamefont {Zhou}\ \emph {et~al.}(2001)\citenamefont {Zhou},
  \citenamefont {Blochwitz}, \citenamefont {Pfeiffer}, \citenamefont {Nollau},
  \citenamefont {Fritz},\ and\ \citenamefont {Leo}}]{Zhou2001}%
  \BibitemOpen
  \bibfield  {author} {\bibinfo {author} {\bibfnamefont {X.}~\bibnamefont
  {Zhou}}, \bibinfo {author} {\bibfnamefont {J.}~\bibnamefont {Blochwitz}},
  \bibinfo {author} {\bibfnamefont {M.}~\bibnamefont {Pfeiffer}}, \bibinfo
  {author} {\bibfnamefont {A.}~\bibnamefont {Nollau}}, \bibinfo {author}
  {\bibfnamefont {T.}~\bibnamefont {Fritz}}, \ and\ \bibinfo {author}
  {\bibfnamefont {K.}~\bibnamefont {Leo}},\ }\href@noop {} {\bibfield
  {journal} {\bibinfo  {journal} {Advanced Functional Materials}\ }\textbf
  {\bibinfo {volume} {11}},\ \bibinfo {pages} {310} (\bibinfo {year}
  {2001})}\BibitemShut {NoStop}%
\bibitem [{\citenamefont {D'Andrade}\ and\ \citenamefont
  {Forrest}(2003)}]{DAndrade2003}%
  \BibitemOpen
  \bibfield  {author} {\bibinfo {author} {\bibfnamefont {B.~W.}\ \bibnamefont
  {D'Andrade}}\ and\ \bibinfo {author} {\bibfnamefont {S.~R.}\ \bibnamefont
  {Forrest}},\ }\href@noop {} {\bibfield  {journal} {\bibinfo  {journal}
  {Journal Of Applied Physics}\ }\textbf {\bibinfo {volume} {94}},\ \bibinfo
  {pages} {3101} (\bibinfo {year} {2003})}\BibitemShut {NoStop}%
\bibitem [{\citenamefont {Goushi}\ \emph {et~al.}(2004)\citenamefont {Goushi},
  \citenamefont {Kwong}, \citenamefont {Brown}, \citenamefont {Sasabe},\ and\
  \citenamefont {Adachi}}]{Goushi2004}%
  \BibitemOpen
  \bibfield  {author} {\bibinfo {author} {\bibfnamefont {K.}~\bibnamefont
  {Goushi}}, \bibinfo {author} {\bibfnamefont {R.}~\bibnamefont {Kwong}},
  \bibinfo {author} {\bibfnamefont {J.~J.}\ \bibnamefont {Brown}}, \bibinfo
  {author} {\bibfnamefont {H.}~\bibnamefont {Sasabe}}, \ and\ \bibinfo {author}
  {\bibfnamefont {C.}~\bibnamefont {Adachi}},\ }\href@noop {} {\bibfield
  {journal} {\bibinfo  {journal} {Journal Of Applied Physics}\ }\textbf
  {\bibinfo {volume} {95}},\ \bibinfo {pages} {7798} (\bibinfo {year}
  {2004})}\BibitemShut {NoStop}%
\bibitem [{\citenamefont {Chin}\ and\ \citenamefont {Lee}(2007)}]{Chin2007}%
  \BibitemOpen
  \bibfield  {author} {\bibinfo {author} {\bibfnamefont {B.~D.}\ \bibnamefont
  {Chin}}\ and\ \bibinfo {author} {\bibfnamefont {C.}~\bibnamefont {Lee}},\
  }\href@noop {} {\bibfield  {journal} {\bibinfo  {journal} {Advanced
  Materials}\ }\textbf {\bibinfo {volume} {19}},\ \bibinfo {pages} {2061}
  (\bibinfo {year} {2007})}\BibitemShut {NoStop}%
\bibitem [{\citenamefont {Adachi}\ \emph
  {et~al.}(2001{\natexlab{a}})\citenamefont {Adachi}, \citenamefont {Baldo},
  \citenamefont {Forrest}, \citenamefont {Lamansky}, \citenamefont {Thompson},\
  and\ \citenamefont {Kwong}}]{Adachi2001}%
  \BibitemOpen
  \bibfield  {author} {\bibinfo {author} {\bibfnamefont {C.}~\bibnamefont
  {Adachi}}, \bibinfo {author} {\bibfnamefont {M.~A.}\ \bibnamefont {Baldo}},
  \bibinfo {author} {\bibfnamefont {S.~R.}\ \bibnamefont {Forrest}}, \bibinfo
  {author} {\bibfnamefont {S.}~\bibnamefont {Lamansky}}, \bibinfo {author}
  {\bibfnamefont {M.~E.}\ \bibnamefont {Thompson}}, \ and\ \bibinfo {author}
  {\bibfnamefont {R.~C.}\ \bibnamefont {Kwong}},\ }\href@noop {} {\bibfield
  {journal} {\bibinfo  {journal} {Applied Physics Letters}\ }\textbf {\bibinfo
  {volume} {78}},\ \bibinfo {pages} {1622} (\bibinfo {year}
  {2001}{\natexlab{a}})}\BibitemShut {NoStop}%
\bibitem [{\citenamefont {He}\ \emph {et~al.}(2004{\natexlab{a}})\citenamefont
  {He}, \citenamefont {Schneider}, \citenamefont {Qin}, \citenamefont {Zhou},
  \citenamefont {Pfeiffer},\ and\ \citenamefont {Leo}}]{He2004a}%
  \BibitemOpen
  \bibfield  {author} {\bibinfo {author} {\bibfnamefont {G.~F.}\ \bibnamefont
  {He}}, \bibinfo {author} {\bibfnamefont {O.}~\bibnamefont {Schneider}},
  \bibinfo {author} {\bibfnamefont {D.~S.}\ \bibnamefont {Qin}}, \bibinfo
  {author} {\bibfnamefont {X.}~\bibnamefont {Zhou}}, \bibinfo {author}
  {\bibfnamefont {M.}~\bibnamefont {Pfeiffer}}, \ and\ \bibinfo {author}
  {\bibfnamefont {K.}~\bibnamefont {Leo}},\ }\href@noop {} {\bibfield
  {journal} {\bibinfo  {journal} {Journal Of Applied Physics}\ }\textbf
  {\bibinfo {volume} {95}},\ \bibinfo {pages} {5773} (\bibinfo {year}
  {2004}{\natexlab{a}})}\BibitemShut {NoStop}%
\bibitem [{\citenamefont {Anthopoulos}\ \emph {et~al.}(2003)\citenamefont
  {Anthopoulos}, \citenamefont {Markham}, \citenamefont {Namdas}, \citenamefont
  {Samuel}, \citenamefont {Lo},\ and\ \citenamefont {Burn}}]{Anthopoulos2003}%
  \BibitemOpen
  \bibfield  {author} {\bibinfo {author} {\bibfnamefont {T.~D.}\ \bibnamefont
  {Anthopoulos}}, \bibinfo {author} {\bibfnamefont {J.~P.~J.}\ \bibnamefont
  {Markham}}, \bibinfo {author} {\bibfnamefont {E.~B.}\ \bibnamefont {Namdas}},
  \bibinfo {author} {\bibfnamefont {I.~D.~W.}\ \bibnamefont {Samuel}}, \bibinfo
  {author} {\bibfnamefont {S.~C.}\ \bibnamefont {Lo}}, \ and\ \bibinfo {author}
  {\bibfnamefont {P.~L.}\ \bibnamefont {Burn}},\ }\href@noop {} {\bibfield
  {journal} {\bibinfo  {journal} {Applied Physics Letters}\ }\textbf {\bibinfo
  {volume} {82}},\ \bibinfo {pages} {4824} (\bibinfo {year}
  {2003})}\BibitemShut {NoStop}%
\bibitem [{\citenamefont {Chuang}\ \emph {et~al.}(2007)\citenamefont {Chuang},
  \citenamefont {Shih}, \citenamefont {Chien}, \citenamefont {Wu},\ and\
  \citenamefont {Shu}}]{Chuang2007}%
  \BibitemOpen
  \bibfield  {author} {\bibinfo {author} {\bibfnamefont {C.~Y.}\ \bibnamefont
  {Chuang}}, \bibinfo {author} {\bibfnamefont {P.~I.}\ \bibnamefont {Shih}},
  \bibinfo {author} {\bibfnamefont {C.~H.}\ \bibnamefont {Chien}}, \bibinfo
  {author} {\bibfnamefont {F.~I.}\ \bibnamefont {Wu}}, \ and\ \bibinfo {author}
  {\bibfnamefont {C.~F.}\ \bibnamefont {Shu}},\ }\href@noop {} {\bibfield
  {journal} {\bibinfo  {journal} {Macromolecules}\ }\textbf {\bibinfo {volume}
  {40}},\ \bibinfo {pages} {247} (\bibinfo {year} {2007})}\BibitemShut
  {NoStop}%
\bibitem [{\citenamefont {Gong}\ \emph {et~al.}(2004)\citenamefont {Gong},
  \citenamefont {Ma}, \citenamefont {Ostrowski}, \citenamefont {Bazan},
  \citenamefont {Moses},\ and\ \citenamefont {Heeger}}]{Gong2004}%
  \BibitemOpen
  \bibfield  {author} {\bibinfo {author} {\bibfnamefont {X.}~\bibnamefont
  {Gong}}, \bibinfo {author} {\bibfnamefont {W.~L.}\ \bibnamefont {Ma}},
  \bibinfo {author} {\bibfnamefont {J.~C.}\ \bibnamefont {Ostrowski}}, \bibinfo
  {author} {\bibfnamefont {G.~C.}\ \bibnamefont {Bazan}}, \bibinfo {author}
  {\bibfnamefont {D.}~\bibnamefont {Moses}}, \ and\ \bibinfo {author}
  {\bibfnamefont {A.~J.}\ \bibnamefont {Heeger}},\ }\href@noop {} {\bibfield
  {journal} {\bibinfo  {journal} {Advanced Materials}\ }\textbf {\bibinfo
  {volume} {16}},\ \bibinfo {pages} {615} (\bibinfo {year} {2004})}\BibitemShut
  {NoStop}%
\bibitem [{\citenamefont {Huang}\ \emph
  {et~al.}(2006{\natexlab{b}})\citenamefont {Huang}, \citenamefont {Hou},
  \citenamefont {Li}, \citenamefont {Li},\ and\ \citenamefont
  {Yang}}]{Huang2006b}%
  \BibitemOpen
  \bibfield  {author} {\bibinfo {author} {\bibfnamefont {J.~S.}\ \bibnamefont
  {Huang}}, \bibinfo {author} {\bibfnamefont {W.~J.}\ \bibnamefont {Hou}},
  \bibinfo {author} {\bibfnamefont {J.~H.}\ \bibnamefont {Li}}, \bibinfo
  {author} {\bibfnamefont {G.}~\bibnamefont {Li}}, \ and\ \bibinfo {author}
  {\bibfnamefont {Y.}~\bibnamefont {Yang}},\ }\href@noop {} {\bibfield
  {journal} {\bibinfo  {journal} {Applied Physics Letters}\ }\textbf {\bibinfo
  {volume} {89}},\ \bibinfo {pages} {133509} (\bibinfo {year}
  {2006}{\natexlab{b}})}\BibitemShut {NoStop}%
\bibitem [{\citenamefont {Huang}\ \emph
  {et~al.}(2006{\natexlab{c}})\citenamefont {Huang}, \citenamefont {Li},
  \citenamefont {Wu}, \citenamefont {Xu},\ and\ \citenamefont
  {Yang}}]{Huang2006a}%
  \BibitemOpen
  \bibfield  {author} {\bibinfo {author} {\bibfnamefont {J.~S.}\ \bibnamefont
  {Huang}}, \bibinfo {author} {\bibfnamefont {G.}~\bibnamefont {Li}}, \bibinfo
  {author} {\bibfnamefont {E.}~\bibnamefont {Wu}}, \bibinfo {author}
  {\bibfnamefont {Q.~F.}\ \bibnamefont {Xu}}, \ and\ \bibinfo {author}
  {\bibfnamefont {Y.}~\bibnamefont {Yang}},\ }\href@noop {} {\bibfield
  {journal} {\bibinfo  {journal} {Advanced Materials}\ }\textbf {\bibinfo
  {volume} {18}},\ \bibinfo {pages} {114} (\bibinfo {year}
  {2006}{\natexlab{c}})}\BibitemShut {NoStop}%
\bibitem [{\citenamefont {Kim}\ \emph {et~al.}(2004)\citenamefont {Kim},
  \citenamefont {Herguth}, \citenamefont {Kang}, \citenamefont {Jen},
  \citenamefont {Tseng},\ and\ \citenamefont {Shu}}]{Kim2004}%
  \BibitemOpen
  \bibfield  {author} {\bibinfo {author} {\bibfnamefont {J.~H.}\ \bibnamefont
  {Kim}}, \bibinfo {author} {\bibfnamefont {P.}~\bibnamefont {Herguth}},
  \bibinfo {author} {\bibfnamefont {M.~S.}\ \bibnamefont {Kang}}, \bibinfo
  {author} {\bibfnamefont {A.~K.~Y.}\ \bibnamefont {Jen}}, \bibinfo {author}
  {\bibfnamefont {Y.~H.}\ \bibnamefont {Tseng}}, \ and\ \bibinfo {author}
  {\bibfnamefont {C.~F.}\ \bibnamefont {Shu}},\ }\href@noop {} {\bibfield
  {journal} {\bibinfo  {journal} {Applied Physics Letters}\ }\textbf {\bibinfo
  {volume} {85}},\ \bibinfo {pages} {1116} (\bibinfo {year}
  {2004})}\BibitemShut {NoStop}%
\bibitem [{\citenamefont {Kim}\ \emph {et~al.}(2006)\citenamefont {Kim},
  \citenamefont {Lee}, \citenamefont {Park}, \citenamefont {Chin},
  \citenamefont {Lee},\ and\ \citenamefont {Kim}}]{Kim2006}%
  \BibitemOpen
  \bibfield  {author} {\bibinfo {author} {\bibfnamefont {T.~H.}\ \bibnamefont
  {Kim}}, \bibinfo {author} {\bibfnamefont {H.~K.}\ \bibnamefont {Lee}},
  \bibinfo {author} {\bibfnamefont {O.~O.}\ \bibnamefont {Park}}, \bibinfo
  {author} {\bibfnamefont {B.~D.}\ \bibnamefont {Chin}}, \bibinfo {author}
  {\bibfnamefont {S.~H.}\ \bibnamefont {Lee}}, \ and\ \bibinfo {author}
  {\bibfnamefont {J.~K.}\ \bibnamefont {Kim}},\ }\href@noop {} {\bibfield
  {journal} {\bibinfo  {journal} {Advanced Functional Materials}\ }\textbf
  {\bibinfo {volume} {16}},\ \bibinfo {pages} {611} (\bibinfo {year}
  {2006})}\BibitemShut {NoStop}%
\bibitem [{\citenamefont {Lee}\ \emph {et~al.}(2005)\citenamefont {Lee},
  \citenamefont {Hwang}, \citenamefont {Jung}, \citenamefont {Cho},
  \citenamefont {Lee}, \citenamefont {Lee},\ and\ \citenamefont
  {Shim}}]{Lee2005}%
  \BibitemOpen
  \bibfield  {author} {\bibinfo {author} {\bibfnamefont {S.~K.}\ \bibnamefont
  {Lee}}, \bibinfo {author} {\bibfnamefont {D.~H.}\ \bibnamefont {Hwang}},
  \bibinfo {author} {\bibfnamefont {B.~J.}\ \bibnamefont {Jung}}, \bibinfo
  {author} {\bibfnamefont {N.~S.}\ \bibnamefont {Cho}}, \bibinfo {author}
  {\bibfnamefont {J.}~\bibnamefont {Lee}}, \bibinfo {author} {\bibfnamefont
  {J.~D.}\ \bibnamefont {Lee}}, \ and\ \bibinfo {author} {\bibfnamefont
  {H.~K.}\ \bibnamefont {Shim}},\ }\href@noop {} {\bibfield  {journal}
  {\bibinfo  {journal} {Advanced Functional Materials}\ }\textbf {\bibinfo
  {volume} {15}},\ \bibinfo {pages} {1647} (\bibinfo {year}
  {2005})}\BibitemShut {NoStop}%
\bibitem [{\citenamefont {Liu}\ \emph {et~al.}(2007{\natexlab{a}})\citenamefont
  {Liu}, \citenamefont {Guo}, \citenamefont {Bu}, \citenamefont {Xie},
  \citenamefont {Cheng}, \citenamefont {Geng}, \citenamefont {Wang},
  \citenamefont {Jing},\ and\ \citenamefont {Wang}}]{Liu2007a}%
  \BibitemOpen
  \bibfield  {author} {\bibinfo {author} {\bibfnamefont {J.}~\bibnamefont
  {Liu}}, \bibinfo {author} {\bibfnamefont {X.}~\bibnamefont {Guo}}, \bibinfo
  {author} {\bibfnamefont {L.~J.}\ \bibnamefont {Bu}}, \bibinfo {author}
  {\bibfnamefont {Z.~Y.}\ \bibnamefont {Xie}}, \bibinfo {author} {\bibfnamefont
  {Y.~X.}\ \bibnamefont {Cheng}}, \bibinfo {author} {\bibfnamefont {Y.~H.}\
  \bibnamefont {Geng}}, \bibinfo {author} {\bibfnamefont {L.~X.}\ \bibnamefont
  {Wang}}, \bibinfo {author} {\bibfnamefont {X.~B.}\ \bibnamefont {Jing}}, \
  and\ \bibinfo {author} {\bibfnamefont {F.~S.}\ \bibnamefont {Wang}},\
  }\href@noop {} {\bibfield  {journal} {\bibinfo  {journal} {Advanced
  Functional Materials}\ }\textbf {\bibinfo {volume} {17}},\ \bibinfo {pages}
  {1917} (\bibinfo {year} {2007}{\natexlab{a}})}\BibitemShut {NoStop}%
\bibitem [{\citenamefont {Liu}\ \emph {et~al.}(2007{\natexlab{b}})\citenamefont
  {Liu}, \citenamefont {Xie}, \citenamefont {Cheng}, \citenamefont {Geng},
  \citenamefont {Wang}, \citenamefont {Jing},\ and\ \citenamefont
  {Wang}}]{Liu2007}%
  \BibitemOpen
  \bibfield  {author} {\bibinfo {author} {\bibfnamefont {J.}~\bibnamefont
  {Liu}}, \bibinfo {author} {\bibfnamefont {Z.~Y.}\ \bibnamefont {Xie}},
  \bibinfo {author} {\bibfnamefont {Y.~X.}\ \bibnamefont {Cheng}}, \bibinfo
  {author} {\bibfnamefont {Y.~H.}\ \bibnamefont {Geng}}, \bibinfo {author}
  {\bibfnamefont {L.~X.}\ \bibnamefont {Wang}}, \bibinfo {author}
  {\bibfnamefont {X.~B.}\ \bibnamefont {Jing}}, \ and\ \bibinfo {author}
  {\bibfnamefont {F.~S.}\ \bibnamefont {Wang}},\ }\href@noop {} {\bibfield
  {journal} {\bibinfo  {journal} {Advanced Materials}\ }\textbf {\bibinfo
  {volume} {19}},\ \bibinfo {pages} {531} (\bibinfo {year}
  {2007}{\natexlab{b}})}\BibitemShut {NoStop}%
\bibitem [{\citenamefont {Liu}\ \emph {et~al.}(2006)\citenamefont {Liu},
  \citenamefont {Zhou}, \citenamefont {Cheng}, \citenamefont {Geng},
  \citenamefont {Wang}, \citenamefont {Ma}, \citenamefont {Jing},\ and\
  \citenamefont {Wang}}]{Liu2006}%
  \BibitemOpen
  \bibfield  {author} {\bibinfo {author} {\bibfnamefont {J.}~\bibnamefont
  {Liu}}, \bibinfo {author} {\bibfnamefont {Q.~G.}\ \bibnamefont {Zhou}},
  \bibinfo {author} {\bibfnamefont {Y.~X.}\ \bibnamefont {Cheng}}, \bibinfo
  {author} {\bibfnamefont {Y.~H.}\ \bibnamefont {Geng}}, \bibinfo {author}
  {\bibfnamefont {L.~X.}\ \bibnamefont {Wang}}, \bibinfo {author}
  {\bibfnamefont {D.~G.}\ \bibnamefont {Ma}}, \bibinfo {author} {\bibfnamefont
  {X.~B.}\ \bibnamefont {Jing}}, \ and\ \bibinfo {author} {\bibfnamefont
  {F.~S.}\ \bibnamefont {Wang}},\ }\href@noop {} {\bibfield  {journal}
  {\bibinfo  {journal} {Advanced Functional Materials}\ }\textbf {\bibinfo
  {volume} {16}},\ \bibinfo {pages} {957} (\bibinfo {year} {2006})}\BibitemShut
  {NoStop}%
\bibitem [{\citenamefont {Luo}\ \emph {et~al.}(2007)\citenamefont {Luo},
  \citenamefont {Li}, \citenamefont {Hou}, \citenamefont {Peng}, \citenamefont
  {Yang},\ and\ \citenamefont {Cao}}]{Luo2007}%
  \BibitemOpen
  \bibfield  {author} {\bibinfo {author} {\bibfnamefont {J.}~\bibnamefont
  {Luo}}, \bibinfo {author} {\bibfnamefont {X.~Z.}\ \bibnamefont {Li}},
  \bibinfo {author} {\bibfnamefont {Q.}~\bibnamefont {Hou}}, \bibinfo {author}
  {\bibfnamefont {J.~B.}\ \bibnamefont {Peng}}, \bibinfo {author}
  {\bibfnamefont {W.}~\bibnamefont {Yang}}, \ and\ \bibinfo {author}
  {\bibfnamefont {Y.}~\bibnamefont {Cao}},\ }\href@noop {} {\bibfield
  {journal} {\bibinfo  {journal} {Advanced Materials}\ }\textbf {\bibinfo
  {volume} {19}},\ \bibinfo {pages} {1113} (\bibinfo {year}
  {2007})}\BibitemShut {NoStop}%
\bibitem [{\citenamefont {Tasch}\ \emph {et~al.}(1997)\citenamefont {Tasch},
  \citenamefont {List}, \citenamefont {Ekström}, \citenamefont {Graupner},
  \citenamefont {Leising}, \citenamefont {Schlichting}, \citenamefont {Rohr},
  \citenamefont {Geerts}, \citenamefont {Scherf},\ and\ \citenamefont
  {Müllen}}]{Tasch1997}%
  \BibitemOpen
  \bibfield  {author} {\bibinfo {author} {\bibfnamefont {S.}~\bibnamefont
  {Tasch}}, \bibinfo {author} {\bibfnamefont {E.~J.~W.}\ \bibnamefont {List}},
  \bibinfo {author} {\bibfnamefont {O.}~\bibnamefont {Ekström}}, \bibinfo
  {author} {\bibfnamefont {W.}~\bibnamefont {Graupner}}, \bibinfo {author}
  {\bibfnamefont {G.}~\bibnamefont {Leising}}, \bibinfo {author} {\bibfnamefont
  {P.}~\bibnamefont {Schlichting}}, \bibinfo {author} {\bibfnamefont
  {U.}~\bibnamefont {Rohr}}, \bibinfo {author} {\bibfnamefont {Y.}~\bibnamefont
  {Geerts}}, \bibinfo {author} {\bibfnamefont {U.}~\bibnamefont {Scherf}}, \
  and\ \bibinfo {author} {\bibfnamefont {K.}~\bibnamefont {Müllen}},\
  }\href@noop {} {\bibfield  {journal} {\bibinfo  {journal} {Applied Physics
  Letters}\ }\textbf {\bibinfo {volume} {71}},\ \bibinfo {pages} {2883}
  (\bibinfo {year} {1997})}\BibitemShut {NoStop}%
\bibitem [{\citenamefont {Tu}\ \emph {et~al.}(2006)\citenamefont {Tu},
  \citenamefont {Mei}, \citenamefont {Zhou}, \citenamefont {Cheng},
  \citenamefont {Geng}, \citenamefont {Wang}, \citenamefont {Ma}, \citenamefont
  {Jing},\ and\ \citenamefont {Wang}}]{Tu2006}%
  \BibitemOpen
  \bibfield  {author} {\bibinfo {author} {\bibfnamefont {G.~L.}\ \bibnamefont
  {Tu}}, \bibinfo {author} {\bibfnamefont {C.~Y.}\ \bibnamefont {Mei}},
  \bibinfo {author} {\bibfnamefont {Q.~G.}\ \bibnamefont {Zhou}}, \bibinfo
  {author} {\bibfnamefont {Y.~X.}\ \bibnamefont {Cheng}}, \bibinfo {author}
  {\bibfnamefont {Y.~H.}\ \bibnamefont {Geng}}, \bibinfo {author}
  {\bibfnamefont {L.~X.}\ \bibnamefont {Wang}}, \bibinfo {author}
  {\bibfnamefont {D.~G.}\ \bibnamefont {Ma}}, \bibinfo {author} {\bibfnamefont
  {X.~B.}\ \bibnamefont {Jing}}, \ and\ \bibinfo {author} {\bibfnamefont
  {F.~S.}\ \bibnamefont {Wang}},\ }\href@noop {} {\bibfield  {journal}
  {\bibinfo  {journal} {Advanced Functional Materials}\ }\textbf {\bibinfo
  {volume} {16}},\ \bibinfo {pages} {101} (\bibinfo {year} {2006})}\BibitemShut
  {NoStop}%
\bibitem [{\citenamefont {Tu}\ \emph {et~al.}(2004)\citenamefont {Tu},
  \citenamefont {Zhou}, \citenamefont {Cheng}, \citenamefont {Wang},
  \citenamefont {Ma}, \citenamefont {Jing},\ and\ \citenamefont
  {Wang}}]{Tu2004}%
  \BibitemOpen
  \bibfield  {author} {\bibinfo {author} {\bibfnamefont {G.~L.}\ \bibnamefont
  {Tu}}, \bibinfo {author} {\bibfnamefont {Q.~G.}\ \bibnamefont {Zhou}},
  \bibinfo {author} {\bibfnamefont {Y.~X.}\ \bibnamefont {Cheng}}, \bibinfo
  {author} {\bibfnamefont {L.~X.}\ \bibnamefont {Wang}}, \bibinfo {author}
  {\bibfnamefont {D.~G.}\ \bibnamefont {Ma}}, \bibinfo {author} {\bibfnamefont
  {X.~B.}\ \bibnamefont {Jing}}, \ and\ \bibinfo {author} {\bibfnamefont
  {F.~S.}\ \bibnamefont {Wang}},\ }\href@noop {} {\bibfield  {journal}
  {\bibinfo  {journal} {Applied Physics Letters}\ }\textbf {\bibinfo {volume}
  {85}},\ \bibinfo {pages} {2172} (\bibinfo {year} {2004})}\BibitemShut
  {NoStop}%
\bibitem [{\citenamefont {Wu}\ \emph {et~al.}(2008)\citenamefont {Wu},
  \citenamefont {Zou}, \citenamefont {Liu}, \citenamefont {Wang}, \citenamefont
  {Mikhailovsky}, \citenamefont {Bazan}, \citenamefont {Yang},\ and\
  \citenamefont {Cao}}]{Wu2008}%
  \BibitemOpen
  \bibfield  {author} {\bibinfo {author} {\bibfnamefont {H.~B.}\ \bibnamefont
  {Wu}}, \bibinfo {author} {\bibfnamefont {J.~H.}\ \bibnamefont {Zou}},
  \bibinfo {author} {\bibfnamefont {F.}~\bibnamefont {Liu}}, \bibinfo {author}
  {\bibfnamefont {L.}~\bibnamefont {Wang}}, \bibinfo {author} {\bibfnamefont
  {A.}~\bibnamefont {Mikhailovsky}}, \bibinfo {author} {\bibfnamefont {G.~C.}\
  \bibnamefont {Bazan}}, \bibinfo {author} {\bibfnamefont {W.}~\bibnamefont
  {Yang}}, \ and\ \bibinfo {author} {\bibfnamefont {Y.}~\bibnamefont {Cao}},\
  }\href@noop {} {\bibfield  {journal} {\bibinfo  {journal} {Advanced
  Materials}\ }\textbf {\bibinfo {volume} {20}},\ \bibinfo {pages} {696}
  (\bibinfo {year} {2008})}\BibitemShut {NoStop}%
\bibitem [{\citenamefont {Wu}\ \emph {et~al.}(2006)\citenamefont {Wu},
  \citenamefont {Lee},\ and\ \citenamefont {Chen}}]{Wu2006}%
  \BibitemOpen
  \bibfield  {author} {\bibinfo {author} {\bibfnamefont {W.~C.}\ \bibnamefont
  {Wu}}, \bibinfo {author} {\bibfnamefont {W.~Y.}\ \bibnamefont {Lee}}, \ and\
  \bibinfo {author} {\bibfnamefont {W.~C.}\ \bibnamefont {Chen}},\ }\href@noop
  {} {\bibfield  {journal} {\bibinfo  {journal} {Macromolecular Chemistry and
  Physics}\ }\textbf {\bibinfo {volume} {207}},\ \bibinfo {pages} {1131}
  (\bibinfo {year} {2006})}\BibitemShut {NoStop}%
\bibitem [{\citenamefont {Xu}\ \emph {et~al.}(2004)\citenamefont {Xu},
  \citenamefont {Duong}, \citenamefont {Wudl},\ and\ \citenamefont
  {Yang}}]{Xu2004}%
  \BibitemOpen
  \bibfield  {author} {\bibinfo {author} {\bibfnamefont {Q.~F.}\ \bibnamefont
  {Xu}}, \bibinfo {author} {\bibfnamefont {H.~M.}\ \bibnamefont {Duong}},
  \bibinfo {author} {\bibfnamefont {F.}~\bibnamefont {Wudl}}, \ and\ \bibinfo
  {author} {\bibfnamefont {Y.}~\bibnamefont {Yang}},\ }\href@noop {} {\bibfield
   {journal} {\bibinfo  {journal} {Applied Physics Letters}\ }\textbf {\bibinfo
  {volume} {85}},\ \bibinfo {pages} {3357} (\bibinfo {year}
  {2004})}\BibitemShut {NoStop}%
\bibitem [{\citenamefont {Gong}\ \emph {et~al.}(2005)\citenamefont {Gong},
  \citenamefont {Wang}, \citenamefont {Moses}, \citenamefont {Bazan},\ and\
  \citenamefont {Heeger}}]{Gong2005}%
  \BibitemOpen
  \bibfield  {author} {\bibinfo {author} {\bibfnamefont {X.}~\bibnamefont
  {Gong}}, \bibinfo {author} {\bibfnamefont {S.}~\bibnamefont {Wang}}, \bibinfo
  {author} {\bibfnamefont {D.}~\bibnamefont {Moses}}, \bibinfo {author}
  {\bibfnamefont {G.~C.}\ \bibnamefont {Bazan}}, \ and\ \bibinfo {author}
  {\bibfnamefont {A.~J.}\ \bibnamefont {Heeger}},\ }\href@noop {} {\bibfield
  {journal} {\bibinfo  {journal} {Advanced Materials}\ }\textbf {\bibinfo
  {volume} {17}},\ \bibinfo {pages} {2053} (\bibinfo {year}
  {2005})}\BibitemShut {NoStop}%
\bibitem [{\citenamefont {Huang}\ \emph {et~al.}(2009)\citenamefont {Huang},
  \citenamefont {Shih}, \citenamefont {Shu}, \citenamefont {Chi},\ and\
  \citenamefont {Jen}}]{Huang2009}%
  \BibitemOpen
  \bibfield  {author} {\bibinfo {author} {\bibfnamefont {F.}~\bibnamefont
  {Huang}}, \bibinfo {author} {\bibfnamefont {P.~I.}\ \bibnamefont {Shih}},
  \bibinfo {author} {\bibfnamefont {C.~F.}\ \bibnamefont {Shu}}, \bibinfo
  {author} {\bibfnamefont {Y.}~\bibnamefont {Chi}}, \ and\ \bibinfo {author}
  {\bibfnamefont {A.~K.~Y.}\ \bibnamefont {Jen}},\ }\href@noop {} {\bibfield
  {journal} {\bibinfo  {journal} {Advanced Materials}\ }\textbf {\bibinfo
  {volume} {21}},\ \bibinfo {pages} {361} (\bibinfo {year} {2009})}\BibitemShut
  {NoStop}%
\bibitem [{\citenamefont {Kawamura}\ \emph {et~al.}(2002)\citenamefont
  {Kawamura}, \citenamefont {Yanagida},\ and\ \citenamefont
  {Forrest}}]{Kawamura2002}%
  \BibitemOpen
  \bibfield  {author} {\bibinfo {author} {\bibfnamefont {Y.}~\bibnamefont
  {Kawamura}}, \bibinfo {author} {\bibfnamefont {S.}~\bibnamefont {Yanagida}},
  \ and\ \bibinfo {author} {\bibfnamefont {S.~R.}\ \bibnamefont {Forrest}},\
  }\href@noop {} {\bibfield  {journal} {\bibinfo  {journal} {Journal of Applied
  Physics}\ }\textbf {\bibinfo {volume} {92}},\ \bibinfo {pages} {87} (\bibinfo
  {year} {2002})}\BibitemShut {NoStop}%
\bibitem [{\citenamefont {Niu}\ \emph {et~al.}(2006)\citenamefont {Niu},
  \citenamefont {Ma}, \citenamefont {Yao}, \citenamefont {Ding}, \citenamefont
  {Tu}, \citenamefont {Xie},\ and\ \citenamefont {Wang}}]{Niu2006}%
  \BibitemOpen
  \bibfield  {author} {\bibinfo {author} {\bibfnamefont {X.~D.}\ \bibnamefont
  {Niu}}, \bibinfo {author} {\bibfnamefont {L.}~\bibnamefont {Ma}}, \bibinfo
  {author} {\bibfnamefont {B.}~\bibnamefont {Yao}}, \bibinfo {author}
  {\bibfnamefont {J.~Q.}\ \bibnamefont {Ding}}, \bibinfo {author}
  {\bibfnamefont {G.~L.}\ \bibnamefont {Tu}}, \bibinfo {author} {\bibfnamefont
  {Z.~Y.}\ \bibnamefont {Xie}}, \ and\ \bibinfo {author} {\bibfnamefont
  {L.~X.}\ \bibnamefont {Wang}},\ }\href@noop {} {\bibfield  {journal}
  {\bibinfo  {journal} {Applied Physics Letters}\ }\textbf {\bibinfo {volume}
  {89}},\ \bibinfo {pages} {213508} (\bibinfo {year} {2006})}\BibitemShut
  {NoStop}%
\bibitem [{\citenamefont {Niu}\ \emph {et~al.}(2007)\citenamefont {Niu},
  \citenamefont {Liu}, \citenamefont {Ka}, \citenamefont {Bardeker},
  \citenamefont {Zin}, \citenamefont {Schofield}, \citenamefont {Chi},\ and\
  \citenamefont {Jen}}]{Niu2007}%
  \BibitemOpen
  \bibfield  {author} {\bibinfo {author} {\bibfnamefont {Y.~H.}\ \bibnamefont
  {Niu}}, \bibinfo {author} {\bibfnamefont {M.~S.}\ \bibnamefont {Liu}},
  \bibinfo {author} {\bibfnamefont {J.~W.}\ \bibnamefont {Ka}}, \bibinfo
  {author} {\bibfnamefont {J.}~\bibnamefont {Bardeker}}, \bibinfo {author}
  {\bibfnamefont {M.~T.}\ \bibnamefont {Zin}}, \bibinfo {author} {\bibfnamefont
  {R.}~\bibnamefont {Schofield}}, \bibinfo {author} {\bibfnamefont
  {Y.}~\bibnamefont {Chi}}, \ and\ \bibinfo {author} {\bibfnamefont {A.~K.~Y.}\
  \bibnamefont {Jen}},\ }\href@noop {} {\bibfield  {journal} {\bibinfo
  {journal} {Advanced Materials}\ }\textbf {\bibinfo {volume} {19}},\ \bibinfo
  {pages} {300} (\bibinfo {year} {2007})}\BibitemShut {NoStop}%
\bibitem [{\citenamefont {Xu}\ \emph {et~al.}(2005)\citenamefont {Xu},
  \citenamefont {Peng}, \citenamefont {Jiang}, \citenamefont {Xu},
  \citenamefont {Yang},\ and\ \citenamefont {Cao}}]{Xu2005}%
  \BibitemOpen
  \bibfield  {author} {\bibinfo {author} {\bibfnamefont {Y.~H.}\ \bibnamefont
  {Xu}}, \bibinfo {author} {\bibfnamefont {J.~B.}\ \bibnamefont {Peng}},
  \bibinfo {author} {\bibfnamefont {J.~X.}\ \bibnamefont {Jiang}}, \bibinfo
  {author} {\bibfnamefont {W.}~\bibnamefont {Xu}}, \bibinfo {author}
  {\bibfnamefont {W.}~\bibnamefont {Yang}}, \ and\ \bibinfo {author}
  {\bibfnamefont {Y.}~\bibnamefont {Cao}},\ }\href@noop {} {\bibfield
  {journal} {\bibinfo  {journal} {Applied Physics Letters}\ }\textbf {\bibinfo
  {volume} {87}},\ \bibinfo {pages} {193502} (\bibinfo {year}
  {2005})}\BibitemShut {NoStop}%
\bibitem [{\citenamefont {Chen}\ \emph {et~al.}(2002)\citenamefont {Chen},
  \citenamefont {Yang}, \citenamefont {Thompson},\ and\ \citenamefont
  {Kido}}]{Chen2002}%
  \BibitemOpen
  \bibfield  {author} {\bibinfo {author} {\bibfnamefont {F.~C.}\ \bibnamefont
  {Chen}}, \bibinfo {author} {\bibfnamefont {Y.}~\bibnamefont {Yang}}, \bibinfo
  {author} {\bibfnamefont {M.~E.}\ \bibnamefont {Thompson}}, \ and\ \bibinfo
  {author} {\bibfnamefont {J.}~\bibnamefont {Kido}},\ }\href@noop {} {\bibfield
   {journal} {\bibinfo  {journal} {Applied Physics Letters}\ }\textbf {\bibinfo
  {volume} {80}},\ \bibinfo {pages} {2308} (\bibinfo {year}
  {2002})}\BibitemShut {NoStop}%
\bibitem [{\citenamefont {Granstrom}\ and\ \citenamefont
  {Inganas}(1996)}]{Granstrom1996}%
  \BibitemOpen
  \bibfield  {author} {\bibinfo {author} {\bibfnamefont {M.}~\bibnamefont
  {Granstrom}}\ and\ \bibinfo {author} {\bibfnamefont {O.}~\bibnamefont
  {Inganas}},\ }\href@noop {} {\bibfield  {journal} {\bibinfo  {journal}
  {Applied Physics Letters}\ }\textbf {\bibinfo {volume} {68}},\ \bibinfo
  {pages} {147} (\bibinfo {year} {1996})}\BibitemShut {NoStop}%
\bibitem [{\citenamefont {Hu}\ and\ \citenamefont {Karasz}(2003)}]{Hu2002}%
  \BibitemOpen
  \bibfield  {author} {\bibinfo {author} {\bibfnamefont {B.}~\bibnamefont
  {Hu}}\ and\ \bibinfo {author} {\bibfnamefont {F.~E.}\ \bibnamefont
  {Karasz}},\ }\href@noop {} {\bibfield  {journal} {\bibinfo  {journal}
  {Journal of Applied Physics}\ }\textbf {\bibinfo {volume} {93}},\ \bibinfo
  {pages} {1995} (\bibinfo {year} {2003})}\BibitemShut {NoStop}%
\bibitem [{\citenamefont {Niu}\ \emph {et~al.}(2004)\citenamefont {Niu},
  \citenamefont {Chen}, \citenamefont {Liu}, \citenamefont {Yip}, \citenamefont
  {Bardecker}, \citenamefont {Jen}, \citenamefont {Kavitha}, \citenamefont
  {Chi}, \citenamefont {Shu}, \citenamefont {Tseng},\ and\ \citenamefont
  {Chien}}]{Niu2004}%
  \BibitemOpen
  \bibfield  {author} {\bibinfo {author} {\bibfnamefont {Y.~H.}\ \bibnamefont
  {Niu}}, \bibinfo {author} {\bibfnamefont {B.~Q.}\ \bibnamefont {Chen}},
  \bibinfo {author} {\bibfnamefont {S.}~\bibnamefont {Liu}}, \bibinfo {author}
  {\bibfnamefont {H.}~\bibnamefont {Yip}}, \bibinfo {author} {\bibfnamefont
  {J.}~\bibnamefont {Bardecker}}, \bibinfo {author} {\bibfnamefont {A.~K.~Y.}\
  \bibnamefont {Jen}}, \bibinfo {author} {\bibfnamefont {J.}~\bibnamefont
  {Kavitha}}, \bibinfo {author} {\bibfnamefont {Y.}~\bibnamefont {Chi}},
  \bibinfo {author} {\bibfnamefont {C.~F.}\ \bibnamefont {Shu}}, \bibinfo
  {author} {\bibfnamefont {Y.~H.}\ \bibnamefont {Tseng}}, \ and\ \bibinfo
  {author} {\bibfnamefont {C.~H.}\ \bibnamefont {Chien}},\ }\href@noop {}
  {\bibfield  {journal} {\bibinfo  {journal} {Applied Physics Letters}\
  }\textbf {\bibinfo {volume} {85}},\ \bibinfo {pages} {1619} (\bibinfo {year}
  {2004})}\BibitemShut {NoStop}%
\bibitem [{\citenamefont {Noh}\ \emph {et~al.}(2003)\citenamefont {Noh},
  \citenamefont {Lee}, \citenamefont {Kim},\ and\ \citenamefont
  {Yase}}]{Noh2003}%
  \BibitemOpen
  \bibfield  {author} {\bibinfo {author} {\bibfnamefont {Y.~Y.}\ \bibnamefont
  {Noh}}, \bibinfo {author} {\bibfnamefont {C.~L.}\ \bibnamefont {Lee}},
  \bibinfo {author} {\bibfnamefont {J.~J.}\ \bibnamefont {Kim}}, \ and\
  \bibinfo {author} {\bibfnamefont {K.}~\bibnamefont {Yase}},\ }\href@noop {}
  {\bibfield  {journal} {\bibinfo  {journal} {Journal Of Chemical Physics}\
  }\textbf {\bibinfo {volume} {118}},\ \bibinfo {pages} {2853} (\bibinfo {year}
  {2003})}\BibitemShut {NoStop}%
\bibitem [{\citenamefont {Pschenitzka}\ and\ \citenamefont
  {Sturm}(1999)}]{Pschenitzka1999}%
  \BibitemOpen
  \bibfield  {author} {\bibinfo {author} {\bibfnamefont {F.}~\bibnamefont
  {Pschenitzka}}\ and\ \bibinfo {author} {\bibfnamefont {J.~C.}\ \bibnamefont
  {Sturm}},\ }\href@noop {} {\bibfield  {journal} {\bibinfo  {journal} {Applied
  Physics Letters}\ }\textbf {\bibinfo {volume} {74}},\ \bibinfo {pages} {1913}
  (\bibinfo {year} {1999})}\BibitemShut {NoStop}%
\bibitem [{\citenamefont {Köhnen}\ \emph {et~al.}(2010)\citenamefont
  {Köhnen}, \citenamefont {Irion}, \citenamefont {Gather}, \citenamefont
  {Rehmann}, \citenamefont {Zacharias},\ and\ \citenamefont
  {Meerholz}}]{Kohnen2010}%
  \BibitemOpen
  \bibfield  {author} {\bibinfo {author} {\bibfnamefont {A.}~\bibnamefont
  {Köhnen}}, \bibinfo {author} {\bibfnamefont {M.}~\bibnamefont {Irion}},
  \bibinfo {author} {\bibfnamefont {M.~C.}\ \bibnamefont {Gather}}, \bibinfo
  {author} {\bibfnamefont {N.}~\bibnamefont {Rehmann}}, \bibinfo {author}
  {\bibfnamefont {P.}~\bibnamefont {Zacharias}}, \ and\ \bibinfo {author}
  {\bibfnamefont {K.}~\bibnamefont {Meerholz}},\ }\href@noop {} {\bibfield
  {journal} {\bibinfo  {journal} {Journal of Materials Chemistry}\ }\textbf
  {\bibinfo {volume} {20}},\ \bibinfo {pages} {3301} (\bibinfo {year}
  {2010})}\BibitemShut {NoStop}%
\bibitem [{\citenamefont {Segal}\ \emph {et~al.}(2003)\citenamefont {Segal},
  \citenamefont {Baldo}, \citenamefont {Holmes}, \citenamefont {Forrest},\ and\
  \citenamefont {Soos}}]{Segal2003}%
  \BibitemOpen
  \bibfield  {author} {\bibinfo {author} {\bibfnamefont {M.}~\bibnamefont
  {Segal}}, \bibinfo {author} {\bibfnamefont {M.~A.}\ \bibnamefont {Baldo}},
  \bibinfo {author} {\bibfnamefont {R.~J.}\ \bibnamefont {Holmes}}, \bibinfo
  {author} {\bibfnamefont {S.~R.}\ \bibnamefont {Forrest}}, \ and\ \bibinfo
  {author} {\bibfnamefont {Z.~G.}\ \bibnamefont {Soos}},\ }\href@noop {}
  {\bibfield  {journal} {\bibinfo  {journal} {Physical Review B}\ }\textbf
  {\bibinfo {volume} {68}},\ \bibinfo {pages} {075211} (\bibinfo {year}
  {2003})}\BibitemShut {NoStop}%
\bibitem [{\citenamefont {Baldo}\ \emph {et~al.}(1999)\citenamefont {Baldo},
  \citenamefont {O'Brien}, \citenamefont {Thompson},\ and\ \citenamefont
  {Forrest}}]{Baldo1999b}%
  \BibitemOpen
  \bibfield  {author} {\bibinfo {author} {\bibfnamefont {M.~A.}\ \bibnamefont
  {Baldo}}, \bibinfo {author} {\bibfnamefont {D.~F.}\ \bibnamefont {O'Brien}},
  \bibinfo {author} {\bibfnamefont {M.~E.}\ \bibnamefont {Thompson}}, \ and\
  \bibinfo {author} {\bibfnamefont {S.~R.}\ \bibnamefont {Forrest}},\
  }\href@noop {} {\bibfield  {journal} {\bibinfo  {journal} {Physical Review
  B}\ }\textbf {\bibinfo {volume} {60}},\ \bibinfo {pages} {14422} (\bibinfo
  {year} {1999})}\BibitemShut {NoStop}%
\bibitem [{\citenamefont {Baldo}\ \emph {et~al.}(1998)\citenamefont {Baldo},
  \citenamefont {O'Brien}, \citenamefont {You}, \citenamefont {Shoustikov},
  \citenamefont {Sibley}, \citenamefont {Thompson},\ and\ \citenamefont
  {Forrest}}]{Baldo1998}%
  \BibitemOpen
  \bibfield  {author} {\bibinfo {author} {\bibfnamefont {M.~A.}\ \bibnamefont
  {Baldo}}, \bibinfo {author} {\bibfnamefont {D.~F.}\ \bibnamefont {O'Brien}},
  \bibinfo {author} {\bibfnamefont {Y.}~\bibnamefont {You}}, \bibinfo {author}
  {\bibfnamefont {A.}~\bibnamefont {Shoustikov}}, \bibinfo {author}
  {\bibfnamefont {S.}~\bibnamefont {Sibley}}, \bibinfo {author} {\bibfnamefont
  {M.~E.}\ \bibnamefont {Thompson}}, \ and\ \bibinfo {author} {\bibfnamefont
  {S.~R.}\ \bibnamefont {Forrest}},\ }\href@noop {} {\bibfield  {journal}
  {\bibinfo  {journal} {Nature}\ }\textbf {\bibinfo {volume} {395}},\ \bibinfo
  {pages} {151} (\bibinfo {year} {1998})}\BibitemShut {NoStop}%
\bibitem [{\citenamefont {Ma}\ \emph {et~al.}(1998)\citenamefont {Ma},
  \citenamefont {Zhang}, \citenamefont {Shen},\ and\ \citenamefont
  {Che}}]{Ma1998}%
  \BibitemOpen
  \bibfield  {author} {\bibinfo {author} {\bibfnamefont {Y.~G.}\ \bibnamefont
  {Ma}}, \bibinfo {author} {\bibfnamefont {H.~Y.}\ \bibnamefont {Zhang}},
  \bibinfo {author} {\bibfnamefont {J.~C.}\ \bibnamefont {Shen}}, \ and\
  \bibinfo {author} {\bibfnamefont {C.~M.}\ \bibnamefont {Che}},\ }\href@noop
  {} {\bibfield  {journal} {\bibinfo  {journal} {Synthetic Metals}\ }\textbf
  {\bibinfo {volume} {94}},\ \bibinfo {pages} {245} (\bibinfo {year}
  {1998})}\BibitemShut {NoStop}%
\bibitem [{\citenamefont {Reineke}\ and\ \citenamefont
  {Baldo}(2012)}]{Reineke2012}%
  \BibitemOpen
  \bibfield  {author} {\bibinfo {author} {\bibfnamefont {S.}~\bibnamefont
  {Reineke}}\ and\ \bibinfo {author} {\bibfnamefont {M.~A.}\ \bibnamefont
  {Baldo}},\ }\href@noop {} {\bibfield  {journal} {\bibinfo  {journal} {Phys.
  Status Solidi A}\ }\textbf {\bibinfo {volume} {209}},\ \bibinfo {pages}
  {2341} (\bibinfo {year} {2012})}\BibitemShut {NoStop}%
\bibitem [{\citenamefont {Yersin}(2004)}]{Yersin2004}%
  \BibitemOpen
  \bibfield  {author} {\bibinfo {author} {\bibfnamefont {H.}~\bibnamefont
  {Yersin}},\ }\href@noop {} {\bibfield  {journal} {\bibinfo  {journal}
  {Transition Metal And Rare Earth Compounds III}\ }\textbf {\bibinfo {volume}
  {241}},\ \bibinfo {pages} {1} (\bibinfo {year} {2004})}\BibitemShut {NoStop}%
\bibitem [{\citenamefont {Yersin}\ \emph {et~al.}(2002)\citenamefont {Yersin},
  \citenamefont {Donges}, \citenamefont {Humbs}, \citenamefont {Strasser},
  \citenamefont {Sitters},\ and\ \citenamefont {Glasbeek}}]{Yersin2002}%
  \BibitemOpen
  \bibfield  {author} {\bibinfo {author} {\bibfnamefont {H.}~\bibnamefont
  {Yersin}}, \bibinfo {author} {\bibfnamefont {D.}~\bibnamefont {Donges}},
  \bibinfo {author} {\bibfnamefont {W.}~\bibnamefont {Humbs}}, \bibinfo
  {author} {\bibfnamefont {J.}~\bibnamefont {Strasser}}, \bibinfo {author}
  {\bibfnamefont {R.}~\bibnamefont {Sitters}}, \ and\ \bibinfo {author}
  {\bibfnamefont {M.}~\bibnamefont {Glasbeek}},\ }\href@noop {} {\bibfield
  {journal} {\bibinfo  {journal} {Inorganic Chemistry}\ }\textbf {\bibinfo
  {volume} {41}},\ \bibinfo {pages} {4915} (\bibinfo {year}
  {2002})}\BibitemShut {NoStop}%
\bibitem [{\citenamefont {Kawamura}\ \emph {et~al.}(2005)\citenamefont
  {Kawamura}, \citenamefont {Goushi}, \citenamefont {Brooks}, \citenamefont
  {Brown}, \citenamefont {Sasabe},\ and\ \citenamefont
  {Adachi}}]{Kawamura2005}%
  \BibitemOpen
  \bibfield  {author} {\bibinfo {author} {\bibfnamefont {Y.}~\bibnamefont
  {Kawamura}}, \bibinfo {author} {\bibfnamefont {K.}~\bibnamefont {Goushi}},
  \bibinfo {author} {\bibfnamefont {J.}~\bibnamefont {Brooks}}, \bibinfo
  {author} {\bibfnamefont {J.~J.}\ \bibnamefont {Brown}}, \bibinfo {author}
  {\bibfnamefont {H.}~\bibnamefont {Sasabe}}, \ and\ \bibinfo {author}
  {\bibfnamefont {C.}~\bibnamefont {Adachi}},\ }\href@noop {} {\bibfield
  {journal} {\bibinfo  {journal} {Applied Physics Letters}\ }\textbf {\bibinfo
  {volume} {86}},\ \bibinfo {pages} {071104} (\bibinfo {year}
  {2005})}\BibitemShut {NoStop}%
\bibitem [{\citenamefont {Kawamura}\ \emph {et~al.}(2006)\citenamefont
  {Kawamura}, \citenamefont {Brooks}, \citenamefont {Brown}, \citenamefont
  {Sasabe},\ and\ \citenamefont {Adachi}}]{Kawamura2006}%
  \BibitemOpen
  \bibfield  {author} {\bibinfo {author} {\bibfnamefont {Y.}~\bibnamefont
  {Kawamura}}, \bibinfo {author} {\bibfnamefont {J.}~\bibnamefont {Brooks}},
  \bibinfo {author} {\bibfnamefont {J.~J.}\ \bibnamefont {Brown}}, \bibinfo
  {author} {\bibfnamefont {H.}~\bibnamefont {Sasabe}}, \ and\ \bibinfo {author}
  {\bibfnamefont {C.}~\bibnamefont {Adachi}},\ }\href@noop {} {\bibfield
  {journal} {\bibinfo  {journal} {Physical Review Letters}\ }\textbf {\bibinfo
  {volume} {96}},\ \bibinfo {pages} {017404} (\bibinfo {year}
  {2006})}\BibitemShut {NoStop}%
\bibitem [{\citenamefont {Baldo}\ \emph
  {et~al.}(2000{\natexlab{a}})\citenamefont {Baldo}, \citenamefont {Adachi},\
  and\ \citenamefont {Forrest}}]{Baldo2000a}%
  \BibitemOpen
  \bibfield  {author} {\bibinfo {author} {\bibfnamefont {M.~A.}\ \bibnamefont
  {Baldo}}, \bibinfo {author} {\bibfnamefont {C.}~\bibnamefont {Adachi}}, \
  and\ \bibinfo {author} {\bibfnamefont {S.~R.}\ \bibnamefont {Forrest}},\
  }\href@noop {} {\bibfield  {journal} {\bibinfo  {journal} {Physical Review
  B}\ }\textbf {\bibinfo {volume} {62}},\ \bibinfo {pages} {10967} (\bibinfo
  {year} {2000}{\natexlab{a}})}\BibitemShut {NoStop}%
\bibitem [{\citenamefont {Reineke}\ \emph
  {et~al.}(2007{\natexlab{a}})\citenamefont {Reineke}, \citenamefont {Walzer},\
  and\ \citenamefont {Leo}}]{Reineke2007}%
  \BibitemOpen
  \bibfield  {author} {\bibinfo {author} {\bibfnamefont {S.}~\bibnamefont
  {Reineke}}, \bibinfo {author} {\bibfnamefont {K.}~\bibnamefont {Walzer}}, \
  and\ \bibinfo {author} {\bibfnamefont {K.}~\bibnamefont {Leo}},\ }\href@noop
  {} {\bibfield  {journal} {\bibinfo  {journal} {Physical Review B}\ }\textbf
  {\bibinfo {volume} {75}},\ \bibinfo {pages} {125328} (\bibinfo {year}
  {2007}{\natexlab{a}})}\BibitemShut {NoStop}%
\bibitem [{\citenamefont {Schwartz}\ \emph {et~al.}(2009)\citenamefont
  {Schwartz}, \citenamefont {Reineke}, \citenamefont {Rosenow}, \citenamefont
  {Walzer},\ and\ \citenamefont {Leo}}]{Schwartz2009}%
  \BibitemOpen
  \bibfield  {author} {\bibinfo {author} {\bibfnamefont {G.}~\bibnamefont
  {Schwartz}}, \bibinfo {author} {\bibfnamefont {S.}~\bibnamefont {Reineke}},
  \bibinfo {author} {\bibfnamefont {T.~C.}\ \bibnamefont {Rosenow}}, \bibinfo
  {author} {\bibfnamefont {K.}~\bibnamefont {Walzer}}, \ and\ \bibinfo {author}
  {\bibfnamefont {K.}~\bibnamefont {Leo}},\ }\href@noop {} {\bibfield
  {journal} {\bibinfo  {journal} {Advanced Functional Materials}\ }\textbf
  {\bibinfo {volume} {19}},\ \bibinfo {pages} {1319} (\bibinfo {year}
  {2009})}\BibitemShut {NoStop}%
\bibitem [{\citenamefont {Sun}\ \emph {et~al.}(2006)\citenamefont {Sun},
  \citenamefont {Giebink}, \citenamefont {Kanno}, \citenamefont {Ma},
  \citenamefont {Thompson},\ and\ \citenamefont {Forrest}}]{Sun2006}%
  \BibitemOpen
  \bibfield  {author} {\bibinfo {author} {\bibfnamefont {Y.~R.}\ \bibnamefont
  {Sun}}, \bibinfo {author} {\bibfnamefont {N.~C.}\ \bibnamefont {Giebink}},
  \bibinfo {author} {\bibfnamefont {H.}~\bibnamefont {Kanno}}, \bibinfo
  {author} {\bibfnamefont {B.~W.}\ \bibnamefont {Ma}}, \bibinfo {author}
  {\bibfnamefont {M.~E.}\ \bibnamefont {Thompson}}, \ and\ \bibinfo {author}
  {\bibfnamefont {S.~R.}\ \bibnamefont {Forrest}},\ }\href@noop {} {\bibfield
  {journal} {\bibinfo  {journal} {Nature}\ }\textbf {\bibinfo {volume} {440}},\
  \bibinfo {pages} {908} (\bibinfo {year} {2006})}\BibitemShut {NoStop}%
\bibitem [{\citenamefont {Reineke}\ \emph
  {et~al.}(2009{\natexlab{b}})\citenamefont {Reineke}, \citenamefont
  {Schwartz}, \citenamefont {Walzer}, \citenamefont {Falke},\ and\
  \citenamefont {Leo}}]{Reineke2009b}%
  \BibitemOpen
  \bibfield  {author} {\bibinfo {author} {\bibfnamefont {S.}~\bibnamefont
  {Reineke}}, \bibinfo {author} {\bibfnamefont {G.}~\bibnamefont {Schwartz}},
  \bibinfo {author} {\bibfnamefont {K.}~\bibnamefont {Walzer}}, \bibinfo
  {author} {\bibfnamefont {M.}~\bibnamefont {Falke}}, \ and\ \bibinfo {author}
  {\bibfnamefont {K.}~\bibnamefont {Leo}},\ }\href@noop {} {\bibfield
  {journal} {\bibinfo  {journal} {Applied Physics Letters}\ }\textbf {\bibinfo
  {volume} {94}},\ \bibinfo {pages} {163305} (\bibinfo {year}
  {2009}{\natexlab{b}})}\BibitemShut {NoStop}%
\bibitem [{\citenamefont {Staroske}\ \emph {et~al.}(2007)\citenamefont
  {Staroske}, \citenamefont {Pfeiffer}, \citenamefont {Leo},\ and\
  \citenamefont {Hoffmann}}]{Staroske2007}%
  \BibitemOpen
  \bibfield  {author} {\bibinfo {author} {\bibfnamefont {W.}~\bibnamefont
  {Staroske}}, \bibinfo {author} {\bibfnamefont {M.}~\bibnamefont {Pfeiffer}},
  \bibinfo {author} {\bibfnamefont {K.}~\bibnamefont {Leo}}, \ and\ \bibinfo
  {author} {\bibfnamefont {M.}~\bibnamefont {Hoffmann}},\ }\href@noop {}
  {\bibfield  {journal} {\bibinfo  {journal} {Physical Review Letters}\
  }\textbf {\bibinfo {volume} {98}},\ \bibinfo {pages} {197402} (\bibinfo
  {year} {2007})}\BibitemShut {NoStop}%
\bibitem [{\citenamefont {Kepler}\ \emph {et~al.}(1963)\citenamefont {Kepler},
  \citenamefont {Avakian}, \citenamefont {Caris},\ and\ \citenamefont
  {Abramson}}]{Kepler1963}%
  \BibitemOpen
  \bibfield  {author} {\bibinfo {author} {\bibfnamefont {R.~G.}\ \bibnamefont
  {Kepler}}, \bibinfo {author} {\bibfnamefont {P.}~\bibnamefont {Avakian}},
  \bibinfo {author} {\bibfnamefont {J.~C.}\ \bibnamefont {Caris}}, \ and\
  \bibinfo {author} {\bibfnamefont {E.}~\bibnamefont {Abramson}},\ }\href@noop
  {} {\bibfield  {journal} {\bibinfo  {journal} {Physical Review Letters}\
  }\textbf {\bibinfo {volume} {10}},\ \bibinfo {pages} {400} (\bibinfo {year}
  {1963})}\BibitemShut {NoStop}%
\bibitem [{\citenamefont {Okumoto}\ \emph {et~al.}(2006)\citenamefont
  {Okumoto}, \citenamefont {Kanno}, \citenamefont {Hamaa}, \citenamefont
  {Takahashi},\ and\ \citenamefont {Shibata}}]{Okumoto2006}%
  \BibitemOpen
  \bibfield  {author} {\bibinfo {author} {\bibfnamefont {K.}~\bibnamefont
  {Okumoto}}, \bibinfo {author} {\bibfnamefont {H.}~\bibnamefont {Kanno}},
  \bibinfo {author} {\bibfnamefont {Y.}~\bibnamefont {Hamaa}}, \bibinfo
  {author} {\bibfnamefont {H.}~\bibnamefont {Takahashi}}, \ and\ \bibinfo
  {author} {\bibfnamefont {K.}~\bibnamefont {Shibata}},\ }\href@noop {}
  {\bibfield  {journal} {\bibinfo  {journal} {Applied Physics Letters}\
  }\textbf {\bibinfo {volume} {89}},\ \bibinfo {pages} {063504} (\bibinfo
  {year} {2006})}\BibitemShut {NoStop}%
\bibitem [{\citenamefont {Kondakov}(2007)}]{Kondakov2007}%
  \BibitemOpen
  \bibfield  {author} {\bibinfo {author} {\bibfnamefont {D.~Y.}\ \bibnamefont
  {Kondakov}},\ }\href@noop {} {\bibfield  {journal} {\bibinfo  {journal}
  {Journal Of Applied Physics}\ }\textbf {\bibinfo {volume} {102}},\ \bibinfo
  {pages} {114504} (\bibinfo {year} {2007})}\BibitemShut {NoStop}%
\bibitem [{\citenamefont {Endo}\ \emph {et~al.}(2009)\citenamefont {Endo},
  \citenamefont {Ogasawara}, \citenamefont {Takahashi}, \citenamefont
  {Yokoyama}, \citenamefont {Kato},\ and\ \citenamefont {Adachi}}]{Endo2009}%
  \BibitemOpen
  \bibfield  {author} {\bibinfo {author} {\bibfnamefont {A.}~\bibnamefont
  {Endo}}, \bibinfo {author} {\bibfnamefont {M.}~\bibnamefont {Ogasawara}},
  \bibinfo {author} {\bibfnamefont {A.}~\bibnamefont {Takahashi}}, \bibinfo
  {author} {\bibfnamefont {D.}~\bibnamefont {Yokoyama}}, \bibinfo {author}
  {\bibfnamefont {Y.}~\bibnamefont {Kato}}, \ and\ \bibinfo {author}
  {\bibfnamefont {C.}~\bibnamefont {Adachi}},\ }\href@noop {} {\bibfield
  {journal} {\bibinfo  {journal} {Advanced Materials}\ }\textbf {\bibinfo
  {volume} {21}},\ \bibinfo {pages} {4802} (\bibinfo {year}
  {2009})}\BibitemShut {NoStop}%
\bibitem [{\citenamefont {Endo}\ \emph {et~al.}(2011)\citenamefont {Endo},
  \citenamefont {Sato}, \citenamefont {Yoshimura}, \citenamefont {Kai},
  \citenamefont {Kawada}, \citenamefont {Miyazaki},\ and\ \citenamefont
  {Adachi}}]{Endo2011}%
  \BibitemOpen
  \bibfield  {author} {\bibinfo {author} {\bibfnamefont {A.}~\bibnamefont
  {Endo}}, \bibinfo {author} {\bibfnamefont {K.}~\bibnamefont {Sato}}, \bibinfo
  {author} {\bibfnamefont {K.}~\bibnamefont {Yoshimura}}, \bibinfo {author}
  {\bibfnamefont {T.}~\bibnamefont {Kai}}, \bibinfo {author} {\bibfnamefont
  {A.}~\bibnamefont {Kawada}}, \bibinfo {author} {\bibfnamefont
  {H.}~\bibnamefont {Miyazaki}}, \ and\ \bibinfo {author} {\bibfnamefont
  {C.}~\bibnamefont {Adachi}},\ }\href@noop {} {\bibfield  {journal} {\bibinfo
  {journal} {Applied Physics Letters}\ }\textbf {\bibinfo {volume} {98}},\
  \bibinfo {pages} {083302} (\bibinfo {year} {2011})}\BibitemShut {NoStop}%
\bibitem [{\citenamefont {Deaton}\ \emph {et~al.}(2010)\citenamefont {Deaton},
  \citenamefont {Switalski}, \citenamefont {Kondakov}, \citenamefont {Young},
  \citenamefont {Pawlik}, \citenamefont {Giesen}, \citenamefont {Harkins},
  \citenamefont {Miller}, \citenamefont {Mickenberg},\ and\ \citenamefont
  {Peters}}]{Deaton2010}%
  \BibitemOpen
  \bibfield  {author} {\bibinfo {author} {\bibfnamefont {J.~C.}\ \bibnamefont
  {Deaton}}, \bibinfo {author} {\bibfnamefont {S.~C.}\ \bibnamefont
  {Switalski}}, \bibinfo {author} {\bibfnamefont {D.~Y.}\ \bibnamefont
  {Kondakov}}, \bibinfo {author} {\bibfnamefont {R.~H.}\ \bibnamefont {Young}},
  \bibinfo {author} {\bibfnamefont {T.~D.}\ \bibnamefont {Pawlik}}, \bibinfo
  {author} {\bibfnamefont {D.~J.}\ \bibnamefont {Giesen}}, \bibinfo {author}
  {\bibfnamefont {S.~B.}\ \bibnamefont {Harkins}}, \bibinfo {author}
  {\bibfnamefont {A.~J.~M.}\ \bibnamefont {Miller}}, \bibinfo {author}
  {\bibfnamefont {S.~F.}\ \bibnamefont {Mickenberg}}, \ and\ \bibinfo {author}
  {\bibfnamefont {J.~C.}\ \bibnamefont {Peters}},\ }\href@noop {} {\bibfield
  {journal} {\bibinfo  {journal} {Journal of the American Chemical Society}\
  }\textbf {\bibinfo {volume} {132}},\ \bibinfo {pages} {9499} (\bibinfo {year}
  {2010})}\BibitemShut {NoStop}%
\bibitem [{\citenamefont {Goushi}\ \emph {et~al.}(2012)\citenamefont {Goushi},
  \citenamefont {Yoshida}, \citenamefont {Sato},\ and\ \citenamefont
  {Adachi}}]{Goushi2012}%
  \BibitemOpen
  \bibfield  {author} {\bibinfo {author} {\bibfnamefont {K.}~\bibnamefont
  {Goushi}}, \bibinfo {author} {\bibfnamefont {K.}~\bibnamefont {Yoshida}},
  \bibinfo {author} {\bibfnamefont {K.}~\bibnamefont {Sato}}, \ and\ \bibinfo
  {author} {\bibfnamefont {C.}~\bibnamefont {Adachi}},\ }\href@noop {}
  {\bibfield  {journal} {\bibinfo  {journal} {Nature Photonics}\ }\textbf
  {\bibinfo {volume} {6}},\ \bibinfo {pages} {253} (\bibinfo {year}
  {2012})}\BibitemShut {NoStop}%
\bibitem [{\citenamefont {Uoyama}\ \emph {et~al.}(2012)\citenamefont {Uoyama},
  \citenamefont {Kenichi~Goushi}, \citenamefont {Shizu}, \citenamefont {H.},\
  and\ \citenamefont {Adachi}}]{Uoyama2012}%
  \BibitemOpen
  \bibfield  {author} {\bibinfo {author} {\bibfnamefont {H.}~\bibnamefont
  {Uoyama}}, \bibinfo {author} {\bibfnamefont {K.}~\bibnamefont
  {Kenichi~Goushi}}, \bibinfo {author} {\bibfnamefont {K.}~\bibnamefont
  {Shizu}}, \bibinfo {author} {\bibfnamefont {N.}~\bibnamefont {H.}}, \ and\
  \bibinfo {author} {\bibfnamefont {C.}~\bibnamefont {Adachi}},\ }\href@noop {}
  {\bibfield  {journal} {\bibinfo  {journal} {Nature}\ }\textbf {\bibinfo
  {volume} {492}},\ \bibinfo {pages} {234} (\bibinfo {year}
  {2012})}\BibitemShut {NoStop}%
\bibitem [{\citenamefont {Segal}\ \emph {et~al.}(2007)\citenamefont {Segal},
  \citenamefont {Singh}, \citenamefont {Rivoire}, \citenamefont {Difley},
  \citenamefont {Van~Voorhis},\ and\ \citenamefont {Baldo}}]{Segal2007}%
  \BibitemOpen
  \bibfield  {author} {\bibinfo {author} {\bibfnamefont {M.}~\bibnamefont
  {Segal}}, \bibinfo {author} {\bibfnamefont {M.}~\bibnamefont {Singh}},
  \bibinfo {author} {\bibfnamefont {K.}~\bibnamefont {Rivoire}}, \bibinfo
  {author} {\bibfnamefont {S.}~\bibnamefont {Difley}}, \bibinfo {author}
  {\bibfnamefont {T.}~\bibnamefont {Van~Voorhis}}, \ and\ \bibinfo {author}
  {\bibfnamefont {M.~A.}\ \bibnamefont {Baldo}},\ }\href@noop {} {\bibfield
  {journal} {\bibinfo  {journal} {Nature Materials}\ }\textbf {\bibinfo
  {volume} {6}},\ \bibinfo {pages} {374} (\bibinfo {year} {2007})}\BibitemShut
  {NoStop}%
\bibitem [{\citenamefont {Coe}\ \emph {et~al.}(2002)\citenamefont {Coe},
  \citenamefont {Woo}, \citenamefont {Bawendi},\ and\ \citenamefont
  {Bulovic}}]{Coe2002}%
  \BibitemOpen
  \bibfield  {author} {\bibinfo {author} {\bibfnamefont {S.}~\bibnamefont
  {Coe}}, \bibinfo {author} {\bibfnamefont {W.~K.}\ \bibnamefont {Woo}},
  \bibinfo {author} {\bibfnamefont {M.}~\bibnamefont {Bawendi}}, \ and\
  \bibinfo {author} {\bibfnamefont {V.}~\bibnamefont {Bulovic}},\ }\href@noop
  {} {\bibfield  {journal} {\bibinfo  {journal} {Nature}\ }\textbf {\bibinfo
  {volume} {420}},\ \bibinfo {pages} {800} (\bibinfo {year}
  {2002})}\BibitemShut {NoStop}%
\bibitem [{\citenamefont {Anikeeva}\ \emph {et~al.}(2007)\citenamefont
  {Anikeeva}, \citenamefont {Halpert}, \citenamefont {Bawendi},\ and\
  \citenamefont {Bulovic}}]{Anikeeva2007}%
  \BibitemOpen
  \bibfield  {author} {\bibinfo {author} {\bibfnamefont {P.~O.}\ \bibnamefont
  {Anikeeva}}, \bibinfo {author} {\bibfnamefont {J.~E.}\ \bibnamefont
  {Halpert}}, \bibinfo {author} {\bibfnamefont {M.~G.}\ \bibnamefont
  {Bawendi}}, \ and\ \bibinfo {author} {\bibfnamefont {V.}~\bibnamefont
  {Bulovic}},\ }\href@noop {} {\bibfield  {journal} {\bibinfo  {journal} {Nano
  Letters}\ }\textbf {\bibinfo {volume} {7}},\ \bibinfo {pages} {2196}
  (\bibinfo {year} {2007})}\BibitemShut {NoStop}%
\bibitem [{\citenamefont {Shirasaki}\ \emph {et~al.}(2013)\citenamefont
  {Shirasaki}, \citenamefont {Supran}, \citenamefont {Bawendi},\ and\
  \citenamefont {Bulovic}}]{Shirasaki2013}%
  \BibitemOpen
  \bibfield  {author} {\bibinfo {author} {\bibfnamefont {Y.}~\bibnamefont
  {Shirasaki}}, \bibinfo {author} {\bibfnamefont {G.}~\bibnamefont {Supran}},
  \bibinfo {author} {\bibfnamefont {M.}~\bibnamefont {Bawendi}}, \ and\
  \bibinfo {author} {\bibfnamefont {V.}~\bibnamefont {Bulovic}},\ }\href@noop
  {} {\bibfield  {journal} {\bibinfo  {journal} {Nature Photonics}\ }\textbf
  {\bibinfo {volume} {7}},\ \bibinfo {pages} {13–} (\bibinfo {year}
  {2013})}\BibitemShut {NoStop}%
\bibitem [{\citenamefont {Caruge}\ \emph {et~al.}(2008)\citenamefont {Caruge},
  \citenamefont {Halpert}, \citenamefont {Wood}, \citenamefont {Bulovic},\ and\
  \citenamefont {Bawendi}}]{Caruge2008}%
  \BibitemOpen
  \bibfield  {author} {\bibinfo {author} {\bibfnamefont {J.~M.}\ \bibnamefont
  {Caruge}}, \bibinfo {author} {\bibfnamefont {J.~E.}\ \bibnamefont {Halpert}},
  \bibinfo {author} {\bibfnamefont {V.}~\bibnamefont {Wood}}, \bibinfo {author}
  {\bibfnamefont {V.}~\bibnamefont {Bulovic}}, \ and\ \bibinfo {author}
  {\bibfnamefont {M.~G.}\ \bibnamefont {Bawendi}},\ }\href@noop {} {\bibfield
  {journal} {\bibinfo  {journal} {Nature Photonics}\ }\textbf {\bibinfo
  {volume} {2}},\ \bibinfo {pages} {247} (\bibinfo {year} {2008})}\BibitemShut
  {NoStop}%
\bibitem [{\citenamefont {Klessinger}\ and\ \citenamefont
  {Michl}(1989)}]{b_Klessinger1989}%
  \BibitemOpen
  \bibfield  {author} {\bibinfo {author} {\bibfnamefont {M.}~\bibnamefont
  {Klessinger}}\ and\ \bibinfo {author} {\bibfnamefont {J.}~\bibnamefont
  {Michl}},\ }\href@noop {} {\emph {\bibinfo {title} {Lichtabsorption und
  Photochemie organischer Moleküle}}}\ (\bibinfo  {publisher} {VCH
  Verlagsgesellschaft mbH, Weinheim},\ \bibinfo {year} {1989})\BibitemShut
  {NoStop}%
\bibitem [{\citenamefont {F\"orster}(1948)}]{Foerster1948}%
  \BibitemOpen
  \bibfield  {author} {\bibinfo {author} {\bibfnamefont {T.}~\bibnamefont
  {F\"orster}},\ }\href@noop {} {\bibfield  {journal} {\bibinfo  {journal}
  {Ann. Physik}\ }\textbf {\bibinfo {volume} {2}},\ \bibinfo {pages} {55}
  (\bibinfo {year} {1948})}\BibitemShut {NoStop}%
\bibitem [{\citenamefont {Dexter}(1953)}]{Dexter1953}%
  \BibitemOpen
  \bibfield  {author} {\bibinfo {author} {\bibfnamefont {D.~L.}\ \bibnamefont
  {Dexter}},\ }\href@noop {} {\bibfield  {journal} {\bibinfo  {journal}
  {Journal of Chemical Physics}\ }\textbf {\bibinfo {volume} {21}},\ \bibinfo
  {pages} {836} (\bibinfo {year} {1953})}\BibitemShut {NoStop}%
\bibitem [{\citenamefont {Braslavsky}\ \emph {et~al.}(2008)\citenamefont
  {Braslavsky}, \citenamefont {Fron}, \citenamefont {Rodriguez}, \citenamefont
  {Roman}, \citenamefont {Scholes}, \citenamefont {Schweitzer}, \citenamefont
  {Valeur},\ and\ \citenamefont {Wirz}}]{Braslavsky2008}%
  \BibitemOpen
  \bibfield  {author} {\bibinfo {author} {\bibfnamefont {S.~E.}\ \bibnamefont
  {Braslavsky}}, \bibinfo {author} {\bibfnamefont {E.}~\bibnamefont {Fron}},
  \bibinfo {author} {\bibfnamefont {H.~B.}\ \bibnamefont {Rodriguez}}, \bibinfo
  {author} {\bibfnamefont {E.~S.}\ \bibnamefont {Roman}}, \bibinfo {author}
  {\bibfnamefont {G.~D.}\ \bibnamefont {Scholes}}, \bibinfo {author}
  {\bibfnamefont {G.}~\bibnamefont {Schweitzer}}, \bibinfo {author}
  {\bibfnamefont {B.}~\bibnamefont {Valeur}}, \ and\ \bibinfo {author}
  {\bibfnamefont {J.}~\bibnamefont {Wirz}},\ }\href@noop {} {\bibfield
  {journal} {\bibinfo  {journal} {Photochemical \& Photobiological Sciences}\
  }\textbf {\bibinfo {volume} {7}},\ \bibinfo {pages} {1444} (\bibinfo {year}
  {2008})}\BibitemShut {NoStop}%
\bibitem [{\citenamefont {Tanaka}\ \emph {et~al.}(2006)\citenamefont {Tanaka},
  \citenamefont {Ise}, \citenamefont {Shiomi}, \citenamefont {Sato},\ and\
  \citenamefont {Takui}}]{Tanaka2006}%
  \BibitemOpen
  \bibfield  {author} {\bibinfo {author} {\bibfnamefont {H.}~\bibnamefont
  {Tanaka}}, \bibinfo {author} {\bibfnamefont {T.}~\bibnamefont {Ise}},
  \bibinfo {author} {\bibfnamefont {D.}~\bibnamefont {Shiomi}}, \bibinfo
  {author} {\bibfnamefont {K.}~\bibnamefont {Sato}}, \ and\ \bibinfo {author}
  {\bibfnamefont {T.}~\bibnamefont {Takui}},\ }\href@noop {} {\bibfield
  {journal} {\bibinfo  {journal} {Journal Of Low Temperature Physics}\ }\textbf
  {\bibinfo {volume} {142}},\ \bibinfo {pages} {601} (\bibinfo {year}
  {2006})}\BibitemShut {NoStop}%
\bibitem [{\citenamefont {Reinhold}(2004)}]{b_Reinhold2004}%
  \BibitemOpen
  \bibfield  {author} {\bibinfo {author} {\bibfnamefont {J.}~\bibnamefont
  {Reinhold}},\ }\href@noop {} {\emph {\bibinfo {title} {Quantentheorie der
  Moleküle: Eine Einführung}}}\ (\bibinfo  {publisher} {B. G. Teubner Verlag
  / GWV Fachverlage GmbH, Wiesbaden},\ \bibinfo {year} {2004})\BibitemShut
  {NoStop}%
\bibitem [{\citenamefont {Wigner}\ and\ \citenamefont
  {Witmer}(1928)}]{Widmer1928}%
  \BibitemOpen
  \bibfield  {author} {\bibinfo {author} {\bibfnamefont {E.}~\bibnamefont
  {Wigner}}\ and\ \bibinfo {author} {\bibfnamefont {E.~E.}\ \bibnamefont
  {Witmer}},\ }\href@noop {} {\bibfield  {journal} {\bibinfo  {journal}
  {Zeitschrift für Physik A Hadrons and Nuclei}\ }\textbf {\bibinfo {volume}
  {51}},\ \bibinfo {pages} {859} (\bibinfo {year} {1928})}\BibitemShut
  {NoStop}%
\bibitem [{\citenamefont {Suna}(1970)}]{Suna1970}%
  \BibitemOpen
  \bibfield  {author} {\bibinfo {author} {\bibfnamefont {A.}~\bibnamefont
  {Suna}},\ }\href@noop {} {\bibfield  {journal} {\bibinfo  {journal} {Physical
  Review B}\ }\textbf {\bibinfo {volume} {1}},\ \bibinfo {pages} {1716}
  (\bibinfo {year} {1970})}\BibitemShut {NoStop}%
\bibitem [{\citenamefont {Murphy}\ \emph {et~al.}(2004)\citenamefont {Murphy},
  \citenamefont {Zhang}, \citenamefont {Troxler}, \citenamefont {Ferry},
  \citenamefont {Martin},\ and\ \citenamefont {Jones}}]{Murphy2004}%
  \BibitemOpen
  \bibfield  {author} {\bibinfo {author} {\bibfnamefont {C.~B.}\ \bibnamefont
  {Murphy}}, \bibinfo {author} {\bibfnamefont {Y.}~\bibnamefont {Zhang}},
  \bibinfo {author} {\bibfnamefont {T.}~\bibnamefont {Troxler}}, \bibinfo
  {author} {\bibfnamefont {V.}~\bibnamefont {Ferry}}, \bibinfo {author}
  {\bibfnamefont {J.~J.}\ \bibnamefont {Martin}}, \ and\ \bibinfo {author}
  {\bibfnamefont {W.~E.}\ \bibnamefont {Jones}},\ }\href@noop {} {\bibfield
  {journal} {\bibinfo  {journal} {Journal Of Physical Chemistry B}\ }\textbf
  {\bibinfo {volume} {108}},\ \bibinfo {pages} {1537} (\bibinfo {year}
  {2004})}\BibitemShut {NoStop}%
\bibitem [{\citenamefont {Fick}(1995)}]{Fick1995}%
  \BibitemOpen
  \bibfield  {author} {\bibinfo {author} {\bibfnamefont {A.}~\bibnamefont
  {Fick}},\ }\href@noop {} {\bibfield  {journal} {\bibinfo  {journal} {Journal
  Of Membrane Science}\ }\textbf {\bibinfo {volume} {100}},\ \bibinfo {pages}
  {33} (\bibinfo {year} {1995})}\BibitemShut {NoStop}%
\bibitem [{\citenamefont {W\"unsche}\ \emph {et~al.}(2010)\citenamefont
  {W\"unsche}, \citenamefont {Reineke}, \citenamefont {L\"ussem},\ and\
  \citenamefont {Leo}}]{Wuensche2010}%
  \BibitemOpen
  \bibfield  {author} {\bibinfo {author} {\bibfnamefont {J.}~\bibnamefont
  {W\"unsche}}, \bibinfo {author} {\bibfnamefont {S.}~\bibnamefont {Reineke}},
  \bibinfo {author} {\bibfnamefont {B.}~\bibnamefont {L\"ussem}}, \ and\
  \bibinfo {author} {\bibfnamefont {K.}~\bibnamefont {Leo}},\ }\href@noop {}
  {\bibfield  {journal} {\bibinfo  {journal} {Physical Review B}\ }\textbf
  {\bibinfo {volume} {81}},\ \bibinfo {pages} {245201} (\bibinfo {year}
  {2010})}\BibitemShut {NoStop}%
\bibitem [{\citenamefont {Rosenow}\ \emph {et~al.}(2010)\citenamefont
  {Rosenow}, \citenamefont {Furno}, \citenamefont {Reineke}, \citenamefont
  {Olthof}, \citenamefont {Lüssem},\ and\ \citenamefont {Leo}}]{Rosenow2010}%
  \BibitemOpen
  \bibfield  {author} {\bibinfo {author} {\bibfnamefont {T.~C.}\ \bibnamefont
  {Rosenow}}, \bibinfo {author} {\bibfnamefont {M.}~\bibnamefont {Furno}},
  \bibinfo {author} {\bibfnamefont {S.}~\bibnamefont {Reineke}}, \bibinfo
  {author} {\bibfnamefont {S.}~\bibnamefont {Olthof}}, \bibinfo {author}
  {\bibfnamefont {B.}~\bibnamefont {Lüssem}}, \ and\ \bibinfo {author}
  {\bibfnamefont {K.}~\bibnamefont {Leo}},\ }\href@noop {} {\bibfield
  {journal} {\bibinfo  {journal} {Journal of Applied Physics}\ }\textbf
  {\bibinfo {volume} {108}},\ \bibinfo {pages} {113113} (\bibinfo {year}
  {2010})}\BibitemShut {NoStop}%
\bibitem [{\citenamefont {Zhou}\ \emph {et~al.}(2007)\citenamefont {Zhou},
  \citenamefont {Ma}, \citenamefont {Zhou}, \citenamefont {Ding},\ and\
  \citenamefont {Hou}}]{Zhou2007}%
  \BibitemOpen
  \bibfield  {author} {\bibinfo {author} {\bibfnamefont {Y.~C.}\ \bibnamefont
  {Zhou}}, \bibinfo {author} {\bibfnamefont {L.~L.}\ \bibnamefont {Ma}},
  \bibinfo {author} {\bibfnamefont {J.}~\bibnamefont {Zhou}}, \bibinfo {author}
  {\bibfnamefont {X.~M.}\ \bibnamefont {Ding}}, \ and\ \bibinfo {author}
  {\bibfnamefont {X.~Y.}\ \bibnamefont {Hou}},\ }\href@noop {} {\bibfield
  {journal} {\bibinfo  {journal} {Physical Review B}\ }\textbf {\bibinfo
  {volume} {75}},\ \bibinfo {pages} {132202} (\bibinfo {year}
  {2007})}\BibitemShut {NoStop}%
\bibitem [{\citenamefont {Giebink}\ \emph {et~al.}(2006)\citenamefont
  {Giebink}, \citenamefont {Sun},\ and\ \citenamefont {Forrest}}]{Giebink2006}%
  \BibitemOpen
  \bibfield  {author} {\bibinfo {author} {\bibfnamefont {N.~C.}\ \bibnamefont
  {Giebink}}, \bibinfo {author} {\bibfnamefont {Y.}~\bibnamefont {Sun}}, \ and\
  \bibinfo {author} {\bibfnamefont {S.~R.}\ \bibnamefont {Forrest}},\
  }\href@noop {} {\bibfield  {journal} {\bibinfo  {journal} {Organic
  Electronics}\ }\textbf {\bibinfo {volume} {7}},\ \bibinfo {pages} {375}
  (\bibinfo {year} {2006})}\BibitemShut {NoStop}%
\bibitem [{\citenamefont {Baldo}\ and\ \citenamefont
  {Forrest}(2000)}]{Baldo2000b}%
  \BibitemOpen
  \bibfield  {author} {\bibinfo {author} {\bibfnamefont {M.~A.}\ \bibnamefont
  {Baldo}}\ and\ \bibinfo {author} {\bibfnamefont {S.~R.}\ \bibnamefont
  {Forrest}},\ }\href@noop {} {\bibfield  {journal} {\bibinfo  {journal}
  {Physical Review B}\ }\textbf {\bibinfo {volume} {62}},\ \bibinfo {pages}
  {10958} (\bibinfo {year} {2000})}\BibitemShut {NoStop}%
\bibitem [{Note4()}]{Note4}%
  \BibitemOpen
  \bibinfo {note} {Because exciton motion is typically isotropic and the
  systems are planar structures, diffusion can be reduced to one dimension in
  space, e.g. $L_x$.}\BibitemShut {Stop}%
\bibitem [{\citenamefont {Meerheim}\ \emph
  {et~al.}(2008{\natexlab{a}})\citenamefont {Meerheim}, \citenamefont {Scholz},
  \citenamefont {Olthof}, \citenamefont {Schwartz}, \citenamefont {Reineke},
  \citenamefont {Walzer},\ and\ \citenamefont {Leo}}]{Meerheim2008}%
  \BibitemOpen
  \bibfield  {author} {\bibinfo {author} {\bibfnamefont {R.}~\bibnamefont
  {Meerheim}}, \bibinfo {author} {\bibfnamefont {S.}~\bibnamefont {Scholz}},
  \bibinfo {author} {\bibfnamefont {S.}~\bibnamefont {Olthof}}, \bibinfo
  {author} {\bibfnamefont {G.}~\bibnamefont {Schwartz}}, \bibinfo {author}
  {\bibfnamefont {S.}~\bibnamefont {Reineke}}, \bibinfo {author} {\bibfnamefont
  {K.}~\bibnamefont {Walzer}}, \ and\ \bibinfo {author} {\bibfnamefont
  {K.}~\bibnamefont {Leo}},\ }\href@noop {} {\bibfield  {journal} {\bibinfo
  {journal} {Journal Of Applied Physics}\ }\textbf {\bibinfo {volume} {104}},\
  \bibinfo {pages} {014510} (\bibinfo {year} {2008}{\natexlab{a}})}\BibitemShut
  {NoStop}%
\bibitem [{\citenamefont {Greiner}(2007)}]{Greiner2007}%
  \BibitemOpen
  \bibfield  {author} {\bibinfo {author} {\bibfnamefont {H.}~\bibnamefont
  {Greiner}},\ }\href@noop {} {\bibfield  {journal} {\bibinfo  {journal}
  {Japanese Journal Of Applied Physics Part 1-Regular Papers Brief
  Communications \& Review Papers}\ }\textbf {\bibinfo {volume} {46}},\
  \bibinfo {pages} {4125} (\bibinfo {year} {2007})}\BibitemShut {NoStop}%
\bibitem [{Note5()}]{Note5}%
  \BibitemOpen
  \bibinfo {note} {Because (i) the organic materials all show different,
  distinct wavelength dependencies and (ii) slight changes in $n$ are
  observable for every two organic materials compared.}\BibitemShut {Stop}%
\bibitem [{\citenamefont {Krummacher}\ \emph {et~al.}(2009)\citenamefont
  {Krummacher}, \citenamefont {Nowy}, \citenamefont {Frischeisen},
  \citenamefont {Klein},\ and\ \citenamefont {Brütting}}]{Krummacher2009}%
  \BibitemOpen
  \bibfield  {author} {\bibinfo {author} {\bibfnamefont {B.~C.}\ \bibnamefont
  {Krummacher}}, \bibinfo {author} {\bibfnamefont {S.}~\bibnamefont {Nowy}},
  \bibinfo {author} {\bibfnamefont {J.}~\bibnamefont {Frischeisen}}, \bibinfo
  {author} {\bibfnamefont {M.}~\bibnamefont {Klein}}, \ and\ \bibinfo {author}
  {\bibfnamefont {W.}~\bibnamefont {Brütting}},\ }\href@noop {} {\bibfield
  {journal} {\bibinfo  {journal} {Organic Electronics}\ }\textbf {\bibinfo
  {volume} {10}},\ \bibinfo {pages} {478} (\bibinfo {year} {2009})}\BibitemShut
  {NoStop}%
\bibitem [{\citenamefont {Meerheim}\ \emph {et~al.}(2010)\citenamefont
  {Meerheim}, \citenamefont {Furno}, \citenamefont {Hofmann}, \citenamefont
  {Lüssem},\ and\ \citenamefont {Leo}}]{Meerheim2010}%
  \BibitemOpen
  \bibfield  {author} {\bibinfo {author} {\bibfnamefont {R.}~\bibnamefont
  {Meerheim}}, \bibinfo {author} {\bibfnamefont {M.}~\bibnamefont {Furno}},
  \bibinfo {author} {\bibfnamefont {S.}~\bibnamefont {Hofmann}}, \bibinfo
  {author} {\bibfnamefont {B.}~\bibnamefont {Lüssem}}, \ and\ \bibinfo
  {author} {\bibfnamefont {K.}~\bibnamefont {Leo}},\ }\href@noop {} {\bibfield
  {journal} {\bibinfo  {journal} {Applied Physics Letters}\ }\textbf {\bibinfo
  {volume} {97}},\ \bibinfo {pages} {253305} (\bibinfo {year}
  {2010})}\BibitemShut {NoStop}%
\bibitem [{\citenamefont {Furno}\ \emph {et~al.}(2010)\citenamefont {Furno},
  \citenamefont {Meerheim}, \citenamefont {Thomschke}, \citenamefont {Hofmann},
  \citenamefont {Lüssem},\ and\ \citenamefont {Leo}}]{Furno2010}%
  \BibitemOpen
  \bibfield  {author} {\bibinfo {author} {\bibfnamefont {M.}~\bibnamefont
  {Furno}}, \bibinfo {author} {\bibfnamefont {R.}~\bibnamefont {Meerheim}},
  \bibinfo {author} {\bibfnamefont {M.}~\bibnamefont {Thomschke}}, \bibinfo
  {author} {\bibfnamefont {S.}~\bibnamefont {Hofmann}}, \bibinfo {author}
  {\bibfnamefont {B.}~\bibnamefont {Lüssem}}, \ and\ \bibinfo {author}
  {\bibfnamefont {K.}~\bibnamefont {Leo}},\ }\href@noop {} {\bibfield
  {journal} {\bibinfo  {journal} {Proc. SPIE}\ }\textbf {\bibinfo {volume}
  {7617}},\ \bibinfo {pages} {761716} (\bibinfo {year} {2010})}\BibitemShut
  {NoStop}%
\bibitem [{\citenamefont {Gärtner}\ and\ \citenamefont
  {Greiner}(2008)}]{Gaertner2008}%
  \BibitemOpen
  \bibfield  {author} {\bibinfo {author} {\bibfnamefont {G.}~\bibnamefont
  {Gärtner}}\ and\ \bibinfo {author} {\bibfnamefont {H.}~\bibnamefont
  {Greiner}},\ }\href@noop {} {\bibfield  {journal} {\bibinfo  {journal}
  {Proceedings of SPIE}\ }\textbf {\bibinfo {volume} {6999}},\ \bibinfo {pages}
  {69992T} (\bibinfo {year} {2008})}\BibitemShut {NoStop}%
\bibitem [{\citenamefont {Adachi}\ \emph
  {et~al.}(2001{\natexlab{b}})\citenamefont {Adachi}, \citenamefont {Baldo},
  \citenamefont {Thompson},\ and\ \citenamefont {Forrest}}]{Adachi2001b}%
  \BibitemOpen
  \bibfield  {author} {\bibinfo {author} {\bibfnamefont {C.}~\bibnamefont
  {Adachi}}, \bibinfo {author} {\bibfnamefont {M.~A.}\ \bibnamefont {Baldo}},
  \bibinfo {author} {\bibfnamefont {M.~E.}\ \bibnamefont {Thompson}}, \ and\
  \bibinfo {author} {\bibfnamefont {S.~R.}\ \bibnamefont {Forrest}},\
  }\href@noop {} {\bibfield  {journal} {\bibinfo  {journal} {Journal Of Applied
  Physics}\ }\textbf {\bibinfo {volume} {90}},\ \bibinfo {pages} {5048}
  (\bibinfo {year} {2001}{\natexlab{b}})}\BibitemShut {NoStop}%
\bibitem [{\citenamefont {Möller}\ and\ \citenamefont
  {Forrest}(2002)}]{Moller2002}%
  \BibitemOpen
  \bibfield  {author} {\bibinfo {author} {\bibfnamefont {S.}~\bibnamefont
  {Möller}}\ and\ \bibinfo {author} {\bibfnamefont {S.~R.}\ \bibnamefont
  {Forrest}},\ }\href@noop {} {\bibfield  {journal} {\bibinfo  {journal}
  {Journal Of Applied Physics}\ }\textbf {\bibinfo {volume} {91}},\ \bibinfo
  {pages} {3324} (\bibinfo {year} {2002})}\BibitemShut {NoStop}%
\bibitem [{\citenamefont {Sun}\ and\ \citenamefont
  {Forrest}(2008{\natexlab{a}})}]{Sun2008}%
  \BibitemOpen
  \bibfield  {author} {\bibinfo {author} {\bibfnamefont {Y.}~\bibnamefont
  {Sun}}\ and\ \bibinfo {author} {\bibfnamefont {S.~R.}\ \bibnamefont
  {Forrest}},\ }\href@noop {} {\bibfield  {journal} {\bibinfo  {journal}
  {Nature Photonics}\ }\textbf {\bibinfo {volume} {2}},\ \bibinfo {pages} {483}
  (\bibinfo {year} {2008}{\natexlab{a}})}\BibitemShut {NoStop}%
\bibitem [{\citenamefont {Sokolik}\ \emph {et~al.}(1996)\citenamefont
  {Sokolik}, \citenamefont {Priestley}, \citenamefont {Walser}, \citenamefont
  {Dorsinville},\ and\ \citenamefont {Tang}}]{Sokolik1996}%
  \BibitemOpen
  \bibfield  {author} {\bibinfo {author} {\bibfnamefont {I.}~\bibnamefont
  {Sokolik}}, \bibinfo {author} {\bibfnamefont {R.}~\bibnamefont {Priestley}},
  \bibinfo {author} {\bibfnamefont {A.~D.}\ \bibnamefont {Walser}}, \bibinfo
  {author} {\bibfnamefont {R.}~\bibnamefont {Dorsinville}}, \ and\ \bibinfo
  {author} {\bibfnamefont {C.~W.}\ \bibnamefont {Tang}},\ }\href@noop {}
  {\bibfield  {journal} {\bibinfo  {journal} {Applied Physics Letters}\
  }\textbf {\bibinfo {volume} {69}},\ \bibinfo {pages} {4168} (\bibinfo {year}
  {1996})}\BibitemShut {NoStop}%
\bibitem [{\citenamefont {Kalinowski}\ \emph {et~al.}(2002)\citenamefont
  {Kalinowski}, \citenamefont {Stampor}, \citenamefont {Mezyk}, \citenamefont
  {Cocchi}, \citenamefont {Virgili}, \citenamefont {Fattori},\ and\
  \citenamefont {Di~Marco}}]{Kalinowski2002}%
  \BibitemOpen
  \bibfield  {author} {\bibinfo {author} {\bibfnamefont {J.}~\bibnamefont
  {Kalinowski}}, \bibinfo {author} {\bibfnamefont {W.}~\bibnamefont {Stampor}},
  \bibinfo {author} {\bibfnamefont {J.}~\bibnamefont {Mezyk}}, \bibinfo
  {author} {\bibfnamefont {M.}~\bibnamefont {Cocchi}}, \bibinfo {author}
  {\bibfnamefont {D.}~\bibnamefont {Virgili}}, \bibinfo {author} {\bibfnamefont
  {V.}~\bibnamefont {Fattori}}, \ and\ \bibinfo {author} {\bibfnamefont
  {P.}~\bibnamefont {Di~Marco}},\ }\href@noop {} {\bibfield  {journal}
  {\bibinfo  {journal} {Physical Review B}\ }\textbf {\bibinfo {volume} {66}},\
  \bibinfo {pages} {235321} (\bibinfo {year} {2002})}\BibitemShut {NoStop}%
\bibitem [{\citenamefont {Reineke}\ \emph
  {et~al.}(2007{\natexlab{b}})\citenamefont {Reineke}, \citenamefont
  {Schwartz}, \citenamefont {Walzer},\ and\ \citenamefont
  {Leo}}]{Reineke2007a}%
  \BibitemOpen
  \bibfield  {author} {\bibinfo {author} {\bibfnamefont {S.}~\bibnamefont
  {Reineke}}, \bibinfo {author} {\bibfnamefont {G.}~\bibnamefont {Schwartz}},
  \bibinfo {author} {\bibfnamefont {K.}~\bibnamefont {Walzer}}, \ and\ \bibinfo
  {author} {\bibfnamefont {K.}~\bibnamefont {Leo}},\ }\href@noop {} {\bibfield
  {journal} {\bibinfo  {journal} {Applied Physics Letters}\ }\textbf {\bibinfo
  {volume} {91}},\ \bibinfo {pages} {123508} (\bibinfo {year}
  {2007}{\natexlab{b}})}\BibitemShut {NoStop}%
\bibitem [{\citenamefont {Su}\ \emph {et~al.}(2008)\citenamefont {Su},
  \citenamefont {Gonmori}, \citenamefont {Sasabe},\ and\ \citenamefont
  {Kido}}]{Su2008}%
  \BibitemOpen
  \bibfield  {author} {\bibinfo {author} {\bibfnamefont {S.~J.}\ \bibnamefont
  {Su}}, \bibinfo {author} {\bibfnamefont {E.}~\bibnamefont {Gonmori}},
  \bibinfo {author} {\bibfnamefont {H.}~\bibnamefont {Sasabe}}, \ and\ \bibinfo
  {author} {\bibfnamefont {J.}~\bibnamefont {Kido}},\ }\href@noop {} {\bibfield
   {journal} {\bibinfo  {journal} {Advanced Materials}\ }\textbf {\bibinfo
  {volume} {20}},\ \bibinfo {pages} {4189} (\bibinfo {year}
  {2008})}\BibitemShut {NoStop}%
\bibitem [{Note6()}]{Note6}%
  \BibitemOpen
  \bibinfo {note} {It is worth mentioning that the luminous efficacy is often
  referred to as \protect \emph {power efficiency} in literature. However,
  strictly speaking, an efficiency should be dimensionless, which is not the
  case for the quantity discussed (cf. [lm\protect \tmspace +\thinmuskip
  {.1667em}W$^{-1}$]).}\BibitemShut {Stop}%
\bibitem [{\citenamefont {Forrest}\ \emph {et~al.}(2003)\citenamefont
  {Forrest}, \citenamefont {Bradley},\ and\ \citenamefont
  {Thompson}}]{Forrest2003}%
  \BibitemOpen
  \bibfield  {author} {\bibinfo {author} {\bibfnamefont {S.~R.}\ \bibnamefont
  {Forrest}}, \bibinfo {author} {\bibfnamefont {D.~D.~C.}\ \bibnamefont
  {Bradley}}, \ and\ \bibinfo {author} {\bibfnamefont {M.~E.}\ \bibnamefont
  {Thompson}},\ }\href@noop {} {\bibfield  {journal} {\bibinfo  {journal}
  {Advanced Materials}\ }\textbf {\bibinfo {volume} {15}},\ \bibinfo {pages}
  {1043} (\bibinfo {year} {2003})}\BibitemShut {NoStop}%
\bibitem [{\citenamefont {Meerheim}\ \emph
  {et~al.}(2008{\natexlab{b}})\citenamefont {Meerheim}, \citenamefont
  {Nitsche},\ and\ \citenamefont {Leo}}]{Meerheim2008a}%
  \BibitemOpen
  \bibfield  {author} {\bibinfo {author} {\bibfnamefont {R.}~\bibnamefont
  {Meerheim}}, \bibinfo {author} {\bibfnamefont {R.}~\bibnamefont {Nitsche}}, \
  and\ \bibinfo {author} {\bibfnamefont {K.}~\bibnamefont {Leo}},\ }\href@noop
  {} {\bibfield  {journal} {\bibinfo  {journal} {Applied Physics Letters}\
  }\textbf {\bibinfo {volume} {93}},\ \bibinfo {pages} {043310} (\bibinfo
  {year} {2008}{\natexlab{b}})}\BibitemShut {NoStop}%
\bibitem [{\citenamefont {Hofmann}\ \emph {et~al.}(2010)\citenamefont
  {Hofmann}, \citenamefont {Thomschke}, \citenamefont {Freitag}, \citenamefont
  {Furno}, \citenamefont {Lüssem},\ and\ \citenamefont {Leo}}]{Hofmann2010}%
  \BibitemOpen
  \bibfield  {author} {\bibinfo {author} {\bibfnamefont {S.}~\bibnamefont
  {Hofmann}}, \bibinfo {author} {\bibfnamefont {M.}~\bibnamefont {Thomschke}},
  \bibinfo {author} {\bibfnamefont {P.}~\bibnamefont {Freitag}}, \bibinfo
  {author} {\bibfnamefont {M.}~\bibnamefont {Furno}}, \bibinfo {author}
  {\bibfnamefont {B.}~\bibnamefont {Lüssem}}, \ and\ \bibinfo {author}
  {\bibfnamefont {K.}~\bibnamefont {Leo}},\ }\href@noop {} {\bibfield
  {journal} {\bibinfo  {journal} {Applied Physics Letters}\ }\textbf {\bibinfo
  {volume} {97}},\ \bibinfo {pages} {253308} (\bibinfo {year}
  {2010})}\BibitemShut {NoStop}%
\bibitem [{\citenamefont {Freitag}\ \emph {et~al.}(2010)\citenamefont
  {Freitag}, \citenamefont {Reineke}, \citenamefont {Olthof}, \citenamefont
  {Furno}, \citenamefont {Lüssem},\ and\ \citenamefont {Leo}}]{Freitag2010}%
  \BibitemOpen
  \bibfield  {author} {\bibinfo {author} {\bibfnamefont {P.}~\bibnamefont
  {Freitag}}, \bibinfo {author} {\bibfnamefont {S.}~\bibnamefont {Reineke}},
  \bibinfo {author} {\bibfnamefont {S.}~\bibnamefont {Olthof}}, \bibinfo
  {author} {\bibfnamefont {M.}~\bibnamefont {Furno}}, \bibinfo {author}
  {\bibfnamefont {B.}~\bibnamefont {Lüssem}}, \ and\ \bibinfo {author}
  {\bibfnamefont {K.}~\bibnamefont {Leo}},\ }\href@noop {} {\bibfield
  {journal} {\bibinfo  {journal} {Organic Electronics}\ }\textbf {\bibinfo
  {volume} {11}},\ \bibinfo {pages} {1676} (\bibinfo {year}
  {2010})}\BibitemShut {NoStop}%
\bibitem [{\citenamefont {Hunt}(1995)}]{Hunt1995}%
  \BibitemOpen
  \bibfield  {author} {\bibinfo {author} {\bibfnamefont {R.}~\bibnamefont
  {Hunt}},\ }\href@noop {} {\emph {\bibinfo {title} {Measuring Color}}}\
  (\bibinfo  {publisher} {Ellis Horwood, London, $2^{\text{nd}}$ edition},\
  \bibinfo {year} {1995})\BibitemShut {NoStop}%
\bibitem [{\citenamefont {Azuma}\ \emph {et~al.}(1995)\citenamefont {Azuma},
  \citenamefont {Einhorn}, \citenamefont {Halstead}, \citenamefont {Jerome},
  \citenamefont {de~Kerf}, \citenamefont {Krtil}, \citenamefont {Münch},
  \citenamefont {Barth}, \citenamefont {Ouweitjes}, \citenamefont {Richter},\
  and\ \citenamefont {Siljeholm}}]{CIE1995}%
  \BibitemOpen
  \bibfield  {author} {\bibinfo {author} {\bibfnamefont {T.}~\bibnamefont
  {Azuma}}, \bibinfo {author} {\bibfnamefont {H.}~\bibnamefont {Einhorn}},
  \bibinfo {author} {\bibfnamefont {M.}~\bibnamefont {Halstead}}, \bibinfo
  {author} {\bibfnamefont {C.}~\bibnamefont {Jerome}}, \bibinfo {author}
  {\bibfnamefont {J.}~\bibnamefont {de~Kerf}}, \bibinfo {author} {\bibfnamefont
  {J.}~\bibnamefont {Krtil}}, \bibinfo {author} {\bibfnamefont
  {W.}~\bibnamefont {Münch}}, \bibinfo {author} {\bibfnamefont
  {E.}~\bibnamefont {Barth}}, \bibinfo {author} {\bibfnamefont
  {J.}~\bibnamefont {Ouweitjes}}, \bibinfo {author} {\bibfnamefont
  {M.}~\bibnamefont {Richter}}, \ and\ \bibinfo {author} {\bibfnamefont
  {G.}~\bibnamefont {Siljeholm}},\ }\href@noop {} {\bibfield  {journal}
  {\bibinfo  {journal} {CIE Publications}\ }\textbf {\bibinfo {volume}
  {13.3}},\ \bibinfo {pages} {1} (\bibinfo {year} {1995})}\BibitemShut
  {NoStop}%
\bibitem [{\citenamefont {Meerheim}\ \emph {et~al.}(2006)\citenamefont
  {Meerheim}, \citenamefont {Walzer}, \citenamefont {Pfeiffer},\ and\
  \citenamefont {Leo}}]{Meerheim2006}%
  \BibitemOpen
  \bibfield  {author} {\bibinfo {author} {\bibfnamefont {R.}~\bibnamefont
  {Meerheim}}, \bibinfo {author} {\bibfnamefont {K.}~\bibnamefont {Walzer}},
  \bibinfo {author} {\bibfnamefont {M.}~\bibnamefont {Pfeiffer}}, \ and\
  \bibinfo {author} {\bibfnamefont {K.}~\bibnamefont {Leo}},\ }\href@noop {}
  {\bibfield  {journal} {\bibinfo  {journal} {Applied Physics Letters}\
  }\textbf {\bibinfo {volume} {89}},\ \bibinfo {pages} {061111} (\bibinfo
  {year} {2006})}\BibitemShut {NoStop}%
\bibitem [{\citenamefont {Zhang}\ \emph {et~al.}(2001)\citenamefont {Zhang},
  \citenamefont {Jiang}, \citenamefont {Zhu}, \citenamefont {Zhang},\ and\
  \citenamefont {Xu}}]{Zhang2001}%
  \BibitemOpen
  \bibfield  {author} {\bibinfo {author} {\bibfnamefont {Z.~L.}\ \bibnamefont
  {Zhang}}, \bibinfo {author} {\bibfnamefont {X.~Y.}\ \bibnamefont {Jiang}},
  \bibinfo {author} {\bibfnamefont {W.~Q.}\ \bibnamefont {Zhu}}, \bibinfo
  {author} {\bibfnamefont {B.~X.}\ \bibnamefont {Zhang}}, \ and\ \bibinfo
  {author} {\bibfnamefont {S.~H.}\ \bibnamefont {Xu}},\ }\href@noop {}
  {\bibfield  {journal} {\bibinfo  {journal} {Journal of Physics D-applied
  Physics}\ }\textbf {\bibinfo {volume} {34}},\ \bibinfo {pages} {3083}
  (\bibinfo {year} {2001})}\BibitemShut {NoStop}%
\bibitem [{\citenamefont {Tsai}\ and\ \citenamefont {Jou}(2006)}]{Tsai2006}%
  \BibitemOpen
  \bibfield  {author} {\bibinfo {author} {\bibfnamefont {Y.~C.}\ \bibnamefont
  {Tsai}}\ and\ \bibinfo {author} {\bibfnamefont {J.~H.}\ \bibnamefont {Jou}},\
  }\href@noop {} {\bibfield  {journal} {\bibinfo  {journal} {Applied Physics
  Letters}\ }\textbf {\bibinfo {volume} {89}},\ \bibinfo {pages} {243521}
  (\bibinfo {year} {2006})}\BibitemShut {NoStop}%
\bibitem [{Note7()}]{Note7}%
  \BibitemOpen
  \bibinfo {note} {As discussed in Section \ref {Rolloff}, the device
  efficiency will drastically decrease from low luminance (where typically the
  maximum luminous efficacy is obtained) to illumination relevant
  levels.}\BibitemShut {Stop}%
\bibitem [{\citenamefont {Mattoussi}\ \emph {et~al.}(1999)\citenamefont
  {Mattoussi}, \citenamefont {Murata}, \citenamefont {Merritt}, \citenamefont
  {Iizumi}, \citenamefont {Kido},\ and\ \citenamefont
  {Kafafi}}]{Mattoussi1999}%
  \BibitemOpen
  \bibfield  {author} {\bibinfo {author} {\bibfnamefont {H.}~\bibnamefont
  {Mattoussi}}, \bibinfo {author} {\bibfnamefont {H.}~\bibnamefont {Murata}},
  \bibinfo {author} {\bibfnamefont {C.~D.}\ \bibnamefont {Merritt}}, \bibinfo
  {author} {\bibfnamefont {Y.}~\bibnamefont {Iizumi}}, \bibinfo {author}
  {\bibfnamefont {J.}~\bibnamefont {Kido}}, \ and\ \bibinfo {author}
  {\bibfnamefont {Z.~H.}\ \bibnamefont {Kafafi}},\ }\href@noop {} {\bibfield
  {journal} {\bibinfo  {journal} {Journal Of Applied Physics}\ }\textbf
  {\bibinfo {volume} {86}},\ \bibinfo {pages} {2642} (\bibinfo {year}
  {1999})}\BibitemShut {NoStop}%
\bibitem [{Note8()}]{Note8}%
  \BibitemOpen
  \bibinfo {note} {Note that other publications [e.g. \cite {Wu2008}] state a
  much higher triplet energy of PVK of 3.0\protect \tmspace +\thinmuskip
  {.1667em}eV.}\BibitemShut {Stop}%
\bibitem [{\citenamefont {Adachi}\ \emph
  {et~al.}(2001{\natexlab{c}})\citenamefont {Adachi}, \citenamefont {Kwong},
  \citenamefont {Djurovich}, \citenamefont {Adamovich}, \citenamefont {Baldo},
  \citenamefont {Thompson},\ and\ \citenamefont {Forrest}}]{Adachi2001a}%
  \BibitemOpen
  \bibfield  {author} {\bibinfo {author} {\bibfnamefont {C.}~\bibnamefont
  {Adachi}}, \bibinfo {author} {\bibfnamefont {R.~C.}\ \bibnamefont {Kwong}},
  \bibinfo {author} {\bibfnamefont {P.}~\bibnamefont {Djurovich}}, \bibinfo
  {author} {\bibfnamefont {V.}~\bibnamefont {Adamovich}}, \bibinfo {author}
  {\bibfnamefont {M.~A.}\ \bibnamefont {Baldo}}, \bibinfo {author}
  {\bibfnamefont {M.~E.}\ \bibnamefont {Thompson}}, \ and\ \bibinfo {author}
  {\bibfnamefont {S.~R.}\ \bibnamefont {Forrest}},\ }\href@noop {} {\bibfield
  {journal} {\bibinfo  {journal} {Applied Physics Letters}\ }\textbf {\bibinfo
  {volume} {79}},\ \bibinfo {pages} {2082} (\bibinfo {year}
  {2001}{\natexlab{c}})}\BibitemShut {NoStop}%
\bibitem [{\citenamefont {Hamada}\ \emph {et~al.}(1992)\citenamefont {Hamada},
  \citenamefont {Adachi}, \citenamefont {Tsutsui},\ and\ \citenamefont
  {Saito}}]{Hamada1992}%
  \BibitemOpen
  \bibfield  {author} {\bibinfo {author} {\bibfnamefont {Y.}~\bibnamefont
  {Hamada}}, \bibinfo {author} {\bibfnamefont {C.}~\bibnamefont {Adachi}},
  \bibinfo {author} {\bibfnamefont {T.}~\bibnamefont {Tsutsui}}, \ and\
  \bibinfo {author} {\bibfnamefont {S.}~\bibnamefont {Saito}},\ }\href@noop {}
  {\bibfield  {journal} {\bibinfo  {journal} {Japanese Journal Of Applied
  Physics Part 1-Regular Papers Short Notes \& Review Papers}\ }\textbf
  {\bibinfo {volume} {31}},\ \bibinfo {pages} {1812} (\bibinfo {year}
  {1992})}\BibitemShut {NoStop}%
\bibitem [{\citenamefont {Cheng}\ \emph {et~al.}(2010)\citenamefont {Cheng},
  \citenamefont {Fei}, \citenamefont {Duan}, \citenamefont {Zhao},
  \citenamefont {Ma},\ and\ \citenamefont {Liu}}]{Cheng2010a}%
  \BibitemOpen
  \bibfield  {author} {\bibinfo {author} {\bibfnamefont {G.}~\bibnamefont
  {Cheng}}, \bibinfo {author} {\bibfnamefont {T.}~\bibnamefont {Fei}}, \bibinfo
  {author} {\bibfnamefont {Y.}~\bibnamefont {Duan}}, \bibinfo {author}
  {\bibfnamefont {Y.}~\bibnamefont {Zhao}}, \bibinfo {author} {\bibfnamefont
  {Y.~G.}\ \bibnamefont {Ma}}, \ and\ \bibinfo {author} {\bibfnamefont {S.~Y.}\
  \bibnamefont {Liu}},\ }\href@noop {} {\bibfield  {journal} {\bibinfo
  {journal} {Optics Letters}\ }\textbf {\bibinfo {volume} {35}},\ \bibinfo
  {pages} {2436} (\bibinfo {year} {2010})}\BibitemShut {NoStop}%
\bibitem [{\citenamefont {Holmes}\ \emph {et~al.}(2003)\citenamefont {Holmes},
  \citenamefont {D'Andrade}, \citenamefont {Forrest}, \citenamefont {Ren},
  \citenamefont {Li},\ and\ \citenamefont {Thompson}}]{Holmes2003}%
  \BibitemOpen
  \bibfield  {author} {\bibinfo {author} {\bibfnamefont {R.~J.}\ \bibnamefont
  {Holmes}}, \bibinfo {author} {\bibfnamefont {B.~W.}\ \bibnamefont
  {D'Andrade}}, \bibinfo {author} {\bibfnamefont {S.~R.}\ \bibnamefont
  {Forrest}}, \bibinfo {author} {\bibfnamefont {X.}~\bibnamefont {Ren}},
  \bibinfo {author} {\bibfnamefont {J.}~\bibnamefont {Li}}, \ and\ \bibinfo
  {author} {\bibfnamefont {M.~E.}\ \bibnamefont {Thompson}},\ }\href@noop {}
  {\bibfield  {journal} {\bibinfo  {journal} {Applied Physics Letters}\
  }\textbf {\bibinfo {volume} {83}},\ \bibinfo {pages} {3818} (\bibinfo {year}
  {2003})}\BibitemShut {NoStop}%
\bibitem [{\citenamefont {Schwartz}\ \emph {et~al.}(2007)\citenamefont
  {Schwartz}, \citenamefont {Pfeiffer}, \citenamefont {Reineke}, \citenamefont
  {Walzer},\ and\ \citenamefont {Leo}}]{Schwartz2007}%
  \BibitemOpen
  \bibfield  {author} {\bibinfo {author} {\bibfnamefont {G.}~\bibnamefont
  {Schwartz}}, \bibinfo {author} {\bibfnamefont {M.}~\bibnamefont {Pfeiffer}},
  \bibinfo {author} {\bibfnamefont {S.}~\bibnamefont {Reineke}}, \bibinfo
  {author} {\bibfnamefont {K.}~\bibnamefont {Walzer}}, \ and\ \bibinfo {author}
  {\bibfnamefont {K.}~\bibnamefont {Leo}},\ }\href@noop {} {\bibfield
  {journal} {\bibinfo  {journal} {Advanced Materials}\ }\textbf {\bibinfo
  {volume} {19}},\ \bibinfo {pages} {3672} (\bibinfo {year}
  {2007})}\BibitemShut {NoStop}%
\bibitem [{\citenamefont {Gather}\ \emph {et~al.}(2011)\citenamefont {Gather},
  \citenamefont {Köhnen},\ and\ \citenamefont {Meerholz}}]{Gather2011}%
  \BibitemOpen
  \bibfield  {author} {\bibinfo {author} {\bibfnamefont {M.~C.}\ \bibnamefont
  {Gather}}, \bibinfo {author} {\bibfnamefont {A.}~\bibnamefont {Köhnen}}, \
  and\ \bibinfo {author} {\bibfnamefont {K.}~\bibnamefont {Meerholz}},\
  }\href@noop {} {\bibfield  {journal} {\bibinfo  {journal} {Advanced
  Materials}\ }\textbf {\bibinfo {volume} {23}},\ \bibinfo {pages} {233}
  (\bibinfo {year} {2011})}\BibitemShut {NoStop}%
\bibitem [{\citenamefont {Berggren}\ \emph {et~al.}(1994)\citenamefont
  {Berggren}, \citenamefont {Inganas}, \citenamefont {Gustafsson},
  \citenamefont {Rasmusson}, \citenamefont {Andersson}, \citenamefont
  {Hjertberg},\ and\ \citenamefont {Wennerstrom}}]{Berggren1994}%
  \BibitemOpen
  \bibfield  {author} {\bibinfo {author} {\bibfnamefont {M.}~\bibnamefont
  {Berggren}}, \bibinfo {author} {\bibfnamefont {O.}~\bibnamefont {Inganas}},
  \bibinfo {author} {\bibfnamefont {G.}~\bibnamefont {Gustafsson}}, \bibinfo
  {author} {\bibfnamefont {J.}~\bibnamefont {Rasmusson}}, \bibinfo {author}
  {\bibfnamefont {M.~R.}\ \bibnamefont {Andersson}}, \bibinfo {author}
  {\bibfnamefont {T.}~\bibnamefont {Hjertberg}}, \ and\ \bibinfo {author}
  {\bibfnamefont {O.}~\bibnamefont {Wennerstrom}},\ }\href@noop {} {\bibfield
  {journal} {\bibinfo  {journal} {Nature}\ }\textbf {\bibinfo {volume} {372}},\
  \bibinfo {pages} {444} (\bibinfo {year} {1994})}\BibitemShut {NoStop}%
\bibitem [{\citenamefont {Thompson}\ \emph {et~al.}(2001)\citenamefont
  {Thompson}, \citenamefont {Blyth}, \citenamefont {Mazzeo}, \citenamefont
  {Anni}, \citenamefont {Gigli},\ and\ \citenamefont
  {Cingolani}}]{Thompson2001}%
  \BibitemOpen
  \bibfield  {author} {\bibinfo {author} {\bibfnamefont {J.}~\bibnamefont
  {Thompson}}, \bibinfo {author} {\bibfnamefont {R.~I.~R.}\ \bibnamefont
  {Blyth}}, \bibinfo {author} {\bibfnamefont {M.}~\bibnamefont {Mazzeo}},
  \bibinfo {author} {\bibfnamefont {M.}~\bibnamefont {Anni}}, \bibinfo {author}
  {\bibfnamefont {G.}~\bibnamefont {Gigli}}, \ and\ \bibinfo {author}
  {\bibfnamefont {R.}~\bibnamefont {Cingolani}},\ }\href@noop {} {\bibfield
  {journal} {\bibinfo  {journal} {Applied Physics Letters}\ }\textbf {\bibinfo
  {volume} {79}},\ \bibinfo {pages} {560} (\bibinfo {year} {2001})}\BibitemShut
  {NoStop}%
\bibitem [{\citenamefont {Castellani}\ and\ \citenamefont
  {Berner}(2007)}]{Castellani2007}%
  \BibitemOpen
  \bibfield  {author} {\bibinfo {author} {\bibfnamefont {M.}~\bibnamefont
  {Castellani}}\ and\ \bibinfo {author} {\bibfnamefont {D.}~\bibnamefont
  {Berner}},\ }\href@noop {} {\bibfield  {journal} {\bibinfo  {journal}
  {Journal Of Applied Physics}\ }\textbf {\bibinfo {volume} {102}},\ \bibinfo
  {pages} {024509} (\bibinfo {year} {2007})}\BibitemShut {NoStop}%
\bibitem [{\citenamefont {Gather}\ \emph
  {et~al.}(2007{\natexlab{b}})\citenamefont {Gather}, \citenamefont {Alle},
  \citenamefont {Becker},\ and\ \citenamefont {Meerholz}}]{Gather2007}%
  \BibitemOpen
  \bibfield  {author} {\bibinfo {author} {\bibfnamefont {M.~C.}\ \bibnamefont
  {Gather}}, \bibinfo {author} {\bibfnamefont {R.}~\bibnamefont {Alle}},
  \bibinfo {author} {\bibfnamefont {H.}~\bibnamefont {Becker}}, \ and\ \bibinfo
  {author} {\bibfnamefont {K.}~\bibnamefont {Meerholz}},\ }\href@noop {}
  {\bibfield  {journal} {\bibinfo  {journal} {Advanced Materials}\ }\textbf
  {\bibinfo {volume} {19}},\ \bibinfo {pages} {4460} (\bibinfo {year}
  {2007}{\natexlab{b}})}\BibitemShut {NoStop}%
\bibitem [{\citenamefont {Zhang}\ \emph {et~al.}(2010)\citenamefont {Zhang},
  \citenamefont {Qin}, \citenamefont {Ding}, \citenamefont {Chen},
  \citenamefont {Xie}, \citenamefont {Cheng},\ and\ \citenamefont
  {Wang}}]{Zhang2010}%
  \BibitemOpen
  \bibfield  {author} {\bibinfo {author} {\bibfnamefont {B.~H.}\ \bibnamefont
  {Zhang}}, \bibinfo {author} {\bibfnamefont {C.~J.}\ \bibnamefont {Qin}},
  \bibinfo {author} {\bibfnamefont {J.~Q.}\ \bibnamefont {Ding}}, \bibinfo
  {author} {\bibfnamefont {L.}~\bibnamefont {Chen}}, \bibinfo {author}
  {\bibfnamefont {Z.~Y.}\ \bibnamefont {Xie}}, \bibinfo {author} {\bibfnamefont
  {Y.~X.}\ \bibnamefont {Cheng}}, \ and\ \bibinfo {author} {\bibfnamefont
  {L.~X.}\ \bibnamefont {Wang}},\ }\href@noop {} {\bibfield  {journal}
  {\bibinfo  {journal} {Advanced Functional Materials}\ }\textbf {\bibinfo
  {volume} {20}},\ \bibinfo {pages} {2951} (\bibinfo {year}
  {2010})}\BibitemShut {NoStop}%
\bibitem [{\citenamefont {Jiang}\ \emph {et~al.}(2006)\citenamefont {Jiang},
  \citenamefont {Xu}, \citenamefont {Yang}, \citenamefont {Guan}, \citenamefont
  {Liu}, \citenamefont {Zhen},\ and\ \citenamefont {Cao}}]{Jiang2006}%
  \BibitemOpen
  \bibfield  {author} {\bibinfo {author} {\bibfnamefont {J.~X.}\ \bibnamefont
  {Jiang}}, \bibinfo {author} {\bibfnamefont {Y.~H.}\ \bibnamefont {Xu}},
  \bibinfo {author} {\bibfnamefont {W.}~\bibnamefont {Yang}}, \bibinfo {author}
  {\bibfnamefont {R.}~\bibnamefont {Guan}}, \bibinfo {author} {\bibfnamefont
  {Z.~Q.}\ \bibnamefont {Liu}}, \bibinfo {author} {\bibfnamefont {H.~Y.}\
  \bibnamefont {Zhen}}, \ and\ \bibinfo {author} {\bibfnamefont
  {Y.}~\bibnamefont {Cao}},\ }\href@noop {} {\bibfield  {journal} {\bibinfo
  {journal} {Advanced Materials}\ }\textbf {\bibinfo {volume} {18}},\ \bibinfo
  {pages} {1769} (\bibinfo {year} {2006})}\BibitemShut {NoStop}%
\bibitem [{Note9()}]{Note9}%
  \BibitemOpen
  \bibinfo {note} {`PL' = photoluminescence}\BibitemShut {NoStop}%
\bibitem [{Note10()}]{Note10}%
  \BibitemOpen
  \bibinfo {note} {To reach color point E, the intensities of the emission
  bands are similar for all incorporated emitters while the shape of a
  multi-emitter spectrum at color point A more looks like a staircase (cf. Fig.
  \ref {_idealwhite}).}\BibitemShut {Stop}%
\bibitem [{\citenamefont {Choukri}\ \emph {et~al.}(2006)\citenamefont
  {Choukri}, \citenamefont {Fischer}, \citenamefont {Forget}, \citenamefont
  {Chenais}, \citenamefont {Castex}, \citenamefont {Ades}, \citenamefont
  {Siove},\ and\ \citenamefont {Geffroy}}]{Choukri2006}%
  \BibitemOpen
  \bibfield  {author} {\bibinfo {author} {\bibfnamefont {H.}~\bibnamefont
  {Choukri}}, \bibinfo {author} {\bibfnamefont {A.}~\bibnamefont {Fischer}},
  \bibinfo {author} {\bibfnamefont {S.}~\bibnamefont {Forget}}, \bibinfo
  {author} {\bibfnamefont {S.}~\bibnamefont {Chenais}}, \bibinfo {author}
  {\bibfnamefont {M.~C.}\ \bibnamefont {Castex}}, \bibinfo {author}
  {\bibfnamefont {D.}~\bibnamefont {Ades}}, \bibinfo {author} {\bibfnamefont
  {A.}~\bibnamefont {Siove}}, \ and\ \bibinfo {author} {\bibfnamefont
  {B.}~\bibnamefont {Geffroy}},\ }\href@noop {} {\bibfield  {journal} {\bibinfo
   {journal} {Applied Physics Letters}\ }\textbf {\bibinfo {volume} {89}},\
  \bibinfo {pages} {183513} (\bibinfo {year} {2006})}\BibitemShut {NoStop}%
\bibitem [{\citenamefont {Duan}\ \emph {et~al.}(2008)\citenamefont {Duan},
  \citenamefont {Mazzeo}, \citenamefont {Maiorano}, \citenamefont {Mariano},
  \citenamefont {Qin}, \citenamefont {Cingolani},\ and\ \citenamefont
  {Gigli}}]{Duan2008}%
  \BibitemOpen
  \bibfield  {author} {\bibinfo {author} {\bibfnamefont {Y.}~\bibnamefont
  {Duan}}, \bibinfo {author} {\bibfnamefont {M.}~\bibnamefont {Mazzeo}},
  \bibinfo {author} {\bibfnamefont {V.}~\bibnamefont {Maiorano}}, \bibinfo
  {author} {\bibfnamefont {F.}~\bibnamefont {Mariano}}, \bibinfo {author}
  {\bibfnamefont {D.}~\bibnamefont {Qin}}, \bibinfo {author} {\bibfnamefont
  {R.}~\bibnamefont {Cingolani}}, \ and\ \bibinfo {author} {\bibfnamefont
  {G.}~\bibnamefont {Gigli}},\ }\href@noop {} {\bibfield  {journal} {\bibinfo
  {journal} {Applied Physics Letters}\ }\textbf {\bibinfo {volume} {92}},\
  \bibinfo {pages} {113304} (\bibinfo {year} {2008})}\BibitemShut {NoStop}%
\bibitem [{\citenamefont {Huang}\ \emph {et~al.}(2002)\citenamefont {Huang},
  \citenamefont {Jou}, \citenamefont {Weng},\ and\ \citenamefont
  {Liu}}]{Huang2002}%
  \BibitemOpen
  \bibfield  {author} {\bibinfo {author} {\bibfnamefont {Y.~S.}\ \bibnamefont
  {Huang}}, \bibinfo {author} {\bibfnamefont {J.~H.}\ \bibnamefont {Jou}},
  \bibinfo {author} {\bibfnamefont {W.~K.}\ \bibnamefont {Weng}}, \ and\
  \bibinfo {author} {\bibfnamefont {J.~M.}\ \bibnamefont {Liu}},\ }\href@noop
  {} {\bibfield  {journal} {\bibinfo  {journal} {Applied Physics Letters}\
  }\textbf {\bibinfo {volume} {80}},\ \bibinfo {pages} {2782} (\bibinfo {year}
  {2002})}\BibitemShut {NoStop}%
\bibitem [{\citenamefont {Kim}\ and\ \citenamefont {Shinar}(2002)}]{Kim2002}%
  \BibitemOpen
  \bibfield  {author} {\bibinfo {author} {\bibfnamefont {C.~H.}\ \bibnamefont
  {Kim}}\ and\ \bibinfo {author} {\bibfnamefont {J.}~\bibnamefont {Shinar}},\
  }\href@noop {} {\bibfield  {journal} {\bibinfo  {journal} {Applied Physics
  Letters}\ }\textbf {\bibinfo {volume} {80}},\ \bibinfo {pages} {2201}
  (\bibinfo {year} {2002})}\BibitemShut {NoStop}%
\bibitem [{\citenamefont {Wu}\ \emph {et~al.}(2005)\citenamefont {Wu},
  \citenamefont {Hwang}, \citenamefont {Chen}, \citenamefont {Lee},
  \citenamefont {Shen},\ and\ \citenamefont {Chen}}]{Wu2005}%
  \BibitemOpen
  \bibfield  {author} {\bibinfo {author} {\bibfnamefont {Y.~S.}\ \bibnamefont
  {Wu}}, \bibinfo {author} {\bibfnamefont {S.~W.}\ \bibnamefont {Hwang}},
  \bibinfo {author} {\bibfnamefont {H.~H.}\ \bibnamefont {Chen}}, \bibinfo
  {author} {\bibfnamefont {M.~T.}\ \bibnamefont {Lee}}, \bibinfo {author}
  {\bibfnamefont {W.~J.}\ \bibnamefont {Shen}}, \ and\ \bibinfo {author}
  {\bibfnamefont {C.~H.}\ \bibnamefont {Chen}},\ }\href@noop {} {\bibfield
  {journal} {\bibinfo  {journal} {Thin Solid Films}\ }\textbf {\bibinfo
  {volume} {488}},\ \bibinfo {pages} {265} (\bibinfo {year}
  {2005})}\BibitemShut {NoStop}%
\bibitem [{\citenamefont {Xie}\ \emph {et~al.}(1999)\citenamefont {Xie},
  \citenamefont {Huang}, \citenamefont {Li}, \citenamefont {Liu}, \citenamefont
  {Wang}, \citenamefont {Li},\ and\ \citenamefont {Shen}}]{Xie1999}%
  \BibitemOpen
  \bibfield  {author} {\bibinfo {author} {\bibfnamefont {Z.~Y.}\ \bibnamefont
  {Xie}}, \bibinfo {author} {\bibfnamefont {J.~S.}\ \bibnamefont {Huang}},
  \bibinfo {author} {\bibfnamefont {C.~N.}\ \bibnamefont {Li}}, \bibinfo
  {author} {\bibfnamefont {S.~Y.}\ \bibnamefont {Liu}}, \bibinfo {author}
  {\bibfnamefont {Y.}~\bibnamefont {Wang}}, \bibinfo {author} {\bibfnamefont
  {Y.~Q.}\ \bibnamefont {Li}}, \ and\ \bibinfo {author} {\bibfnamefont {J.~C.}\
  \bibnamefont {Shen}},\ }\href@noop {} {\bibfield  {journal} {\bibinfo
  {journal} {Applied Physics Letters}\ }\textbf {\bibinfo {volume} {74}},\
  \bibinfo {pages} {641} (\bibinfo {year} {1999})}\BibitemShut {NoStop}%
\bibitem [{\citenamefont {Chuen}\ and\ \citenamefont {Tao}(2002)}]{Chuen2002}%
  \BibitemOpen
  \bibfield  {author} {\bibinfo {author} {\bibfnamefont {C.~H.}\ \bibnamefont
  {Chuen}}\ and\ \bibinfo {author} {\bibfnamefont {Y.~T.}\ \bibnamefont
  {Tao}},\ }\href@noop {} {\bibfield  {journal} {\bibinfo  {journal} {Applied
  Physics Letters}\ }\textbf {\bibinfo {volume} {81}},\ \bibinfo {pages} {4499}
  (\bibinfo {year} {2002})}\BibitemShut {NoStop}%
\bibitem [{\citenamefont {Yang}\ \emph {et~al.}(2011)\citenamefont {Yang},
  \citenamefont {Peng}, \citenamefont {Ye}, \citenamefont {Wu}, \citenamefont
  {Liu},\ and\ \citenamefont {Wang}}]{Yang2011}%
  \BibitemOpen
  \bibfield  {author} {\bibinfo {author} {\bibfnamefont {Y.}~\bibnamefont
  {Yang}}, \bibinfo {author} {\bibfnamefont {T.}~\bibnamefont {Peng}}, \bibinfo
  {author} {\bibfnamefont {K.~Q.}\ \bibnamefont {Ye}}, \bibinfo {author}
  {\bibfnamefont {Y.}~\bibnamefont {Wu}}, \bibinfo {author} {\bibfnamefont
  {Y.}~\bibnamefont {Liu}}, \ and\ \bibinfo {author} {\bibfnamefont
  {Y.}~\bibnamefont {Wang}},\ }\href@noop {} {\bibfield  {journal} {\bibinfo
  {journal} {Organic Electronics}\ }\textbf {\bibinfo {volume} {12}},\ \bibinfo
  {pages} {29} (\bibinfo {year} {2011})}\BibitemShut {NoStop}%
\bibitem [{\citenamefont {Jou}\ \emph {et~al.}(2006)\citenamefont {Jou},
  \citenamefont {Chiu}, \citenamefont {Wang}, \citenamefont {Wang},\ and\
  \citenamefont {Hu}}]{Jou2006a}%
  \BibitemOpen
  \bibfield  {author} {\bibinfo {author} {\bibfnamefont {J.~H.}\ \bibnamefont
  {Jou}}, \bibinfo {author} {\bibfnamefont {Y.~S.}\ \bibnamefont {Chiu}},
  \bibinfo {author} {\bibfnamefont {C.~P.}\ \bibnamefont {Wang}}, \bibinfo
  {author} {\bibfnamefont {R.~Y.}\ \bibnamefont {Wang}}, \ and\ \bibinfo
  {author} {\bibfnamefont {C.}~\bibnamefont {Hu}},\ }\href@noop {} {\bibfield
  {journal} {\bibinfo  {journal} {Applied Physics Letters}\ }\textbf {\bibinfo
  {volume} {88}},\ \bibinfo {pages} {193501} (\bibinfo {year}
  {2006})}\BibitemShut {NoStop}%
\bibitem [{\citenamefont {Swanson}\ \emph {et~al.}(2003)\citenamefont
  {Swanson}, \citenamefont {Wallraff}, \citenamefont {Chen}, \citenamefont
  {Zhang}, \citenamefont {Bozano}, \citenamefont {Carter}, \citenamefont
  {Salem}, \citenamefont {Villa},\ and\ \citenamefont {Scott}}]{Swanson2003}%
  \BibitemOpen
  \bibfield  {author} {\bibinfo {author} {\bibfnamefont {S.~A.}\ \bibnamefont
  {Swanson}}, \bibinfo {author} {\bibfnamefont {G.~M.}\ \bibnamefont
  {Wallraff}}, \bibinfo {author} {\bibfnamefont {J.~P.}\ \bibnamefont {Chen}},
  \bibinfo {author} {\bibfnamefont {W.~J.}\ \bibnamefont {Zhang}}, \bibinfo
  {author} {\bibfnamefont {L.~D.}\ \bibnamefont {Bozano}}, \bibinfo {author}
  {\bibfnamefont {K.~R.}\ \bibnamefont {Carter}}, \bibinfo {author}
  {\bibfnamefont {J.~R.}\ \bibnamefont {Salem}}, \bibinfo {author}
  {\bibfnamefont {R.}~\bibnamefont {Villa}}, \ and\ \bibinfo {author}
  {\bibfnamefont {J.~C.}\ \bibnamefont {Scott}},\ }\href@noop {} {\bibfield
  {journal} {\bibinfo  {journal} {Chemistry Of Materials}\ }\textbf {\bibinfo
  {volume} {15}},\ \bibinfo {pages} {2305} (\bibinfo {year}
  {2003})}\BibitemShut {NoStop}%
\bibitem [{\citenamefont {Xie}\ \emph {et~al.}(2003)\citenamefont {Xie},
  \citenamefont {Liu},\ and\ \citenamefont {Zhao}}]{Xie2003}%
  \BibitemOpen
  \bibfield  {author} {\bibinfo {author} {\bibfnamefont {W.~F.}\ \bibnamefont
  {Xie}}, \bibinfo {author} {\bibfnamefont {S.~Y.}\ \bibnamefont {Liu}}, \ and\
  \bibinfo {author} {\bibfnamefont {Y.}~\bibnamefont {Zhao}},\ }\href@noop {}
  {\bibfield  {journal} {\bibinfo  {journal} {Journal of Physics D-Applied
  Physics}\ }\textbf {\bibinfo {volume} {36}},\ \bibinfo {pages} {1246}
  (\bibinfo {year} {2003})}\BibitemShut {NoStop}%
\bibitem [{\citenamefont {Baldo}\ \emph
  {et~al.}(2000{\natexlab{b}})\citenamefont {Baldo}, \citenamefont {Thompson},\
  and\ \citenamefont {Forrest}}]{Baldo2000}%
  \BibitemOpen
  \bibfield  {author} {\bibinfo {author} {\bibfnamefont {M.~A.}\ \bibnamefont
  {Baldo}}, \bibinfo {author} {\bibfnamefont {M.~E.}\ \bibnamefont {Thompson}},
  \ and\ \bibinfo {author} {\bibfnamefont {S.~R.}\ \bibnamefont {Forrest}},\
  }\href@noop {} {\bibfield  {journal} {\bibinfo  {journal} {Nature}\ }\textbf
  {\bibinfo {volume} {403}},\ \bibinfo {pages} {750} (\bibinfo {year}
  {2000}{\natexlab{b}})}\BibitemShut {NoStop}%
\bibitem [{\citenamefont {Jordan}\ \emph {et~al.}(1996)\citenamefont {Jordan},
  \citenamefont {Dodabalapur}, \citenamefont {Strukelj},\ and\ \citenamefont
  {Miller}}]{Jordan1996}%
  \BibitemOpen
  \bibfield  {author} {\bibinfo {author} {\bibfnamefont {R.~H.}\ \bibnamefont
  {Jordan}}, \bibinfo {author} {\bibfnamefont {A.}~\bibnamefont {Dodabalapur}},
  \bibinfo {author} {\bibfnamefont {M.}~\bibnamefont {Strukelj}}, \ and\
  \bibinfo {author} {\bibfnamefont {T.~M.}\ \bibnamefont {Miller}},\
  }\href@noop {} {\bibfield  {journal} {\bibinfo  {journal} {Applied Physics
  Letters}\ }\textbf {\bibinfo {volume} {68}},\ \bibinfo {pages} {1192}
  (\bibinfo {year} {1996})}\BibitemShut {NoStop}%
\bibitem [{\citenamefont {Strukelj}\ \emph {et~al.}(1996)\citenamefont
  {Strukelj}, \citenamefont {Jordan},\ and\ \citenamefont
  {Dodabalapur}}]{Strukelj1996}%
  \BibitemOpen
  \bibfield  {author} {\bibinfo {author} {\bibfnamefont {M.}~\bibnamefont
  {Strukelj}}, \bibinfo {author} {\bibfnamefont {R.~H.}\ \bibnamefont
  {Jordan}}, \ and\ \bibinfo {author} {\bibfnamefont {A.}~\bibnamefont
  {Dodabalapur}},\ }\href@noop {} {\bibfield  {journal} {\bibinfo  {journal}
  {Journal of the American Chemical Society}\ }\textbf {\bibinfo {volume}
  {118}},\ \bibinfo {pages} {1213} (\bibinfo {year} {1996})}\BibitemShut
  {NoStop}%
\bibitem [{\citenamefont {Hamada}\ \emph {et~al.}(1996)\citenamefont {Hamada},
  \citenamefont {Sang}, \citenamefont {Fujii}, \citenamefont {Nishio},
  \citenamefont {Takanashi},\ and\ \citenamefont {Shibata}}]{Hamada1996}%
  \BibitemOpen
  \bibfield  {author} {\bibinfo {author} {\bibfnamefont {Y.}~\bibnamefont
  {Hamada}}, \bibinfo {author} {\bibfnamefont {T.}~\bibnamefont {Sang}},
  \bibinfo {author} {\bibfnamefont {H.}~\bibnamefont {Fujii}}, \bibinfo
  {author} {\bibfnamefont {Y.}~\bibnamefont {Nishio}}, \bibinfo {author}
  {\bibfnamefont {H.}~\bibnamefont {Takanashi}}, \ and\ \bibinfo {author}
  {\bibfnamefont {K.}~\bibnamefont {Shibata}},\ }\href@noop {} {\bibfield
  {journal} {\bibinfo  {journal} {Japanese Journal of Applied Physics Part
  2-letters}\ }\textbf {\bibinfo {volume} {35}},\ \bibinfo {pages} {L1339}
  (\bibinfo {year} {1996})}\BibitemShut {NoStop}%
\bibitem [{\citenamefont {Feng}\ \emph {et~al.}(2001)\citenamefont {Feng},
  \citenamefont {Li}, \citenamefont {Gao}, \citenamefont {Liu}, \citenamefont
  {Liu},\ and\ \citenamefont {Wang}}]{Feng2001}%
  \BibitemOpen
  \bibfield  {author} {\bibinfo {author} {\bibfnamefont {J.}~\bibnamefont
  {Feng}}, \bibinfo {author} {\bibfnamefont {F.}~\bibnamefont {Li}}, \bibinfo
  {author} {\bibfnamefont {W.~B.}\ \bibnamefont {Gao}}, \bibinfo {author}
  {\bibfnamefont {S.~Y.}\ \bibnamefont {Liu}}, \bibinfo {author} {\bibfnamefont
  {Y.}~\bibnamefont {Liu}}, \ and\ \bibinfo {author} {\bibfnamefont
  {Y.}~\bibnamefont {Wang}},\ }\href@noop {} {\bibfield  {journal} {\bibinfo
  {journal} {Applied Physics Letters}\ }\textbf {\bibinfo {volume} {78}},\
  \bibinfo {pages} {3947} (\bibinfo {year} {2001})}\BibitemShut {NoStop}%
\bibitem [{\citenamefont {Mazzeo}\ \emph {et~al.}(2003)\citenamefont {Mazzeo},
  \citenamefont {Pisignano}, \citenamefont {Della~Sala}, \citenamefont
  {Thompson}, \citenamefont {Blyth}, \citenamefont {Gigli}, \citenamefont
  {Cingolani}, \citenamefont {Sotgiu},\ and\ \citenamefont
  {Barbarella}}]{Mazzeo2003}%
  \BibitemOpen
  \bibfield  {author} {\bibinfo {author} {\bibfnamefont {M.}~\bibnamefont
  {Mazzeo}}, \bibinfo {author} {\bibfnamefont {D.}~\bibnamefont {Pisignano}},
  \bibinfo {author} {\bibfnamefont {F.}~\bibnamefont {Della~Sala}}, \bibinfo
  {author} {\bibfnamefont {J.}~\bibnamefont {Thompson}}, \bibinfo {author}
  {\bibfnamefont {R.~I.~R.}\ \bibnamefont {Blyth}}, \bibinfo {author}
  {\bibfnamefont {G.}~\bibnamefont {Gigli}}, \bibinfo {author} {\bibfnamefont
  {R.}~\bibnamefont {Cingolani}}, \bibinfo {author} {\bibfnamefont
  {G.}~\bibnamefont {Sotgiu}}, \ and\ \bibinfo {author} {\bibfnamefont
  {G.}~\bibnamefont {Barbarella}},\ }\href@noop {} {\bibfield  {journal}
  {\bibinfo  {journal} {Applied Physics Letters}\ }\textbf {\bibinfo {volume}
  {82}},\ \bibinfo {pages} {334} (\bibinfo {year} {2003})}\BibitemShut
  {NoStop}%
\bibitem [{\citenamefont {Tong}\ \emph {et~al.}(2007)\citenamefont {Tong},
  \citenamefont {Lai}, \citenamefont {Chan}, \citenamefont {Tang},
  \citenamefont {Kwong}, \citenamefont {Lee},\ and\ \citenamefont
  {Lee}}]{Tong2007}%
  \BibitemOpen
  \bibfield  {author} {\bibinfo {author} {\bibfnamefont {Q.~X.}\ \bibnamefont
  {Tong}}, \bibinfo {author} {\bibfnamefont {S.~L.}\ \bibnamefont {Lai}},
  \bibinfo {author} {\bibfnamefont {M.~Y.}\ \bibnamefont {Chan}}, \bibinfo
  {author} {\bibfnamefont {J.~X.}\ \bibnamefont {Tang}}, \bibinfo {author}
  {\bibfnamefont {H.~L.}\ \bibnamefont {Kwong}}, \bibinfo {author}
  {\bibfnamefont {C.~S.}\ \bibnamefont {Lee}}, \ and\ \bibinfo {author}
  {\bibfnamefont {S.~T.}\ \bibnamefont {Lee}},\ }\href@noop {} {\bibfield
  {journal} {\bibinfo  {journal} {Applied Physics Letters}\ }\textbf {\bibinfo
  {volume} {91}},\ \bibinfo {pages} {023503} (\bibinfo {year}
  {2007})}\BibitemShut {NoStop}%
\bibitem [{\citenamefont {Seidler}\ \emph {et~al.}(2010)\citenamefont
  {Seidler}, \citenamefont {Reineke}, \citenamefont {Walzer}, \citenamefont
  {Lüssem}, \citenamefont {Tomkeviciene}, \citenamefont {Grazulevicius},\ and\
  \citenamefont {Leo}}]{Seidler2010}%
  \BibitemOpen
  \bibfield  {author} {\bibinfo {author} {\bibfnamefont {N.}~\bibnamefont
  {Seidler}}, \bibinfo {author} {\bibfnamefont {S.}~\bibnamefont {Reineke}},
  \bibinfo {author} {\bibfnamefont {K.}~\bibnamefont {Walzer}}, \bibinfo
  {author} {\bibfnamefont {B.}~\bibnamefont {Lüssem}}, \bibinfo {author}
  {\bibfnamefont {A.}~\bibnamefont {Tomkeviciene}}, \bibinfo {author}
  {\bibfnamefont {J.~V.}\ \bibnamefont {Grazulevicius}}, \ and\ \bibinfo
  {author} {\bibfnamefont {K.}~\bibnamefont {Leo}},\ }\href@noop {} {\bibfield
  {journal} {\bibinfo  {journal} {Applied Physics Letters}\ }\textbf {\bibinfo
  {volume} {96}},\ \bibinfo {pages} {093304} (\bibinfo {year}
  {2010})}\BibitemShut {NoStop}%
\bibitem [{Note11()}]{Note11}%
  \BibitemOpen
  \bibinfo {note} {Even more complicated is the fact that the fluorescent blue
  system can either be a neat film or a host-guest system. In the latter case,
  one would need to include exciton transfers from host to guest for both
  singlets and triplets.}\BibitemShut {Stop}%
\bibitem [{\citenamefont {Reineke}\ \emph
  {et~al.}(2009{\natexlab{c}})\citenamefont {Reineke}, \citenamefont
  {Schwartz}, \citenamefont {Walzer},\ and\ \citenamefont {Leo}}]{Reineke2009}%
  \BibitemOpen
  \bibfield  {author} {\bibinfo {author} {\bibfnamefont {S.}~\bibnamefont
  {Reineke}}, \bibinfo {author} {\bibfnamefont {G.}~\bibnamefont {Schwartz}},
  \bibinfo {author} {\bibfnamefont {K.}~\bibnamefont {Walzer}}, \ and\ \bibinfo
  {author} {\bibfnamefont {K.}~\bibnamefont {Leo}},\ }\href@noop {} {\bibfield
  {journal} {\bibinfo  {journal} {Physica Status Solidi RRL}\ }\textbf
  {\bibinfo {volume} {3}},\ \bibinfo {pages} {67} (\bibinfo {year}
  {2009}{\natexlab{c}})}\BibitemShut {NoStop}%
\bibitem [{\citenamefont {Baek}\ and\ \citenamefont {Lee}(2008)}]{Baek2008}%
  \BibitemOpen
  \bibfield  {author} {\bibinfo {author} {\bibfnamefont {H.~I.}\ \bibnamefont
  {Baek}}\ and\ \bibinfo {author} {\bibfnamefont {C.~H.}\ \bibnamefont {Lee}},\
  }\href@noop {} {\bibfield  {journal} {\bibinfo  {journal} {Journal of Physics
  D-applied Physics}\ }\textbf {\bibinfo {volume} {41}},\ \bibinfo {pages}
  {105101} (\bibinfo {year} {2008})}\BibitemShut {NoStop}%
\bibitem [{\citenamefont {Schwartz}\ \emph
  {et~al.}(2008{\natexlab{a}})\citenamefont {Schwartz}, \citenamefont {Ke},
  \citenamefont {Wu}, \citenamefont {Walzer},\ and\ \citenamefont
  {Leo}}]{Schwartz2008a}%
  \BibitemOpen
  \bibfield  {author} {\bibinfo {author} {\bibfnamefont {G.}~\bibnamefont
  {Schwartz}}, \bibinfo {author} {\bibfnamefont {T.~H.}\ \bibnamefont {Ke}},
  \bibinfo {author} {\bibfnamefont {C.~C.}\ \bibnamefont {Wu}}, \bibinfo
  {author} {\bibfnamefont {K.}~\bibnamefont {Walzer}}, \ and\ \bibinfo {author}
  {\bibfnamefont {K.}~\bibnamefont {Leo}},\ }\href@noop {} {\bibfield
  {journal} {\bibinfo  {journal} {Applied Physics Letters}\ }\textbf {\bibinfo
  {volume} {93}},\ \bibinfo {pages} {073304} (\bibinfo {year}
  {2008}{\natexlab{a}})}\BibitemShut {NoStop}%
\bibitem [{\citenamefont {Leem}\ \emph {et~al.}(2010)\citenamefont {Leem},
  \citenamefont {Kim}, \citenamefont {Jung}, \citenamefont {Kim}, \citenamefont
  {Kim}, \citenamefont {Kim}, \citenamefont {Kim}, \citenamefont {Kwon},\ and\
  \citenamefont {Kim}}]{Leem2010}%
  \BibitemOpen
  \bibfield  {author} {\bibinfo {author} {\bibfnamefont {D.~S.}\ \bibnamefont
  {Leem}}, \bibinfo {author} {\bibfnamefont {J.~W.}\ \bibnamefont {Kim}},
  \bibinfo {author} {\bibfnamefont {S.~O.}\ \bibnamefont {Jung}}, \bibinfo
  {author} {\bibfnamefont {S.~O.}\ \bibnamefont {Kim}}, \bibinfo {author}
  {\bibfnamefont {S.~H.}\ \bibnamefont {Kim}}, \bibinfo {author} {\bibfnamefont
  {K.~Y.}\ \bibnamefont {Kim}}, \bibinfo {author} {\bibfnamefont {Y.~H.}\
  \bibnamefont {Kim}}, \bibinfo {author} {\bibfnamefont {S.~K.}\ \bibnamefont
  {Kwon}}, \ and\ \bibinfo {author} {\bibfnamefont {J.~J.}\ \bibnamefont
  {Kim}},\ }\href@noop {} {\bibfield  {journal} {\bibinfo  {journal} {Journal
  of Physics D-Applied Physics}\ }\textbf {\bibinfo {volume} {43}},\ \bibinfo
  {pages} {405102} (\bibinfo {year} {2010})}\BibitemShut {NoStop}%
\bibitem [{\citenamefont {Ho}\ \emph {et~al.}(2008{\natexlab{a}})\citenamefont
  {Ho}, \citenamefont {Wong}, \citenamefont {Wang}, \citenamefont {Ma},
  \citenamefont {Wang},\ and\ \citenamefont {Lin}}]{Ho2008a}%
  \BibitemOpen
  \bibfield  {author} {\bibinfo {author} {\bibfnamefont {C.~L.}\ \bibnamefont
  {Ho}}, \bibinfo {author} {\bibfnamefont {W.~Y.}\ \bibnamefont {Wong}},
  \bibinfo {author} {\bibfnamefont {Q.}~\bibnamefont {Wang}}, \bibinfo {author}
  {\bibfnamefont {D.~G.}\ \bibnamefont {Ma}}, \bibinfo {author} {\bibfnamefont
  {L.~X.}\ \bibnamefont {Wang}}, \ and\ \bibinfo {author} {\bibfnamefont
  {Z.~Y.}\ \bibnamefont {Lin}},\ }\href@noop {} {\bibfield  {journal} {\bibinfo
   {journal} {Advanced Functional Materials}\ }\textbf {\bibinfo {volume}
  {18}},\ \bibinfo {pages} {928} (\bibinfo {year}
  {2008}{\natexlab{a}})}\BibitemShut {NoStop}%
\bibitem [{\citenamefont {Yan}\ \emph {et~al.}(2007)\citenamefont {Yan},
  \citenamefont {Cheung}, \citenamefont {Kui}, \citenamefont {Xiang},
  \citenamefont {Roy}, \citenamefont {Xu},\ and\ \citenamefont
  {Che}}]{Yan2007}%
  \BibitemOpen
  \bibfield  {author} {\bibinfo {author} {\bibfnamefont {B.~P.}\ \bibnamefont
  {Yan}}, \bibinfo {author} {\bibfnamefont {C.~C.~C.}\ \bibnamefont {Cheung}},
  \bibinfo {author} {\bibfnamefont {S.~C.~F.}\ \bibnamefont {Kui}}, \bibinfo
  {author} {\bibfnamefont {H.~F.}\ \bibnamefont {Xiang}}, \bibinfo {author}
  {\bibfnamefont {V.~A.~L.}\ \bibnamefont {Roy}}, \bibinfo {author}
  {\bibfnamefont {S.~J.}\ \bibnamefont {Xu}}, \ and\ \bibinfo {author}
  {\bibfnamefont {C.~M.}\ \bibnamefont {Che}},\ }\href@noop {} {\bibfield
  {journal} {\bibinfo  {journal} {Advanced Materials}\ }\textbf {\bibinfo
  {volume} {19}},\ \bibinfo {pages} {3599} (\bibinfo {year}
  {2007})}\BibitemShut {NoStop}%
\bibitem [{\citenamefont {Ho}\ \emph {et~al.}(2008{\natexlab{b}})\citenamefont
  {Ho}, \citenamefont {Lin}, \citenamefont {Wong}, \citenamefont {Wong},\ and\
  \citenamefont {Chen}}]{Ho2008}%
  \BibitemOpen
  \bibfield  {author} {\bibinfo {author} {\bibfnamefont {C.~L.}\ \bibnamefont
  {Ho}}, \bibinfo {author} {\bibfnamefont {M.~F.}\ \bibnamefont {Lin}},
  \bibinfo {author} {\bibfnamefont {W.~Y.}\ \bibnamefont {Wong}}, \bibinfo
  {author} {\bibfnamefont {W.~K.}\ \bibnamefont {Wong}}, \ and\ \bibinfo
  {author} {\bibfnamefont {C.~H.}\ \bibnamefont {Chen}},\ }\href@noop {}
  {\bibfield  {journal} {\bibinfo  {journal} {Applied Physics Letters}\
  }\textbf {\bibinfo {volume} {92}},\ \bibinfo {pages} {083301} (\bibinfo
  {year} {2008}{\natexlab{b}})}\BibitemShut {NoStop}%
\bibitem [{\citenamefont {Seo}\ \emph {et~al.}(2007)\citenamefont {Seo},
  \citenamefont {Seo}, \citenamefont {Park}, \citenamefont {Kim}, \citenamefont
  {Kim}, \citenamefont {Hyung}, \citenamefont {Lee},\ and\ \citenamefont
  {Yoon}}]{Seo2007}%
  \BibitemOpen
  \bibfield  {author} {\bibinfo {author} {\bibfnamefont {J.~H.}\ \bibnamefont
  {Seo}}, \bibinfo {author} {\bibfnamefont {J.~H.}\ \bibnamefont {Seo}},
  \bibinfo {author} {\bibfnamefont {J.~H.}\ \bibnamefont {Park}}, \bibinfo
  {author} {\bibfnamefont {Y.~K.}\ \bibnamefont {Kim}}, \bibinfo {author}
  {\bibfnamefont {J.~H.}\ \bibnamefont {Kim}}, \bibinfo {author} {\bibfnamefont
  {G.~W.}\ \bibnamefont {Hyung}}, \bibinfo {author} {\bibfnamefont {K.~H.}\
  \bibnamefont {Lee}}, \ and\ \bibinfo {author} {\bibfnamefont {S.~S.}\
  \bibnamefont {Yoon}},\ }\href@noop {} {\bibfield  {journal} {\bibinfo
  {journal} {Applied Physics Letters}\ }\textbf {\bibinfo {volume} {90}},\
  \bibinfo {pages} {203507} (\bibinfo {year} {2007})}\BibitemShut {NoStop}%
\bibitem [{\citenamefont {Zhang}\ \emph {et~al.}(2008)\citenamefont {Zhang},
  \citenamefont {Hua}, \citenamefont {Wu}, \citenamefont {Wang},\ and\
  \citenamefont {Yin}}]{Zhang2008}%
  \BibitemOpen
  \bibfield  {author} {\bibinfo {author} {\bibfnamefont {L.~J.}\ \bibnamefont
  {Zhang}}, \bibinfo {author} {\bibfnamefont {Y.~L.}\ \bibnamefont {Hua}},
  \bibinfo {author} {\bibfnamefont {X.~M.}\ \bibnamefont {Wu}}, \bibinfo
  {author} {\bibfnamefont {Y.}~\bibnamefont {Wang}}, \ and\ \bibinfo {author}
  {\bibfnamefont {S.~G.}\ \bibnamefont {Yin}},\ }\href@noop {} {\bibfield
  {journal} {\bibinfo  {journal} {Chinese Physics B}\ }\textbf {\bibinfo
  {volume} {17}},\ \bibinfo {pages} {3097} (\bibinfo {year}
  {2008})}\BibitemShut {NoStop}%
\bibitem [{\citenamefont {Deaton}\ \emph {et~al.}(2008)\citenamefont {Deaton},
  \citenamefont {Kondakova}, \citenamefont {Giesen}, \citenamefont {Begley},\
  and\ \citenamefont {Kondakov}}]{Deaton2008}%
  \BibitemOpen
  \bibfield  {author} {\bibinfo {author} {\bibfnamefont {J.~C.}\ \bibnamefont
  {Deaton}}, \bibinfo {author} {\bibfnamefont {M.~E.}\ \bibnamefont
  {Kondakova}}, \bibinfo {author} {\bibfnamefont {D.~J.}\ \bibnamefont
  {Giesen}}, \bibinfo {author} {\bibfnamefont {W.~J.}\ \bibnamefont {Begley}},
  \ and\ \bibinfo {author} {\bibfnamefont {D.~Y.}\ \bibnamefont {Kondakov}},\
  }\href@noop {} {\bibfield  {journal} {\bibinfo  {journal} {US patent:
  US020080286610A1}\ } (\bibinfo {year} {2008})}\BibitemShut {NoStop}%
\bibitem [{\citenamefont {Adachi}\ \emph {et~al.}(2000)\citenamefont {Adachi},
  \citenamefont {Baldo}, \citenamefont {Forrest},\ and\ \citenamefont
  {Thompson}}]{Adachi2000}%
  \BibitemOpen
  \bibfield  {author} {\bibinfo {author} {\bibfnamefont {C.}~\bibnamefont
  {Adachi}}, \bibinfo {author} {\bibfnamefont {M.~A.}\ \bibnamefont {Baldo}},
  \bibinfo {author} {\bibfnamefont {S.~R.}\ \bibnamefont {Forrest}}, \ and\
  \bibinfo {author} {\bibfnamefont {M.~E.}\ \bibnamefont {Thompson}},\
  }\href@noop {} {\bibfield  {journal} {\bibinfo  {journal} {Applied Physics
  Letters}\ }\textbf {\bibinfo {volume} {77}},\ \bibinfo {pages} {904}
  (\bibinfo {year} {2000})}\BibitemShut {NoStop}%
\bibitem [{\citenamefont {Cheng}\ \emph {et~al.}(2003)\citenamefont {Cheng},
  \citenamefont {Li}, \citenamefont {Duan}, \citenamefont {Feng}, \citenamefont
  {Liu}, \citenamefont {Qiu}, \citenamefont {Lin}, \citenamefont {Ma},\ and\
  \citenamefont {Lee}}]{Cheng2003a}%
  \BibitemOpen
  \bibfield  {author} {\bibinfo {author} {\bibfnamefont {G.}~\bibnamefont
  {Cheng}}, \bibinfo {author} {\bibfnamefont {F.}~\bibnamefont {Li}}, \bibinfo
  {author} {\bibfnamefont {Y.}~\bibnamefont {Duan}}, \bibinfo {author}
  {\bibfnamefont {J.}~\bibnamefont {Feng}}, \bibinfo {author} {\bibfnamefont
  {S.~Y.}\ \bibnamefont {Liu}}, \bibinfo {author} {\bibfnamefont
  {S.}~\bibnamefont {Qiu}}, \bibinfo {author} {\bibfnamefont {D.}~\bibnamefont
  {Lin}}, \bibinfo {author} {\bibfnamefont {Y.~G.}\ \bibnamefont {Ma}}, \ and\
  \bibinfo {author} {\bibfnamefont {S.~T.}\ \bibnamefont {Lee}},\ }\href@noop
  {} {\bibfield  {journal} {\bibinfo  {journal} {Applied Physics Letters}\
  }\textbf {\bibinfo {volume} {82}},\ \bibinfo {pages} {4224} (\bibinfo {year}
  {2003})}\BibitemShut {NoStop}%
\bibitem [{\citenamefont {Cheng}\ \emph
  {et~al.}(2006{\natexlab{a}})\citenamefont {Cheng}, \citenamefont {Zhang},
  \citenamefont {Zhao}, \citenamefont {Liu},\ and\ \citenamefont
  {Ma}}]{Cheng2006}%
  \BibitemOpen
  \bibfield  {author} {\bibinfo {author} {\bibfnamefont {G.}~\bibnamefont
  {Cheng}}, \bibinfo {author} {\bibfnamefont {Y.~F.}\ \bibnamefont {Zhang}},
  \bibinfo {author} {\bibfnamefont {Y.}~\bibnamefont {Zhao}}, \bibinfo {author}
  {\bibfnamefont {S.~Y.}\ \bibnamefont {Liu}}, \ and\ \bibinfo {author}
  {\bibfnamefont {Y.~G.}\ \bibnamefont {Ma}},\ }\href@noop {} {\bibfield
  {journal} {\bibinfo  {journal} {Applied Physics Letters}\ }\textbf {\bibinfo
  {volume} {88}},\ \bibinfo {pages} {083512} (\bibinfo {year}
  {2006}{\natexlab{a}})}\BibitemShut {NoStop}%
\bibitem [{\citenamefont {Kanno}\ \emph
  {et~al.}(2006{\natexlab{c}})\citenamefont {Kanno}, \citenamefont {Sun},\ and\
  \citenamefont {Forrest}}]{Kanno2006b}%
  \BibitemOpen
  \bibfield  {author} {\bibinfo {author} {\bibfnamefont {H.}~\bibnamefont
  {Kanno}}, \bibinfo {author} {\bibfnamefont {Y.~R.}\ \bibnamefont {Sun}}, \
  and\ \bibinfo {author} {\bibfnamefont {S.~R.}\ \bibnamefont {Forrest}},\
  }\href@noop {} {\bibfield  {journal} {\bibinfo  {journal} {Applied Physics
  Letters}\ }\textbf {\bibinfo {volume} {89}},\ \bibinfo {pages} {143516}
  (\bibinfo {year} {2006}{\natexlab{c}})}\BibitemShut {NoStop}%
\bibitem [{\citenamefont {Xue}\ \emph {et~al.}(2010)\citenamefont {Xue},
  \citenamefont {Xie}, \citenamefont {Chen}, \citenamefont {Lu}, \citenamefont
  {Zhang}, \citenamefont {Tang}, \citenamefont {Zhao}, \citenamefont {Hou},\
  and\ \citenamefont {Liu}}]{Xue2010}%
  \BibitemOpen
  \bibfield  {author} {\bibinfo {author} {\bibfnamefont {Q.}~\bibnamefont
  {Xue}}, \bibinfo {author} {\bibfnamefont {G.~H.}\ \bibnamefont {Xie}},
  \bibinfo {author} {\bibfnamefont {P.}~\bibnamefont {Chen}}, \bibinfo {author}
  {\bibfnamefont {J.~H.}\ \bibnamefont {Lu}}, \bibinfo {author} {\bibfnamefont
  {D.~D.}\ \bibnamefont {Zhang}}, \bibinfo {author} {\bibfnamefont {Y.~N.}\
  \bibnamefont {Tang}}, \bibinfo {author} {\bibfnamefont {Y.}~\bibnamefont
  {Zhao}}, \bibinfo {author} {\bibfnamefont {J.~Y.}\ \bibnamefont {Hou}}, \
  and\ \bibinfo {author} {\bibfnamefont {S.~Y.}\ \bibnamefont {Liu}},\
  }\href@noop {} {\bibfield  {journal} {\bibinfo  {journal} {Synthetic Metals}\
  }\textbf {\bibinfo {volume} {160}},\ \bibinfo {pages} {829} (\bibinfo {year}
  {2010})}\BibitemShut {NoStop}%
\bibitem [{\citenamefont {Lei}\ \emph {et~al.}(2004)\citenamefont {Lei},
  \citenamefont {Wang},\ and\ \citenamefont {Qiu}}]{Lei2004}%
  \BibitemOpen
  \bibfield  {author} {\bibinfo {author} {\bibfnamefont {G.~T.}\ \bibnamefont
  {Lei}}, \bibinfo {author} {\bibfnamefont {L.~D.}\ \bibnamefont {Wang}}, \
  and\ \bibinfo {author} {\bibfnamefont {Y.}~\bibnamefont {Qiu}},\ }\href@noop
  {} {\bibfield  {journal} {\bibinfo  {journal} {Applied Physics Letters}\
  }\textbf {\bibinfo {volume} {85}},\ \bibinfo {pages} {5403} (\bibinfo {year}
  {2004})}\BibitemShut {NoStop}%
\bibitem [{\citenamefont {Zhou}\ \emph {et~al.}(2002)\citenamefont {Zhou},
  \citenamefont {Qin}, \citenamefont {Pfeiffer}, \citenamefont
  {Blochwitz-Nimoth}, \citenamefont {Werner}, \citenamefont {Drechsel},
  \citenamefont {Maennig}, \citenamefont {Leo}, \citenamefont {Bold},
  \citenamefont {Erk},\ and\ \citenamefont {Hartmann}}]{Zhou2002}%
  \BibitemOpen
  \bibfield  {author} {\bibinfo {author} {\bibfnamefont {X.}~\bibnamefont
  {Zhou}}, \bibinfo {author} {\bibfnamefont {D.~S.}\ \bibnamefont {Qin}},
  \bibinfo {author} {\bibfnamefont {M.}~\bibnamefont {Pfeiffer}}, \bibinfo
  {author} {\bibfnamefont {J.}~\bibnamefont {Blochwitz-Nimoth}}, \bibinfo
  {author} {\bibfnamefont {A.}~\bibnamefont {Werner}}, \bibinfo {author}
  {\bibfnamefont {J.}~\bibnamefont {Drechsel}}, \bibinfo {author}
  {\bibfnamefont {B.}~\bibnamefont {Maennig}}, \bibinfo {author} {\bibfnamefont
  {K.}~\bibnamefont {Leo}}, \bibinfo {author} {\bibfnamefont {M.}~\bibnamefont
  {Bold}}, \bibinfo {author} {\bibfnamefont {P.}~\bibnamefont {Erk}}, \ and\
  \bibinfo {author} {\bibfnamefont {H.}~\bibnamefont {Hartmann}},\ }\href@noop
  {} {\bibfield  {journal} {\bibinfo  {journal} {Applied Physics Letters}\
  }\textbf {\bibinfo {volume} {81}},\ \bibinfo {pages} {4070} (\bibinfo {year}
  {2002})}\BibitemShut {NoStop}%
\bibitem [{\citenamefont {Hofmann}\ \emph {et~al.}(2012)\citenamefont
  {Hofmann}, \citenamefont {Rosenow}, \citenamefont {Gather}, \citenamefont
  {Luessem},\ and\ \citenamefont {Leo}}]{Hofmann2012}%
  \BibitemOpen
  \bibfield  {author} {\bibinfo {author} {\bibfnamefont {S.}~\bibnamefont
  {Hofmann}}, \bibinfo {author} {\bibfnamefont {T.~C.}\ \bibnamefont
  {Rosenow}}, \bibinfo {author} {\bibfnamefont {M.~C.}\ \bibnamefont {Gather}},
  \bibinfo {author} {\bibfnamefont {B.}~\bibnamefont {Luessem}}, \ and\
  \bibinfo {author} {\bibfnamefont {K.}~\bibnamefont {Leo}},\ }\href@noop {}
  {\bibfield  {journal} {\bibinfo  {journal} {Physical Review B}\ }\textbf
  {\bibinfo {volume} {85}},\ \bibinfo {pages} {245209} (\bibinfo {year}
  {2012})}\BibitemShut {NoStop}%
\bibitem [{Note12()}]{Note12}%
  \BibitemOpen
  \bibinfo {note} {Note that the EQE values in two-unit stacked OLEDs can
  theoretically be doubled, because for every electron that is injected, two
  photons can be emitted.}\BibitemShut {Stop}%
\bibitem [{\citenamefont {Schwartz}\ \emph
  {et~al.}(2008{\natexlab{b}})\citenamefont {Schwartz}, \citenamefont
  {Reineke}, \citenamefont {Walzer},\ and\ \citenamefont {Leo}}]{Schwartz2008}%
  \BibitemOpen
  \bibfield  {author} {\bibinfo {author} {\bibfnamefont {G.}~\bibnamefont
  {Schwartz}}, \bibinfo {author} {\bibfnamefont {S.}~\bibnamefont {Reineke}},
  \bibinfo {author} {\bibfnamefont {K.}~\bibnamefont {Walzer}}, \ and\ \bibinfo
  {author} {\bibfnamefont {K.}~\bibnamefont {Leo}},\ }\href@noop {} {\bibfield
  {journal} {\bibinfo  {journal} {Applied Physics Letters}\ }\textbf {\bibinfo
  {volume} {92}},\ \bibinfo {pages} {053311} (\bibinfo {year}
  {2008}{\natexlab{b}})}\BibitemShut {NoStop}%
\bibitem [{\citenamefont {Kondakova}\ \emph {et~al.}(2010)\citenamefont
  {Kondakova}, \citenamefont {Deaton}, \citenamefont {Pawlik}, \citenamefont
  {Giesen}, \citenamefont {Kondakov}, \citenamefont {Young}, \citenamefont
  {Royster}, \citenamefont {Comfort},\ and\ \citenamefont
  {Shore}}]{Kondakova2010}%
  \BibitemOpen
  \bibfield  {author} {\bibinfo {author} {\bibfnamefont {M.~E.}\ \bibnamefont
  {Kondakova}}, \bibinfo {author} {\bibfnamefont {J.~C.}\ \bibnamefont
  {Deaton}}, \bibinfo {author} {\bibfnamefont {T.~D.}\ \bibnamefont {Pawlik}},
  \bibinfo {author} {\bibfnamefont {D.~J.}\ \bibnamefont {Giesen}}, \bibinfo
  {author} {\bibfnamefont {D.~Y.}\ \bibnamefont {Kondakov}}, \bibinfo {author}
  {\bibfnamefont {R.~H.}\ \bibnamefont {Young}}, \bibinfo {author}
  {\bibfnamefont {T.~L.}\ \bibnamefont {Royster}}, \bibinfo {author}
  {\bibfnamefont {D.~L.}\ \bibnamefont {Comfort}}, \ and\ \bibinfo {author}
  {\bibfnamefont {J.~D.}\ \bibnamefont {Shore}},\ }\href@noop {} {\bibfield
  {journal} {\bibinfo  {journal} {Journal of Applied Physics}\ }\textbf
  {\bibinfo {volume} {107}},\ \bibinfo {pages} {014515} (\bibinfo {year}
  {2010})}\BibitemShut {NoStop}%
\bibitem [{\citenamefont {Hung}\ \emph {et~al.}(2010)\citenamefont {Hung},
  \citenamefont {Chi}, \citenamefont {Chen}, \citenamefont {Chen},
  \citenamefont {Chou},\ and\ \citenamefont {Wong}}]{Hung2010}%
  \BibitemOpen
  \bibfield  {author} {\bibinfo {author} {\bibfnamefont {W.~Y.}\ \bibnamefont
  {Hung}}, \bibinfo {author} {\bibfnamefont {L.~C.}\ \bibnamefont {Chi}},
  \bibinfo {author} {\bibfnamefont {W.~J.}\ \bibnamefont {Chen}}, \bibinfo
  {author} {\bibfnamefont {Y.~M.}\ \bibnamefont {Chen}}, \bibinfo {author}
  {\bibfnamefont {S.~H.}\ \bibnamefont {Chou}}, \ and\ \bibinfo {author}
  {\bibfnamefont {K.~T.}\ \bibnamefont {Wong}},\ }\href@noop {} {\bibfield
  {journal} {\bibinfo  {journal} {Journal of Materials Chemistry}\ }\textbf
  {\bibinfo {volume} {20}},\ \bibinfo {pages} {10113} (\bibinfo {year}
  {2010})}\BibitemShut {NoStop}%
\bibitem [{\citenamefont {Kobayashi}\ \emph {et~al.}(2005)\citenamefont
  {Kobayashi}, \citenamefont {Ide}, \citenamefont {Matsusue},\ and\
  \citenamefont {Naito}}]{Kobayashi2005}%
  \BibitemOpen
  \bibfield  {author} {\bibinfo {author} {\bibfnamefont {T.}~\bibnamefont
  {Kobayashi}}, \bibinfo {author} {\bibfnamefont {N.}~\bibnamefont {Ide}},
  \bibinfo {author} {\bibfnamefont {N.}~\bibnamefont {Matsusue}}, \ and\
  \bibinfo {author} {\bibfnamefont {H.}~\bibnamefont {Naito}},\ }\href@noop {}
  {\bibfield  {journal} {\bibinfo  {journal} {Japanese Journal Of Applied
  Physics Part 1-Regular Papers Brief Communications \& Review Papers}\
  }\textbf {\bibinfo {volume} {44}},\ \bibinfo {pages} {1966} (\bibinfo {year}
  {2005})}\BibitemShut {NoStop}%
\bibitem [{\citenamefont {Tokito}\ \emph {et~al.}(2005)\citenamefont {Tokito},
  \citenamefont {Tsuzuki}, \citenamefont {Sato},\ and\ \citenamefont
  {Iijima}}]{Tokito2005}%
  \BibitemOpen
  \bibfield  {author} {\bibinfo {author} {\bibfnamefont {S.}~\bibnamefont
  {Tokito}}, \bibinfo {author} {\bibfnamefont {T.}~\bibnamefont {Tsuzuki}},
  \bibinfo {author} {\bibfnamefont {F.}~\bibnamefont {Sato}}, \ and\ \bibinfo
  {author} {\bibfnamefont {T.}~\bibnamefont {Iijima}},\ }\href@noop {}
  {\bibfield  {journal} {\bibinfo  {journal} {Current Applied Physics}\
  }\textbf {\bibinfo {volume} {5}},\ \bibinfo {pages} {331} (\bibinfo {year}
  {2005})}\BibitemShut {NoStop}%
\bibitem [{\citenamefont {Tokito}\ \emph {et~al.}(2003)\citenamefont {Tokito},
  \citenamefont {Iijima}, \citenamefont {Suzuri}, \citenamefont {Kita},
  \citenamefont {Tsuzuki},\ and\ \citenamefont {Sato}}]{Tokito2003}%
  \BibitemOpen
  \bibfield  {author} {\bibinfo {author} {\bibfnamefont {S.}~\bibnamefont
  {Tokito}}, \bibinfo {author} {\bibfnamefont {T.}~\bibnamefont {Iijima}},
  \bibinfo {author} {\bibfnamefont {Y.}~\bibnamefont {Suzuri}}, \bibinfo
  {author} {\bibfnamefont {H.}~\bibnamefont {Kita}}, \bibinfo {author}
  {\bibfnamefont {T.}~\bibnamefont {Tsuzuki}}, \ and\ \bibinfo {author}
  {\bibfnamefont {F.}~\bibnamefont {Sato}},\ }\href@noop {} {\bibfield
  {journal} {\bibinfo  {journal} {Applied Physics Letters}\ }\textbf {\bibinfo
  {volume} {83}},\ \bibinfo {pages} {569} (\bibinfo {year} {2003})}\BibitemShut
  {NoStop}%
\bibitem [{\citenamefont {Lai}\ \emph {et~al.}(2010)\citenamefont {Lai},
  \citenamefont {Tao}, \citenamefont {Chan}, \citenamefont {Ng}, \citenamefont
  {Lo}, \citenamefont {Lee}, \citenamefont {Zhang},\ and\ \citenamefont
  {Lee}}]{Lai2010}%
  \BibitemOpen
  \bibfield  {author} {\bibinfo {author} {\bibfnamefont {S.~L.}\ \bibnamefont
  {Lai}}, \bibinfo {author} {\bibfnamefont {S.~L.}\ \bibnamefont {Tao}},
  \bibinfo {author} {\bibfnamefont {M.~Y.}\ \bibnamefont {Chan}}, \bibinfo
  {author} {\bibfnamefont {T.~W.}\ \bibnamefont {Ng}}, \bibinfo {author}
  {\bibfnamefont {M.~F.}\ \bibnamefont {Lo}}, \bibinfo {author} {\bibfnamefont
  {C.~S.}\ \bibnamefont {Lee}}, \bibinfo {author} {\bibfnamefont {X.~H.}\
  \bibnamefont {Zhang}}, \ and\ \bibinfo {author} {\bibfnamefont {S.~T.}\
  \bibnamefont {Lee}},\ }\href@noop {} {\bibfield  {journal} {\bibinfo
  {journal} {Organic Electronics}\ }\textbf {\bibinfo {volume} {11}},\ \bibinfo
  {pages} {1511} (\bibinfo {year} {2010})}\BibitemShut {NoStop}%
\bibitem [{\citenamefont {Tsai}\ \emph {et~al.}(2006)\citenamefont {Tsai},
  \citenamefont {Lin}, \citenamefont {Su}, \citenamefont {Ke}, \citenamefont
  {Wu}, \citenamefont {Fang}, \citenamefont {Liao}, \citenamefont {Wong},\ and\
  \citenamefont {Wu}}]{Tsai2006a}%
  \BibitemOpen
  \bibfield  {author} {\bibinfo {author} {\bibfnamefont {M.~H.}\ \bibnamefont
  {Tsai}}, \bibinfo {author} {\bibfnamefont {H.~W.}\ \bibnamefont {Lin}},
  \bibinfo {author} {\bibfnamefont {H.~C.}\ \bibnamefont {Su}}, \bibinfo
  {author} {\bibfnamefont {T.~H.}\ \bibnamefont {Ke}}, \bibinfo {author}
  {\bibfnamefont {C.~C.}\ \bibnamefont {Wu}}, \bibinfo {author} {\bibfnamefont
  {F.~C.}\ \bibnamefont {Fang}}, \bibinfo {author} {\bibfnamefont {Y.~L.}\
  \bibnamefont {Liao}}, \bibinfo {author} {\bibfnamefont {K.~T.}\ \bibnamefont
  {Wong}}, \ and\ \bibinfo {author} {\bibfnamefont {C.~I.}\ \bibnamefont
  {Wu}},\ }\href@noop {} {\bibfield  {journal} {\bibinfo  {journal} {Advanced
  Materials}\ }\textbf {\bibinfo {volume} {18}},\ \bibinfo {pages} {1216}
  (\bibinfo {year} {2006})}\BibitemShut {NoStop}%
\bibitem [{\citenamefont {Yeh}\ \emph {et~al.}(2005)\citenamefont {Yeh},
  \citenamefont {Wu}, \citenamefont {Chen}, \citenamefont {Song}, \citenamefont
  {Chi}, \citenamefont {Ho}, \citenamefont {Hsu},\ and\ \citenamefont
  {Chen}}]{Yeh2005}%
  \BibitemOpen
  \bibfield  {author} {\bibinfo {author} {\bibfnamefont {S.~J.}\ \bibnamefont
  {Yeh}}, \bibinfo {author} {\bibfnamefont {M.~F.}\ \bibnamefont {Wu}},
  \bibinfo {author} {\bibfnamefont {C.~T.}\ \bibnamefont {Chen}}, \bibinfo
  {author} {\bibfnamefont {Y.~H.}\ \bibnamefont {Song}}, \bibinfo {author}
  {\bibfnamefont {Y.}~\bibnamefont {Chi}}, \bibinfo {author} {\bibfnamefont
  {M.~H.}\ \bibnamefont {Ho}}, \bibinfo {author} {\bibfnamefont {S.~F.}\
  \bibnamefont {Hsu}}, \ and\ \bibinfo {author} {\bibfnamefont {C.~H.}\
  \bibnamefont {Chen}},\ }\href@noop {} {\bibfield  {journal} {\bibinfo
  {journal} {Advanced Materials}\ }\textbf {\bibinfo {volume} {17}},\ \bibinfo
  {pages} {285} (\bibinfo {year} {2005})}\BibitemShut {NoStop}%
\bibitem [{\citenamefont {Lei}\ \emph {et~al.}(2006)\citenamefont {Lei},
  \citenamefont {Wang},\ and\ \citenamefont {Qiu}}]{Lei2006}%
  \BibitemOpen
  \bibfield  {author} {\bibinfo {author} {\bibfnamefont {G.~T.}\ \bibnamefont
  {Lei}}, \bibinfo {author} {\bibfnamefont {L.~D.}\ \bibnamefont {Wang}}, \
  and\ \bibinfo {author} {\bibfnamefont {Y.}~\bibnamefont {Qiu}},\ }\href@noop
  {} {\bibfield  {journal} {\bibinfo  {journal} {Applied Physics Letters}\
  }\textbf {\bibinfo {volume} {88}},\ \bibinfo {pages} {103508} (\bibinfo
  {year} {2006})}\BibitemShut {NoStop}%
\bibitem [{\citenamefont {Wang}\ \emph
  {et~al.}(2009{\natexlab{a}})\citenamefont {Wang}, \citenamefont {Ding},
  \citenamefont {Ma}, \citenamefont {Cheng}, \citenamefont {Wang},
  \citenamefont {Jing},\ and\ \citenamefont {Wang}}]{Wang2009}%
  \BibitemOpen
  \bibfield  {author} {\bibinfo {author} {\bibfnamefont {Q.}~\bibnamefont
  {Wang}}, \bibinfo {author} {\bibfnamefont {J.~Q.}\ \bibnamefont {Ding}},
  \bibinfo {author} {\bibfnamefont {D.~G.}\ \bibnamefont {Ma}}, \bibinfo
  {author} {\bibfnamefont {Y.~X.}\ \bibnamefont {Cheng}}, \bibinfo {author}
  {\bibfnamefont {L.~X.}\ \bibnamefont {Wang}}, \bibinfo {author}
  {\bibfnamefont {X.~B.}\ \bibnamefont {Jing}}, \ and\ \bibinfo {author}
  {\bibfnamefont {F.~S.}\ \bibnamefont {Wang}},\ }\href@noop {} {\bibfield
  {journal} {\bibinfo  {journal} {Advanced Functional Materials}\ }\textbf
  {\bibinfo {volume} {19}},\ \bibinfo {pages} {84} (\bibinfo {year}
  {2009}{\natexlab{a}})}\BibitemShut {NoStop}%
\bibitem [{\citenamefont {Kim}\ \emph {et~al.}(2007)\citenamefont {Kim},
  \citenamefont {Jang},\ and\ \citenamefont {Lee}}]{Kim2007a}%
  \BibitemOpen
  \bibfield  {author} {\bibinfo {author} {\bibfnamefont {S.~H.}\ \bibnamefont
  {Kim}}, \bibinfo {author} {\bibfnamefont {J.}~\bibnamefont {Jang}}, \ and\
  \bibinfo {author} {\bibfnamefont {J.~Y.}\ \bibnamefont {Lee}},\ }\href@noop
  {} {\bibfield  {journal} {\bibinfo  {journal} {Applied Physics Letters}\
  }\textbf {\bibinfo {volume} {91}},\ \bibinfo {pages} {123509} (\bibinfo
  {year} {2007})}\BibitemShut {NoStop}%
\bibitem [{\citenamefont {Chang}\ \emph
  {et~al.}(2010{\natexlab{a}})\citenamefont {Chang}, \citenamefont {Chen},
  \citenamefont {Wu}, \citenamefont {Chang}, \citenamefont {Hung},\ and\
  \citenamefont {Chi}}]{Chang2010a}%
  \BibitemOpen
  \bibfield  {author} {\bibinfo {author} {\bibfnamefont {C.~H.}\ \bibnamefont
  {Chang}}, \bibinfo {author} {\bibfnamefont {C.~C.}\ \bibnamefont {Chen}},
  \bibinfo {author} {\bibfnamefont {C.~C.}\ \bibnamefont {Wu}}, \bibinfo
  {author} {\bibfnamefont {S.~Y.}\ \bibnamefont {Chang}}, \bibinfo {author}
  {\bibfnamefont {J.~Y.}\ \bibnamefont {Hung}}, \ and\ \bibinfo {author}
  {\bibfnamefont {Y.}~\bibnamefont {Chi}},\ }\href@noop {} {\bibfield
  {journal} {\bibinfo  {journal} {Organic Electronics}\ }\textbf {\bibinfo
  {volume} {11}},\ \bibinfo {pages} {266} (\bibinfo {year}
  {2010}{\natexlab{a}})}\BibitemShut {NoStop}%
\bibitem [{Note13()}]{Note13}%
  \BibitemOpen
  \bibinfo {note} {Estimated to 1000\protect \tmspace +\thinmuskip
  {.1667em}cd\protect \tmspace +\thinmuskip {.1667em}m$^{-2}$.}\BibitemShut
  {Stop}%
\bibitem [{\citenamefont {Sun}\ and\ \citenamefont
  {Forrest}(2008{\natexlab{b}})}]{Sun2008a}%
  \BibitemOpen
  \bibfield  {author} {\bibinfo {author} {\bibfnamefont {Y.~R.}\ \bibnamefont
  {Sun}}\ and\ \bibinfo {author} {\bibfnamefont {S.~R.}\ \bibnamefont
  {Forrest}},\ }\href@noop {} {\bibfield  {journal} {\bibinfo  {journal}
  {Organic Electronics}\ }\textbf {\bibinfo {volume} {9}},\ \bibinfo {pages}
  {994} (\bibinfo {year} {2008}{\natexlab{b}})}\BibitemShut {NoStop}%
\bibitem [{\citenamefont {Wang}\ \emph
  {et~al.}(2009{\natexlab{b}})\citenamefont {Wang}, \citenamefont {Ding},
  \citenamefont {Ma}, \citenamefont {Cheng}, \citenamefont {Wang},\ and\
  \citenamefont {Wang}}]{Wang2009a}%
  \BibitemOpen
  \bibfield  {author} {\bibinfo {author} {\bibfnamefont {Q.}~\bibnamefont
  {Wang}}, \bibinfo {author} {\bibfnamefont {J.~Q.}\ \bibnamefont {Ding}},
  \bibinfo {author} {\bibfnamefont {D.~G.}\ \bibnamefont {Ma}}, \bibinfo
  {author} {\bibfnamefont {Y.~X.}\ \bibnamefont {Cheng}}, \bibinfo {author}
  {\bibfnamefont {L.~X.}\ \bibnamefont {Wang}}, \ and\ \bibinfo {author}
  {\bibfnamefont {F.~S.}\ \bibnamefont {Wang}},\ }\href@noop {} {\bibfield
  {journal} {\bibinfo  {journal} {Advanced Materials}\ }\textbf {\bibinfo
  {volume} {21}},\ \bibinfo {pages} {2397} (\bibinfo {year}
  {2009}{\natexlab{b}})}\BibitemShut {NoStop}%
\bibitem [{\citenamefont {Seo}\ \emph {et~al.}(2010)\citenamefont {Seo},
  \citenamefont {Lee}, \citenamefont {Seo}, \citenamefont {Moon}, \citenamefont
  {Lee}, \citenamefont {Park}, \citenamefont {Yoon},\ and\ \citenamefont
  {Kim}}]{Seo2010}%
  \BibitemOpen
  \bibfield  {author} {\bibinfo {author} {\bibfnamefont {J.~H.}\ \bibnamefont
  {Seo}}, \bibinfo {author} {\bibfnamefont {S.~J.}\ \bibnamefont {Lee}},
  \bibinfo {author} {\bibfnamefont {B.~M.}\ \bibnamefont {Seo}}, \bibinfo
  {author} {\bibfnamefont {S.~J.}\ \bibnamefont {Moon}}, \bibinfo {author}
  {\bibfnamefont {K.~H.}\ \bibnamefont {Lee}}, \bibinfo {author} {\bibfnamefont
  {J.~K.}\ \bibnamefont {Park}}, \bibinfo {author} {\bibfnamefont {S.~S.}\
  \bibnamefont {Yoon}}, \ and\ \bibinfo {author} {\bibfnamefont {Y.~K.}\
  \bibnamefont {Kim}},\ }\href@noop {} {\bibfield  {journal} {\bibinfo
  {journal} {Organic Electronics}\ }\textbf {\bibinfo {volume} {11}},\ \bibinfo
  {pages} {1759} (\bibinfo {year} {2010})}\BibitemShut {NoStop}%
\bibitem [{\citenamefont {Sasabe}\ \emph {et~al.}(2010)\citenamefont {Sasabe},
  \citenamefont {Takamatsu}, \citenamefont {Motoyama}, \citenamefont
  {Watanabe}, \citenamefont {Wagenblast}, \citenamefont {Langer}, \citenamefont
  {Molt}, \citenamefont {Fuchs}, \citenamefont {Lennartz},\ and\ \citenamefont
  {Kido}}]{Sasabe2010}%
  \BibitemOpen
  \bibfield  {author} {\bibinfo {author} {\bibfnamefont {H.}~\bibnamefont
  {Sasabe}}, \bibinfo {author} {\bibfnamefont {J.}~\bibnamefont {Takamatsu}},
  \bibinfo {author} {\bibfnamefont {T.}~\bibnamefont {Motoyama}}, \bibinfo
  {author} {\bibfnamefont {S.}~\bibnamefont {Watanabe}}, \bibinfo {author}
  {\bibfnamefont {G.}~\bibnamefont {Wagenblast}}, \bibinfo {author}
  {\bibfnamefont {N.}~\bibnamefont {Langer}}, \bibinfo {author} {\bibfnamefont
  {O.}~\bibnamefont {Molt}}, \bibinfo {author} {\bibfnamefont {E.}~\bibnamefont
  {Fuchs}}, \bibinfo {author} {\bibfnamefont {C.}~\bibnamefont {Lennartz}}, \
  and\ \bibinfo {author} {\bibfnamefont {J.}~\bibnamefont {Kido}},\ }\href@noop
  {} {\bibfield  {journal} {\bibinfo  {journal} {Advanced Materials}\ }\textbf
  {\bibinfo {volume} {22}},\ \bibinfo {pages} {5003} (\bibinfo {year}
  {2010})}\BibitemShut {NoStop}%
\bibitem [{\citenamefont {Chang}\ \emph
  {et~al.}(2010{\natexlab{b}})\citenamefont {Chang}, \citenamefont {Tien},
  \citenamefont {Chen}, \citenamefont {Lin}, \citenamefont {Cheng},
  \citenamefont {Liu}, \citenamefont {Wu}, \citenamefont {Hung}, \citenamefont
  {Chiu},\ and\ \citenamefont {Chi}}]{Chang2010}%
  \BibitemOpen
  \bibfield  {author} {\bibinfo {author} {\bibfnamefont {C.~H.}\ \bibnamefont
  {Chang}}, \bibinfo {author} {\bibfnamefont {K.~C.}\ \bibnamefont {Tien}},
  \bibinfo {author} {\bibfnamefont {C.~C.}\ \bibnamefont {Chen}}, \bibinfo
  {author} {\bibfnamefont {M.~S.}\ \bibnamefont {Lin}}, \bibinfo {author}
  {\bibfnamefont {H.~C.}\ \bibnamefont {Cheng}}, \bibinfo {author}
  {\bibfnamefont {S.~H.}\ \bibnamefont {Liu}}, \bibinfo {author} {\bibfnamefont
  {C.~C.}\ \bibnamefont {Wu}}, \bibinfo {author} {\bibfnamefont {J.~Y.}\
  \bibnamefont {Hung}}, \bibinfo {author} {\bibfnamefont {Y.~C.}\ \bibnamefont
  {Chiu}}, \ and\ \bibinfo {author} {\bibfnamefont {Y.}~\bibnamefont {Chi}},\
  }\href@noop {} {\bibfield  {journal} {\bibinfo  {journal} {Organic
  Electronics}\ }\textbf {\bibinfo {volume} {11}},\ \bibinfo {pages} {412}
  (\bibinfo {year} {2010}{\natexlab{b}})}\BibitemShut {NoStop}%
\bibitem [{\citenamefont {D'Andrade}\ \emph
  {et~al.}(2002{\natexlab{b}})\citenamefont {D'Andrade}, \citenamefont
  {Thompson},\ and\ \citenamefont {Forrest}}]{DAndrade2002a}%
  \BibitemOpen
  \bibfield  {author} {\bibinfo {author} {\bibfnamefont {B.~W.}\ \bibnamefont
  {D'Andrade}}, \bibinfo {author} {\bibfnamefont {M.~E.}\ \bibnamefont
  {Thompson}}, \ and\ \bibinfo {author} {\bibfnamefont {S.~R.}\ \bibnamefont
  {Forrest}},\ }\href@noop {} {\bibfield  {journal} {\bibinfo  {journal}
  {Advanced Materials}\ }\textbf {\bibinfo {volume} {14}},\ \bibinfo {pages}
  {147} (\bibinfo {year} {2002}{\natexlab{b}})}\BibitemShut {NoStop}%
\bibitem [{\citenamefont {Cheng}\ \emph
  {et~al.}(2006{\natexlab{b}})\citenamefont {Cheng}, \citenamefont {Zhang},
  \citenamefont {Zhao}, \citenamefont {Lin}, \citenamefont {Ruan},
  \citenamefont {Liu}, \citenamefont {Fei}, \citenamefont {Ma},\ and\
  \citenamefont {Cheng}}]{Cheng2006a}%
  \BibitemOpen
  \bibfield  {author} {\bibinfo {author} {\bibfnamefont {G.}~\bibnamefont
  {Cheng}}, \bibinfo {author} {\bibfnamefont {Y.~F.}\ \bibnamefont {Zhang}},
  \bibinfo {author} {\bibfnamefont {Y.}~\bibnamefont {Zhao}}, \bibinfo {author}
  {\bibfnamefont {Y.~Y.}\ \bibnamefont {Lin}}, \bibinfo {author} {\bibfnamefont
  {C.~Y.}\ \bibnamefont {Ruan}}, \bibinfo {author} {\bibfnamefont {S.~Y.}\
  \bibnamefont {Liu}}, \bibinfo {author} {\bibfnamefont {T.}~\bibnamefont
  {Fei}}, \bibinfo {author} {\bibfnamefont {Y.~G.}\ \bibnamefont {Ma}}, \ and\
  \bibinfo {author} {\bibfnamefont {Y.~X.}\ \bibnamefont {Cheng}},\ }\href@noop
  {} {\bibfield  {journal} {\bibinfo  {journal} {Applied Physics Letters}\
  }\textbf {\bibinfo {volume} {89}},\ \bibinfo {pages} {043504} (\bibinfo
  {year} {2006}{\natexlab{b}})}\BibitemShut {NoStop}%
\bibitem [{Note14()}]{Note14}%
  \BibitemOpen
  \bibinfo {note} {Upon a further increase of emitter concentration, the PLQY
  decreases again as dominated by concentration quenching [\cite
  {Kawamura2006}].}\BibitemShut {Stop}%
\bibitem [{\citenamefont {Su}\ \emph {et~al.}(2010)\citenamefont {Su},
  \citenamefont {Sasabe}, \citenamefont {Pu}, \citenamefont {Nakayama},\ and\
  \citenamefont {Kido}}]{Su2010}%
  \BibitemOpen
  \bibfield  {author} {\bibinfo {author} {\bibfnamefont {S.~J.}\ \bibnamefont
  {Su}}, \bibinfo {author} {\bibfnamefont {H.}~\bibnamefont {Sasabe}}, \bibinfo
  {author} {\bibfnamefont {Y.~J.}\ \bibnamefont {Pu}}, \bibinfo {author}
  {\bibfnamefont {K.}~\bibnamefont {Nakayama}}, \ and\ \bibinfo {author}
  {\bibfnamefont {J.}~\bibnamefont {Kido}},\ }\href@noop {} {\bibfield
  {journal} {\bibinfo  {journal} {Advanced Materials}\ }\textbf {\bibinfo
  {volume} {22}},\ \bibinfo {pages} {3311} (\bibinfo {year}
  {2010})}\BibitemShut {NoStop}%
\bibitem [{\citenamefont {He}\ \emph {et~al.}(2004{\natexlab{b}})\citenamefont
  {He}, \citenamefont {Pfeiffer}, \citenamefont {Leo}, \citenamefont {Hofmann},
  \citenamefont {Birnstock}, \citenamefont {Pudzich},\ and\ \citenamefont
  {Salbeck}}]{He2004}%
  \BibitemOpen
  \bibfield  {author} {\bibinfo {author} {\bibfnamefont {G.~F.}\ \bibnamefont
  {He}}, \bibinfo {author} {\bibfnamefont {M.}~\bibnamefont {Pfeiffer}},
  \bibinfo {author} {\bibfnamefont {K.}~\bibnamefont {Leo}}, \bibinfo {author}
  {\bibfnamefont {M.}~\bibnamefont {Hofmann}}, \bibinfo {author} {\bibfnamefont
  {J.}~\bibnamefont {Birnstock}}, \bibinfo {author} {\bibfnamefont
  {R.}~\bibnamefont {Pudzich}}, \ and\ \bibinfo {author} {\bibfnamefont
  {J.}~\bibnamefont {Salbeck}},\ }\href@noop {} {\bibfield  {journal} {\bibinfo
   {journal} {Applied Physics Letters}\ }\textbf {\bibinfo {volume} {85}},\
  \bibinfo {pages} {3911} (\bibinfo {year} {2004}{\natexlab{b}})}\BibitemShut
  {NoStop}%
\bibitem [{\citenamefont {Weichsel}\ \emph {et~al.}(2012)\citenamefont
  {Weichsel}, \citenamefont {Reineke}, \citenamefont {Furno}, \citenamefont
  {Luessem},\ and\ \citenamefont {Leo}}]{Weichsel2012}%
  \BibitemOpen
  \bibfield  {author} {\bibinfo {author} {\bibfnamefont {C.}~\bibnamefont
  {Weichsel}}, \bibinfo {author} {\bibfnamefont {S.}~\bibnamefont {Reineke}},
  \bibinfo {author} {\bibfnamefont {M.}~\bibnamefont {Furno}}, \bibinfo
  {author} {\bibfnamefont {B.}~\bibnamefont {Luessem}}, \ and\ \bibinfo
  {author} {\bibfnamefont {K.}~\bibnamefont {Leo}},\ }\href@noop {} {\bibfield
  {journal} {\bibinfo  {journal} {Journal of Applied Physics}\ }\textbf
  {\bibinfo {volume} {111}},\ \bibinfo {pages} {033102} (\bibinfo {year}
  {2012})}\BibitemShut {NoStop}%
\bibitem [{\citenamefont {Zheng}\ and\ \citenamefont
  {Rillema}(1998)}]{Zheng1998}%
  \BibitemOpen
  \bibfield  {author} {\bibinfo {author} {\bibfnamefont {G.~Y.}\ \bibnamefont
  {Zheng}}\ and\ \bibinfo {author} {\bibfnamefont {D.~P.}\ \bibnamefont
  {Rillema}},\ }\href@noop {} {\bibfield  {journal} {\bibinfo  {journal}
  {Inorganic Chemistry}\ }\textbf {\bibinfo {volume} {37}},\ \bibinfo {pages}
  {1392} (\bibinfo {year} {1998})}\BibitemShut {NoStop}%
\bibitem [{\citenamefont {Connick}\ \emph {et~al.}(1996)\citenamefont
  {Connick}, \citenamefont {Henling}, \citenamefont {Marsh},\ and\
  \citenamefont {Gray}}]{Connick1996}%
  \BibitemOpen
  \bibfield  {author} {\bibinfo {author} {\bibfnamefont {W.~B.}\ \bibnamefont
  {Connick}}, \bibinfo {author} {\bibfnamefont {L.~M.}\ \bibnamefont
  {Henling}}, \bibinfo {author} {\bibfnamefont {R.~E.}\ \bibnamefont {Marsh}},
  \ and\ \bibinfo {author} {\bibfnamefont {H.~B.}\ \bibnamefont {Gray}},\
  }\href@noop {} {\bibfield  {journal} {\bibinfo  {journal} {Inorganic
  Chemistry}\ }\textbf {\bibinfo {volume} {35}},\ \bibinfo {pages} {6261}
  (\bibinfo {year} {1996})}\BibitemShut {NoStop}%
\bibitem [{\citenamefont {Ma}\ \emph {et~al.}(2006)\citenamefont {Ma},
  \citenamefont {Djurovich}, \citenamefont {Garon}, \citenamefont {Alleyne},\
  and\ \citenamefont {Thompson}}]{Ma2006}%
  \BibitemOpen
  \bibfield  {author} {\bibinfo {author} {\bibfnamefont {B.~W.}\ \bibnamefont
  {Ma}}, \bibinfo {author} {\bibfnamefont {P.~I.}\ \bibnamefont {Djurovich}},
  \bibinfo {author} {\bibfnamefont {S.}~\bibnamefont {Garon}}, \bibinfo
  {author} {\bibfnamefont {B.}~\bibnamefont {Alleyne}}, \ and\ \bibinfo
  {author} {\bibfnamefont {M.~E.}\ \bibnamefont {Thompson}},\ }\href@noop {}
  {\bibfield  {journal} {\bibinfo  {journal} {Advanced Functional Materials}\
  }\textbf {\bibinfo {volume} {16}},\ \bibinfo {pages} {2438} (\bibinfo {year}
  {2006})}\BibitemShut {NoStop}%
\bibitem [{\citenamefont {Koshiyama}\ \emph {et~al.}(2004)\citenamefont
  {Koshiyama}, \citenamefont {Omura},\ and\ \citenamefont
  {Kato}}]{Koshiyama2004}%
  \BibitemOpen
  \bibfield  {author} {\bibinfo {author} {\bibfnamefont {T.}~\bibnamefont
  {Koshiyama}}, \bibinfo {author} {\bibfnamefont {A.}~\bibnamefont {Omura}}, \
  and\ \bibinfo {author} {\bibfnamefont {M.}~\bibnamefont {Kato}},\ }\href@noop
  {} {\bibfield  {journal} {\bibinfo  {journal} {Chemistry Letters}\ }\textbf
  {\bibinfo {volume} {33}},\ \bibinfo {pages} {1386} (\bibinfo {year}
  {2004})}\BibitemShut {NoStop}%
\bibitem [{\citenamefont {Cocchi}\ \emph {et~al.}(2009)\citenamefont {Cocchi},
  \citenamefont {Kalinowski}, \citenamefont {Fattori}, \citenamefont
  {Williams},\ and\ \citenamefont {Murphy}}]{Cocchi2009}%
  \BibitemOpen
  \bibfield  {author} {\bibinfo {author} {\bibfnamefont {M.}~\bibnamefont
  {Cocchi}}, \bibinfo {author} {\bibfnamefont {J.}~\bibnamefont {Kalinowski}},
  \bibinfo {author} {\bibfnamefont {V.}~\bibnamefont {Fattori}}, \bibinfo
  {author} {\bibfnamefont {J.~A.~G.}\ \bibnamefont {Williams}}, \ and\ \bibinfo
  {author} {\bibfnamefont {L.}~\bibnamefont {Murphy}},\ }\href@noop {}
  {\bibfield  {journal} {\bibinfo  {journal} {Applied Physics Letters}\
  }\textbf {\bibinfo {volume} {94}},\ \bibinfo {pages} {073309} (\bibinfo
  {year} {2009})}\BibitemShut {NoStop}%
\bibitem [{\citenamefont {Cocchi}\ \emph {et~al.}(2010)\citenamefont {Cocchi},
  \citenamefont {Kalinowski}, \citenamefont {Murphy}, \citenamefont
  {Williams},\ and\ \citenamefont {Fattori}}]{Cocchi2010}%
  \BibitemOpen
  \bibfield  {author} {\bibinfo {author} {\bibfnamefont {M.}~\bibnamefont
  {Cocchi}}, \bibinfo {author} {\bibfnamefont {J.}~\bibnamefont {Kalinowski}},
  \bibinfo {author} {\bibfnamefont {L.}~\bibnamefont {Murphy}}, \bibinfo
  {author} {\bibfnamefont {J.~A.~G.}\ \bibnamefont {Williams}}, \ and\ \bibinfo
  {author} {\bibfnamefont {V.}~\bibnamefont {Fattori}},\ }\href@noop {}
  {\bibfield  {journal} {\bibinfo  {journal} {Organic Electronics}\ }\textbf
  {\bibinfo {volume} {11}},\ \bibinfo {pages} {388} (\bibinfo {year}
  {2010})}\BibitemShut {NoStop}%
\bibitem [{\citenamefont {Kalinowski}\ \emph {et~al.}(2010)\citenamefont
  {Kalinowski}, \citenamefont {Cocchi}, \citenamefont {Fattori}, \citenamefont
  {Murphy},\ and\ \citenamefont {Williams}}]{Kalinowski2010}%
  \BibitemOpen
  \bibfield  {author} {\bibinfo {author} {\bibfnamefont {J.}~\bibnamefont
  {Kalinowski}}, \bibinfo {author} {\bibfnamefont {M.}~\bibnamefont {Cocchi}},
  \bibinfo {author} {\bibfnamefont {V.}~\bibnamefont {Fattori}}, \bibinfo
  {author} {\bibfnamefont {L.}~\bibnamefont {Murphy}}, \ and\ \bibinfo {author}
  {\bibfnamefont {J.~A.~G.}\ \bibnamefont {Williams}},\ }\href@noop {}
  {\bibfield  {journal} {\bibinfo  {journal} {Organic Electronics}\ }\textbf
  {\bibinfo {volume} {11}},\ \bibinfo {pages} {724} (\bibinfo {year}
  {2010})}\BibitemShut {NoStop}%
\bibitem [{\citenamefont {Zhou}\ \emph {et~al.}(2009)\citenamefont {Zhou},
  \citenamefont {Wang}, \citenamefont {Ho}, \citenamefont {Wong}, \citenamefont
  {Ma},\ and\ \citenamefont {Wang}}]{Zhou2009}%
  \BibitemOpen
  \bibfield  {author} {\bibinfo {author} {\bibfnamefont {G.~J.}\ \bibnamefont
  {Zhou}}, \bibinfo {author} {\bibfnamefont {Q.}~\bibnamefont {Wang}}, \bibinfo
  {author} {\bibfnamefont {C.~L.}\ \bibnamefont {Ho}}, \bibinfo {author}
  {\bibfnamefont {W.~Y.}\ \bibnamefont {Wong}}, \bibinfo {author}
  {\bibfnamefont {D.~G.}\ \bibnamefont {Ma}}, \ and\ \bibinfo {author}
  {\bibfnamefont {L.~X.}\ \bibnamefont {Wang}},\ }\href@noop {} {\bibfield
  {journal} {\bibinfo  {journal} {Chemical Communications}\ ,\ \bibinfo {pages}
  {3574}} (\bibinfo {year} {2009})}\BibitemShut {NoStop}%
\bibitem [{\citenamefont {Lu}\ and\ \citenamefont {Sturm}(2002)}]{Lu2002}%
  \BibitemOpen
  \bibfield  {author} {\bibinfo {author} {\bibfnamefont {M.~H.}\ \bibnamefont
  {Lu}}\ and\ \bibinfo {author} {\bibfnamefont {J.~C.}\ \bibnamefont {Sturm}},\
  }\href@noop {} {\bibfield  {journal} {\bibinfo  {journal} {Journal Of Applied
  Physics}\ }\textbf {\bibinfo {volume} {91}},\ \bibinfo {pages} {595}
  (\bibinfo {year} {2002})}\BibitemShut {NoStop}%
\bibitem [{\citenamefont {Furno}\ \emph {et~al.}(2012)\citenamefont {Furno},
  \citenamefont {Meerheim}, \citenamefont {Hofmann}, \citenamefont {Luessem},\
  and\ \citenamefont {Leo}}]{Furno2012}%
  \BibitemOpen
  \bibfield  {author} {\bibinfo {author} {\bibfnamefont {M.}~\bibnamefont
  {Furno}}, \bibinfo {author} {\bibfnamefont {R.}~\bibnamefont {Meerheim}},
  \bibinfo {author} {\bibfnamefont {S.}~\bibnamefont {Hofmann}}, \bibinfo
  {author} {\bibfnamefont {B.}~\bibnamefont {Luessem}}, \ and\ \bibinfo
  {author} {\bibfnamefont {K.}~\bibnamefont {Leo}},\ }\href@noop {} {\bibfield
  {journal} {\bibinfo  {journal} {Physical Review B}\ }\textbf {\bibinfo
  {volume} {85}},\ \bibinfo {pages} {115205} (\bibinfo {year}
  {2012})}\BibitemShut {NoStop}%
\bibitem [{\citenamefont {Neyts}(1998)}]{Neyts1998}%
  \BibitemOpen
  \bibfield  {author} {\bibinfo {author} {\bibfnamefont {K.}~\bibnamefont
  {Neyts}},\ }\href@noop {} {\bibfield  {journal} {\bibinfo  {journal} {J. Opt.
  Soc. Am. A}\ }\textbf {\bibinfo {volume} {15}},\ \bibinfo {pages} {962}
  (\bibinfo {year} {1998})}\BibitemShut {NoStop}%
\bibitem [{\citenamefont {Nowy}\ \emph {et~al.}(2008)\citenamefont {Nowy},
  \citenamefont {Krummacher}, \citenamefont {Frischeisen}, \citenamefont
  {Reinke},\ and\ \citenamefont {Brütting}}]{Nowy2008}%
  \BibitemOpen
  \bibfield  {author} {\bibinfo {author} {\bibfnamefont {S.}~\bibnamefont
  {Nowy}}, \bibinfo {author} {\bibfnamefont {B.~C.}\ \bibnamefont
  {Krummacher}}, \bibinfo {author} {\bibfnamefont {J.}~\bibnamefont
  {Frischeisen}}, \bibinfo {author} {\bibfnamefont {N.~A.}\ \bibnamefont
  {Reinke}}, \ and\ \bibinfo {author} {\bibfnamefont {W.}~\bibnamefont
  {Brütting}},\ }\href@noop {} {\bibfield  {journal} {\bibinfo  {journal}
  {Journal Of Applied Physics}\ }\textbf {\bibinfo {volume} {104}},\ \bibinfo
  {pages} {123109} (\bibinfo {year} {2008})}\BibitemShut {NoStop}%
\bibitem [{\citenamefont {D'Andrade}\ and\ \citenamefont
  {Brown}(2006)}]{DAndrade2006}%
  \BibitemOpen
  \bibfield  {author} {\bibinfo {author} {\bibfnamefont {B.~W.}\ \bibnamefont
  {D'Andrade}}\ and\ \bibinfo {author} {\bibfnamefont {J.~J.}\ \bibnamefont
  {Brown}},\ }\href@noop {} {\bibfield  {journal} {\bibinfo  {journal} {Applied
  Physics Letters}\ }\textbf {\bibinfo {volume} {88}},\ \bibinfo {pages}
  {192908} (\bibinfo {year} {2006})}\BibitemShut {NoStop}%
\bibitem [{\citenamefont {Madigan}\ \emph {et~al.}(2000)\citenamefont
  {Madigan}, \citenamefont {Lu},\ and\ \citenamefont {Sturm}}]{Madigan2000}%
  \BibitemOpen
  \bibfield  {author} {\bibinfo {author} {\bibfnamefont {C.~F.}\ \bibnamefont
  {Madigan}}, \bibinfo {author} {\bibfnamefont {M.~H.}\ \bibnamefont {Lu}}, \
  and\ \bibinfo {author} {\bibfnamefont {J.~C.}\ \bibnamefont {Sturm}},\
  }\href@noop {} {\bibfield  {journal} {\bibinfo  {journal} {Applied Physics
  Letters}\ }\textbf {\bibinfo {volume} {76}},\ \bibinfo {pages} {1650}
  (\bibinfo {year} {2000})}\BibitemShut {NoStop}%
\bibitem [{\citenamefont {Nakamura}\ \emph {et~al.}(2005)\citenamefont
  {Nakamura}, \citenamefont {Tsutsumi}, \citenamefont {Juni},\ and\
  \citenamefont {Fujii}}]{Nakamura2005}%
  \BibitemOpen
  \bibfield  {author} {\bibinfo {author} {\bibfnamefont {T.}~\bibnamefont
  {Nakamura}}, \bibinfo {author} {\bibfnamefont {N.}~\bibnamefont {Tsutsumi}},
  \bibinfo {author} {\bibfnamefont {N.}~\bibnamefont {Juni}}, \ and\ \bibinfo
  {author} {\bibfnamefont {H.}~\bibnamefont {Fujii}},\ }\href@noop {}
  {\bibfield  {journal} {\bibinfo  {journal} {Journal Of Applied Physics}\
  }\textbf {\bibinfo {volume} {97}},\ \bibinfo {pages} {054505} (\bibinfo
  {year} {2005})}\BibitemShut {NoStop}%
\bibitem [{\citenamefont {Koh}\ \emph {et~al.}(2010)\citenamefont {Koh},
  \citenamefont {Choi}, \citenamefont {Lee},\ and\ \citenamefont
  {Yoo}}]{Koh2010}%
  \BibitemOpen
  \bibfield  {author} {\bibinfo {author} {\bibfnamefont {T.~W.}\ \bibnamefont
  {Koh}}, \bibinfo {author} {\bibfnamefont {J.~M.}\ \bibnamefont {Choi}},
  \bibinfo {author} {\bibfnamefont {S.}~\bibnamefont {Lee}}, \ and\ \bibinfo
  {author} {\bibfnamefont {S.}~\bibnamefont {Yoo}},\ }\href@noop {} {\bibfield
  {journal} {\bibinfo  {journal} {Advanced Materials}\ }\textbf {\bibinfo
  {volume} {22}},\ \bibinfo {pages} {1849} (\bibinfo {year}
  {2010})}\BibitemShut {NoStop}%
\bibitem [{\citenamefont {Koo}\ \emph {et~al.}(2010)\citenamefont {Koo},
  \citenamefont {Jeong}, \citenamefont {Araoka}, \citenamefont {Ishikawa},
  \citenamefont {Nishimura}, \citenamefont {Toyooka},\ and\ \citenamefont
  {Takezoe}}]{Koo2010}%
  \BibitemOpen
  \bibfield  {author} {\bibinfo {author} {\bibfnamefont {W.~H.}\ \bibnamefont
  {Koo}}, \bibinfo {author} {\bibfnamefont {S.~M.}\ \bibnamefont {Jeong}},
  \bibinfo {author} {\bibfnamefont {F.}~\bibnamefont {Araoka}}, \bibinfo
  {author} {\bibfnamefont {K.}~\bibnamefont {Ishikawa}}, \bibinfo {author}
  {\bibfnamefont {S.}~\bibnamefont {Nishimura}}, \bibinfo {author}
  {\bibfnamefont {T.}~\bibnamefont {Toyooka}}, \ and\ \bibinfo {author}
  {\bibfnamefont {H.}~\bibnamefont {Takezoe}},\ }\href@noop {} {\bibfield
  {journal} {\bibinfo  {journal} {Nature Photonics}\ }\textbf {\bibinfo
  {volume} {4}},\ \bibinfo {pages} {222} (\bibinfo {year} {2010})}\BibitemShut
  {NoStop}%
\bibitem [{\citenamefont {Fehse}\ \emph {et~al.}(2007)\citenamefont {Fehse},
  \citenamefont {Walzer}, \citenamefont {Leo}, \citenamefont {Loevenich},\ and\
  \citenamefont {Elschner}}]{Fehse2007}%
  \BibitemOpen
  \bibfield  {author} {\bibinfo {author} {\bibfnamefont {K.}~\bibnamefont
  {Fehse}}, \bibinfo {author} {\bibfnamefont {K.}~\bibnamefont {Walzer}},
  \bibinfo {author} {\bibfnamefont {K.}~\bibnamefont {Leo}}, \bibinfo {author}
  {\bibfnamefont {W.}~\bibnamefont {Loevenich}}, \ and\ \bibinfo {author}
  {\bibfnamefont {A.}~\bibnamefont {Elschner}},\ }\href@noop {} {\bibfield
  {journal} {\bibinfo  {journal} {Advanced Materials}\ }\textbf {\bibinfo
  {volume} {19}},\ \bibinfo {pages} {441} (\bibinfo {year} {2007})}\BibitemShut
  {NoStop}%
\bibitem [{Note15()}]{Note15}%
  \BibitemOpen
  \bibinfo {note} {The transport layer thickness of the ETL and HTL slightly
  differ to meet the field antinode, accounting for the different optical
  properties of the two substrate types.}\BibitemShut {Stop}%
\bibitem [{\citenamefont {Frischeisen}\ \emph {et~al.}(2010)\citenamefont
  {Frischeisen}, \citenamefont {Yokoyama}, \citenamefont {Adachi},\ and\
  \citenamefont {Bruetting}}]{Frischeisen2010}%
  \BibitemOpen
  \bibfield  {author} {\bibinfo {author} {\bibfnamefont {J.}~\bibnamefont
  {Frischeisen}}, \bibinfo {author} {\bibfnamefont {D.}~\bibnamefont
  {Yokoyama}}, \bibinfo {author} {\bibfnamefont {C.}~\bibnamefont {Adachi}}, \
  and\ \bibinfo {author} {\bibfnamefont {W.}~\bibnamefont {Bruetting}},\
  }\href@noop {} {\bibfield  {journal} {\bibinfo  {journal} {Applied Physics
  Letters}\ }\textbf {\bibinfo {volume} {96}},\ \bibinfo {pages} {073302}
  (\bibinfo {year} {2010})}\BibitemShut {NoStop}%
\bibitem [{\citenamefont {Taneda}\ \emph {et~al.}(2011)\citenamefont {Taneda},
  \citenamefont {Yasuda},\ and\ \citenamefont {Adachi}}]{Taneda2011}%
  \BibitemOpen
  \bibfield  {author} {\bibinfo {author} {\bibfnamefont {M.}~\bibnamefont
  {Taneda}}, \bibinfo {author} {\bibfnamefont {T.}~\bibnamefont {Yasuda}}, \
  and\ \bibinfo {author} {\bibfnamefont {C.}~\bibnamefont {Adachi}},\
  }\href@noop {} {\bibfield  {journal} {\bibinfo  {journal} {Applied Physics
  Express}\ }\textbf {\bibinfo {volume} {4}},\ \bibinfo {pages} {071602}
  (\bibinfo {year} {2011})}\BibitemShut {NoStop}%
\bibitem [{\citenamefont {Flaemmich}\ \emph {et~al.}(2011)\citenamefont
  {Flaemmich}, \citenamefont {Frischeisen}, \citenamefont {Setz}, \citenamefont
  {Michaelis}, \citenamefont {Krummacher}, \citenamefont {Schmidt},
  \citenamefont {Bruetting},\ and\ \citenamefont {Danz}}]{Flaemmich2011}%
  \BibitemOpen
  \bibfield  {author} {\bibinfo {author} {\bibfnamefont {M.}~\bibnamefont
  {Flaemmich}}, \bibinfo {author} {\bibfnamefont {J.}~\bibnamefont
  {Frischeisen}}, \bibinfo {author} {\bibfnamefont {D.~S.}\ \bibnamefont
  {Setz}}, \bibinfo {author} {\bibfnamefont {D.}~\bibnamefont {Michaelis}},
  \bibinfo {author} {\bibfnamefont {B.~C.}\ \bibnamefont {Krummacher}},
  \bibinfo {author} {\bibfnamefont {T.~D.}\ \bibnamefont {Schmidt}}, \bibinfo
  {author} {\bibfnamefont {W.}~\bibnamefont {Bruetting}}, \ and\ \bibinfo
  {author} {\bibfnamefont {N.}~\bibnamefont {Danz}},\ }\href@noop {} {\bibfield
   {journal} {\bibinfo  {journal} {Organic Electronics}\ }\textbf {\bibinfo
  {volume} {12}},\ \bibinfo {pages} {1663} (\bibinfo {year}
  {2011})}\BibitemShut {NoStop}%
\bibitem [{\citenamefont {Schmidt}\ \emph {et~al.}(2011)\citenamefont
  {Schmidt}, \citenamefont {Setz}, \citenamefont {Flaemmich}, \citenamefont
  {Frischeisen}, \citenamefont {Michaelis}, \citenamefont {Krummacher},
  \citenamefont {Danz},\ and\ \citenamefont {Bruetting}}]{Schmidt2011}%
  \BibitemOpen
  \bibfield  {author} {\bibinfo {author} {\bibfnamefont {T.~D.}\ \bibnamefont
  {Schmidt}}, \bibinfo {author} {\bibfnamefont {D.~S.}\ \bibnamefont {Setz}},
  \bibinfo {author} {\bibfnamefont {M.}~\bibnamefont {Flaemmich}}, \bibinfo
  {author} {\bibfnamefont {J.}~\bibnamefont {Frischeisen}}, \bibinfo {author}
  {\bibfnamefont {D.}~\bibnamefont {Michaelis}}, \bibinfo {author}
  {\bibfnamefont {B.~C.}\ \bibnamefont {Krummacher}}, \bibinfo {author}
  {\bibfnamefont {N.}~\bibnamefont {Danz}}, \ and\ \bibinfo {author}
  {\bibfnamefont {W.}~\bibnamefont {Bruetting}},\ }\href@noop {} {\bibfield
  {journal} {\bibinfo  {journal} {Applied Physics Letters}\ }\textbf {\bibinfo
  {volume} {99}},\ \bibinfo {pages} {163302} (\bibinfo {year}
  {2011})}\BibitemShut {NoStop}%
\bibitem [{\citenamefont {Brütting}\ \emph {et~al.}(2012)\citenamefont
  {Brütting}, \citenamefont {Frischeisen}, \citenamefont {Schmidt},
  \citenamefont {Scholz},\ and\ \citenamefont {Mayr}}]{Bruetting2012}%
  \BibitemOpen
  \bibfield  {author} {\bibinfo {author} {\bibfnamefont {W.}~\bibnamefont
  {Brütting}}, \bibinfo {author} {\bibfnamefont {J.}~\bibnamefont
  {Frischeisen}}, \bibinfo {author} {\bibfnamefont {T.~D.}\ \bibnamefont
  {Schmidt}}, \bibinfo {author} {\bibfnamefont {B.~J.}\ \bibnamefont {Scholz}},
  \ and\ \bibinfo {author} {\bibfnamefont {C.}~\bibnamefont {Mayr}},\ }\href
  {\doibase 10.1002/pssa.201228320} {\bibfield  {journal} {\bibinfo  {journal}
  {Phys. Status Solidi A}\ ,\ \bibinfo {pages} {doi: 10.1002/pssa.201228320}}
  (\bibinfo {year} {2012})}\BibitemShut {NoStop}%
\bibitem [{\citenamefont {Frischeisen}\ \emph {et~al.}(2011)\citenamefont
  {Frischeisen}, \citenamefont {Yokoyama}, \citenamefont {Endo}, \citenamefont
  {Adachi},\ and\ \citenamefont {Bruetting}}]{Frischeisen2011}%
  \BibitemOpen
  \bibfield  {author} {\bibinfo {author} {\bibfnamefont {J.}~\bibnamefont
  {Frischeisen}}, \bibinfo {author} {\bibfnamefont {D.}~\bibnamefont
  {Yokoyama}}, \bibinfo {author} {\bibfnamefont {A.}~\bibnamefont {Endo}},
  \bibinfo {author} {\bibfnamefont {C.}~\bibnamefont {Adachi}}, \ and\ \bibinfo
  {author} {\bibfnamefont {W.}~\bibnamefont {Bruetting}},\ }\href@noop {}
  {\bibfield  {journal} {\bibinfo  {journal} {Organic Electronics}\ }\textbf
  {\bibinfo {volume} {12}},\ \bibinfo {pages} {809} (\bibinfo {year}
  {2011})}\BibitemShut {NoStop}%
\bibitem [{\citenamefont {Shen}\ \emph {et~al.}(1997)\citenamefont {Shen},
  \citenamefont {Burrows}, \citenamefont {Bulovic}, \citenamefont {Forrest},\
  and\ \citenamefont {Thompson}}]{Shen1997}%
  \BibitemOpen
  \bibfield  {author} {\bibinfo {author} {\bibfnamefont {Z.~L.}\ \bibnamefont
  {Shen}}, \bibinfo {author} {\bibfnamefont {P.~E.}\ \bibnamefont {Burrows}},
  \bibinfo {author} {\bibfnamefont {V.}~\bibnamefont {Bulovic}}, \bibinfo
  {author} {\bibfnamefont {S.~R.}\ \bibnamefont {Forrest}}, \ and\ \bibinfo
  {author} {\bibfnamefont {M.~E.}\ \bibnamefont {Thompson}},\ }\href@noop {}
  {\bibfield  {journal} {\bibinfo  {journal} {Science}\ }\textbf {\bibinfo
  {volume} {276}},\ \bibinfo {pages} {2009} (\bibinfo {year}
  {1997})}\BibitemShut {NoStop}%
\bibitem [{\citenamefont {Chen}\ \emph {et~al.}(2010)\citenamefont {Chen},
  \citenamefont {Deng}, \citenamefont {Xie}, \citenamefont {Peng},
  \citenamefont {Xie}, \citenamefont {Fan},\ and\ \citenamefont
  {Huang}}]{Chen2010}%
  \BibitemOpen
  \bibfield  {author} {\bibinfo {author} {\bibfnamefont {S.~F.}\ \bibnamefont
  {Chen}}, \bibinfo {author} {\bibfnamefont {L.~L.}\ \bibnamefont {Deng}},
  \bibinfo {author} {\bibfnamefont {J.}~\bibnamefont {Xie}}, \bibinfo {author}
  {\bibfnamefont {L.}~\bibnamefont {Peng}}, \bibinfo {author} {\bibfnamefont
  {L.~H.}\ \bibnamefont {Xie}}, \bibinfo {author} {\bibfnamefont {Q.~L.}\
  \bibnamefont {Fan}}, \ and\ \bibinfo {author} {\bibfnamefont
  {W.}~\bibnamefont {Huang}},\ }\href@noop {} {\bibfield  {journal} {\bibinfo
  {journal} {Advanced Materials}\ }\textbf {\bibinfo {volume} {22}},\ \bibinfo
  {pages} {5227} (\bibinfo {year} {2010})}\BibitemShut {NoStop}%
\bibitem [{\citenamefont {Hung}\ \emph {et~al.}(2001)\citenamefont {Hung},
  \citenamefont {Tang}, \citenamefont {Mason}, \citenamefont {Raychaudhuri},\
  and\ \citenamefont {Madathil}}]{Hung2001}%
  \BibitemOpen
  \bibfield  {author} {\bibinfo {author} {\bibfnamefont {L.~S.}\ \bibnamefont
  {Hung}}, \bibinfo {author} {\bibfnamefont {C.~W.}\ \bibnamefont {Tang}},
  \bibinfo {author} {\bibfnamefont {M.~G.}\ \bibnamefont {Mason}}, \bibinfo
  {author} {\bibfnamefont {P.}~\bibnamefont {Raychaudhuri}}, \ and\ \bibinfo
  {author} {\bibfnamefont {J.}~\bibnamefont {Madathil}},\ }\href@noop {}
  {\bibfield  {journal} {\bibinfo  {journal} {Applied Physics Letters}\
  }\textbf {\bibinfo {volume} {78}},\ \bibinfo {pages} {544} (\bibinfo {year}
  {2001})}\BibitemShut {NoStop}%
\bibitem [{\citenamefont {Thomschke}\ \emph {et~al.}(2009)\citenamefont
  {Thomschke}, \citenamefont {Nitsche}, \citenamefont {Furno},\ and\
  \citenamefont {Leo}}]{Thomschke2009}%
  \BibitemOpen
  \bibfield  {author} {\bibinfo {author} {\bibfnamefont {M.}~\bibnamefont
  {Thomschke}}, \bibinfo {author} {\bibfnamefont {R.}~\bibnamefont {Nitsche}},
  \bibinfo {author} {\bibfnamefont {M.}~\bibnamefont {Furno}}, \ and\ \bibinfo
  {author} {\bibfnamefont {K.}~\bibnamefont {Leo}},\ }\href@noop {} {\bibfield
  {journal} {\bibinfo  {journal} {Applied Physics Letters}\ }\textbf {\bibinfo
  {volume} {94}},\ \bibinfo {pages} {083303} (\bibinfo {year}
  {2009})}\BibitemShut {NoStop}%
\bibitem [{\citenamefont {Hsu}\ \emph {et~al.}(2005)\citenamefont {Hsu},
  \citenamefont {Lee}, \citenamefont {Hwang},\ and\ \citenamefont
  {Chen}}]{Hsu2005}%
  \BibitemOpen
  \bibfield  {author} {\bibinfo {author} {\bibfnamefont {S.~F.}\ \bibnamefont
  {Hsu}}, \bibinfo {author} {\bibfnamefont {C.~C.}\ \bibnamefont {Lee}},
  \bibinfo {author} {\bibfnamefont {S.~W.}\ \bibnamefont {Hwang}}, \ and\
  \bibinfo {author} {\bibfnamefont {C.~H.}\ \bibnamefont {Chen}},\ }\href@noop
  {} {\bibfield  {journal} {\bibinfo  {journal} {Applied Physics Letters}\
  }\textbf {\bibinfo {volume} {86}},\ \bibinfo {pages} {253508} (\bibinfo
  {year} {2005})}\BibitemShut {NoStop}%
\bibitem [{\citenamefont {Thomschke}\ \emph {et~al.}(2012)\citenamefont
  {Thomschke}, \citenamefont {Reineke}, \citenamefont {Luessem},\ and\
  \citenamefont {Leo}}]{Thomschke2012}%
  \BibitemOpen
  \bibfield  {author} {\bibinfo {author} {\bibfnamefont {M.}~\bibnamefont
  {Thomschke}}, \bibinfo {author} {\bibfnamefont {S.}~\bibnamefont {Reineke}},
  \bibinfo {author} {\bibfnamefont {B.}~\bibnamefont {Luessem}}, \ and\
  \bibinfo {author} {\bibfnamefont {K.}~\bibnamefont {Leo}},\ }\href@noop {}
  {\bibfield  {journal} {\bibinfo  {journal} {Nano Letters}\ }\textbf {\bibinfo
  {volume} {12}},\ \bibinfo {pages} {424} (\bibinfo {year} {2012})}\BibitemShut
  {NoStop}%
\bibitem [{Note16()}]{Note16}%
  \BibitemOpen
  \bibinfo {note} {Note that this is a two-unit stacked device potentially
  having 200\protect \tmspace +\thinmuskip {.1667em}\% internal quantum
  efficiency.}\BibitemShut {Stop}%
\bibitem [{Note17()}]{Note17}%
  \BibitemOpen
  \bibinfo {note} {Note that the absolute numbers of $K_\protect \text r$
  differ for the spectra shown in Fig. \ref {_idealwhite} and for the values
  displayed in the table of Fig. \ref {white_OLED_perspective}. While the
  latter are based on OLED spectra, the simulation of Fig. \ref {_idealwhite}
  only takes the PL spectra of the emitters into account, giving rise to slight
  differences.}\BibitemShut {Stop}%
\bibitem [{Note18()}]{Note18}%
  \BibitemOpen
  \bibinfo {note} {Note that the reference device is an optimized OLED
  employing high refractive index substrates and thick electron transport
  layers.}\BibitemShut {Stop}%
\bibitem [{\citenamefont {Tanaka}\ \emph
  {et~al.}(2007{\natexlab{a}})\citenamefont {Tanaka}, \citenamefont {Sasabe},
  \citenamefont {Li}, \citenamefont {Su}, \citenamefont {Takeda},\ and\
  \citenamefont {Kido}}]{Tanaka2007}%
  \BibitemOpen
  \bibfield  {author} {\bibinfo {author} {\bibfnamefont {D.}~\bibnamefont
  {Tanaka}}, \bibinfo {author} {\bibfnamefont {H.}~\bibnamefont {Sasabe}},
  \bibinfo {author} {\bibfnamefont {Y.~J.}\ \bibnamefont {Li}}, \bibinfo
  {author} {\bibfnamefont {S.~J.}\ \bibnamefont {Su}}, \bibinfo {author}
  {\bibfnamefont {T.}~\bibnamefont {Takeda}}, \ and\ \bibinfo {author}
  {\bibfnamefont {J.}~\bibnamefont {Kido}},\ }\href@noop {} {\bibfield
  {journal} {\bibinfo  {journal} {Japanese Journal Of Applied Physics Part
  2-Letters \& Express Letters}\ }\textbf {\bibinfo {volume} {46}},\ \bibinfo
  {pages} {L10} (\bibinfo {year} {2007}{\natexlab{a}})}\BibitemShut {NoStop}%
\bibitem [{\citenamefont {Tanaka}\ \emph
  {et~al.}(2007{\natexlab{b}})\citenamefont {Tanaka}, \citenamefont {Agata},
  \citenamefont {Takeda}, \citenamefont {Watanabe},\ and\ \citenamefont
  {Kido}}]{Tanaka2007a}%
  \BibitemOpen
  \bibfield  {author} {\bibinfo {author} {\bibfnamefont {D.}~\bibnamefont
  {Tanaka}}, \bibinfo {author} {\bibfnamefont {Y.}~\bibnamefont {Agata}},
  \bibinfo {author} {\bibfnamefont {T.}~\bibnamefont {Takeda}}, \bibinfo
  {author} {\bibfnamefont {S.}~\bibnamefont {Watanabe}}, \ and\ \bibinfo
  {author} {\bibfnamefont {J.}~\bibnamefont {Kido}},\ }\href@noop {} {\bibfield
   {journal} {\bibinfo  {journal} {Japanese Journal Of Applied Physics Part
  2-Letters \& Express Letters}\ }\textbf {\bibinfo {volume} {46}},\ \bibinfo
  {pages} {L117} (\bibinfo {year} {2007}{\natexlab{b}})}\BibitemShut {NoStop}%
\bibitem [{Note19()}]{Note19}%
  \BibitemOpen
  \bibinfo {note} {This increase with respect to the flat device was obtained
  in the LE using the pyramid pattern.}\BibitemShut {Stop}%
\bibitem [{\citenamefont {Namdas}\ \emph {et~al.}(2005)\citenamefont {Namdas},
  \citenamefont {Ruseckas}, \citenamefont {Samuel}, \citenamefont {Lo},\ and\
  \citenamefont {Burn}}]{Namdas2005}%
  \BibitemOpen
  \bibfield  {author} {\bibinfo {author} {\bibfnamefont {E.~B.}\ \bibnamefont
  {Namdas}}, \bibinfo {author} {\bibfnamefont {A.}~\bibnamefont {Ruseckas}},
  \bibinfo {author} {\bibfnamefont {I.~D.~W.}\ \bibnamefont {Samuel}}, \bibinfo
  {author} {\bibfnamefont {S.~C.}\ \bibnamefont {Lo}}, \ and\ \bibinfo {author}
  {\bibfnamefont {P.~L.}\ \bibnamefont {Burn}},\ }\href@noop {} {\bibfield
  {journal} {\bibinfo  {journal} {Applied Physics Letters}\ }\textbf {\bibinfo
  {volume} {86}},\ \bibinfo {pages} {091104} (\bibinfo {year}
  {2005})}\BibitemShut {NoStop}%
\bibitem [{\citenamefont {Reineke}\ \emph {et~al.}(2010)\citenamefont
  {Reineke}, \citenamefont {Rosenow}, \citenamefont {L\"ussem},\ and\
  \citenamefont {Leo}}]{Reineke2010}%
  \BibitemOpen
  \bibfield  {author} {\bibinfo {author} {\bibfnamefont {S.}~\bibnamefont
  {Reineke}}, \bibinfo {author} {\bibfnamefont {T.~C.}\ \bibnamefont
  {Rosenow}}, \bibinfo {author} {\bibfnamefont {B.}~\bibnamefont {L\"ussem}}, \
  and\ \bibinfo {author} {\bibfnamefont {K.}~\bibnamefont {Leo}},\ }\href@noop
  {} {\bibfield  {journal} {\bibinfo  {journal} {Advanced Materials}\ }\textbf
  {\bibinfo {volume} {22}},\ \bibinfo {pages} {3189} (\bibinfo {year}
  {2010})}\BibitemShut {NoStop}%
\bibitem [{\citenamefont {Kang}\ \emph {et~al.}(2007)\citenamefont {Kang},
  \citenamefont {Lee}, \citenamefont {Park}, \citenamefont {Jeong},
  \citenamefont {Yoo}, \citenamefont {Park},\ and\ \citenamefont
  {Kim}}]{Kang2007}%
  \BibitemOpen
  \bibfield  {author} {\bibinfo {author} {\bibfnamefont {J.~W.}\ \bibnamefont
  {Kang}}, \bibinfo {author} {\bibfnamefont {S.~H.}\ \bibnamefont {Lee}},
  \bibinfo {author} {\bibfnamefont {H.~D.}\ \bibnamefont {Park}}, \bibinfo
  {author} {\bibfnamefont {W.~I.}\ \bibnamefont {Jeong}}, \bibinfo {author}
  {\bibfnamefont {K.~M.}\ \bibnamefont {Yoo}}, \bibinfo {author} {\bibfnamefont
  {Y.~S.}\ \bibnamefont {Park}}, \ and\ \bibinfo {author} {\bibfnamefont
  {J.~J.}\ \bibnamefont {Kim}},\ }\href@noop {} {\bibfield  {journal} {\bibinfo
   {journal} {Applied Physics Letters}\ }\textbf {\bibinfo {volume} {90}},\
  \bibinfo {pages} {223508} (\bibinfo {year} {2007})}\BibitemShut {NoStop}%
\bibitem [{\citenamefont {Han}\ \emph {et~al.}(2008)\citenamefont {Han},
  \citenamefont {Yang}, \citenamefont {Li}, \citenamefont {Chu}, \citenamefont
  {Chen}, \citenamefont {Su}, \citenamefont {Zhang}, \citenamefont {Yan},
  \citenamefont {Hu},\ and\ \citenamefont {Zhang}}]{Han2008}%
  \BibitemOpen
  \bibfield  {author} {\bibinfo {author} {\bibfnamefont {L.~L.}\ \bibnamefont
  {Han}}, \bibinfo {author} {\bibfnamefont {D.~F.}\ \bibnamefont {Yang}},
  \bibinfo {author} {\bibfnamefont {W.~L.}\ \bibnamefont {Li}}, \bibinfo
  {author} {\bibfnamefont {B.}~\bibnamefont {Chu}}, \bibinfo {author}
  {\bibfnamefont {Y.}~\bibnamefont {Chen}}, \bibinfo {author} {\bibfnamefont
  {Z.~S.}\ \bibnamefont {Su}}, \bibinfo {author} {\bibfnamefont {D.~Y.}\
  \bibnamefont {Zhang}}, \bibinfo {author} {\bibfnamefont {F.}~\bibnamefont
  {Yan}}, \bibinfo {author} {\bibfnamefont {Z.~Z.}\ \bibnamefont {Hu}}, \ and\
  \bibinfo {author} {\bibfnamefont {Z.~Q.}\ \bibnamefont {Zhang}},\ }\href@noop
  {} {\bibfield  {journal} {\bibinfo  {journal} {Applied Physics Letters}\
  }\textbf {\bibinfo {volume} {93}},\ \bibinfo {pages} {153303} (\bibinfo
  {year} {2008})}\BibitemShut {NoStop}%
\bibitem [{\citenamefont {Sasabe}\ and\ \citenamefont
  {Kido}(2011)}]{Sasabe2011}%
  \BibitemOpen
  \bibfield  {author} {\bibinfo {author} {\bibfnamefont {H.}~\bibnamefont
  {Sasabe}}\ and\ \bibinfo {author} {\bibfnamefont {J.}~\bibnamefont {Kido}},\
  }\href@noop {} {\bibfield  {journal} {\bibinfo  {journal} {Chemistry of
  Materials}\ }\textbf {\bibinfo {volume} {23}},\ \bibinfo {pages} {621}
  (\bibinfo {year} {2011})}\BibitemShut {NoStop}%
\bibitem [{\citenamefont {Sasabe}\ \emph {et~al.}(2011)\citenamefont {Sasabe},
  \citenamefont {Tanaka}, \citenamefont {Yokoyama}, \citenamefont {Chiba},
  \citenamefont {Pu}, \citenamefont {Nakayama}, \citenamefont {Yokoyama},\ and\
  \citenamefont {Kido}}]{Sasabe2011a}%
  \BibitemOpen
  \bibfield  {author} {\bibinfo {author} {\bibfnamefont {H.}~\bibnamefont
  {Sasabe}}, \bibinfo {author} {\bibfnamefont {D.}~\bibnamefont {Tanaka}},
  \bibinfo {author} {\bibfnamefont {D.}~\bibnamefont {Yokoyama}}, \bibinfo
  {author} {\bibfnamefont {T.}~\bibnamefont {Chiba}}, \bibinfo {author}
  {\bibfnamefont {Y.-J.}\ \bibnamefont {Pu}}, \bibinfo {author} {\bibfnamefont
  {K.-i.}\ \bibnamefont {Nakayama}}, \bibinfo {author} {\bibfnamefont
  {M.}~\bibnamefont {Yokoyama}}, \ and\ \bibinfo {author} {\bibfnamefont
  {J.}~\bibnamefont {Kido}},\ }\href@noop {} {\bibfield  {journal} {\bibinfo
  {journal} {Advanced Functional Materials}\ }\textbf {\bibinfo {volume}
  {21}},\ \bibinfo {pages} {336} (\bibinfo {year} {2011})}\BibitemShut
  {NoStop}%
\bibitem [{\citenamefont {Harada}\ \emph {et~al.}(2005)\citenamefont {Harada},
  \citenamefont {Werner}, \citenamefont {Pfeiffer}, \citenamefont {Bloom},
  \citenamefont {Elliott},\ and\ \citenamefont {Leo}}]{Harada2005}%
  \BibitemOpen
  \bibfield  {author} {\bibinfo {author} {\bibfnamefont {K.}~\bibnamefont
  {Harada}}, \bibinfo {author} {\bibfnamefont {A.~G.}\ \bibnamefont {Werner}},
  \bibinfo {author} {\bibfnamefont {M.}~\bibnamefont {Pfeiffer}}, \bibinfo
  {author} {\bibfnamefont {C.~J.}\ \bibnamefont {Bloom}}, \bibinfo {author}
  {\bibfnamefont {C.~M.}\ \bibnamefont {Elliott}}, \ and\ \bibinfo {author}
  {\bibfnamefont {K.}~\bibnamefont {Leo}},\ }\href@noop {} {\bibfield
  {journal} {\bibinfo  {journal} {Physical Review Letters}\ }\textbf {\bibinfo
  {volume} {94}},\ \bibinfo {pages} {036601} (\bibinfo {year}
  {2005})}\BibitemShut {NoStop}%
\bibitem [{\citenamefont {Cai}\ \emph {et~al.}(2011{\natexlab{a}})\citenamefont
  {Cai}, \citenamefont {Su}, \citenamefont {Chiba}, \citenamefont {Sasabe},
  \citenamefont {Pu}, \citenamefont {Nakayama},\ and\ \citenamefont
  {Kido}}]{Cai2011}%
  \BibitemOpen
  \bibfield  {author} {\bibinfo {author} {\bibfnamefont {C.}~\bibnamefont
  {Cai}}, \bibinfo {author} {\bibfnamefont {S.-J.}\ \bibnamefont {Su}},
  \bibinfo {author} {\bibfnamefont {T.}~\bibnamefont {Chiba}}, \bibinfo
  {author} {\bibfnamefont {H.}~\bibnamefont {Sasabe}}, \bibinfo {author}
  {\bibfnamefont {Y.-J.}\ \bibnamefont {Pu}}, \bibinfo {author} {\bibfnamefont
  {K.}~\bibnamefont {Nakayama}}, \ and\ \bibinfo {author} {\bibfnamefont
  {J.}~\bibnamefont {Kido}},\ }\href@noop {} {\bibfield  {journal} {\bibinfo
  {journal} {Organic Electronics}\ }\textbf {\bibinfo {volume} {12}},\ \bibinfo
  {pages} {843} (\bibinfo {year} {2011}{\natexlab{a}})}\BibitemShut {NoStop}%
\bibitem [{\citenamefont {Cai}\ \emph {et~al.}(2011{\natexlab{b}})\citenamefont
  {Cai}, \citenamefont {Su}, \citenamefont {Chiba}, \citenamefont {Sasabe},
  \citenamefont {Pu}, \citenamefont {Nakayama},\ and\ \citenamefont
  {Kido}}]{Cai2011a}%
  \BibitemOpen
  \bibfield  {author} {\bibinfo {author} {\bibfnamefont {C.}~\bibnamefont
  {Cai}}, \bibinfo {author} {\bibfnamefont {S.-J.}\ \bibnamefont {Su}},
  \bibinfo {author} {\bibfnamefont {T.}~\bibnamefont {Chiba}}, \bibinfo
  {author} {\bibfnamefont {H.}~\bibnamefont {Sasabe}}, \bibinfo {author}
  {\bibfnamefont {Y.-J.}\ \bibnamefont {Pu}}, \bibinfo {author} {\bibfnamefont
  {K.}~\bibnamefont {Nakayama}}, \ and\ \bibinfo {author} {\bibfnamefont
  {J.}~\bibnamefont {Kido}},\ }\href@noop {} {\bibfield  {journal} {\bibinfo
  {journal} {Japanese Journal of Applied Physics}\ }\textbf {\bibinfo {volume}
  {50}},\ \bibinfo {pages} {040204} (\bibinfo {year}
  {2011}{\natexlab{b}})}\BibitemShut {NoStop}%
\end{thebibliography}%

\end{document}